\documentclass[a4paper,openright,twoside,11pt,italian]{book}

\usepackage[italian]{babel}
\usepackage{amsfonts}
\usepackage{latexsym}
\usepackage{epsfig}
\usepackage{graphicx}
\usepackage{fancyhdr}

\author{Giovanni Ricco}
\def\notD{{\slash\kern-7pt D}}

\newcommand{\dd}{\partial}

\newcommand{\tr}{{\rm tr \,}}

\newcommand{\diag}{{\rm diag \,}}

\newcommand{\emp}{}

\newcommand{\be}{\begin{equation}}
\newcommand{\ee}{\end{equation}}
\newcommand{\ben}{\begin{displaymath}}
\newcommand{\een}{\end{displaymath}}
\newcommand{\ba}{\begin{eqnarray}}
\newcommand{\ea}{\end{eqnarray}}

\newcommand{\bean}{\begin{eqnarray*}}
\newcommand{\eean}{\end{eqnarray*}}

\begin{document}

\pagestyle{empty}

\newpage

              %


\thispagestyle{empty}

\begin{flushright}
ROM2F-05/14
\end{flushright}

\begin{center}

\hspace*{-0.6cm}\makebox[\textwidth]{
    \includegraphics[height=1.45truecm]{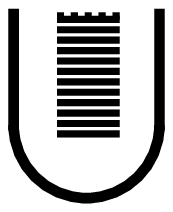}
    \raisebox{0.68cm}{
    \begin{minipage}[h]{13truecm}
        \hspace*{0.2cm}{\LARGE UNIVERSIT\`A DEGLI STUDI DI ROMA
        \hspace*{3.0cm}``TOR \rule{0pt}{24pt}VERGATA''}
    \end{minipage}}}
 
\end{center}

\vspace{0.8cm}

\begin{center}
FACOLT\`A DI SCIENZE MATEMATICHE, FISICHE E NATURALI \\
Dipartimento di Fisica \\
\end{center}

\vspace{1cm}

\begin{center}
{\LARGE {\bf Superstringhe oltre un loop}}
\end{center}
\vspace{0.5cm}
\begin{center}
Tesi di Laurea in Fisica
\end{center}

\vspace{2.5cm}
\noindent
Candidato \\
\rule{0pt}{20pt}{{\em Giovanni RICCO} \\
Relatore \\
\rule{0pt}{20pt}{\large Prof. {\em Augusto SAGNOTTI} \\

\vspace{2cm}
\begin{center}
Anno Accademico 2004-2005
\end{center}

\pagebreak
\newpage
\cleardoublepage
\pagestyle{fancy}\fancyhf{}
\renewcommand{\chaptermark}[1]{\markboth{#1}{}}
\renewcommand{\sectionmark}[1]{\markright{\thesection.\ #1}{}}
\fancyhead[LE,RO]{\bfseries\thepage}
\fancypagestyle{plain}{\fancyhead{} \renewcommand{\headrulewidth}{0pt}}
\oddsidemargin +0.5cm \evensidemargin +0.5cm
\fancyhead[RE]{\bfseries\leftmark}
\fancyhead[LO]{\slshape\rightmark}
\setlength{\headwidth}{15cm}

\pagenumbering{roman}
\tableofcontents
\newpage
\cleardoublepage
\vfill\eject
\pagenumbering{arabic}

\chapter*{Ringraziamenti \markboth{Ringraziamenti}{Ringraziamenti}}
\addcontentsline{toc}{chapter}{Ringraziamenti}

Questa lavoro di Tesi \`e stato svolto presso il Dipartimento di Fisica dell'Universit\`a di Roma ``Tor Vergata'', e presso il CPHT dell'Ecole Polytechnique di Parigi, in cui ho avuto la fortuna di trovare ambienti di ricerca estremamente stimolanti. Vorrei ringraziare in particolare il Prof. A. Sagnotti per essere essere stato una guida autorevole e disponibile nel vasto mondo della Teoria delle Stringhe e per avermi offerto la preziosa possibilit\`a di collaborare a questo progetto di ricerca. Vorrei ringraziare anche il Prof. Emilian Dudas, con cui ho avuto modo di lavorare a Parigi, per la disponibilit\`a e l'istruttiva collaborazione. 

Infine vorrei esprimere la mia riconoscenza ai dottorandi del Dipartimento di Fisica dell'Universit\`a di Roma ``Tor Vergata'' per l'ambiente piacevole e sereno che mi hanno fatto trovare in questi mesi di studio e di ricerca nella mia ``nuova'' Universit\`a. Un particolare ringraziamento va al Dr. Marco Nicolosi per le utili discussioni e consigli. 

{\it This work was supported in part by INFN, by the MIUR-COFIN Contract 2003-023852, by the EU contracts
MRTN-CT-2004-503369 and MRTN-CT-2004-512194, by the INTAS contract 03-51-6346, and the NATO grant PST.CLG.978785. The author would like to thank the CPhT of the Ecole Poytechnique, and in particular Prof. E. Dudas, for the kind hospitality extended to him while this Thesis was being completed.}

\chapter*{Introduzione \markboth{Introduzione}{Introduzione}}
\addcontentsline{toc}{chapter}{Introduzione}

\section*{Il cammino dell'unificazione}

La nostra attuale conoscenza delle leggi fondamentali dell'universo poggia su due teorie, il Modello Standard e la Relativit\`a Generale. Entrambe risultano descrivere l'universo correttamente nei limiti delle verifiche empiriche oggi permesse, ma allo stesso tempo esse sono due teorie inconciliabili. 

Nel corso degli ultimi cento anni la fisica \`e avanzata superando di volta in volta le contraddizioni emerse fra le teorie esistenti. Il tentativo di rendere l'elettromagnetismo di Maxwell compatibile con la relativit\`a Galileana port\`o Einstein alla formulazione della Relativit\`a Ristretta. In seguito, il tentativo di conciliare gravitazione Newtoniana e Relativit\`a Speciale lo port\`o a sviluppare la Teoria della Relativit\`a Generale. Infine, la Teoria Quantistica dei Campi nacque dal bisogno di rendere compatibili la Meccanica Quantistica e la Relativit\`a Speciale.

Allo stesso tempo lo sviluppo storico della fisica \`e stato segnato da una straordinaria unificazione nella descrizione di fenomeni apparentemente molto diversi. 
Il primo grande passo fu l'unificazione dei fenomeni elettrici e magnetici. La teoria elettrostatica, formulata da Cavedish nel periodo dal 1771 al 1773 e completata da Coulomb nel 1785, fu messa in rapporto con i fenomeni magnetici dai lavori di Oersted, che osserv\`o la deflessione dell'ago di una bussola a causa della presenza di correnti (1819), di Biot-Savart e Amp\`ere, che stabilirono le regole con cui le correnti elettriche producono campi magnetici (1820-25), e di Faraday, che mostr\`o che campi magnetici variabili generano campi elettrici (1873). L'unificazione dei fenomeni elettrici e magnetici fu portata a termine da Maxwell, che costru\`i un sistema coerente di equazioni in grado di descrivere tutti i fenomeni noti (1831). Le equazioni di Maxwell portarono alla previsione delle onde elettromagnetiche. 
Una seconda unificazione di tipo differente, ma di grandissima portata, fu prodotta dalla teoria della Relativit\`a Ristretta, formulata nel 1905 da Albert Einstein, che mostr\`o come lo spazio e il tempo non siano entit\`a separate e assolute, come nella Meccanica Newtoniana, ma formino piuttosto un continuo spazio-temporale che \`e l'arena in cui i fenomeni fisici si dipanano. Allo stesso modo la teoria, portava all'unificazione dei concetti di massa ed energia, fino ad allora ben distinti. 
Sulla strada dell'unificazione, un radicale cambiamento di paradigma fu introdotto dalla formulazione della Meccanica Quantistica. I lavori di molti scienziati e in particolare di Planck, Schr\"odinger, Dirac, Heinsenberg e Einstein, mostrarono come una corretta descrizione dei fenomeni microscopici necessitasse dell'unificazione dei concetti di onda e particella, in alcuni casi la quantizzazione dei possibili valori degli osservabili, l'individuazione di osservabili che non possono essere misurati simultaneamente per i quali valgono relazioni di indeterminazione, e l'abbandono di una meccanica deterministica in favore di leggi intrinsecamente probabilistiche. La Meccanica Quantistica ha inoltre portato all'introduzione di un momento intrinseco delle particelle, detto \emph{spin} e  alla distinzione delle particelle in due famiglie con differenti propriet\`a statistiche, i \emph{fermioni} e i \emph{bosoni}, rispettivamente di spin semintero e intero.

La nostra attuale visione del mondo poggia sull'identificazione di quattro forze fondamentali e di diverse profonde relazioni fra tre di queste. La forza fondamentale che storicamente \`e stata identificata per prima \`e la \emph{forza di gravit\`a}, descritta prima dalla  Gravitazione Newtoniana e poi dalla Relativit\`a Generale. 
 
La seconda forza fondamentale \`e la \emph{forza elettromagnetica}. La descrizione delle forze elettromagnetiche in termini della teoria classica di campo di Maxwell \`e stata superata con la formulazione della prima teoria quantistica relativistica di campo, l'Elettrodinamica Quantistica (QED). Lo schema concettuale della Teoria Quantistica dei Campi (QFT), realizza pienamente la dualit\`a onda-particella, associando le particelle a \emph{quanti di energia} di corrispondenti campi d'onda (nel caso della QED i quanti del campo sono i fotoni), in maniera tale da rendere evidente l'indistinguibilit\`a di tutte le particelle di uno stesso tipo. Nella visione tradizionale della QFT, le particelle fondamentali sono i quanti di energia dei campi che entrano nella lagrangiana della teoria fondamentale. La descrizione in termini di campi assegna alle particelle un numero limitato di attributi (\emph{numeri quantici}), tra i quali la massa, lo spin e uno o pi\`u tipi di carica. Le interazioni tra particelle sono mediate da scambi di altre particelle: ad esempio le interazioni elettromagnetiche sono associate allo scambio di fotoni.  

La terza forza fondamentale \`e la \emph{forza debole}, responsabile dei processi di decadimento $\beta$, fra i quali il pi\`u noto \`e il decadimento di un protone in un neutrone, un elettrone e un antineutrino. I decadimenti $\beta$ sono noti sin dalla fine del diciannovesimo secolo, ma l'identificazione di questa forza fondamentale \`e avvenuta solo nella prima met\`a del ventesimo secolo. Le interazioni deboli sono sensibilmente pi\`u flebili delle interazioni elettromagnetiche, e la loro prima descrizione \`e dovuta a Fermi. Negli anni '60, i lavori di Glashow, Weinberg e Salam hanno portato a formulare una teoria in grado di descrivere le forze deboli e elettromagnetiche all'interno di uno schema unificato. La teoria quantistica di campo che descrive le forze deboli e elettromagnetiche \`e nota come \emph{teoria elettrodebole}.

Infine la quarta forza fondamentale \`e detta \emph{forza forte}, o forza di colore. La forza forte \`e responsabile di diversi stati aggregati come i neutroni, i protoni, i pioni e di molte altre particelle subnucleari. I costituenti elementari di questi stati vengono identificati nei \emph{quarks}, che non possono propagarsi liberamente, ma solo in stati aggregati, a causa delle peculiari propriet\`a della forza forte. La teoria quantistica di campo che descrive questa forza \`e nota come Cromodinamica Quantistica (QCD). La teoria elettrodebole e la QCD formano insieme il Modello Standard della fisica delle particelle.

Il Modello Standard \`e quindi una teoria quantistica di campo, e costituisce una straordinaria sintesi delle nostre conoscenze di tre delle quattro forze fondamentali: le forze elettromagnetiche, le forze deboli e le forze forti. La descrizione di queste interazioni avviene in termini di una teoria di gauge di Yang-Mills con gruppo di gauge $SU(3)\times SU(2)\times U(1)$. Una teoria di gauge \`e una generalizzazione della Teoria di Maxwell dell'elettromagnetismo, i cui pontenziali sono in generale matrici che soddisfano equazioni di campo non lineari, anche in assenza di materia. 

In particolare, la simmetria di gauge $SU(2)\times U(1)$ \`e associata alle interazioni deboli, mediate dai bosoni massivi di spin $1$, $W^\pm$ e $Z^0$, e alle interazioni elettromagnetiche mediate dai fotoni, bosoni di massa nulla di spin $1$; il gruppo di gauge $SU(3)$ \`e invece associato alle interazioni forti, mediate da otto bosoni di massa nulla, detti gluoni. Ci sono poi diverse particelle fondamentali di ``materia'', fermioni di spin $1/2$, divisi in due gruppi: i \emph{leptoni} e i \emph{quarks}. I leptoni includono elettroni $e^-$, muoni $\mu^-$, e tauoni $\tau^-$ con i rispettivi neutrini di massa quasi nulla $\nu_e$, $\nu_\mu$, $\nu_\tau$, e sono organizzati in tre doppietti,
\be
Leptoni: \qquad (\nu_e, e^-)\ , \qquad (\nu_\mu, \mu^-)\ , \qquad (\nu_\tau, \tau^-) \ .
\ee
Considerando anche le rispettive antiparticelle, si ha quindi un totale di dodici leptoni.

I quarks hanno carica forte (colore), debole e elettrica. Esistono sei differenti tipi di quarks, detti \emph{sapori}:  \emph{up} e \emph{down}, \emph{charmed} e \emph{strange}, e \emph{top} e \emph{bottom}. I quarks sono organizzati in tre doppietti, che sono la ripetizione di massa crescente del primo doppietto, formato dai costituenti della materia ordinaria, i quarks $u$ e $d$,
\be
Quarks: \qquad (u, d)\ , \qquad (c, s)\ , \qquad (t, b)\ .
\ee
La carica forte si presenta in tre differenti colori. Considerando anche le antiparticelle, si hanno $6\times 3\times 2=36$ quarks, e il Modello Standard ha quindi un totale di 48 particelle di materia fermioniche e 12 bosoni di gauge. 

Il Modello Standard \`e completato dal meccanismo di \emph{rottura spontanea della simmetria di gauge}, detto comunemente meccanismo di Higgs, ma dovuto a R. Brout, F. Englert e P. Higgs, che permette di avere una teoria rinormalizzabile con bosoni di gauge massivi, come richiesto dalle interazioni deboli. La rottura di simmetria ha luogo La lagrangiana elettrodebole descrive con quattro bosoni di gauge di massa nulla, una propriet\`a essenziale ai fini della sua rinormalizzabilit\`a, mentre il processo di rottura di simmetria genera le masse dei bosoni vettoriali responsabili delle interazioni deboli: $W^+$, $W^-$, e $Z^0$. La particella che rimane di massa nulla \`e il fotone.
La rottura di simmetria di gauge produce un ulteriore effetto fisicamente: la comparsa di uno scalare massivo, il bosone di Higgs. Ad oggi, la verifica sperimentale dell'esistenza di questo bosone rimane il tassello mancante nella verifica delle previsioni del Modello Standard. 

Il comportamento delle forze elettromagnetiche e di quelle deboli e forti \`e estremamente diverso. I fotoni non portano carica elettromagnetica, e l'interazione fra particelle cariche risentir\`a quindi solo delle correzioni radiative originate dalla creazione di coppie di particella-antiparticella nel vuoto. Al contrario, le forze deboli e forti sono associate a bosoni di gauge carichi, e l'autointerazione dei bosoni di gauge genera un effetto di anti-schermo. Per questa ragione, mentre le forze elettromagnetiche sono libere per piccoli impulsi e grandi distanze, e quindi possono essere studiate perturbativamente, al contrario le forze forti sono \emph{asintoticamente libere} ad alte energie. Questo comportamento \`e alla base del fenomeno del confinamento dei quarks, suggerito dalla crescita dell'interazione al diminuire dell'energia, che spiega come mai non esistano apparentemente particelle libere con carica di colore. A causa del meccanismo di Higgs, che da massa ai bosoni di gauge, le forze deboli risultano inoltre effettivamente deboli per energie minori di $M_W\simeq 100 GeV$.

Come si \`e detto, il Modello Standard incorpora tutte le propriet\`a conosciute delle interazioni forti, deboli e elettromagnetiche e risulta in straordinario accordo con i dati sperimentali. D'altra parte, esso \`e per molti aspetti una teoria poco soddisfacente: ha un numero molto alto di parametri arbitrari (circa una ventina) che devono essere opportunamente fissati e non incorpora in alcun modo le interazioni gravitazionali. Queste due osservazioni danno ragione degli innumerevoli sforzi profusi nel corso degli ultimi trenta anni per individuare una teoria pi\`u fondamentale che incorpori la gravit\`a e che possibilmente non contenga parametri liberi.

\`E importante anche notare come il Modello Standard non sia realmente una teoria di unificazione delle forze fondamentali, sebbene i settori elettrodebole e forte non siano del tutto separati (esistono molte particelle soggette ad entrambe le forze). Negli anni ci sono stati molti tentativi di formulare una \emph{Teoria di Grande Unificazione} (GUT) in grado di unire questi due settori del Modello Standard. Ad oggi, nonostante i molti indizi incoraggianti, specie, come si vedr\`a, per le estensioni supersimmetriche del Modello Standard, il programma di unificazione \`e tutt'altro che completo. 

Prima di discutere le forze gravitazionali, \`e utile accennare ai problemi di rinormalizzabilit\`a delle teorie quantistiche di campo. Sin dalla nascita della Teoria Quantistica dei Campi \`e stato chiaro che lo sviluppo perturbativo, sin dalle prime correzioni quantistiche all'ordine ad un loop, porta in generale ad avere quantit\`a divergenti per gli ordinari osservabili fisici. La ragione di queste divergenze \`e da ricercare nel passaggio da un numero finito di gradi di libert\`a nella Meccanica Quantistica, ad un numero infinito nella Teoria Quantistica dei Campi. Questo porta a una somma continua su un numero infinito di modi interni negli integrali di loop, che genera quantit\`a divergenti. Evitando i dettagli tecnici, l'idea chiave della rinormalizzazione \`e che i parametri ``nudi'' che compaiono nella lagrangiana di campo, come le costanti di accoppiamento e le masse, siano divergenti ma comunque non misurabili. Inoltre le divergenze di questi parametri possono essere scelte in modo da cancellare gli infiniti negli osservabili della teoria. Assorbendo questi infiniti nei parametri ``nudi'' della teoria, \`e possibile definire nuovi parametri ``vestiti'', finiti e misurabili. Una teoria si dice rinormalizzabile se \`e resa non divergente da un numero finito di queste ridefinizioni. 

La forza gravitazionale \`e estremamente debole rispetto alle altre forze fondamentali ma, a differenza di queste, \`e solo attrattiva e quindi risulta dominante nella dinamica su grande scala dell'universo. La forza gravitazionale \`e descritta, in termini geometrici, dalla Relativit\`a Generale di Einstein. In questa teoria, a differenza della Relativit\`a Speciale, la metrica \`e una struttura dinamica e le forze gravitazionali nascono dalla curvatura dello spazio-tempo. La Relativit\`a Generale \`e una teoria di campo classica e, ad oggi, non esiste alcuna sua formulazione quantistica consistente.

Come si \`e visto, una delle idee guida della Teoria Quantistica dei Campi associa una particella ad ogni campo, sia di forza che di materia. Le particelle che mediano la trasmissione delle interazioni si muovono nello spazio tempo fra oggetti che portano le cariche dell'interazione. Nel caso della gravit\`a, l'idea va incontro ad alcune difficolt\`a concettuali. La forza di gravit\`a \`e infatti associata alla dinamica dello stesso spazio-tempo, mentre si vorrebbe che le forze gravitazionali fossero mediate da particelle (i \emph{gravitoni}) che si propaghino su un background spazio-temporale definito. Il modo pi\`u naturale per ottenere questa descrizione \`e linearizzare la teoria, separando il tensore metrico in una parte di background, ad esempio dato da una metrica Minkowskiana $\eta_{\mu \nu}\equiv \diag (-,+,+,+, \dots)$, e una fluttuazione dipendente dalla posizione $h_{\mu \nu}(x)$, che deve essere piccola , $|h_{\mu \nu}(x)| \ll 1$,
\be
g_{\mu \nu}=\eta_{\mu \nu}+h_{\mu \nu}(x) \ .
\ee
In questo modo $h_{\mu \nu}(x)$ pu\`o essere associata ad una particella di spin due, il gravitone che si propaga su un background di riferimento. Questo \`e il punto di partenza necessario per formulare una teoria quantistica della gravit\`a. Le difficolt\`a per\`o non si limitano a questo.

La difficolt\`a maggiore nel definire una teoria quantistica di campo che descriva tutte le forze fondamentali risiede nell'apparente impossibilit\`a nello scrivere una teoria quantistica della gravit\`a che sia rinormalizzabile \cite{Rovelli:1999hz}. 

\begin{figure}
\begin{center}
\includegraphics[width=13cm,height=6cm]{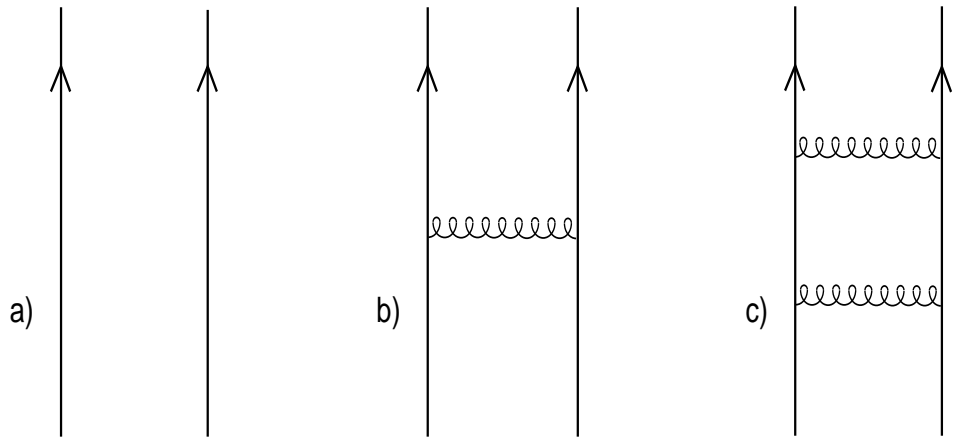}
\end{center}
\caption{a) Due particelle che si propagano liberamente. b) Correzione dovuta allo scambio di un gravitone. c) Correzione per lo scambio di due gravitoni.}
\label{divergenzaqg}
\end{figure}

Per dare un'idea del problema possiamo considerare un processo di scambio di gravitoni. In analogia con la teoria elettromagnetica in cui le interazioni sono pesate dalla costante di struttura fine $\alpha=q^2/\hbar c$, \`e possibile definire un accoppiamento adimensionale per le interazioni gravitazionali come $\alpha_G=G_NE^2/\hbar c^5$, dove $G_N$ \`e la costante di Newton. In unit\`a naturali $\hbar=c=1$ si ha $G_N = M_P^{-2}$, dove la massa di Planck $M_P= 1.2 \times 10^{19}GeV$. La figura \ref{divergenzaqg} mostra due particelle che si propagano liberamente e le correzioni al processo dovute allo scambio di uno o due gravitoni. La prima correzione sar\`a proporzionale a $\alpha_G$ e quindi a $E^2/M_P^{2}$, dove $E$ \`e naturalmente l'energia caratteristica del processo. Questa correzione diventa quindi rilevante per alte energie con $E>M_P$. A queste energie i diagrammi di scambio di due gravitoni sono di ordine
\be
M_P^{-4}\int^\infty dE'E'^3 \ ,
\ee
dove $E'$ e l'energia degli stati virtuali intermedi, e quindi la teoria presenta delle divergenze. Agli ordini successivi le divergenze, come si capisce facilmente, sono pi\`u gravi e questo rende la teoria perturbativa inservibile. Questo problema di \emph{divergenze ad alte energie} rende la teoria non rinormalizzabile. In realt\`a la teoria di Einstein \`e priva di divergenze ad un loop per ragioni di simmetria non evidenti, mentre le prime divergenze si manifestano a due loop \cite{Goroff:1986th}.

Per spiegare le divergenze che si trovano in gravit\`a quantistica vengono avanzate alcune ipotesi, tra le quali quelle comunemente considerate pi\`u ragionevoli sono due. La prima \`e che la teoria abbia un punto fisso non banale ultravioletto, ovvero che le divergenze abbiano origine dall'espansione perturbativa in potenze dell'accoppiamento, mentre una risoluzione esatta della teoria risulterebbe perfettamente consistente. La seconda \`e che alla scala di Planck sia presente ``nuova fisica'' e che pertanto la Relativit\`a Generale sia soltanto una teoria effettiva di bassa energia di un teoria pi\`u profonda. Questa seconda ipotesi \`e storicamente fondata nel passaggio dalla teoria di Fermi delle interazioni deboli alla teoria elettrodebole.

Ci si aspetta che le divergenze debbano scomparire in una teoria fondamentale che sia in grado di distribuire nello spazio tempo l'interazione. Il modo pi\`u ovvio di farlo \`e operare una discretizzazione dello spazio-tempo, ma questo distrugge l'invarianza di Lorentz al di sotto di una certa scala, e rende molto difficile mantenere l'invarianza per trasformazioni locali di coordinate, che \`e propria della Relativit\`a Generale. Come si vedr\`a, l'alternativa fornita dalla Teoria delle Stringhe \`e associare le particelle ad oggetti unidimensionali, stringhe, piuttosto che puntiformi come nella Teoria Quantistica dei Campi.

Nel discutere di possibili generalizzazioni delle teorie che oggi conosciamo, \`e utile accennare a altri due problemi non risolti della Fisica Teorica. Uno di questi \`e il \emph{problema di gerarchia}. La nostra conoscenza delle leggi fondamentali non \`e, infatti, in grado di spiegare i diversi ordini di grandezza che riscontriamo in natura. Un primo problema di gerarchia \`e la grande differenza fra le scale degli accoppiamenti delle diverse forze. Ad esempio, il rapporto fra la costante di Fermi $G_F$ e la costante di Newton, che determinano le scale delle interazioni deboli e gravitazionali a bassa energia, \`e  $G_F/G_N \sim 10^{35}\hbar^2c^{-2}$. Allo stesso modo l'attuale fisica teorica pu\`o solo registrare ma non spiegare le gerarchie presenti nei parametri del Modello Standard, ad esempio fra le masse dei fermioni.

Un secondo problema aperto \`e il \emph{problema di costante cosmologica}, ovvero la determinazione teorica del suo valore in accordo con i dati sperimentali. La costante cosmologica \`e legata alla densit\`a dell'energia di vuoto e determina la curvatura media dell'universo. Le stime teoriche prodotte dalla QFT ($\rho_{th}\sim M_{P}^4 c^5/\hbar^3$) sono in forte disaccordo con i valori osservati in natura ($\rho_{exp}\sim H^2 c^2 /G_N$), dove $H$ \`e la costante di Hubble. La stima teorica risulta di 120 ordini di grandezze pi\`u grande di quella sperimentale, e questo \`e il pi\`u grande disaccordo tra teoria ed esperimenti nella storia della scienza. 

\section*{Oltre il Modello Standard e la Relativit\`a Generale}

Nella ricerca di completamenti del Modello Standard una delle idee teoriche pi\`u affascinati \`e la \emph{supersimmetria}. La supersimmetria \`e una simmetria spazio-temporale fra particelle bosoniche e fermioniche apparentemente non realizzata alle energie ordinarie. La supersimmetria \`e stata inizialmente sviluppata alla fine degli anni '60 nei tentativi di individuazione di un \emph{Master Group} che combinasse i gruppi di simmetria interni e il gruppo di Lorentz in maniera non banale (Myazawa, 1966). ed \`e stata riscoperta nel 1971 a partire da due diversi filoni di ricerca: da un lato nell'ambito della teoria delle stringhe, dall'altro nella ricerca di generalizzazioni dell'usuale algebra spazio-temporale. La prima azione supersimmetrica in quattro dimensioni \`e stata proposta nel 1974 da Wess e Zumino.

Nonostante la supersimmetria non risulti in apparenza verificata, sono molte le ragioni che inducono allo studio di possibili estensioni supersimmetriche della fisica che conosciamo. La prima \`e sicuramente legata alle difficolt\`a che si incontrano nel costruire modelli di teorie di campo unificate. Il problema \`e che, nel costruire un gruppo unificato che combini il gruppo di Lorentz e un gruppo di Lie compatto, si incorre nel Teorema di Coleman-Mandula che stabilisce che un gruppo di questo tipo non pu\`o avere rappresentazioni finito-dimensionali. Questo indica che non \`e possibile costruire un \emph{master group} che combini sia la gravit\`a che lo spettro di particelle. 

La supersimmetria \`e un modo per evitare il Teorema di Coleman-Mandula, che non riguarda appunto una simmetria non banale che mescoli campi fermionici e bosonici e che ponga entrambi in uno stesso multipletto.

In teorie supersimmetriche \`e presente un operatore $Q$ che trasforma stati bosonici $|B \rangle$ in stati fermionici $|F \rangle$,
\be
Q|B \rangle = |F \rangle \ ,
\ee
e una sua conseguenza diretta \`e la possibilit\`a di avere multipletti in cui compaiano fermioni e bosoni.

Un altro aspetto di estremo interesse di questa teoria \`e che rendendo la supersimmetria una simmetria locale, si giunge necessariamente ad una teoria della gravit\`a. La nuova teoria chiamata Supergravit\`a, e scoperta da Ferrara, Freedman e van Nieuwenhuizen nel 1976, anche se non finita, risulta meno divergente della gravit\`a ordinaria. Come si vedr\`a essa ha un ruolo centrale nelle ricerche in corso attualmente.

\begin{figure}
\begin{center}
\includegraphics{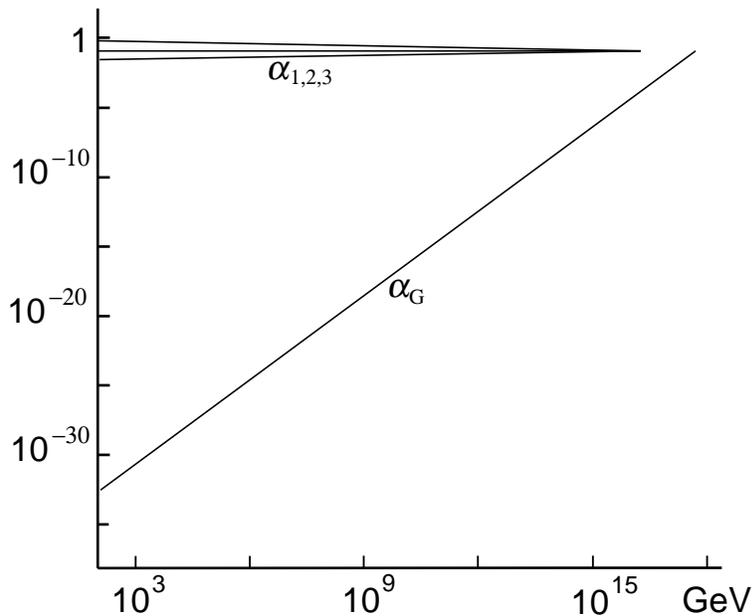}
\end{center}
\caption{I tre accoppiamenti di gauge e l'accoppiamento gravitazionale $\alpha_G$ in funzione dell'energia in teorie supersimmetriche.}
\label{unificazione}
\end{figure}

Uno degli aspetti di maggiore interesse nello studio della supersimmetria \`e che questa conduce naturalmente a tentativi di unificazione della fisica delle particelle e della gravit\`a. In teorie supersimmetriche gli accoppiamenti di gauge delle tre forze del Modello Standard realizzano l'unificazione, con una buona approssimazione, a energie intorno a $2\times 10^{16}GeV$. L'accoppiamento adimensionale della gravit\`a, come si \`e visto, a differenza di quanto avviene per le altre interazioni, dipende fortemente dalla scala di energia. La figura \ref{unificazione} mostra come in teorie supersimmetriche ci siano incoraggianti indicazioni di un'unificazione delle quattro forze fondamentali. 

Infine la supersimmetria sembra avere un ruolo per il problema della costante cosmologica. Fermioni e bosoni contribuiscono infatti all'energia di vuoto con segni opposti, e una teoria supersimmetrica nello spazio piatto da quindi luogo a una costante cosmologica nulla. Un teoria realistica e semplicemente verificabile dovrebbe prevedere la rottura della supersimmetria a scale di energie prossime a quelle dei nuovi acceleratori. Modelli di questo tipo portano per\`o a stime della costante cosmologica migliori di quelle della QFT standard, ma che restano lontane dai valori osservati.

Un'altra idea di grande interesse, introdotta nei tentativi di realizzazione di una teoria di grande unificazione \`e quella delle \emph{dimensioni extra}. \`E possibile che, su scale dell'ordine della lunghezza di Planck ($\ell_P=10^{-33}cm$) o anche maggiori, lo spazio-tempo evidenzi oltre alle quattro dimensioni estese ordinarie (le tre dimensioni spaziali e il tempo), diverse altre dimensioni spaziali piccole (\emph{compattificate}). Naturalmente queste dimensioni sfuggirebbero ad una rilevazione con ogni tipo di sonde di lunghezze d'onda grandi rispetto alla scala delle dimensioni compatte.

Quest'idea, a prima vista singolare, risulta invece piuttosto naturale. Il numero di dimensioni spazio-temporali non \`e in alcun modo vincolato dalla Teoria dei Campi, mentre una teoria pi\`u fondamentale potrebbe avere il pregio di fissarlo univocamente per ragioni di consistenza interna (come si vedr\`a nel caso della Teoria delle Stringhe).

Un primo argomento sulla ragionevolezza delle dimensioni extra \`e di tipo cosmologico. L'attuale modello cosmologico contempla una fase iniziale di espansione dell'universo. Quelle che oggi sono dimensioni estese sono state, nelle fasi iniziali dell'universo, piccole e fortemente curvate. Non sembrerebbe pertanto irragionevole se le dimensioni fossero pi\`u di quattro e se solo alcune di queste avessero subito il processo di espansione, fino alla struttura dello spazio-tempo che conosciamo, dal momento che non c'\`e alcuna ragione per supporre un processo  di espansione isotropo. In natura sono presenti numerose simmetrie spontaneamente rotte, e un principio di rottura di simmetria simile potrebbe quindi interessare anche le simmetrie spazio-temporali. Il gruppo di simmetria ordinario di Lorentz $SO(3, 1)$, potrebbe essere, in questo caso, il gruppo di simmetria residua di un gruppo pi\`u grande $SO(d, 1)$, spontaneamente rotto, $d>3$. La simmetria verrebbe rotta dalla compattificazione delle dimensioni extra. 

L'interesse per la possibile esistenza per ulteriori dimensioni \`e giustificato dal fatto che queste potrebbero essere responsabili di alcuni aspetti della fisica osservata. A titolo di esempio, si po\`o considerare un modello molto simile al quello proposto nel 1919 da T. Kaluza, che diede orgine alle teorie con dimensioni extra, anche note come \emph{teorie di Kaluza-Klein}. L'idea \`e che, oltre alle 4 dimensioni ordinarie, esista una quinta dimensione, compattificata nel modo pi\`u semplice possibile, e cio\`e su un cerchio, tramite l'identificazione
\be
\phi(x_5) \simeq \phi(x_5 + 2 \pi R) \ ,
\ee
dove $R$ \`e il raggio di compattificazione della quinta dimensione. Espandendo il campo $\phi$,
\be
\phi(x)= \sum_n \phi_n e^{ipx} \ ,
\ee
si trova che la relazione di periodicit\`a impone la quantizzazione dei momenti $p_5=n/r$ in termini di un intero $n$. Per raggi di compattificazione dell'ordine della lunghezza di Planck, i modi pi\`u alti con $n \neq 0$ sono particolarmente massivi $m \sim 10^{19}GeV$. Nel limite di bassa energia questi modi possono essere ignorati, tenendo conto solo del modo $n=0$. Questo significa che in questa approssimazione, il campo $\phi(x)$ perde la sua dipendenza dalla quinta coordinata,
\be
\partial_5 \phi(x)\sim 0 \ .
\ee
Si pu\`o cos\`i decomporre la Relativit\`a Generale in 5-dimensioni in campi quadridimensionali, scrivendo il tensore metrico nella forma
\begin{equation}
g_{AB}=
\left[
\begin{array}{ccc}
&\vrule&\\[-5pt]
\hspace{5pt}g_{\mu\nu}\hspace{5pt}&\vrule&\hspace{-7pt}A_\mu\hspace{-3pt}
\\[-5pt]
&\vrule&\\
\hline&\vrule&\\[-15pt]
A_\nu&\vrule&\hspace{-9pt}\phi'\hspace{-4pt}\\[-15pt]&\vrule&
\end{array}
\right] \ .
\label{kk}
\end{equation}
Considerando le equazioni di Einstein in cinque dimensioni e scrivendole in termini delle componenti (\ref{kk}), si trovano sia le equazioni di Einstein per la metrica quadridimensionale $g_{\mu \nu}$ che le equazioni di Maxwell per il potenziale vettore $A_\mu$. L'ultimo elemento della matrice \`e un campo scalare $\phi'$, noto come dilatone, che si accoppia sia al campo elettromagnetico che alla metrica e che avremo modo di incontrare spesso nel seguito di questa Tesi.

In maniera simile, per spazi con dimensioni pi\`u alte, dall'equazione di Dirac \`e possibile ritrovare differenti generazioni di quarks e leptoni: un singolo spinore nello spazio con molte dimensioni porta a molti campi spinoriali in quattro dimensioni. Il problema \`e per\`o, in generale, realizzare la struttura chirale delle interazioni deboli.

\section*{La Teoria delle Stringhe come teoria di unificazione}

La Teoria delle Stringhe, nata alla fine degli anni '60 dal tentativo di descrivere le forze nucleari forti, \`e oggi il candidato pi\`u promettente per una teoria unificata di tutte le forze fondamentali \cite{Polchinski:1994mb}. Dal momento che per le interazioni forti appariva impossibile far ricorso alla teoria perturbativa dei campi, si cercavano al tempo esempi concreti di ``matrici S'', ovvero  espressioni in grado di determinare direttamente le probabilit\`a di diverse interazioni sotto condizioni assegnate. Nel 1968 Veneziano propose un'ampiezza che sembrava appunto descrivere un'interazione fra particelle scalari risultante dallo scambio di infinite particelle di massa e spin crescente \cite{veneziano}. L'ampiezza aveva anche sorprendenti propriet\`a di simmetria fra le variabili di Mandelstam $s$ e $t$ legate alla descrizione degli impulsi in gioco. Poco dopo il lavoro di Veneziano, Shapiro e Virasoro proposero una nuova ``matrice S'' con una simmetria pi\`u ampia tra le variabili che descrivono gli impulsi \cite{shapiro, sv}. Nel 1970 le ampiezze furono reinterpretate, principalmente grazie ai lavori di Nambu e Susskind, come ampiezze d'interazione di oggetti unidimensionali, stringhe appunto. Nel 1971 l'inclusione dei gradi di libert\`a fermionici port\`o alla scoperta delle stringhe supersimmetriche \cite{nsr}.

Lo studio di questi modelli, detti allora ``modelli duali'', continu\`o fino allo sviluppo della QCD, che si dimostr\`o presto la corretta teoria delle interazioni forti. Nel 1974 per\`o Scherk e Schwarz, e indipendentemente Yoneya proposero che la Teoria delle Stringhe poteva descrivere le interazioni gravitazionali e poteva essere per questo un candidato come teoria di unificazione \cite{Scherk:1974ca, Yoneya:1974jg}.

Il periodo 1984-85 vide importanti risultati che convinsero un'ampia parte della comunit\`a scientifica del potenziale della Teoria delle Stringhe in relaziona al problema dell'unificazione. In particolare, nel 1984 Green e Schwarz mostrarono che la teoria di superstringhe aperte e chiuse di Tipo I \`e priva di anomalie, e quindi quantisticamente consistente, grazie ad un meccanismo di cancellazione del tutto nuovo, se il suo gruppo di gauge \`e SO(32) \cite{gs}.

Nella Teoria delle Stringhe le forze sono unificate in maniera molto profonda, dal momento che le particelle sono unificate. Mentre nella Teoria Quantistica dei Campi le particelle fondamentali sono considerate puntiformi, in Teoria delle Stringhe esse sono identificate con i modi vibrazionali di oggetti fondamentali unidimensionali, le stringhe. 

Lo studio di una teoria quantistica relativistica di oggetti unidimensionali rivela una ricchezza sorprendente, e molte delle caratteristiche attese da una teoria di unificazione:
\begin{itemize}
\item[-] La Teoria di Stringhe include naturalmente la gravit\`a. Ogni teoria consistente di stringhe contiene uno stato vibrazionale di massa nulla e spin 2 che pu\`o essere identificato con il gravitone, dal momento che a basse energie la sua dinamica \`e descritta dalle equazioni di Einstein. 
\item[-] Almeno in teoria delle perturbazioni, la teoria \`e una teoria quantistica della gravit\`a priva di divergenze.
\item[-] E' possibile in maniera molto naturale introdurre grandi gruppi di gauge in grado di contenere, almeno in linea di principio, il Modello Standard, come atteso dalle Teorie di Grande Unificazione.
\item[-] La quantizzazione della Teoria delle Stringhe porta a fissare univocamente la dimensione dello spazio tempo, per ragioni di consistenza. La dimensionalit\`a dello spazio-tempo si trova essere per la stringa bosonica $D=26$, e per la superstringa $D=10$. 
\item[-] La supersimmetria pu\`o essere inclusa in maniera molto naturale nella Teorie delle Stringhe, e questo produce notevoli semplificazioni.
\item[-] La Teoria non ha parametri liberi. In particolare, la costante di accoppiamento di stringa, $g_s$, \`e determinata dinamicamente dal campo del dilatone, $g_s=e^{\phi}$. 
\end{itemize}

Le stringhe hanno una scala dimensionale naturale che pu\`o essere stimata con l'analisi dimensionale. Dal momento che la teoria delle stringhe \`e una teoria quantistica che descrive anche la gravit\`a, essa deve coinvolgere le costanti fondamentali $c$ (velocit\`a della luce), $\hbar$ (costante di Planck), e $G_N$ (costante di Newton). Da queste costanti naturali, si pu\`o formare una lunghezza, la lunghezza di Plack, a cui si \`e gi\`a fatto cenno,
\be
\ell_P =\sqrt{\alpha'}= \left(\frac{\hbar G_N}{c^3}\right)^{3/2}= 1.6 \times 10^{-33} \, cm \ ,
\ee
in maniera simile si definisce la massa di Planck come
\be
m_P=\left(\frac{\hbar c}{G_N}\right)^{1/2} = 1.2 \times 10^{19}\, GeV/c^2 \ .
\ee
Esperimenti ad energie molto inferiori all'energia di Planck, come quelli possibili attualmente, non possono risolvere distanze dell'ordine della lunghezza di Planck. Alle energie accessibili le stringhe possono quindi essere approssimate efficacemente da particelle puntiformi. Questo giustifica, dal punto di vista della Teoria delle Stringhe, il successo della QFT nel descrivere la fisica che conosciamo.

Nella sua evoluzione temporale, una stringa disegna una superficie bidimensionale nello spazio-tempo, detta \emph{superficie d'universo} (\emph{world-sheet}), che pu\`o essere parametrizzata con coordinate ($\sigma$, $\tau$). La superficie d'universo \`e l'equivalente per una stringa della traiettoria descritta da una particella puntiforme. La ``storia'' di una stringa nello spazio-tempo $D$-dimensionale \`e descritta dalla sua coordinata $X^\mu(\sigma, \tau)$, e l'azione generica di stringa \`e della forma
\be
S=-T\int dA \ ,
\ee
dove $T$ indica la tensione di stringa e $dA$ \`e l'elemento d'area spazzato dalla stringa nel suo cammino. Come si vedr\`a in dettaglio nel prossimo capitolo, si tratta della naturale generalizzazione dell'azione di particella relativistica. 

\begin{figure}
\begin{center}
\includegraphics[width=11cm,height=6cm]{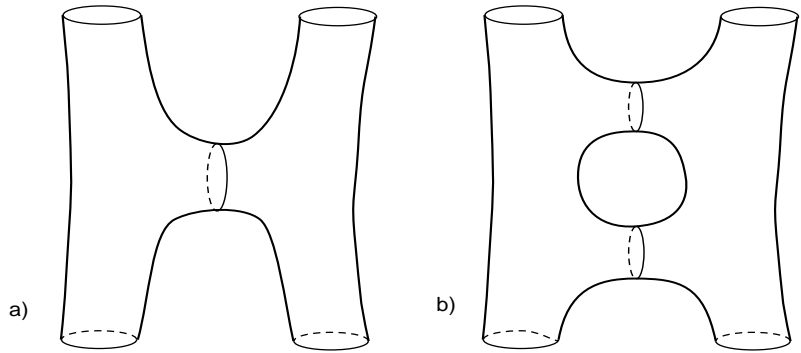}
\end{center}
\caption{L'equivalente in Teoria delle Stringhe dei diagrammi in figura \ref{divergenzaqg}. a) Interazione fra stringhe chiuse con scambio di una stringa. b) Interazione con scambio di due stringhe. }
\label{stringdiag}
\end{figure}

Nella formulazione perturbativa della Teoria Quantistica dei Campi i contributi delle ampiezze sono associati ai diagrammi di Feynman, che raffigurano tutte le possibile configurazioni delle traiettorie. In particolare, le interazioni corrispondono alle giunzioni delle traiettorie. Allo stesso modo, la teoria perturbativa di stringa coinvolge superfici di universo di diverse topologie. L'esistenza delle interazioni \`e per\`o una conseguenza della topologia della superficie d'universo piuttosto che di una singolarit\`a locale (vedi figura \ref{stringdiag}). Questa differenza rispetto alla teoria dei campi, ha due importanti implicazioni. La prima \`e che in Teoria delle Stringhe la struttura delle interazioni \`e unicamente determinata dalla teoria libera, mentre la seconda \`e che, dal momento che le superfici di universo sono lisce, le ampiezze in Teoria delle Stringhe non hanno divergenze ultraviolette (almeno nel caso di stringhe chiuse). La comparsa di divergenze nella Teoria dei Campi Quantistica \`e invece riconducibile al fatto che le interazioni sono localizzate in punti dello spazio-tempo.

Esistono due tipi possibili di stringhe, aperte e chiuse. Le stringhe aperte hanno due estremit\`a, mentre quelle chiuse non ne hanno alcuna. Si possono avere teorie di sole stringhe chiuse o di stringhe chiuse e aperte. La ragione intuitiva dell'impossibilit\`a di scrivere teorie di sole stringhe aperte e quindi senza gravit\`a (che ha sempre origine da stringhe chiuse), \`e che le stringhe aperte possono chiudersi dando luogo a stringhe chiuse. 

Una seconda divisione \`e tra stringhe bosoniche e superstringhe. Le stringhe bosoniche descrivono solo particelle di spin intero, e per questo motivo non appaiono realistiche. Al contrario, le superstringhe descrivono anche fermioni, e ci si aspetta che una descrizione realistica unificata delle forze fondamentali possa sorgere dalle teorie di superstringa.

Un aspetto cruciale della teoria delle stringhe \`e legato allo studio delle anomalie. Un'anomalia \`e una violazione quantistica di una simmetria posseduta da una teoria a livello classico. Le anomalie sono ben note anche nello studio della Teoria dei Campi. Nel caso la violazione riguardi una simmetria globale, questa non \`e dannosa ma comporta la comparsa di effetti nella fenomenologia della teoria. Ad esempio, nel Modello Standard le anomalie globali sono importanti nella determinazione della vita media del $\pi^0$ e della massa della particella $\eta'$.  Al contrario, se l'anomalia riguarda una simmetria di gauge questo  porta in generale ad una inconsistenza delle teoria. Infatti le simmetrie di gauge sono necessarie per la cancellazione dei gradi di libert\`a non fisici, e l'impossibilit\`a di farlo porta ad una perdita di unitariet\`a della teoria. 
In realt\`a, per una teoria bidimensionale come il modello chirale di Schwinger (una teoria di gauge $U(1)$ accoppiata ad un fermione di massa nulla), le anomalie di gauge non sono dannose e i gradi di libert\`a aggiuntivi possono essere inclusi ottenendo una teoria consistente, ma non si \`e ancora in grado di generalizzare questa procedura in dimensioni pi\`u alte.
Per formulare una teoria delle stringhe consistente \`e quindi necessario che tutte le anomalie locali (di gauge, gravitazionali, miste) si cancellino.
In presenza di stringhe aperte la cancellazione delle anomalie di gauge si ottiene, come si vedr\`a in dettaglio, imponendo un'opportuna cancellazione nel settore di Ramond-Ramond (R-R), che \`e uno dei settori dello spettro di superstringa. Questa condizione equivale a richiedere che lo spazio-tempo compattificato sia globalmente neutro, rispetto alle cariche corrispondenti, e fissa necessariamente il gruppo di gauge della superstringa di Tipo I in dieci dimensioni. La cancellazione delle anomalie del diagramma di esagono all'ordine ad un loop, l'analogo del diagramma a triangolo in quattro dimensioni, avviene grazie al contributo del diagramma di propagazione al livello ad albero della 2-forma: questo \`e noto come meccanismo di Green e Schwarz \cite{gs}. 
Infine, la cancellazione dell'anomalia di Weyl porta a determinare univocamente la dimensione $D$ dello spazio-tempo. 

La nostra comprensione della Teoria delle Stringhe \`e stata per molti anni limitata agli aspetti perturbativi. Dalla teoria dei campi \`e per\`o noto che molti importanti effetti dinamici sorgono quando si hanno molti gradi di libert\`a in accoppiamento forte, quali il confinamento e il meccanismo di Higgs, che sono necessari alla comprensione della fisica della teoria. Fortunatamente, a partire dalla met\`a degli anni '90, si \`e ottenuto un notevole progresso nella comprensione delle teorie supersimmetriche nel limite di accoppiamento forte, che ha avuto come effetto una sorprendente unificazione delle teorie di superstringa note.

\section*{Relazioni di dualit\`a e M-teoria}

La famiglia delle teorie di superstringa contiene cinque differenti modelli con supersimmetria spazio-temporale in $10$ dimensioni: Tipo I $SO(32)$, Tipo IIA, Tipo IIB, eterotica $SO(32)$ (HO), eterotica $E_8\times E_8$ (HE), oltre ad alcuni altri modelli non supersimmetrici. Negli ultimi dieci anni molti sforzi sono stati spesi nello studio delle relazioni che collegano i differenti modelli. Si \`e cos\`i giuti a comprendere che le diverse teorie supersimmetriche sono legate fra loro da alcune trasformazioni, dette di \emph{dualit\`a}, e che esse possono essere interpretate come differenti limiti di un'unica teoria sottostante in $11$ dimensioni, detta M-teoria, i cui gradi di libert\`a fondamentali non sono per\`o stringhe \cite{Schwarz:1996du, Sen:1997yy, Witten:1995ex, Hull:1995mz}. 

Le teorie di Tipo II A e IIB sono teorie di superstringhe chiuse orientate, mentre la Tipo I \`e una teoria di superstringhe aperte e chiuse non orientate. Nel limite di basse energie, in cui la teoria di stringhe pu\`o essere interpretata come una teoria di campo, queste tre teorie descrivono altrettanti modelli di supergravit\`a in $10$ dimensioni, rispettivamente i modelli di Tipo IIA, IIB, di Tipo I, quest'ultima accoppiata ad una teoria con supersimmetrica di Yang-Mills con gruppo di gauge $SO(32)$.

La stringa eterotica \`e un modello di sole stringhe chiuse. Dal momento che, come si vedr\`a, per una stringa chiusa i modi destri e sinistri sono indipendenti, \`e possibile considerare modelli i cui modi destri sono i modi di una superstringa in 10 dimensioni, mentre quelli sinistri sono i modi di una stringa bosonica in 26 dimensioni. Naturalmente occorre compattificare le dimensioni bosoniche chirali in eccesso. Questa costruzione, a prima vista singolare, porta alla formulazione di una teoria supersimmetrica consistente e introduce in modo naturale i gradi di libert\`a interni di una teoria di gauge senza dover considerare stringhe aperte. Le condizioni di consistenza selezionano solo due possibili gruppi di gauge, $SO(32)$ e $E_8 \times E_8$.

Della M-teoria si conosce la teoria di basse energie, trovata alla fine degli anni '70 da Cremmer, Julia e Scherk, che risulta essere l'unica teoria di supergravit\`a in 11 dimensioni \cite{Cremmer:1978km}. In particolare, si vede che uno spazio-tempo 11-dimensionale \`e il pi\`u grande in cui sia possibile formulare una teoria di supergravit\`a. La supergravit\`a in undici dimensioni \`e una teoria non rinormalizzabile, e quindi non pu\`o essere considerata come una teoria fondamentale. Al contrario, la corretta interpretazione sembra essere quella di una teoria effettiva di bassa energia della M-teoria. La M-teoria non \`e direttamente legata ad una teoria di stringa, dal momento che il suo limite di bassa energia non contiene il potenziale di oggetti unidimensionali.

Fra le cinque teorie di superstringa e la M-teoria esiste una fitta rete di relazioni, nota come ``rete di dualit\`a''. In generale una dualit\`a \`e una relazione di equivalenza fra sistemi fisici apparentemente distinti, che connette gli stati di una teoria con quelli di un'altra (o diversi stati della stessa teoria di partenza) preservando le interazioni e le simmetrie.  

Le relazioni di dualit\`a sono ben note in teoria dei campi. Un esempio particolarmente importante di questo tipo di relazioni \`e la dualit\`a elettro-magnetica. Una sorprendente caratteristica delle equazioni di Maxwell \`e la simmetria dei termini sinistri delle equazioni sotto gli scambi del campo elettrico e del campo magnetico $\mathbf{E} \to - \mathbf{B}$ e $\mathbf{B} \to \mathbf{E}$. Questa simmetria porta all'idea suggestiva che possano esistere cariche magnetiche oltre che cariche elettriche, e una conseguenza di questa ipotesi \`e una relazione di quantizzazione del prodotto tra cariche elettriche e magnetiche trovata da Dirac. La ricerca di evidenze sperimentali di questa congettura, che ha solide basi nelle teorie di grande unificazione e di superstringa, \`e oggi un attivo campo di ricerca.

\begin{figure}
\begin{center}
\includegraphics{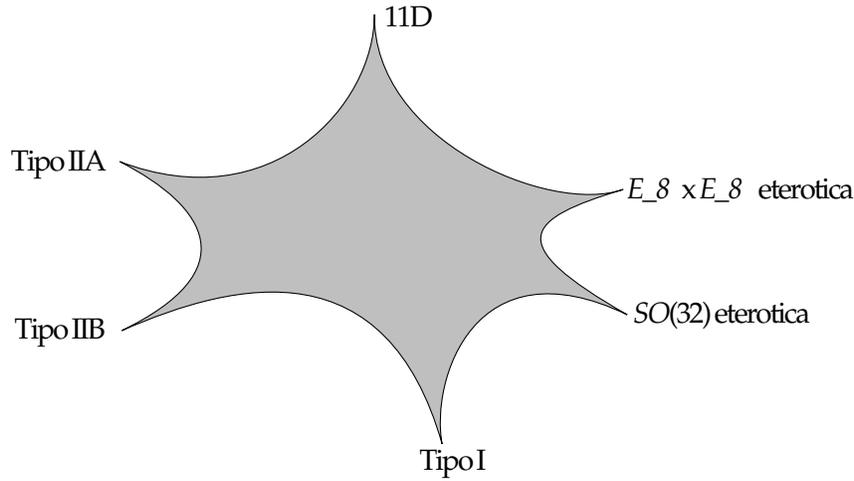}
\end{center}
\caption{Le cinque teorie di stringa e la M-teoria come limiti di un'unica teoria.}
\label{esagono}
\end{figure}

Come si \`e detto, le relazioni di dualit\`a sono di straordinario interesse anche in teoria delle stringhe. Un primo esempio \`e la S-dualit\`a \cite{Sduality}  che collega settori di una teoria con accoppiamento forte a settori con accoppiamento debole 
\be
S \ : \ g_s \ \longleftrightarrow \ \frac{1}{g_s} \ .
\ee
L'importanza di questa dualit\`a risiede nel fatto che, come si \`e gi\`a avuto modo di sottolineare, mentre per accoppiamento forte non \`e possibile lo studio perturbativo della teoria, al contrario questo diventa possibile in accoppiamento debole.

Si trova che la S-dualit\`a identifica il limite di accoppiamento debole della teoria di stringa eterotica $SO(32)$, con il limite di accoppiamento forte della Tipo I $SO(32)$. Inoltre si trova che la teoria Tipo IIB \`e auto-duale sotto questa trasformazione. La simmetria tra accoppiamento forte e debole per S-dualit\`a \`e esplicita nelle teorie di basse energie, mentre rimane una congettura per le teorie complete, anche se sono molti gli indizi che spingono a ritenerla vera.

In teoria delle stringhe la S-dualit\`a scambia fra di loro diversi tipi di oggetti, in particolare stringhe e alcuni oggetti estesi conosciuti come \emph{$D$-brane} \cite{pol95}, che risultano fondamentali per avere il giusto conteggio  dei gradi di libert\`a della teoria prima e dopo la dualit\`a. Le $D$-brane corrispondono a configurazioni solitoniche la cui tensione \`e proporzionale all'inverso della costante d'accoppiamento di stringa, e sono mappate dalla S-dualit\`a in stati di stringa. Nel limite perturbativo ($g_s$ piccolo) le $D$-brane si presentano come oggetti essenzialmente rigidi e privi di dinamica, mentre al contrario nel limite di accoppiamento forte le $D$-brane si rivelano oggetti dinamici che possono muoversi e curvarsi. Una $D$-brana \`e caratterizzata, oltre che dalla tensione, anche dalla sua carica di Ramond-Ramond. Naturalmente esisteranno anche anti-brane, con carica opposta e stessa tensione.

Una seconda relazione di dualit\`a \`e la T-dualit\`a, che identifica teorie compattificate su un cerchio di raggio $R$, con teorie compattificate su un raggio $1/R$,
\be
T \ : \ R \ \longleftrightarrow \ \frac{\alpha'}{R} \ .
\ee
La T-dualit\`a identifica le due teorie di tipo II, e le due teorie di stringa eterotica. A differenza della S-dualit\`a, la T-dualit\`a \`e una simmetria perturbativa \cite{dlp,horava2,green}.

Esiste una ulteriore relazione che unifica la teoria di Tipo I con la IIB, conosciuta come proiezione di orientifold $\Omega$ \cite{cargese}. L'azione di $\Omega$ scambia modi sinistri e destri di una stringa chiusa, e i punti fissi di questa proiezione corrispondono nello spazio-tempo a oggetti estesi non dinamici detti piani di orientifold o $O$-piani. Gli $O$-piani risultano essere carichi e avere tensione che, a differenza di quella delle $D$-brane, pu\`o avere valori anche negativi.

Le ultime relazioni utili a completare il diagramma delle dualit\`a sono quelle che collegano la M-teoria alle teorie IIA e HE. Si pu\`o vedere che, partendo dalla teoria di supergravit\`a 11-dimensionale, e compattificando l'undicesima dimensione su un cerchio $S^1$, si trova la supergravit\`a di Tipo IIA. Al contrario, compattificandola su un segmento $S^1/\mathbb Z_2$, si ritrova la teoria di bassa energia della stringa eterotica $E_8\times E_8$ \cite{hwi}. 

Complessivamente le relazioni di dualit\`a ricostruiscono i sei lati dell'esagono di dualit\`a in figura \ref{esagono}. Questa mostra la profonda unit\`a delle teorie di superstringa in 10 dimensioni, che possono essere pensate come limiti di una teoria sottostante che viene identificata con la M-teoria. La teoria delle stringhe rivela quindi, sorprendentemente, di non essere realmente una teoria di oggetti unidimensionali.

\section*{Teorie realistiche e problemi legati alla loro verifica}

Si \`e visto che la Teoria delle Stringhe sembra essere buon candidato per una teoria di unificazione. Ci si aspetta quindi che sia possibile far emergere, insieme alla gravitazione quantistica, il Modello Standard dalla teoria delle stringhe nel limite di bassa energia. Si \`e visto che la teoria delle stringhe possiede strutture in grado di accogliere tutte le particelle e interazioni del Modello Standard, ma allo stato attuale della ricerca non si \`e ancora riusciti ad identificare un modello che nel limite di bassa energia che includa tutti i dettagli del Modello Standard.

I modelli di superstringa, che contengono sia bosoni che fermioni, sono definiti in $D=10$. La costruzione di modelli realistici in quattro dimensioni, in grado di descrivere l'universo che conosciamo, deve necessariamente affrontare la compattificazione delle sei dimensioni extra e la rottura della supersimmetria, che come abbiamo detto non \`e osservata in natura. Come si vedr\`a i due problemi sono in parte collegati.

Il modo pi\`u semplice di compattificare le ulteriori dimensioni \`e considerare un toro $T^n$, dove $n$ indica la sua  dimensione. Si tratta di un'estensione molto semplice del modello di Kaluza-Klein, che nel caso di stringhe chiuse porta ad una fisica pi\`u ricca, grazie alla possibilit\`a che le stringhe si avvolgano intorno alle dimensioni compattificate. Una generalizzazione di questo tipo di compattificazioni sono le compattificazioni su orbifold \cite{dhvw}, che si ottengono identificando punti della variet\`a interna, sotto l'azione di un gruppo discreto. Queste identificazioni lasciano, in generale, dei punti fissi che in teoria delle stringhe non producono singolarit\`a. Per questo tipo di compattificazioni \`e possibile determinare analiticamente sia lo spettro al livello ad albero che le interazioni.  Infine, una classe di compattificazioni di grande interesse sono gli spazi di Calabi-Yau che, a differenza delle compattificazioni di orbifold, sono variet\`a lisce ma che in alcuni limiti si riducono ad orbifold. Un Calabi-Yau \`e, in generale, una variet\`a $n$-dimensionale complessa, compatta, con metrica di K\"ahler Ricci-piatta, e con gruppo di olonomia $SU(d)$, con $d=n/2$ \cite{Greene:1996cy}. La riduzione del gruppo di olonomia da $SO(n)$ a $SU(d)$ \`e una conseguenza della richiesta che una supersimmetria sia presente in $D=4$. 

La rottura delle supersimmetria pu\`o essere ottenuta anche in compattificazioni toroidali o su orbifold. Nel caso di compattificazioni toroidali \`e possibile generalizzare alle stringhe il meccanismo di Scherk-Schwarz \cite{ss}, che per teorie supersimmetriche di campo con compattificazioni consiste nell'introdurre degli shift dei momenti di Kaluza-Klein dei vari campi proporzionali alle loro cariche, introducendo in generale differenze di massa fra fermioni e bosoni che rompono la supersimmetria. Questo meccanismo in Teorie delle Stringhe si arricchisce della possibilit\`a di introdurre shift non solo nei momenti ma anche nei winding \cite{ads1}, che sono i numeri quantici associati al numero di avvolgimenti della stringa intorno alla direzione compatta. La supersimmetria \`e conseguentemente rotta alla scala $1/R$ dove $R$ \`e la scala tipica della dimensione compatta. Si tratta di un meccanismo di rottura spontanea di simmetria, regolato da un parametro continuo, e nel limite di decompattificazione la supersimmetria viene ripristinata.
Nel caso di stringhe chiuse l'introduzione degli shift nei momenti o nei winding porta essenzialmente allo stesso fenomeno. Questo \`e dovuto alle propriet\`a delle teorie di stringa chiusa sotto T-dualit\`a. Al contrario, per la teoria di Tipo I l'effetto dei due tipi di shift sulla fisica del modello \`e molto diverso, e i due meccanismi risultanti sono detti \emph{Scherk-Schwarz supersymmetry breaking} ed \emph{M-theory breaking} \cite{ads1,bsb}. Mentre nel primo caso la supersimmetria \`e rotta sia nello spazio tempo che sulle brane, per il modello di M-theory breaking si ha  un interessante fenomeno detto \emph{brane supersymmetry}, ovvero le eccitazioni di bassa energia di brane immerse in uno spazio-tempo non supersimmetrico possono essere supersimmetriche. Questo \`e vero a livello classico, ma la supersimmetria delle brane viene rotta da correzioni radiative.

In modelli con compattificazioni su orbifold \`e possibile introdurre la rottura della supersimmetria sulle brane in due diversi modi. Il primo modo detto \emph{brane supersymmetry breaking} si realizza in modelli che richiedono configurazioni con la presenza simultanea di brane e antibrane di diversi tipi. In questo caso il settore chiuso \`e supersimmetrico, anche se in generale \`e diverso dal settore chiuso di una teoria non compattificata, mentre nel settore aperto la supersimmetria \`e rotta alla scala di stringa. Questo tipo di configurazioni, che si studieranno in dettaglio nei prossimi capitoli, risultano stabili ovvero privi di tachioni. Il secondo modo consiste nel deformare uno spettro aperto supersimmetrico con un sistema di coppie separate di brane e antibrane dello stesso tipo. Questo tipo di configurazione risultano instabili a causa delle forze attrattive fra brane e antibrane.

L'ultimo metodo conosciuto per rompere la supersimmetria in Teoria delle Stringhe consiste nell'introdurre campi magnetici all'interno delle dimensioni compatte che equivale, nella rappresentazione che si ottiene con una T-dualit\`a, all'introduzione di brane ruotate \cite{magnetic, bachasmag, Larosa}. La rottura di simmetria avviene perch\`e gli estremi delle stringhe aperte sono carichi e possono accoppiarsi ai campi magnetici, dando luogo a degli shift nelle masse degli stati di stringa differenti a seconda dei loro spin. Le differenze di massa che si producono in questo modo possono portare alla rottura della simmetria fra fermioni e bosoni (un'opportuna scelta dei campi magnetici pu\`o anche preservare la supersimmetria).

Uno dei principali problemi nello studio di teorie di stringa realistiche \`e che la teoria ha un enorme numero di vuoti approssimativamente stabili, che corrispondono alle diverse scelte per possibili per le dimensioni compattificate. La fisica osservata risente in modo cruciale dalla scelta del vuoto, dal momento che i parametri di bassa energia in quattro dimensioni dipendono da alcuni \emph{moduli} continui e discreti che codificano il tipo di compattificazione. 
D'altra parte la Teoria delle Stringhe condivide apparentemente con la Relativit\`a di Einstein la mancanza di un principio globale di minimo che permetta di selezionare globalmente le configurazioni energeticamente pi\`u favorevoli. In cosmologia solitamente si fa ricorso ad argomenti basati sulle condizioni iniziali, le simmetrie, la semplicit\`a. 
In  Teoria delle Stringhe questo significa che non \`e possibile selezionare uno dei vuoti stabili fissando arbitrariamente i moduli. I moduli si presentano nella teoria di bassa energia come campi scalari con accoppiamenti esclusivamente derivativi e quindi con potenziali piatti e valori di vuoto indeterminati. In questo modo la Teoria delle Stringhe, la cui formulazione \`e priva di parametri liberi, viene a generare un numero molto alto di parametri discreti e continui attraverso le sue soluzioni. 
Una teoria \`e naturalmente tanto pi\`u predittiva quanto pi\`u basso \`e il numero delle sue soluzioni. In Teoria delle Stringhe, la presenza di molti vuoti \`e oggi il maggiore ostacolo alla possibilit\`a di estrarre i parametri del Modello Standard. Questo problema \`e noto come \emph{problema dei moduli}.

La ricerca di uno (o pi\`u) modelli di stringa in grado di prevedere tutti i parametri del Modello Standard  \`e uno dei principali obiettivi oggi perseguiti al fine di giustificare l'adozione della Teoria delle Stringhe come teoria delle interazioni fondamentali. D'altra parte \`e di grande interesse anche l'esplorazione della teoria per la gravit\`a quantistica. Grandi progressi sono stati compiuti nello studio delle situazioni in cui gli effetti quantistici della gravit\`a diventano rilevanti. Il pi\`u celebre di questi \`e forse la precisa interpretazione statistica della termodinamica di Bekenstein dei buchi neri, per una vasta classe di questi oggetti legati in modo diretto alla supersimmetria. Lo studio semi-classico della gravit\`a quantistica, dove campi quantistici vengono studiati in un background classico di buco nero, ha portato Hawking all'inizio degli anni '70 ad ipotizzare l'emissione di radiazione termica da parte dei buchi neri. Precedentemente Bekenstein aveva proposto una formula per l'entropia dei buchi neri  associata alla loro area. Questa formula, che va sotto il nome di entropia di buco nero di Bekenstein-Hawking, \`e semplicemente 
\be
S=\frac{A}{4G_N} \ ,
\ee
costituisce il cuore della delle leggi termodinamiche dei buchi neri. Quando fu formulata, questa legge non era sostenuta da nessuna teoria della gravit\`a quantistica che spiegasse in termini statistici la relazione fra l'entropia e le propriet\`a dei buchi neri. La teoria delle stringhe ha permesso, almeno per una classe di buchi neri, di giustificare questa formula in funzione di gradi di libert\`a microscopici, eccitazioni di $D$-brane \cite{Strominger:1996sh}.
Naturalmente una teoria completa della gravitazione quantistica sarebbe di straordinaria importanza nello studio della cosmologia delle fasi iniziali di formazione dell'universo. In queste fasi infatti la Relativit\`a Generale non \`e assolutamente in grado di fornire informazioni e la nostra attuale conoscenza della Teoria delle Stringhe appare insufficiente. 

\`E utile infine accennare a due campi di ricerca sperimentale che potrebbero fornire utili indizi sulla Teoria delle Stringhe, la ricerca delle ulteriori dimensioni e della supersimmetria. Si \`e precedentemente indicata la lunghezza di Planck, $\ell_P \sim 10^{-33}$ cm, come scala naturale delle ulteriori dimensioni. Questa scelta porta le ulteriori dimensioni a scale lontane dalle nostre possibilit\`a di verifica sperimentale, in quanto gli attuali acceleratori sono infatti in grado di esplorare distanze solo fino a $10^{-16}$ cm. Anche se questo scenario appare probabile, la Teoria delle Stringhe non esclude la presenza di dimensioni extra ``grandi'', dell'ordine, ad esempio, di $10^{-18}$ cm. In questo caso una verifica sperimentale sarebbe possibile e qualora avvenisse costituirebbe una spettacolare evidenza per la plausibilit\`a della Teoria. Allo stesso modo, se si dovessero trovare evidenze sperimentali della realizzazione della supersimmetria ad alte energie, questo costituirebbe un indizio del fatto che la Teoria delle Stringhe si sta muovendo nella giusta direzione.

\section*{Oltre un loop}

Come si \`e detto, lo studio
sistematico delle propriet\`a di vuoti con supersimmetria rotta in quattro dimensioni \`e attualmente uno dei principali ostacoli da superare per giungere ad una comprensione soddisfacente del legame tra la Teoria
delle Stringhe e il Modello Standard delle interazioni fondamentali. Questa Tesi tratta di alcuni aspetti tecnici legati a questo confronto, e in particolare discute la struttura di alcune correzioni radiative in
modelli con supersimmetria rotta.

Al momento esistono diverse tecniche per il calcolo di diagrammi ad un loop nella Teoria delle Stringhe, e il loro studio ha rivelato una serie di propriet\`a sorprendenti, tra cui nuovi meccanismi per la cancellazione delle anomalie e per la rottura della supersimmetria. Inoltre, alcuni dei risultati di questa analisi (correzioni di soglia, l'analogo nella Teoria delle Stringhe
delle funzioni beta del gruppo di rinormalizzazione) sono alla base dei tentativi di confronto con la Fisica delle Particelle Elementari. Non esistono invece tecniche generali oltre un loop, e per molto tempo la
stessa definizione dei diagrammi ha presentato difficolt\`a ritenute  insormontabili. Recentemente E. D'Hoker e D.H. Phong hanno proposto una definizione operativa per il contributo a due loop in una classe di superstringhe chiuse e orientate in D=10. Oggetto di questo lavoro di Tesi \`e l'estensione di questo risultato ad altri tipi di superstringhe in D=10, in vista anche della costruzione di un'espressione generale per le correzioni di soglia a due loops. 

Nel primo capitolo vengono discusse le propriet\`a fondamentali e le lagrangiane delle stringhe relativistiche aperte e chiuse, sia nel caso bosonico che in quello di superstringa, lo spettro delle loro eccitazioni e i metodi di quantizzazione canonica e nel cono di luce.

Il secondo capitolo contiene una breve introduzione alla formulazione funzionale della teoria delle stringhe (integrale di Polyakov). In questa formulazione le ampiezze di stringa sono definite da una somma su tutte le possibili ``storie'' di stringa che interpolino tra stati iniziali e finali. L'integrale di Polyakov d\`a quindi luogo ad una
espansione perturbativa della teoria in potenze della ``costante di accoppiamento di stringa'', ordinata dalla caratteristica di Eulero delle superfici di Riemann coinvolte. 

Il terzo capitolo contiene una breve rassegna sulle ampiezze di stringa bosonica, e mostra in dettaglio alcune ampiezze di stringa aperta e chiusa all'ordine ad albero e ad un loop.

Il quarto capitolo contiene i risultati principali dello studio delle ampiezze di vuoto ad un loop, che in Teoria delle Stringhe sono interpretabili come funzioni di partizioni e permettono di estrarre un gran numero di informazioni sulla struttura dei modelli e di definirne le regole costitutive.

Il quinto capitolo \`e dedicato ai pi\`u semplici meccanismi di compattificazione delle dimensioni aggiuntive della teoria, su tori e orbifold. Queste compattificazioni rivelano una simmetria perturbativa  molto profonda, nota come T-dualit\`a, che gioca un ruolo centrale anche nella definizione di altri oggetti estesi, le  $D$-brane e gli $O$-piani. 

Il sesto capitolo \`e dedicato all'introduzione dei modelli con supersimmetria spazio-temporale rotta sia in maniera esplicita (Tipo $0$) che spontaneamente. 

L'ultimo capitolo \`e dedicato alle ampiezze di superstringa e ai problemi relativi alla loro definizione a genere pi\`u alto del primo. Il lavoro originale di questa Tesi consiste nella generalizzazione di
risultati ottenuti recentemente da D'Hoker e Phong per le ampiezze di genere due per i modelli di superstringa chiusa ed orientata ad altri casi con supersimmetria rotta (modelli di Tipo $0$ e modelli con
``brane supersymmetry breaking'' in dieci dimensioni), e in risultati preliminari sulle loro correzioni di soglia. Questo studio rappresenta il punto di partenza per lo studio sistematico delle ridefinizioni di vuoto introdotte a due loops dalla rottura della supersimmetria. Il contenuto di questo capitolo \`e frutto di una collaborazione in corso con il Dr. Carlo Angelantonj dell'Universit\`a di Torino, il Prof. Emilian Dudas dell'Ecole Polytechnique di Parigi, e il mio relatore di Tesi.


\chapter{Stringhe Relativistiche}

\section{Particella relativistica}

Una particella relativistica puntiforme che viva in uno spazio $D$-dimensionale, che consideriamo con metrica $\eta_{\mu \nu}=\diag (-, +, +, \dots )$, descrive nel suo moto una traiettoria (\emph{una superficie d'universo} 1-dimensionale) parametrizzata dal tempo proprio $\tau$. L'azione pi\`u semplice per una particella massiva \`e data dalla lunghezza del cammino percorso nello spazio-tempo. La lunghezza infinitesima \`e
\be
d{\ell}=(-ds^2)^{1/2}=(-dX^\mu dX^\nu \eta_{\mu \nu})^{1/2}=(-dX^\mu dX_\mu)^{1/2} \ ,
\ee
dove si ha $(-ds^2) > 0$ per una particella massiva e l'azione \`e quindi
\be
\label{azionepr}
S_{pr} = -m\int d\ell = -m \int d\tau \sqrt{-\eta_{\mu \nu}\dot X^\mu \dot X^\nu} \ ,
\ee
dove il punto indica la derivata rispetto a $\tau$. Variando l'azione si trova
\be
\delta S_{pr}=m \int d\tau \left( \frac{\dot X^\mu \delta \dot X_\mu}{\sqrt{-\dot X^\mu \dot X_\mu}}\right)=
- m \int d\tau \left( \frac{\ddot X^\nu}{\sqrt{-\dot X^\mu \dot X_\mu}}\right)\delta X_\nu \ ,
\ee
e pertanto per una variazione $\delta X_\nu$ arbitraria, si ottiene l'equazione del moto
\be
\frac{d^2 X^\mu}{d\tau^2}=0 \ .
\ee

L'azione (\ref{azionepr}), oltre a contenere una radice che rende complicata la quantizzazione, non pu\`o  descrivere particelle di massa nulla. \`E possibile scrivere una nuova azione equivalente alla (\ref{azionepr}) nel caso massivo, ma in grado di descrivere anche il caso di particelle di massa nulla,
\be
\label{azionepr2}
S'_{pr}= \frac{1}{2}\int d\tau \left(e ^{-1}\dot X^\mu \dot X_\mu -em^2\right) \ ,
\ee
dove $e(\tau)$ \`e un moltiplicatore di Lagrange, un termine che non ha dinamica. Da un punto di vista fisico si pu\`o immaginare $e(\tau)$ come legato ad una possibile `metrica' $\gamma_{\tau \tau}$ definita sulla traiettoria della particella
\be
e(\tau)=\sqrt{-\gamma_{\tau \tau}(\tau)} \ , \qquad ds^2= \gamma_{\tau \tau}d\tau d\tau \ .
\ee
Si tratta naturalmente di una descrizione ridondante nel caso di una particella puntiforme, che descrive traiettorie unidimensionali, ma nel caso di stringa sar\`a particolarmente utile. Variando la (\ref{azionepr2}) rispetto ad $e$ si ottiene
\be
\delta S'_{pr} = \frac{1}{2}\int d\tau \left[-e^{-2}\dot X^\mu \dot X_\mu - m^2 \right]\delta e \ ,
\ee
e considerando variazioni arbitrarie $\delta e$ si trova l'equazione del moto per $e$
\be
\dot X^\mu \dot X_\mu + e^2m^2=0 \ ,
\ee
che costituisce un vincolo per la teoria. Risolvendo l'equazione trovata si ottiene
\be
e=\frac{1}{m}\sqrt{- \dot X^\mu \dot X_\mu} \ ,
\ee
che sostituita nell'azione (\ref{azionepr2}) da
\be
S'_{pr} = \frac{1}{2}\int d\tau \left[m\sqrt{- \dot X^\mu \dot X_\mu} 
+ \sqrt{- \dot X^\mu \dot X_\mu}m^{-1}m^2 \right] = S_{pr} \ ,
\ee
mostrando l'equivalenza a livello classico della due azioni. 

La nuova azione ha due simmetrie interessanti. La prima \`e l'invarianza sotto trasformazioni di Poincar\'e dello spazio-tempo ambiente,
\be
X^\mu \to X'^\mu=\Lambda^\mu_{\phantom{\mu}\nu}X^\nu+a^\mu \ ,
\ee
dove $\Lambda^\mu_{\phantom{\mu}\nu}$ \`e una matrice di $SO(1, D-1)$ e $a^\mu$ un vettore arbitrario $D$-dimensionale. Questa \`e una simmetria globale comune ad entrambe le azioni scritte. La seconda simmetria \`e propria solo della seconda azione, ed \`e una simmetria locale o \emph{di gauge}, definita sulla linea di universo e dovuta alla presenza di $e(\tau)$. L'azione \`e infatti invariante sotto riparametrizzazioni della forma
\be
\delta X = \zeta(\tau)\frac{dX(\tau)}{d\tau} \ , \qquad \delta e = \frac{d}{d\tau}[\zeta(\tau)e(\tau)] \ ,
\ee
per un arbitrario parametro $\zeta(\tau)$. 

\section{La Stringa Bosonica}

Una stringa \`e un oggetto esteso unidimensionale che descrive nel suo moto una superficie di universo (\emph{world-sheet}) bidimensionale che pu\`o essere parametrizzata con coordinate $(\tau, \sigma)$.  La prima pu\`o essere vista come il tempo proprio e la seconda come la coordinata spaziale che corre lungo la stringa, scegliendo $0 \leq \sigma \leq \pi$. L'evoluzione della stringa nello spazio-tempo \`e descritta dalle funzioni $X^\mu(\tau, \sigma)$, con $\mu = 0, \dots, D-1$, che descrivono l'immersione della superficie di universo nello spazio-tempo.

\subsection{Azione di Stringa Bosonica}

L'azione e tutte le quantit\`a fisiche, come nel caso di particella puntiforme, devono essere indipendenti dalla parametrizzazione della superficie d'universo. La pi\`u semplice azione che possiamo scrivere \`e proporzionale all'area del world-sheet spazzato dalla stringa. Per esprimere l'azione in termini di $X^\mu(\tau, \sigma)$, definiamo la metrica indotta $h_{ab}$, i cui indici $a, b$ corrono sui valori $(\tau, \sigma)$:
\be
h_{ab}=\partial_aX^\mu\partial_bX_\mu \ .
\ee
L'azione di stringa che si ottiene \`e l'azione di Nambu-Goto, ovvero
\be
S_{NG}=-\frac{1}{2\pi \alpha'}\int_M d\tau d\sigma \sqrt{-\det h_{ab}} \ ,
\ee
dove M indica la superficie di universo, e $\alpha'$ \`e la \emph{pendenza di Regge}. Non \`e difficile rendersi conto che l'integrale \`e l'area spazzata dalla stringa nel suo moto, mentre la costante moltiplicativa
\be
T=\frac{1}{2\pi \alpha'}
\ee
\`e dimensionalmente una forza fratto una lunghezza, e pu\`o essere interpretata come la tensione di stringa. In forma pi\`u esplicita l'azione si scrive
\be
S_{NG}=-\frac{1}{2\pi \alpha'}\int_M d\tau d\sigma \sqrt{\left( \frac{\partial X^\mu}{\partial \sigma}\frac{\partial X^\mu}{\partial \tau}\right)^2-\left(\frac{\partial X^\mu}{\partial \sigma}\right)^2\left(\frac{\partial X_\mu}{\partial \tau}\right)^2} \ .
\ee

Come nel caso di particella puntiforme, \`e possibile scrivere un'azione classicamente equivalente introducendo una metrica indipendente $\gamma_{ab}(\sigma, \tau)$, che scegliamo con segnatura Lorentziana $(-,+)$, sulla superficie d'universo $M$. La nuova azione, derivata in origine da Brink, Di Vecchia, Howe, Deser e Zumino \cite{bdhdz}, ma conosciuta come azione di Polyakov \cite{pol1}, che ne ha studiato in dettaglio la quantizzazione, \`e
\ba
\label{azionep}
S_P &=& -\frac{1}{4\pi\alpha'}\int_M d\sigma d\tau \, (-\gamma)^{1/2}\gamma^{ab}\partial_aX^\mu\partial_bX^\nu \eta_{\mu \nu} \nonumber \\
&=& -\frac{1}{4\pi \alpha'}\int_M d\tau d\sigma (-\gamma)^{1/2}\gamma^{ab}h_{ab}  \ ,
\ea
dove $\gamma=\det \gamma_{ab}$.  Per verificare l'equivalenza delle azioni a livello classico si utilizzano, come nel caso di particella puntiforme, le equazioni del moto che si ottengono variando la metrica. La variazione dell'azione \`e
\be
\delta_\gamma S_P = -\frac{1}{4\pi \alpha'}\int_M d\tau d\sigma (-\gamma)^{1/2}\left\{-\frac{1}{2}\delta \gamma \gamma^{ab}h_{ab} + \delta \gamma^{ab}h_{ab} \right\} \ ,
\ee
ed utilizzando la variazione del determinante $\delta \gamma = \gamma \gamma^{ab}\delta \gamma_{ab}= -\gamma \gamma_{ab}\delta \gamma^{ab}$, diventa
\be
\delta_\gamma S_P = -\frac{1}{4\pi \alpha'}\int_M d\tau d\sigma (-\gamma)^{1/2} \delta \gamma^{ab}\left\{h_{ab} - \frac{1}{2}\gamma_{ab} \gamma^{cd}h_{cd} \right\} \ .
\ee
da cui si trova l'equazione del moto
\be
h_{ab} = \frac{1}{2}\gamma_{ab} \gamma^{cd}h_{cd} \ .
\ee
Dividendo ciscun membro di questa equazione per la radice quadrata di meno il suo determinante si ottiene 
\be
\label{prop}
h_{ab}(-h)^{-1/2}=\gamma_{ab}(-\gamma)^{-1/2} \ ,
\ee
una relazione di proporzionalit\`a fra $\gamma_{ab}$ e la metrica indotta, e sostituendo infine nell'azione di Polyakov si ritrova l'azione di Nambu-Goto.

Le azioni di Polyakov e di Nambu-Goto hanno diverse simmetrie. In particolare, entrambe sono invarianti sotto le trasformazioni dello spazio tempo del gruppo Poincar\'e 
\ba
X'^\mu(\tau, \sigma) &=& \Lambda^\mu_{\phantom{\mu}\nu}X^\nu(\tau, \sigma)+a^\mu \ , \nonumber \\ 
\gamma'_{ab}(\tau, \sigma) &=& \gamma_{ab}(\tau, \sigma)
\ea
e sotto i diffeomorfismi, trasformazioni generali delle coordinate sul world-sheet,
\ba
X'^\mu(\tau', \sigma') &=& X^\mu(\tau, \sigma) \ , \nonumber \\
\frac{\partial \sigma'^c}{\partial \sigma^a}\frac{\partial \sigma'^d}{\partial \sigma^b}\gamma'_{cd}(\tau', \sigma')
&=& \gamma_{ab}(\tau, \sigma) \ ,
\ea
con $(\tau'(\tau, \sigma), \sigma'(\tau, \sigma))$ le nuove coordinate. L'azione di Polyakov ha per\`o un'ulteriore invarianza sotto trasformazioni bidimensionali di Weyl
\ba
X'^\mu(\tau, \sigma) &=& X^\mu(\tau, \sigma) \ , \nonumber \\
\gamma'_{ab}(\tau, \sigma) &=& e^{2\omega (\tau, \sigma)}\gamma_{ab}(\tau, \sigma) \ ,
\ea
per $\omega(\tau, \sigma)$ arbitrario. Si pu\`o comprendere questa ulteriore simmetria osservando che l'equazione del moto (\ref{prop}) che lega l'azione di Polyakov a quella di Nambu-Goto determina $\gamma_{ab}$ solo a meno di un riscalamento. Tutte le metriche collegate da trasformazioni di Weyl corrispondono quindi alla medesima metrica indotta, e quindi alla stessa descrizione nello spazio-tempo in termini di $X^\mu(\tau, \sigma)$. 

L'azione di Polyakov pu\`o essere generalizzata aggiungendo termini polinomiali nelle derivate che abbiano tutte le simmetrie dell'azione scritta. L'unico termine invariante sotto trasformazioni di Poincar\'e, diffeomorfismi e trasformazioni di Weyl \`e l'azione di Einstein-Hilbert in due dimensioni
\begin{equation}
\chi=\frac{1}{4\pi}\int_M d\tau d{\sigma}\, (-\gamma)^{1/2}R + \frac{1}{2\pi}\int_{\partial M}ds\, k \,,
\end{equation}
dove $R$ \`e lo scalare di Ricci costruito da $\gamma_{ab}$ e $k$ \`e la traccia del tensore di curvatura geodesica sul bordo della superficie d'universo. Sotto una trasformazione di Weyl  $(-\gamma)^{1/2}\to e^{2\omega}(-\gamma)^{1/2}$
e $R\to e^{-2\omega}(R-2\nabla^2 \omega)$ l'azione scritta \`e invariante, mentre un termine di costante cosmologica 
\be
\Theta = \frac{1}{4\pi \alpha'} \int_M d\tau d{\sigma}(-\gamma)^{1/2} \ ,
\ee
non \`e al contrario invariante sotto trasformazioni di Weyl. 

L'azione di Polyakov generalizzata \`e quindi
\ba
S'_P&=&S_P- \lambda \chi \nonumber \\
&=& -\int_M d\sigma d\tau \, (-\gamma)^{1/2}\left\{\frac{1}{4\pi\alpha'} \gamma^{ab}\partial_aX^\mu\partial_bX^\nu \eta_{\mu \nu}+\frac{\lambda}{4\pi}R \right\} - \frac{\lambda}{2\pi}\int_{\partial M}ds\, k \ .
\ea
Il termine di Einstein-Hilbert in due dimensioni e il corrispondente temine contenente $k$ non portano per\`o dinamica alla metrica, dal momento che risultano essere derivate totali, e il loro contributo ad $S$ dipende per questo solo dalla topologia del world-sheet.

\subsection{Tensore energia-impulso ed equazioni del moto}

La variazione dell'azione (\ref{azionep}) rispetto alla metrica definisce il tensore energia-impulso,
\ba
T^{ab}(\tau, \sigma)&=&-4\pi (-\gamma)^{-1/2}\frac{\delta}{\delta \gamma_{ab}}S_P \nonumber \\
&=&-\frac{1}{\alpha'}(\partial^aX^\mu \partial^bX_\mu - \frac{1}{2}\gamma^{ab}\partial^cX^\mu \partial_cX_\mu) \ ,
\ea
che per l'equazione del moto (\ref{prop}) \`e identicamente nullo, ovvero $T^{ab}=0$. Dall'invarianza sotto trasformazioni di Weyl si ottiene anche $T^a_a=\gamma_{ab}T^{ab}=0$.

Per ottenere le equazioni del moto dei campi occorre variare l'azione rispetto ad $X^\mu$ 
\ba
\delta S_P &=& \frac{1}{2\pi \alpha'}\int d\tau d\sigma \partial_a\left\{ \sqrt{-\gamma}\gamma^{ab}\partial_bX_\mu \right\} \delta X^\mu \nonumber \\
&-& \frac{1}{2\pi \alpha'}\int d\tau \sqrt{-\gamma}\partial_\sigma X_\mu \delta X^\mu \Big \arrowvert_{\sigma=0}^{\sigma=\pi} \ ,
\ea
che porta all'equazione
\be
\label{moto}
\partial_a\left( \sqrt{-\gamma}\gamma^{ab}\partial_bX_\mu \right)= \sqrt{-\gamma}\nabla^2X^\mu=0 \ ,
\ee
dove $\nabla$ denota la derivata covariante. L'equazione trovata deve essere accompagnata da condizioni che cancellino i termini di bordo e che siano consistenti con l'invarianza sotto trasformazioni di Poincar\`e. Si possono imporre condizioni al bordo di Neumann
\be
\label{condizioni1}
Stringhe \, Aperte:\,
\left\{
\begin{array}{c}
X'^\mu(\tau, 0)=0 \nonumber \\
X'^\mu(\tau, \pi)=0 
\end{array}
\right.
\ ,
\ee
che portano a definire stringhe aperte con estremi liberi di muoversi nello spazio-tempo, oppure condizioni di periodicit\`a 
\be
\label{condizioni2}
Stringhe \, Chiuse:\,
\left\{
\begin{array}{c}
X'^\mu(\tau, 0)=X'^\mu(\tau, \pi) \nonumber \\
X^\mu(\tau, 0)=X^\mu(\tau, \pi) \nonumber \\
\gamma_{ab}(\tau, 0)=\gamma_{ab}(\tau, \pi) 
\end{array}
\right.
\ .
\ee
che portano a definire stringhe chiuse.

Per risolvere le equazioni del moto (\ref{moto}) in maniera diretta si pu\`o semplificare il problema utilizzando le simmetrie di gauge dell'azione. La metrica $\gamma_{ab}$ \`e una matrice simmetrica $2\times 2$ ed \`e specificata da tre funzioni indipendenti. Fissando le due riparametrizzazioni delle coordinate sulla superficie di universo e l'invarianza di Weyl, si sceglie 
\be
\label{gaugeconforme}
\gamma_{ab}=\eta_{ab}e^{\phi}=
\left(
\begin{array}{cc}
-1 & 0 \\
0 & 1 
\end{array}
\right)e^{\phi} \ .
\ee
Questa scelta della gauge, in cui la metrica bidimensionale \`e piatta a meno di un fattore conforme, \`e detta \emph{gauge conforme}. Le equazioni del moto divengono quindi
\be
\left(\frac{\partial^2}{\partial \sigma^2} - \frac{\partial^2}{\partial \tau^2} \right)X^\mu (\tau, \sigma)= 0 \ ,
\ee
che si riconosce essere l'equazione delle onde in due dimensioni, la cui soluzione pi\`u generale \`e
\be
X^\mu(\tau, \sigma)=X_L^\mu(\tau + \sigma) + X_R^\mu(\tau - \sigma) \ .
\ee
Imponendo sulla soluzione generale dell'equazione del moto le condizioni al bordo (\ref{condizioni1}) si ottiene, nel caso di stringa aperta, l'espansione nei modi di vibrazione 
\be
\label{modes_exp_bos1}
X^\mu \ = \ x^\mu \ + \ 2\alpha'p^\mu\tau \ + \
i\sqrt{2\alpha'} \ \sum_{n\neq 0}\frac{\alpha^\mu_n}{n} \ e^{-in\tau}\cos{n\sigma} \ ,
\ee
mentre imponendo le condizioni di periodicit\`a (\ref{condizioni2}) si trova per una stringa chiusa
\be
\label{modes_exp_bos2}
X^\mu \ = \ x^\mu \ + \ 2\alpha'p^\mu\tau \ + \
i\frac{\sqrt{2\alpha'}}{2} \ \sum_{n\neq
  0}\left(\frac{\alpha^\mu_n}{n} \ e^{-2in(\tau-\sigma)}+
\frac{\tilde\alpha^\mu_n}{n} \ e^{-2in(\tau+\sigma)}\right) \ ,
\ee
dove, per avere una soluzione reale, si impone $\alpha^\mu_{-n}=(\alpha^\mu_{n})^\ast$ e $\tilde\alpha^\mu_{-n}=(\tilde\alpha^\mu_n)^\ast$. Si osserva che $x^\mu$ e $p^\mu$ sono rispettivamente la posizione e l'impulso del centro di massa della stringa. Si pu\`o identificare $p^\mu$ come il modo zero dell'espansione:
\ba
\label{zeromodi}
stringa \, aperta: \qquad \alpha_0^\mu &=&(2\alpha')^{1/2}p^\mu \ , \nonumber \\
stringa \, chiusa: \qquad \alpha_0^\mu &=&\left(\frac{\alpha'}{2}\right)^{1/2}p^\mu \ .
\ea
Dal punto di vista fisico l'espansione nei modi della stringa chiusa \`e quella di coppie di onde indipendenti che viaggiano lungo la stringa in direzioni opposte mentre, nel caso aperto si hanno onde stazionarie, dal momento che le condizioni al bordo impongono ai modi destri e sinistri di riflettersi gli uni negli altri. I vincoli sul tensore energia impulso sono
\ba
T_{\tau \sigma}&=&T_{\sigma \tau}=\frac{1}{\alpha'}\dot X^\mu X'_\mu=0 \ ,\nonumber \\
T_{\sigma \sigma}&=&T_{\tau \tau}=\frac{1}{2\alpha'}\left( \dot X^\mu \dot X_\mu + X'^\mu X'_\mu\right)=0 \ ,
\ea
che possono anche essere scritti come
\be
(\dot X \pm X')^2=0
\ee

Introducendo le nuove coordinate $\sigma^\pm = \tau \pm \sigma$ si ha $X^\mu(\tau, \sigma)=X_L^\mu(\sigma^+) + X_R^\mu(\sigma^-)$. La metrica diventa $ds^2=-d\tau^2 + d\sigma^2 \to -d\sigma^+ d\sigma^-$ e quindi $\eta_{-+}=\eta_{+-}=-1/2$, $\eta^{+-}=\eta^{-+}=-2$ e $\eta_{++}=\eta_{--}=\eta^{++}=\eta^{--}=0$. Le derivate vengono scomposte come $\partial_\tau=\partial_+ + \partial_-$ e $\partial_\sigma=\partial_+ - \partial_-$. I vincoli sul tensore energia impulso nelle nuove coordinate sono
\ba
\label{TT}
T_{++}&=& \frac{1}{2}(T_{\tau \tau}+T_{\tau \sigma})=\frac{1}{\alpha'}\partial_+X^{\mu} \partial_+X_\mu =0 \ ,\nonumber \\
T_{--}&=& \frac{1}{2}(T_{\tau \tau}-T_{\tau \sigma})= \frac{1}{\alpha'}\partial_-X^\mu \partial_-X_\mu  =0 \ .
\ea
Nelle nuove coordinate \`e immediato rendersi conto che la scelta della gauge conforme (\ref{gaugeconforme}) non fissa completamente la simmetria locale. Infatti, per due trasformazioni indipendenti delle coordinate sulla superficie d'universo del tipo 
\be
\sigma^+ \to \sigma'^+=f(\sigma^+) \ , \qquad \sigma^- \to \sigma'^-=g(\sigma^-) \ ,
\ee
si ha una trasformazione della metrica
\be
\gamma'_{+-}=\left(\frac{\partial f(\sigma^+)}{\partial \sigma^+}\frac{\partial g(\sigma^-)}{\partial \sigma^-}\right)^{-1}\gamma_{+-}
\ee
che pu\`o essere facilmente rissorbita con una trasformazione di Weyl della forma
\be
\gamma'_{+-}=e^{2\omega_L(\sigma^+)+2\omega_R(\sigma^-)}\gamma_{+-}
\ee
per $e^{2\omega_L(\sigma^+)}=\partial_+ f(\sigma^+)$ e $e^{2\omega_R(\sigma^-)}=\partial_- g(\sigma^-)$. La teoria di stringhe, nella gauge conforme, definisce una teoria di campo conforme bidimensionale \cite{Cft,Ginsp,bpz}.

\subsection{Dinamica Hamiltoniana}

La densit\`a lagrangiana, per la scelta fatta della metrica, \`e
\be
{\mathcal L} = -\frac{1}{4 \pi \alpha'}(\partial_\sigma X^\mu \partial_\sigma X_\mu - \partial_\tau X^\mu \partial_\tau X_\mu) \ , 
\ee
da cui si pu\`o derivare il momento coniugato di $X^\mu$
\be
\Pi^\mu=\frac{\delta{\mathcal L}}{\delta(\partial_\tau X^\mu)}=\frac{1}{2 \pi \alpha'}\dot X^\mu \ .
\ee
Si hanno classicamente le parentesi di Poisson a tempi uguali:
\ba
\{ X^\mu(\sigma), \Pi^\nu(\sigma')\}_{PB}&=&\eta^{\mu \nu}\delta(\sigma - \sigma') \ , \nonumber \\
\{ X^\mu(\sigma), X^\nu(\sigma')\}_{PB}&=&0 \ , \nonumber \\
\{ \Pi^\mu(\sigma), \Pi^\nu(\sigma')\}_{PB}&=&0 \ .
\ea
Da cui si derivano facilmente le relazioni sugli oscillatori, sull'impulso e sulla coordinata del centro di massa
\ba
\label{parentesipoisson}
\{\alpha_m^\mu, \alpha_n^\nu \}_{PB}&=&\{\tilde \alpha_m^\mu, \tilde \alpha_n^\nu \}_{PB} = im\delta_{m+n}\eta^{\mu \nu} \ , \nonumber \\
\{x^\mu, p^\nu\}_{PB}&=&\eta^{\mu \nu} \ , \nonumber \\
\{\alpha_m^\mu, \tilde \alpha_n^\nu \}_{PB} &=& 0 \ .
\ea 
La densit\`a hamiltoniana \`e
\be
{\mathcal H}=\dot X^\mu \Pi_\mu - {\mathcal L}=\frac{1}{4 \pi \alpha'}(\partial_\sigma X^\mu \partial_\sigma X_\mu + \partial_\tau X^\mu \partial_\tau X_\mu) \ ,
\ee
da cui si ricava l'Hamiltoniana integrando lungo la stringa
\ba
H&=&\int_0^\pi d\sigma {\mathcal H}(\sigma)=\frac{1}{2}\sum_{-\infty}^\infty \alpha_{-n}\cdot \alpha_n \qquad stringhe \, aperte \ , \nonumber \\
H&=&\int_0^{2\pi} d\sigma {\mathcal H}(\sigma)=\frac{1}{2}\sum_{-\infty}^\infty (\alpha_{-n}\cdot \alpha_n + \tilde \alpha_{-n}\cdot \tilde \alpha_n ) \qquad stringhe \, chiuse \ .
\ea
Dalle (\ref{TT}) possiamo definire gli operatori di Virasoro come i modi di Fourier del tensore energia-impulso. Per la stringa chiusa essi sono
\ba
L_m&=&\frac{1}{\pi \alpha'}\int_0^{2\pi}d\sigma T_{--}e^{im(\tau - \sigma)}=\frac{1}{2}\sum_{-\infty}^\infty \alpha_{m-n}\cdot \alpha_n \ , \nonumber \\
\bar L_m&=&\frac{1}{\pi \alpha'}\int_0^{2\pi}d\sigma T_{++}e^{im(\tau + \sigma)}=\frac{1}{2}\sum_{-\infty}^\infty \tilde \alpha_{m-n}\cdot \tilde \alpha_n \ ,
\ea
e soddisfano le condizioni di realit\`a
\be
L_m^\ast=L_{-m} \qquad e \qquad \bar L_m^\ast=\bar L_{-m}
\ee
Nel caso di stringa aperta non c'\`e differenza fra oscillatori destri e sinistri, e
\ba
L_m&=&\frac{1}{\pi \alpha'}\int_0^{\pi}d\sigma \left\{T_{--}e^{im(\tau - \sigma)}+ T_{++}e^{im(\tau + \sigma)}\right\} \nonumber \\
&=&\frac{1}{2}\sum_{-\infty}^\infty \alpha_{m-n}\cdot \alpha_n \ .
\ea
L'Hamiltoniana si pu\`o riscrivere in termini di operatori di Virasoro:
\ba
H&=&L_0 \qquad stringhe \, aperte \ , \nonumber \\
H&=&L_0 + \bar L_0 \qquad stringhe \, chiuse \ .
\ea
I vincoli (\ref{TT}), in questo formalismo, equivalgono a richiedere per tutti i modi $L_m=0$, $\bar L_m=0$, per ogni $m$. Usando le parentesi di Poisson per gli oscillatori (\ref{parentesipoisson}) si trova l'algebra di Virasoro:
\ba
\{L_m, L_n\}_{PB}&=&-i(m-n)L_{m+n} \ , \nonumber \\
\{\bar L_m, \bar L_n\}_{PB}&=&-i(m-n)\bar L_{m+n} \ , \nonumber \\
\{L_m, \bar L_n\}_{PB}&=&0 \ .
\ea

\section{Quantizzazione della Stringa Bosonica}

Ci sono diversi approcci alla quantizzazione delle stringhe. Come in teoria dei campi \`e possibile quantizzare la teoria classica sia in maniera canonica che nel formalismo dell'integrale sui cammini. La quantizzazione canonica consiste nel sostituire le variabili classiche con operatori quantistici e nel sostituire le parentesi di Poisson con relazioni di commutazione fra operatori
\be
\label{quant}
\{ \, \ , \,  \}_{PB} \to -i[\, \ , \,\, ] \ .
\ee
Come si \`e visto si sono introdotti dei vincoli nello studio della stringa classica. Nella teoria quantistica, l'introduzuone di questi vincoli \`e possibile con due approcci differenti. Il primo, conosciuto come \emph{quantizzazione canonica covariante},  consiste nel quantizzare le varibili classiche senza considerare i vincoli per poi imporli sullo spazio di Hilbert degli stati: in questo modo si preserva l'invarianza di Lorentz esplicita della teoria. Il secondo approccio, la \emph{quantizzazione nel cono di luce}, consiste nel risolvere esplicitamente i vincoli al livello della teoria classica, nella \emph{gauge del cono di luce}, per poi quantizzare. In questo caso l'invarianza di Lorentz esplicita viene persa.

Nel formalismo funzionale dell'integrale sui cammini il metodo di Faddeev-Popov viene associato a tecniche BRST e si trova uno spazio degli stati manifestamente Lorentz invariante che per\`o contiene anche dei campi aggiuntivi non fisici chiamati campi di \emph{ghost}. Il formalismo dell'integrale funzionale, che in teoria delle stringhe \`e conosciuto come integrale di Polyakov, sar\`a studiato in dettaglio nel prossimo capitolo.

\subsection{Quantizzazione canonica covariante}

La quantizzazione (\ref{quant}) conduce alle relazioni di commutazione
\ba
\label{comm}
\left[ X^\mu(\sigma), \Pi^\nu(\sigma') \right] &=& i\eta^{\mu \nu}\delta(\sigma - \sigma') \ , \nonumber \\
\left[ \alpha_m^\mu, \alpha_n^\nu \right] &=& \left[ \tilde \alpha_m^\mu, \tilde \alpha_n^\nu \right] = m\delta_{m+n}\eta^{\mu \nu} \ , \nonumber \\
\left[x^\nu ,  p^\mu \right] &=& i\eta^{\mu \nu} \ , \qquad \left[ \alpha_m^\mu, \tilde \alpha_n^\nu \right] = 0 \ ,
\ea 
mentre le condizioni di realt\`a divengono condizioni di hermiticit\`a degli operatori. Definendo nuovi operatori $\sqrt m \alpha_{\pm m}^\mu$ si trovano le relazioni di commutazione di $D$ coppie di operatori di creazione e di distruzione di un oscillatore quantistico.  Gli operatori di Virasoro $L_m$ sono stati costruiti dagli operatori di creazione e distruzione, come in Teoria dei Campi, e la quantizzazione richiede che siano \emph{normalmente ordinati} 
\be
L_m=\frac{1}{2}\sum_{-\infty}^\infty :\alpha_{m-n}\cdot \alpha_n: + a\delta_{m, 0}\ ,
\ee
cio\'e che tutti gli operatori di distruzione siano sulla destra. 

In Teoria dei Campi, la prescrizione dell'ordinamento normale \`e necessaria per ottenere operatori quantistici correttamente definiti, ovvero che abbiano autovalori finiti sugli stati fisici. In generale un operatore in una teoria di campo quantistica \`e un prodotto a punti uguali di campi fondamentali. Non \`e difficile accorgersi che un prodotto di campi a punti uguali \`e formalmente divergente dal momento che si hanno delle somme infinite di prodotti di operatori di creazione di distruzione. La prescrizione di spostare gli operatori di creazione a sinistra e quelli di distruzione a destra elimina le divergenze.

Nel nostro caso, l'ordinamento normale non comporta problemi, date le relazioni di commutazione (\ref{comm}), eccetto che per $L_0$, dal momento che $\alpha_n^\mu$ e $\alpha_{-m}^\mu$ non commutano. Si trova
\be
\label{Lzero}
L_0=\frac{1}{2}\alpha_0^2+\sum_{n=1}^\infty \alpha_{-n}\cdot \alpha_{n}+D\sum_{n=1}^\infty n \ ,
\ee
il secondo termine \`e una costante divergente che, come si vedr\`a in dettaglio nel prossimo paragrafo, pu\`o essere regolata lasciando un termine finito che corrisponde all'energia di punto zero degli oscillatori, che indichiamo con $a$. Calcolando l'algebra di Virasoro per la teoria quantistica, facendo attenzione all'ordinamento normale, si trova che l'algebra classica \`e modificata dalla presenza di un termine centrale:
\ba
\left[L_m, L_n\right]&=&(m-n)L_{m+n}+ \frac{D}{12}m(m^2-1)\delta_{m+n} \ , \nonumber  \\
\left[\bar L_m, \bar L_n\right]&=&(m-n) \bar L_{m+n}+ \frac{D}{12}m(m^2-1)\delta_{m+n} \ , \nonumber \\
\left[L_m, \bar L_n\right]&=&0 \ .
\ea

Lo spazio di Hilbert degli stati pu\`o essere costruito a partire dallo stato di vuoto $|0; k \rangle$, che definiamo come lo stato annichilato da tutti gli operatori di distruzione, e dove $k$ indica l'impulso del centro di massa della stringa. Uno stato generico $|\phi \rangle$ sar\`a costruito agendo con gli operatori di creazione sul vuoto. Non \`e difficile rendersi conto che lo spazio degli stati contiene anche stati non fisici, a norma negativa, dal momento che per $\mu=0$ si hanno relazioni di commutazione (\ref{comm}) con segno opposto alle altre. Ad esempio
\be
|\alpha^0_{-1}|0; k \rangle |^2= \langle k; 0|\alpha^0_{1}\alpha^0_{-1}|0; k \rangle = -1 \ . 
\ee
Si devono pertanto imporre dei vincoli sullo spazio degli stati:
\ba
\label{qvincoli}
(L_0-a)|\phi \rangle &=& 0 \ , \qquad L_m |\phi \rangle \qquad {\it per} \, m>0 \nonumber \\
(\bar L_0-a)|\phi \rangle &=& 0 \ , \qquad \bar L_m |\phi \rangle \qquad {\it per} \, m>0 \ .
\ea
Si noti che i vincoli sono imposti in maniera ``debole'' (in  modo tale che il valore di aspettazione sugli stati sia zero), ovvero solo per $m>0$, dal momento che, a causa dell'estensione centrale dell'algebra di Virasoro, si avrebbe altrimenti un'inconsistenza:
\be
0=\langle \phi|[L_m, L_{-m}]|\phi \rangle=2m \langle \phi|L_0|\phi \rangle + \frac{D}{12}m(m^2-1)\langle \phi|\phi \rangle \ne 0 \ .
\ee
La massa degli stati viene studiata definendo un operatore di massa, $M^2=-p^\mu p_\mu$. Ricordano le definizioni (\ref{zeromodi}), dalla (\ref{Lzero}) si ha, nel caso di stringa chiusa,
\ba
\label{mass1}
L_0&=&\frac{\alpha'}{4}p^\mu p_\mu+\sum_{n=1}^\infty \alpha_{-n}\cdot \alpha_{n}= \nonumber \\
&=&-\frac{\alpha'}{4}M^2+\sum_{n=1}^\infty \alpha_{-n}\cdot \alpha_{n} \ , \nonumber \\
\bar L_0&=&\frac{\alpha'}{4}p^\mu p_\mu+\sum_{n=1}^\infty \tilde \alpha_{-n}\cdot \tilde \alpha_{n}= \nonumber \\
&=&-\frac{\alpha'}{4} M^2+\sum_{n=1}^\infty \tilde \alpha_{-n}\cdot \tilde \alpha_{n} \ , 
\ea
da cui utilizzando i vincoli (\ref{qvincoli}) si trovano le formule di massa
\ba
M^2&=&\frac{4}{\alpha'}\left(\sum_{n=1}^\infty \alpha_{-n}\cdot \alpha_{n} -a \right) \ , \nonumber \\
M^2&=&\frac{4}{\alpha'}\left(\sum_{n=1}^\infty \tilde \alpha_{-n}\cdot \tilde \alpha_{n} -a \right) \ .
\ea
Possiamo riscrivere in forma simmetrica la formula di massa come semisomma e semidifferenza delle due formule trovate. Si ottiene
\ba
M^2=\frac{2}{\alpha'}\left( \sum_{n=1}^\infty \alpha_{-n}\cdot \alpha_{n}+\sum_{n=1}^\infty \tilde \alpha_{-n}\cdot \tilde \alpha_{n} -2a \right) \ , \nonumber \\
\sum_{n=1}^\infty \alpha_{-n}\cdot \alpha_{n}=\sum_{n=1}^\infty \tilde \alpha_{-n}\cdot \tilde \alpha_{n} \ ,
\ea
dove la seconda condizione \`e un vincolo sullo spettro di massa (level matching condition). Per la stringa aperta si trova pi\`u semplicemente
\be
M^2=\frac{1}{\alpha'}\left(\sum_{n=1}^\infty \alpha_{-n}\cdot \alpha_{n} -a \right) \ .
\ee
Definiamo due operatori numero che contino il numero pesato degli oscillatori sinistri e destri: $N=\sum \alpha_{-n}\cdot \alpha_{n}= \sum n N_n$ e $\bar N=\sum \tilde \alpha_{-n}\cdot \tilde \alpha_{n}\sum n \bar N_n$, nei termini dei quali le formule di massa acquistano una forma particolarmente compatta:
\ba
\label{massachiusa}
M^2=\frac{2}{\alpha'}\left(N+\bar N -2a \right), \qquad N=\bar N \ ,\qquad stringhe \, chiuse \\
\label{massaperta}
M^2=\frac{1}{\alpha'}\left(N -a \right) \ .\qquad stringhe \, aperte 
\ea

Gli stati fisici della teoria sono gli stati costruiti con gli operatori di creazione che rispettino le condizioni (\ref{qvincoli}). Oltre a questi stati si trovano degli stati 
\be
|{\rm spur} \, \rangle = L_{-n}| \, \rangle \ , 
\ee
detti \emph{spuri}, ortogonali a tutti gli stati fisici. Esistono anche stati che sono sia fisici che spuri, ma che si possono eliminare dallo spazio di Hilbert fisico dal momento che si pu\`o vedere che corrispondono a stati nulli.  Lo studio dettagliato dello spettro culmina in un celebre teorema (\emph{no ghost theorem}) che stabilisce che per $D=26$ lo spettro fisico ottenuto dalle (\ref{qvincoli}) contiene solo stati a norma positiva. 

\`E utile, a questo punto, proseguire lo studio dello spettro della stringa bosonica introducendo la quantizzazione nel cono di luce.

\subsection{Quantizzazione nel cono di luce}

Le coordinate del cono di luce sono definite come
\be
X^{\pm}=\frac{X^0\pm X^{1}}{\sqrt 2} \ ,
\ee
mentre le coordinate $X^i$ per $i \ne 0, 1$ rimangono invariate. Nelle nuove coordinate la metrica si vede facilmente essere
\ba
a^\mu b_\mu = -a^+b^--a^-b^++a^ib^i \ , \nonumber \\
a_-=-a^+ \ , \qquad a_+=-a^- \ , \qquad a_i=a^i \ .
\ea
La simmetria locale residua nella gauge conforme permette di fissare
\be
X^+(\sigma, \tau)=x^+ + 2\alpha' p^+\tau \ ,
\ee
che \`e detta gauge del cono di luce. I vincoli (\ref{TT}) possono essere risolti esprimendo le coordinate $X^-$ in funzione di quelle trasverse, scrivendo
\ba
\partial_+X^{\mu} \partial_+X_\mu = -2\partial_+X^{+} \partial_+X_- + (\partial_+X^i)^2 = 0 \ ,\nonumber \\
\partial_-X^{\mu} \partial_-X_\mu = -2\partial_-X^{+} \partial_-X_- + (\partial_-X^i)^2 = 0 \ .
\ea
Dal momento che
\be
\partial_{\pm}X^+=\alpha' p^+ \ ,
\ee
sostituendo si trova
\ba
2\alpha' p^+ \partial_+X_-=(\partial_+X^i)^2 \ , \nonumber\\
2\alpha' p^+ \partial_-X_-=(\partial_+X^i)^2 \ .
\ea
Sostituendo l'espansione nei modi delle coordinate $X^\mu$ si trova per gli oscillatori (stringa chiusa)
\ba
\label{oscillatorimeno}
\alpha_m^-=\frac{1}{\sqrt{2 \alpha'}p^+}\sum_{n\in \mathbb Z} \alpha_{m-n}^i\alpha_n^i  \ ,
\ea
e un'espressione analoga per $\tilde \alpha^-$, mentre nel caso di stringa aperta cambiano solo le normalizzazioni.

Dall'espressione trovata si giunge facilmente alla formula di massa. Vediamolo in dettaglio per il caso di stringa chiusa, dal momento che il caso di stringa aperta cambia solo per le definizioni degli oscillatori. Ricordando le espressioni degli zero modi (\ref{zeromodi}), dall'espressione (\ref{oscillatorimeno}) si ottiene
\be
\alpha' p^+p^-=\sum_{n\in \mathbb Z} \alpha_{-n}^i\alpha_n^i=\frac{\alpha'}{2}(p^i)^2+\sum_{n\in \mathbb Z-\{0\}} \alpha_{-n}^i\alpha_n^i \ ,
\ee
insieme all'analoga espressione in termini di $\tilde \alpha_n^i$. Dal momento che $M^2=p^\mu p_\mu$, si trovano l'espressione classica della formula di massa e la level matching condition in funzione dei soli coefficienti degli oscillatori trasversi
\ba
M^2=\frac{1}{\alpha'}\left(\sum_{n\in \mathbb Z-\{0\}} \alpha_{-n}^i\alpha_n^i + \sum_{n\in \mathbb Z-\{0\}} \tilde \alpha_{-n}^i \tilde \alpha_n^i \right) \ , \nonumber \\
\sum_{n\in \mathbb Z-\{0\}} \alpha_{-n}^i\alpha_n^i = \sum_{n\in \mathbb Z-\{0\}} \tilde \alpha_{-n}^i \tilde \alpha_n^i \ .
\ea

Per quantizzare occorre imporre le relazioni di commutazione canoniche (\ref{comm}), e l'ordinamento normale che porta, come si \`e visto nel paragrafo precedente, ad un termine costante infinito nella formula di massa
\be
\frac{D-2}{2}\sum_{n}^\infty n \ ,
\ee
dove il fattore $(D-2)$ viene dalla somma sulle direzioni trasverse. L'energia di punto zero pu\`o essere regolata cancellando con un contro-termine la parte divergente. Inseriamo nella sommatoria un regolatore (\emph{cutoff}) esponenziale in modo da valutarne il termine finito come
\ba
\frac{D-2}{2}\sum_{n}^\infty n e^{-\epsilon n}&=&-\frac{D-2}{2}\frac{d}{d\epsilon}\sum_{n}^\infty e^{-\epsilon n}
=-\frac{D-2}{2}\frac{d}{d\epsilon}\left(\frac{1}{1-e^{-\epsilon}}\right) \nonumber \\
&=&\frac{D-2}{2}\left(\frac{1}{\epsilon^2}-\frac{1}{12}+O(\epsilon)\right) \ .
\ea
Il primo termine pu\`o essere cancellato con un controtermine del tipo $\int d^2\sigma (-\gamma)^{1/2}$, lasciando nel limite $\epsilon \to 0$, la costante
\be
a=\frac{D-2}{24}  \ .
\ee
Introducendo gli operatori numero trasversi $N^\bot$, $\bar N^\bot$ in cui le somme coninvolgono solo operatori relativi alle dimensioni trasverse, si ha 
\ba
\label{massachiusa2}
M^2=\frac{2}{\alpha'}\left(N^\bot+\bar N^\bot -2a \right), \qquad N^\bot=\bar N^\bot \ ,\qquad stringhe \, chiuse \\
\label{massaperta2}
M^2=\frac{1}{\alpha'}\left(N^\bot -a \right) \ . \qquad stringhe \, aperte 
\ea

A questo punto possiamo studiare lo spettro di massa. Iniziamo dal caso di stringa aperta, un cui stato generico pu\`o essere costruito agendo sul vuoto $|0; k\rangle$ con gli operatori di creazione,
\be
|N; k\rangle = \left[\prod_{i=2}^{D-1}\prod_{n=1}^{\infty}\frac{(\alpha^i_{-n})^{N_{in}}}{(n^{N_{in}}N_{in}!)^{1/2}} \right]|0; k\rangle \ ,
\ee
dove $k$ \`e il momento del centro di massa e $N_{in}$ sono i numeri di occupazione di ciascun modo. Ogni scelta dei numeri d'occupazione rappresenta, dal punto di vista dello spazio-tempo, una diversa particella o stato di spin. Lo stato pi\`u basso in massa \`e
\be
|0; k\rangle \ , \qquad M^2=\frac{2-D}{24\alpha'} \ ,
\ee
la cui massa quadrata \`e negativa per $D>2$: si tratta quindi di un tachione. Poich\'e in Teoria dei Campi l'energia potenziale di un campo scalare libero \`e $\frac{1}{2}m^2\phi$, un valore di massa quadrata negativo indica che il vuoto \`e uno stato instabile, come ad esempio il vuoto simmetrico nel caso di una teoria con rottura spontanea di simmetria. La presenza di un tachione in teoria delle stringhe \`e quindi un'indicazione dell'instabilit\`a del vuoto. Si tratta di un problema di definizione della stringa bosonica non del tutto risolto allo stato attuale.

I primi stati eccitati si ottengono eccitando uno solo dei modi $n=1$ 
\be
\alpha^i_{-1}|0; k\rangle  \ , \qquad M^2=\frac{26-D}{24\alpha'} \ .
\ee
Per uno stato massivo ci si pu\`o sempre porre in un sistema di riferimento in cui la particella sia a riposto $p^\mu=(m, 0, \dots, 0)$. Gli stati interni formano una rappresentazione del gruppo delle rotazioni spaziali $SO(D-1)$. Per una particella di massa nulla non esiste un sistema di riferimento in cui sia a riposo, ma si pu\`o scegliere un sistema di riferimento in cui $p^\mu=(E, E, \dots, 0)$, e gli stati corrispondenti formano rappresentazioni del gruppo $SO(D-2)$. L'invarianza di Lorentz richiede quindi che si abbiano, in $D$ dimensioni, $D-1$ stati di spin per una particella vettoriale massiva e $D-2$ per una particella vettoriale a massa nulla. Dal momento che il primo stato eccitato ha solo $D-2$ stati di spin, deve essere un vettore a massa nulla $A_\mu$. Si trova pertanto che l'invarianza di Lorentz fissa la dimensione dello spazio-tempo della teoria di stringa bosonica a $D=26$. 

Nel caso della stringa chiusa, a partire dallo stato pi\`u basso in massa $|0,0; k\rangle$, si pu\`o costruire, con gli operatori di creazione dei modi destri e sinistri, uno stato generico 
\be
|N,\bar N; k\rangle = \left[\prod_{i=2}^{D-1}\prod_{n=1}^{\infty}\frac{(\alpha^i_{-n})^{N_{in}}(\tilde \alpha^i_{-n})^{\bar N_{in}}}{(n^{N_{in}}N_{in}!n^{\bar N_{in}}\bar N_{in}!)^{1/2}} \right]|0,0; k\rangle \ ,
\ee
dove i numeri di occupazione devono essere tali da rispettare il vincolo $N=\bar N$. Lo stato pi\`u basso in massa \`e ancora un tachione
\be
|0,0; k\rangle \ , \qquad M^2=\frac{2-D}{6\alpha'} \ .
\ee
I primi stati eccitati sono
\be
\alpha^i_{-1}\tilde \alpha^j_{-1}|0,0; k\rangle  \ , \qquad M^2=\frac{26-D}{6\alpha'} \ ,
\ee
come nel caso aperto questi stati non completano una rappresentazione di $SO(D-1)$ e devono essere a massa nulla, fissando ancora $D=26$, e quindi $a=1$. Lo stato risultante a massa nulla \`e una rappresentazione tensoriale di $SO(D-2)$, pu\`o essere decomposto in un tensore simmetrico a traccia nulla, in un tensore antisimmetrico, e uno scalare. Infatti, in generale, un tensore pu\`o essere scritto come
\be
T^{ij}=\frac{1}{2}\left(T^{ij}+T^{ji}-\frac{2}{D-2}\delta^{ij}T^{kk}\right)+\frac{1}{2}\left(T^{ij}-T^{ji}\right)+
\frac{1}{D-2}\delta^{ij}T^{kk} \ ,
\ee
in cui i tre termini non si mischiano sotto rotazione. I tre stati possono essere interpretati come un gravitone $G_{\mu \nu}$, un tensore antisimmetrico $B_{\mu \nu}$ e un dilatone $\phi$. 

\subsection{Azione di Stringa in un background curvo}

Per giustificare l'identificazione degli stati a massa nulla, occorre fare una piccola digressione discutendo una generalizzazione dell'azione di Polyakov che descrive il moto di una stringa chiusa in un background coerente dei propri stati a massa nulla,
\be
\label{sigmamodel}
S_\sigma = \frac{1}{4\pi \alpha'}\int_M d^2\sigma g^{1/2}\left[\left(g^{ab}G_{\mu \nu}(X)+i\epsilon^{ab}B_{\mu \nu}(X)\right)\partial_a X^\mu \partial_b X^\nu+ \alpha'R\Phi(X)\right] \ .
\ee
L'azione (\ref{sigmamodel}) \`e essenzialmente una approssimazione di bassa energia della descrizione di una stringa che si propaghi in un background dovuto alla presenza di un campo di stringa, di cui si manifestano solo gli stati a massa nulla. Questo tipo di azione \`e ben nota anche per alcuni modelli di Teoria dei Campi ed \`e conosciuta come \emph{modello sigma non lineare}.

La presenza della metrica $G_{\mu \nu}(X)$ in sostituzione della metrica piatta $\eta_{\mu \nu}$ pu\`o essere interpretata come l'introduzione di un background curvo che correttamente risulta essere uno ``stato coerente'' di gravitoni. Infatti come sar\`a pi\`u chiaro nel formalismo funzionale dell'integrale di Polyakov, la presenza di $G_{\mu \nu}(X)$ nell'azione equivale ad inserire un operatore di vertice del gravitone. A questo punto dovrebbe risultare abbastanza naturale, nel linguaggio di stringa, l'inclusione nell'azione (\ref{sigmamodel}) anche degli altri campi a massa nulla nel background, $B_{\mu \nu}(X)$ e $\Phi(X)$. 

Nella nuova teoria gli accoppiamenti sono quindi tutti determinati dagli stati a massa nulla della teoria stessa. In particolare, come si avr\`a modo di vedere in dettaglio nel prossimo capitolo, la costante di accoppiamento di stringa \`e essenzialmente data dall'esponenziale del termine di Eulero, che nella (\ref{sigmamodel}) \`e moltiplicato per il campo del dilatone $\Phi(X)$. Si trova cos\`i un risultato di grande interesse, il fatto che la costante d'accoppiamento di stringa \`e determinata dalla parte costante del campo del dilatone (valore di aspettazione),
\be
g_s = e^{<\Phi>} \ .
\ee

Perch\`e l'azione (\ref{sigmamodel}) definisca una teoria quantistica di stringa consistente occorre imporre la cancellazione di eventuali anomalie per l'invarianza di Weyl che equivale a richiedere che il tensore bidimensionale energia-impulso sia a traccia nulla. Questo porta ad imporre l'annullamento di tre funzioni $\beta$, legate ai tre campi a massa nulla di background, che risultano essere
\begin{eqnarray}
\beta_{\mu\nu}^G
&=&\alpha^\prime\left(R_{\mu\nu}+2\nabla_\mu\nabla_\nu\Phi
-{1\over4}H_{\mu\kappa\sigma}H_\nu^{\phantom{\nu}\kappa\sigma}\right)
+O(\alpha^{\prime2}) \ ,\nonumber\\ 
\beta_{\mu\nu}^B
&=&\alpha^\prime\left(-{1\over2}\nabla^\kappa
H_{\kappa\mu\nu}+\nabla^\kappa \Phi H_{\kappa\mu\nu}
\right)+O(\alpha^{\prime2})\ ,\nonumber\\
\beta^\Phi
&=&\alpha^\prime\left({D-26\over6\alpha^\prime}
-{1\over2}\nabla^2\Phi+\nabla_\kappa\Phi\nabla^\kappa \Phi 
-{1\over24}H_{\kappa\mu\nu}H^{\kappa\mu\nu}
\right)+O(\alpha^{\prime2})\ ,
\label{eqnmtn}
\end{eqnarray}
con $H_{\mu\nu\kappa}\equiv\partial_{\mu} B_{\nu\kappa}+\partial_{\nu} B_{\kappa\mu}+\partial_{\kappa} B_{\mu\nu}$.
Queste funzioni sono essenzialmente le funzioni beta del gruppo di rinormalizzazione degli accoppiamenti. Le equazioni
\be
\label{backmoto}
\beta^G_{\mu \nu}=\beta^B_{\mu \nu}=\beta^\Phi=0 \ .
\ee
hanno la struttura di equazioni del moto per i campi di background, e possono essere infatti derivate da un'azione di bassa energia nello spazio-tempo,
\begin{eqnarray}
{\rm S}&=&{1\over2\kappa^2_0}\int
d^DX(-G)^{1/2}e^{-2\Phi}\left[R+4\nabla_\mu
\Phi\nabla^\mu\Phi-{1\over12}H_{\mu\nu\lambda}
H^{\mu\nu\lambda}\right.\nonumber\\
&&\hskip5cm
\left.-{2(D-26)\over3\alpha^\prime}+O(\alpha^\prime)\right]\ .
\label{stringfrm}
\end{eqnarray}

In particolare, al primo ordine perturbativo, l'equazione $\beta^G_{\mu \nu}=0$ \`e l'equazione di Einstein con termini di sorgente dovuti al tensore antisimmetrico e al dilatone, mentre l'equazione $\beta^B_{\mu \nu}=0$ \`e la generalizzazione delle equazioni di Maxwell per una sorgente unidimensionale. Agli ordini successivi compaiono delle correzioni di stringa alle equazioni di bassa energia. 

\`E abbastanza sorprendente che l'equazione di Einstein compaia come condizione dell'invarianza di Weyl della teoria bidimensionale. In questo modo la gravit\`a emerge naturalmente dalla richiesta di consistenza della teoria, giustificando l'identificazione di $G_{\mu \nu}$ con il gravitone.

\section{La Superstringa}

La teoria di stringa bosonica, discussa fin qui, ha il pregio di mettere in luce, in un contesto relativamente semplice,  alcuni dei punti di forza di una reinterpretazione delle particelle fondamentali in termini di oscillazioni di oggetti unidimensionali, ma d'altra parte non descrive fermioni e presenta nello spettro il tachione come indice dell'instabilit\`a del vuoto. Si \`e quindi portati a cercare generalizzazioni dell'azione bosonica a partire dall'inclusione di gradi di libert\`a fermionici. 

Un'idea particolarmente feconda \`e quella di introdurre una supersimmetria sul world-sheet che leghi le coordinate spazio-temporali $X^\mu(\tau, \sigma)$, che si sono viste essere campi bosonici sul world-sheet, ad un partner fermionico $\psi^\mu_\alpha(\tau, \sigma)$. L'indice $\mu$ indica che, dal punto di vista dello spazio-tempo, la coordinata fermionica trasforma come un vettore, le cui componenti sono spinori sul world-sheet. La teoria ottenuta in questo modo \`e detta teoria di superstringa, e l'azione corrispondente \`e
\ba
\label{susylocale}
S & = & -\frac{1}{4\pi \alpha'}  \int d\sigma d\tau \ \sqrt{-\gamma} 
\bigg[\gamma^{ab} \dd_a X^{\mu} \dd_b X^{\nu} -i  \bar\psi^\mu
  \Gamma^a \nabla_a \psi^\nu\nonumber\\
  &&- i \bar\chi_a \Gamma^b \Gamma^a \psi^\mu
  \left( \dd_b X^\nu -\frac{i}{4}\bar\chi_b \psi^\nu \right)
  \bigg]\eta_{\mu\nu} \ ,
\ea
dove $\chi_a$ \`e un gravitino di Majorana e, come $\sqrt{-\gamma} \ \gamma^{ab}$, \`e un moltiplicatore di Lagrange senza dinamica. Infatti mentre, come si \`e gia detto, il termine dinamico della metrica bidimensionale di Einstein-Hilbert \`e un termine topologico, il termine cinetico del gravitino dovrebbe essere l'azione di Rarita-Schiwinger che, essendo proporzionale ad un tensore a tre indici $\gamma_{abc}$, in due dimensioni \`e identicamente nulla. Per accoppiare i campi spinoriali alla gravit\`a bidimensionale si \`e introdotto il \emph{formalismo del vielbein}, ovvero una base nello spazio tangente della variet\`a definita dalla metrica $\gamma_{ab}$,
\ba
\gamma_{ab}=\eta_{mn}e^m_ae^n_b \ , \nonumber \\
\eta^{mn}=\gamma^{ab}e^m_ae^n_b \ .
\ea
Le matrici $\Gamma^a$ sono definite come $\Gamma^a \equiv e^a_m\Gamma^m$, dove le $\Gamma^m$ sono matrici di Dirac ordinarie, mentre la derivata covariante per i campi spinoriali, $\nabla_a$, \`e definita come $\nabla_a \equiv \dd_a-\frac{i}{4}\omega^{\phantom{a}mn}_a\sigma_{mn}$.

Oltre a possedere le simmetrie note l'azione scritta \`e invariante sotto trasformazioni locali di supersimmetria
\ba
\delta \gamma_{ab}&=&2i\Gamma_a \chi_b \ , \nonumber \\
\delta \chi_a &=& 2 \nabla_a \epsilon \ , \nonumber \\
\delta \psi^\mu &=& \Gamma^a(\dd_aX^\mu -\frac{i}{2}\chi_a \psi^\mu)\epsilon \ , \nonumber \\
\delta X^\mu &=&i \epsilon \psi^\mu \ ,
\ea
dove il parametro $\epsilon$ \`e uno spinore di Majorana. L'azione pu\`o essere semplificata per un'opportuna scelta della gauge. Scegliendo l'equivalente della gauge conforme del caso bosonico, detta gauge superconforme, 
\be
\gamma_{ab}=\eta_{ab}e^\phi \ , \qquad \chi_a=\Gamma_a\zeta \ ,
\ee
ed utilizzando l'identit\`a delle matrici gamma bidimensionali $\Gamma_a \Gamma^b \Gamma^a = 0$, l'azione si riduce a 
\be
\label{susyglobale}
S = -\frac{1}{4\pi \alpha'}  \int d\sigma d\tau \left( \eta^{ab} \dd_a X^{\mu} \dd_b X^{\nu}
-i  \bar\psi^\mu \Gamma^a \dd_a \psi^\nu \right)\eta_{\mu \nu} \ ,
\ee
l'azione di D campi scalari e D campi fermionici liberi. Procedendo come nel caso della stringa bosonica si trovano due correnti conservate, il tensore energia-impulso, ottenuto dalla variazione della metrica sulla superficie d'universo 
\ba
\label{constraint1}
T_{ab}  &=& \ -\frac{1}{\alpha'}\left( \dd_a X^\mu \dd_b X_{\mu} \
+ \ \frac{i}{4} \ \bar\psi^\mu \ (\Gamma_a\dd_b  \ + \
\Gamma_b\dd_a) \ \psi_\mu \right) \nonumber\\
&& + \ \frac{1}{2\alpha'} \ \eta_{ab} \ \left(\dd^c X^\mu
\dd_c X_\mu \ + \ \frac{1}{2} \ \bar\psi^\mu\Gamma\cdot\dd \
\psi_\mu \right)  \ = \ 0 \ ,
\ea
e la supercorrente, ottenuta variando il gravitino,
\be
\label{constraint2}
J^a \ = \ \frac{1}{2\alpha'} \ \Gamma^b\Gamma^a \ \psi^\mu \
\dd_b X_\mu \ = \ 0 \ .
\ee
Come nel caso bosonico, le due equazioni sono vincoli della teoria, che in questo caso prendendo il nome di vincoli di super-Virasoro. 

Variando l'azione rispetto ai campi $X^\mu$ e $\psi^\mu$ e imponendo l'annullamento dei termini di bordo
\be
\delta {\mathcal L}_{bordo} \sim \left[X'_\mu \delta X^\mu + i(\psi_+\delta \psi_+ - \psi_-\delta \psi_-) \right]_0^\pi = 0 \ ,
\ee
si trovano le equazioni di Klein-Gordon e di Dirac:
\ba
\label{motosuper}
\left(\frac{\partial^2}{\partial \sigma^2} - \frac{\partial^2}{\partial \tau^2} \right)X^\mu (\tau, \sigma) &=& 0 \ , \nonumber \\
i\Gamma^a\dd_a \psi^\mu &=& 0 \ .
\ea
Le matrici $\Gamma^a$ in due dimensioni possono essere scelte puramente immaginarie 
\be
\Gamma^0=\sigma_2=
\left(
\begin{array}{cc}
0 & -i \\
i & 0 
\end{array}
\right) \ , \qquad 
\Gamma^1=i\sigma_1=
\left(
\begin{array}{cc}
0 & i \\
i & 0 
\end{array}
\right) \ ,
\ee
e l'operatore di chiralit\`a  \`e
\be
\Gamma^3=\Gamma^0\Gamma^1=\sigma_3=
\left(
\begin{array}{cc}
1 & 0 \\
0 & -1 
\end{array}
\right) \ .
\ee
L'operatore di Dirac in questa base \`e reale, 
\be
i\Gamma^a\dd_a \psi^\mu = 
\left(
\begin{array}{cc}
0 & (\dd_\tau-\dd_\sigma) \\
-(\dd_\tau+\dd_\sigma) & 0 
\end{array}
\right)
\left(
{\psi_+^\mu \atop \psi_-^\mu}
\right)
\ee
e quindi le componenti di $\psi_a^{\mu}$ sul world-sheet possono essere scelte reali. Dalle equazioni del moto si vede che i campi $\psi_\pm^{\mu}$ sono funzioni delle coordinate $\sigma^\pm= \tau \pm \sigma$. Infine, la matrice di chiralit\`a mostra come  $\psi_\pm^{\mu}$ siano due spinori di Majorana-Weyl.
Come nel caso bosonico le condizioni di annullamento dei termini di bordo, compatibili con l'invarianza di Lorentz, definiscono le stringhe aperte e chiuse. Per le stringhe aperte insieme alle condizioni (\ref{condizioni1}) occorre richiedere che
\ba
\psi^\mu_+(0)&=&\psi^\mu_-(0) \ , \nonumber \\
\psi^\mu_+(\pi)&=&\pm \psi^\mu_-(\pi) \ .
\ea
Le condizioni che si ottengono in corrispondenza del segno relativo positivo sono dette di Ramond, mentre quelle corrispondenti al segno negativo sono dette di Neveu-Schwarz \cite{nsr}. Allo stesso modo per la superstringa chiusa le condizioni possibili insieme alle (\ref{condizioni2}) sono
\ba
\psi^\mu_i(\pi)&=&+\psi^\mu_i(0) \ , \qquad condizioni \, di \, Ramond \, (R) \nonumber \\
\psi^\mu_i(\pi)&=&-\psi^\mu_i(0) \ . \qquad condizioni \, di \, Neveu-Schwarz \, (NS)
\ea
Per la stringa chiusa si hanno quindi quattro possibili combinazioni di condizioni al bordo, che definiscono quattro distiniti settori: R-R, NS-R, R-NS, NS-NS.

Le equazioni del moto (\ref{motosuper}) possono essere ora risolte imponendo sulle soluzioni le condizioni al bordo trovate. Per il campo bosonico si ottengono le soluzioni gi\`a note (\ref{modes_exp_bos1}, \ref{modes_exp_bos2}). Le soluzioni dell'equazione di Dirac, nel caso di stringa chiusa, sono per i modi destri e sinistri:
\ba
\psi_+ \ &=& \ \sqrt{2\alpha'}\sum_{r} \ \tilde{\psi}_r^\mu \
e^{-2ir(\tau+\sigma)} \qquad r \in {\mathbb Z} \, {\rm oppure} \, r \in {\mathbb Z} + \frac{1}{2}\nonumber\\
\psi_- \ &=& \ \sqrt{2\alpha'}\sum_{r}
\ \psi_r^\mu \ e^{-2ir(\tau-\sigma)} \qquad r \in {\mathbb Z} \, {\rm oppure} \, r \in {\mathbb Z} + \frac{1}{2}\ ,
\ea
dove l'indice $r$ corre sui seminteri per condizioni di Ramond, e sugli interi per condizioni di Neveu-Schwarz. Le soluzioni di stringa aperta che si ottengono sono della stessa forma di quelle di stringa chiusa, ma con le frequenze di oscillazione dimezzate, normalizzazione $\sqrt{\alpha'}$ e coefficienti dei modi destri e sinistri identificati.

\`E utile riscrivere, nelle nuove coordinate $\sigma^\pm=\tau \pm \sigma$, il tensore energia impulso e la supercorrente separando le parti olomorfe, che dipendono da $\sigma^+$, da quelle antiolomorfe, che dipendono da $\sigma^-$:
\ba
\label{A.a.10}
T_{\pm\pm} \ &=& \ \dd_\pm X^\mu\dd_\pm X_\mu \ + \ \frac{i}{2} \
\psi_\pm\dd_\pm\psi_\pm \ , \nonumber \\
J_\pm(\xi^\pm) \ &=& \ \psi_\pm^\mu\dd_\pm X_\mu \ .
\ea

\subsection{Quantizzazione canonica}

La quantizzazione canonica della superstringa si ottiene imponendo le relazioni di commutazione usuali per campi di spin intero e semintero:
\ba
[\dot{X}^\mu(\sigma,\tau),X^\nu(\sigma',\tau)] \ &=& \ -i\delta(\sigma-\sigma')\eta^{\mu\nu}
\ , \nonumber\\
\{\psi^\mu_\pm(\sigma,\tau),\psi^\nu_\pm(\sigma',\tau)\} \ &=& \ \delta(\sigma-\sigma')\eta^{\mu\nu} \ ,
\ea
da cui si ottengono per gli oscillatori le relazioni di commutazione
\ba
\label{algebra_oscillator}
[\alpha^\mu_m,\alpha^\nu_n] \ &=& \ m\delta_{m+n}\eta^{\mu\nu} \ , \nonumber\\
\{\psi_r^\mu, \psi_s^\nu\} \ &=& \ \eta^{\mu\nu}\delta_{r+s}  \quad  \ ,
\ea
e le medesime relazioni per gli oscillatori destri. Come nel caso bosonico si riconoscono, dopo un riscalamento, le regole di commutazione degli operatori di creazione e distruzione. Uno stato generico $|\phi \rangle$ sar\`a ottenuto dall'azione degli operatori bosonici e fermionici di creazione sul vuoto. 

La quantizzazione covariante nel caso della superstringa richiede che si impongano i vincoli
\be
\label{mode_constraints}
T_{++} \ = \ T_{--} \ = \ J_+ \ = \ J_- \ = \ 0 \ ,
\ee
che, come nel caso bosonico, possono essere decomposti in modi normali. Per la stringa chiusa si ha
\ba
\label{supervincoli}
L_m|\phi \rangle \ &=& \ {\bar L}_m|\phi \rangle \ = \ 0 \ , m > 0 \ , \nonumber \\
(L_0-a)|\phi \rangle \ &=& \ ({\bar L}_0 - \bar a)|\phi \rangle \ = \ 0 \ ,  \nonumber \\
G_r|\phi \rangle \ &=& \ {\bar G}_r|\phi \rangle \ = \ 0 \ , r > 0 \ ,
\ea
dove
\ba
L_m \ &=& \ \frac{1}{\pi \alpha'} \ \int_0^{2\pi} d\sigma \ T_{--} \ e^{-im\sigma} = \frac{1}{2}\sum_{n}:\alpha^\mu_{m-n}\alpha_{\mu, n}: + \frac{1}{4}\sum_r(2r-m):\psi_{m-r}^\mu \psi_{\mu, r}: + a\delta_{m, 0}  \ , \nonumber\\
G_r \ &=& \ \frac{2}{\pi \alpha'} \ \int_0^{2\pi} d\sigma J_- \ \ e^{-ir\sigma}= \sum_n :\alpha_n \psi_{r-n}: \ ,
\ea
In maniera analoga si trovano gli operatori di super-Virasoro nel caso di stringa aperta. L'algebra degli operatori \`e
\ba
\left[L_m, L_n \right] &=& (m-n)L_{m+n}+\frac{c}{12}(m^3-m)\delta_{m+n} \ , \nonumber \\
\{G_r, G_s\} &=& 2L_{r+s} + \frac{c}{12}(4r^2-1)\delta_{r+s} \ , \nonumber \\
\left[L_m, G_r \right] &=& \frac{1}{2}(m-2r)G_{m+r} \ ,
\ea
dove $c=D+D/2$ con $D$ il contributo dei gradi di libert\`a bosonici e $D/2$ il contributo dei gradi di libert\`a fermionici. La super-algebra \`e nota come algebra di Ramond per $r, s$ interi e di Neveu-Schwarz per $r, s$ seminteri. I campi antiolomorfi portano una seconda copia di questa algebra.

Prima di discutere in dettaglio lo spettro, vediamo brevemente la quantizzazione nel formalismo del cono di luce. Le coordinate di cono di luce sono $X^\pm = (X^0\pm X^{1})/\sqrt{2}$ e in maniera analoga definiamo $\psi_\pm$. La guage di cono di luce \`e
\ba
X^+ \ &=& \ x^+ +2\alpha'p^+\tau\nonumber\\
\psi_\pm^+ \ &=& \ 0 \ ,
\ea
e risolvendo come nel caso bosonico i vincoli (\ref{mode_constraints}), si ottiene per gli oscillatori
\ba
\label{solve_constr}
\alpha^-_n \ &=& \frac{2}{\sqrt{2\alpha'} \ p^+} \ \Bigg[\sum_{i=1}^{D-2}
\ \frac{1}{2}\sum_{m \ \in \ \mathbb{Z}}:\alpha^i_{n-m}\alpha^i_m:
\nonumber\\
&&+ \ \sum_{i=1}^{D-2} \ \frac{1}{2}\sum_{r}(r-\frac{n}{2}):\psi^i_{n-r}\psi^i_r: +a
\ \delta_n\Bigg] \ , \nonumber\\
\psi_r^- \ &=& \ \frac{2}{\sqrt{2\alpha'} \ p^+} \ \sum_{i=1}^{D-2} \sum_{s} \
\alpha^i_{r-s}\psi^i_s \  \ \ ,
\ea
dove $r, s \in \mathbb{Z}$ nel settore di Ramond e $r, s \in \mathbb{Z}+ 1/2$ nel settore di Neveu-Schwarz mentre $a$ \`e la parte finita del termine divergente dovuto all'ordinamento normale che varia a seconda del settore. Le energie di punto zero possono essere determinate, sul modello di quanto gi\`a fatto, studiando l'andamento della funzione $\zeta$ di Riemann
\be
\zeta_\alpha(-1,\epsilon) \ = \ \sum_{n=1}^\infty(n+\alpha) \ e^{-(n+\alpha)\epsilon} \ \stackrel{\epsilon \sim 0}{\longrightarrow} \ -\frac{1}{12}[6\alpha(\alpha-1)+1] + O(\frac{1}{\epsilon})
\ ,
\ee
da cui si ottiene che ogni grado di libert\`a bosonico contribuisce all'energia di vuoto con $a=-\frac{1}{24}$, mentre ogni grado di libert\`a fermionico con $a=+\frac{1}{24}$ per condizioni periodiche e con $a=-\frac{1}{48}$ per condizioni antiperiodiche. Si ha quindi $a^R=0$ e $a^{NS}=-\frac{D-2}{16}$, e anche per la superstringa la dimensione dello spazio tempo \`e fissata dall'invarianza di Lorentz, in questo caso a $D=10$, per cui $a^R=0$ e $a^{NS}=-\frac{1}{2}$.

Definendo gli operatori numero bosonici e fermionici
\ba
\ N_B^\bot&=&\sum_{i=1}^{D-2}\sum_{n=1}^\infty\alpha_{-n}^i\alpha_n^i \ , \quad \nonumber \\
\ N_F^\bot&=&\sum_{i=1}^{D-2}\sum_{r}^\infty r \ \psi_{-r}^i \psi_r^i \ , 
\ea
dagli zero modi si trova per la formula di massa nel caso chiuso
\be
\label{supermassachiusa}
M^2 \ = \ \frac{2}{\alpha'} \ \left[N_B^\bot + \bar N_B^\bot + N_F^\bot + \bar N_F^\bot + a + \bar a \right], \qquad N_B^\bot + N_F^\bot+a \ = \ \bar N_B^\bot+ \bar N_F^\bot+ \bar a \ ,
\ee
mentre per la superstringa aperta
\be
\label{supermassaperta}
M^2 \ = \ \frac{2}{\alpha'} \ \left[N_B^\bot + N_F^\bot + a \right]\ .
\ee

\subsection{Spettro di Superstringa}

La teoria descritta fin qui, \`e in realt\`a inconsistente come teoria quantistica anche per $D=10$; per ottenere una teoria consistente lo spettro deve essere opportunamente troncato come proposto inizialmente da Gliozzi, Scherk e Olive \cite{gso} (proiezione GSO). Una prima ragione \`e la presenza nello spettro di un tachione nel settore NS dovuto al contributo negativo all'energia di vuoto di $a^{NS}$. Un seconda ragione risiede nella particolarit\`a dello spettro di avere operatori anticommutanti $\psi^\mu_r$ che mappano bosoni in bosoni: dato uno stato bosonico $|\phi \rangle$, lo stato ottenuto applicando un operatore anticommuntante $\psi^\mu_r|\phi \rangle$, \`e ancora uno stato bosonico. Questo viola  il teorema di spin-statistica, si tratta di un effetto poco naturale e quindi indesiderabile. La terza ragione che ci spinge a definire una proiezione dello spettro \`e che in questo modo \`e possibile ottenere uno spettro supersimmetrico. Infatti, se la supersimmetria sulla superficie bidimensionale d'universo \`e gi\`a presente nella teoria, dal punto di vista dello spazio-tempo 10 dimensionale, la supersimmetria va implementata con un opportuno troncamento dello spettro.

Nel settore NS, il tachione scalare $\mid 0\rangle_{NS}$ pu\`o essere eliminato defininendo un operatore di proiezione $P_{GSO}^{(NS)}=\frac{1-(-)^F}{2}$, dove $F$ conta il numero di operatori fermionici. Lo stato fondamentale dello spettro proiettato diventa $\psi_{-\frac{1}{2}}^i\mid 0\rangle_{NS}$. Nel settore R, dal momento che gli operatori $\psi_0^i$ soddisfano l'algebra di Clifford (a meno di un riscalamento)
\be
\{\psi_0^i,\psi_0^j\}=\delta^{ij} \ ,
\ee
lo stato fondamentale \`e una rappresentazione spinoriale dell'algebra di $SO(8)$, in particolare uno spinore di Majorana a massa nulla. Introducendo l'operatore di proiezione $P^{(R)}_{GSO}=\frac{1+(-1)^F\Gamma_9}{2}$, dove $\Gamma_9$ \`e la matrice di chiralit\`a definita nello spazio trasverso, lo stato fondamentale diventa uno spinore di Majorana-Weyl di chiralit\`a definita e gli stati eccitati costruiti da questo con operatori di creazione fermionici avranno, alternativamente, chiralit\`a destra e sinistra.

Lo spettro di bassa energia diella superstringa aperta GSO proiettato (teoria di Tipo I), contiene un vettore a massa nulla nel settore NS e un fermione di Majorana-Weyl nel settore R, che insieme formano un multipletto di super Yang-Mills in $D=10$. La teoria di tipo I risulta priva di anomalie \cite{gs}.

Nelle stringhe chiuse la proiezione GSO porta ad uno spettro di bassa energia di ${\mathcal N}=2$ supergravit\`a. La teoria che si ottiene scegliendo le chiralit\`a dei vuoti destro e sinistro concordi, \`e detta di Tipo $IIB$, mentre la teoria con vuoti discordi \`e detta di Tipo $IIA$. Dal momento che gli spin sono additivi, i settori NS-NS e R-R dello spettro conterranno bosoni e i settori NS-R e R-NS fermioni. In termini di rappresentazioni di massa nulla si ha:
\ba
Tipo \, IIA: \, (\bold{8_v  \oplus 8_s})\otimes(\bold{8_v  \oplus 8_c}) \ ,\nonumber \\
Tipo \, IIB: \, (\bold{8_v  \oplus 8_s})\otimes(\bold{8_v  \oplus 8_s}) \ .
\ea
Decomponendo questi prodotti in termini di rappresentazioni del gruppo di Lorentz $SO(8)$, si trova il contenuto in termini di stati. Nel settore NS-NS si ottiene
\be
(\bold{8_v \otimes 8_v}) =\bold{1 \oplus 28 \oplus 35} = \Phi + B_{\mu \nu}+ h_{\mu \nu} \ ,
\ee
che identifichiamo con il dilatone, il tensore antisimmetrico e il gravitone. Allo stesso modo nel settore R-R si trova, per la $IIB$, una $2$-forma e una $4$-forma con una curvatura autoduale e, per la $IIA$ un vettore e una $3$-forma. Nei settori misti si hanno due gravitini e due spinori (detti dilatini). Nella $IIA$ i due gravitini sono di chiralit\`a opposta e cos\`i anche i due dilatini mentre, nella $IIB$ i due gravitini hanno la stessa chiralit\`a e chiralit\`a opposta rispetto ai due dilatini. Si pu\`o vedere che, sebbene la IIB abbia spettro chirale, la teoria \`e priva di anomalie gravitazionali \cite{alvgaum}.


\chapter{Integrale di Polyakov}

\section{L'integrale funzionale di Polyakov}

L'integrale sui cammini di Feynman \`e un metodo molto naturale per definire una teoria di campo.  Anche in Teoria delle Stringhe \`e possibile definire un integrale sui ``cammini'' di stringa, ovvero una somma su tutte le possibile \emph{storie} della stringa (world-sheet o superfici d'universo) che interpolino tra stati iniziali e finali. Le storie sono pesate da un fattore
\begin{equation}
e^{(iS/\hbar)} \ ,
\end{equation}
dove $S$ indica l'azione corrispondente. L'unico modo di implementare le interazioni in teoria di stringhe in maniera consistente con le simmetrie, \`e considerare come permesse le sole interazioni gi\`a implicite nella somma sui world-sheet. In questo modo si pu\`o ottenere una teoria consistente, priva di divergenze e unitaria. \`E interessante osservare come le interazioni siano indotte dalla topologia globale del world-sheet, mentre localmente le propriet\`e del world-sheet sono indistinguibili da quelle di una teoria libera. Si pu\`o osservare come, data una sezione di un diagramma di interazione, questa sembra aver luogo in un dato punto dello spazio-tempo, ma in realt\`a la sezione \`e arbitraria. \`E infatti possibile definire una nuova sezione del diagramma, che corrisponde ad effettuare una trasformazione di Lorentz, il cui risultato \`e spostare l'interazione in un punto differente dello spazio-tempo.

A seconda delle topologie che si considerano nella somma sui world-sheet, si definiscono differenti teorie di stringa. Disegnando un diagramma di stringa si hanno due tipi di ``bordi'': i bordi definiti dalle stringhe iniziali e finali e quelli che corrispondono alle world-line degli eventuali estremi di stringa aperta. Limitiamoci a considerare i secondi per le definizioni seguenti. Esistono quattro possibili scelte per definire la somma sui world-sheet, che corrispondono a quattro distinte teorie di stringa:
\begin{enumerate}
\item stringhe chiuse orientate: tutti i world-sheet sono orientabili senza bordi;
\item stringhe chiuse non orientate: tutti i world-sheet sono senza bordi;
\item stringhe aperte e chiuse orientate: tutti i world sheet sono orientabili con bordi;
\item strighe aperte e chiuse non orientate: tutti i world sheet sono ammessi.
\end{enumerate}
Non \`e possibile definire, invece, teorie di sole stringhe aperte. Consideriamo infatti un world-sheet con topologia di anello: questo \`e un diagramma di vuoto di stringa aperta, ma \`e anche interpretabile come diagramma di propagazione di stringa chiusa (figura). Quindi, partendo da una teoria di sole stringhe aperte, la somma sui world-sheet porter\`a ad includere necessariamente processi in cui lo scattering di stringhe aperte produce stringhe chiuse.

Sviluppiamo l'idea della somma sui world-sheet \cite{pol1, pol2}. Il punto di partenza \`e la teoria di stringa Euclidea, in cui la metrica Minkowkiana sul world-sheet $\gamma_{ab}$ \`e sostituita da una metrica Euclidea $g_{ab}(\sigma^1 ,\sigma^2)$.Consideriamo per semplicit\`a una teoria bosonica. L'integrale funzionale che definisce la teoria \`e su tutte le metriche Euclidee e su tutte le immersioni $X^\mu(\sigma^1 ,\sigma^2)$ del world-sheet nello spazio-tempo di Minkowski:
\begin{equation}
\label{intfunzionale}
\int[dX][dg]e^{(-S)} \ .
\end{equation}
L'azione euclidea \`e
\begin{equation}
S=S_X + \lambda\chi \ ,
\end{equation}
dove 
\begin{equation}
S_X=\frac{1}{4\pi\alpha'}\int_M d{\sigma^2}\, g^{1/2}g^{ab}\partial_aX^\mu\partial_bX_\mu \ ,
\end{equation}
\begin{equation}
\label{chi}
\chi=\frac{1}{4\pi}\int_M d{\sigma^2}\, g^{1/2}R + \frac{1}{2\pi}\int_{\partial M}ds\, k \ ,
\end{equation}
e $k$ \`e la curvatura geodetica dei bordi. Il vantaggio dell'integrale funzionale Euclideo \`e che l'integrale sulle metriche \`e ben definito. Sui world-sheet non banali a cui abbiamo accennato si possono avere metriche Euclidee non singolari, ma non sempre metriche Minkowskiane non singolari. Si pu\`o dimostrare inoltre che in due dimensioni il passaggio all'euclideo \`e ben giustificato e, la teoria \`e equivalente a quella Minkowkiana.

Il termine $\chi$ \`e localmente una derivata totale in due dimensioni e, quindi, dipende solo dalla topologia del world-sheet; \`e detto \emph{numero di Eulero} del world-sheet. Il fattore $e^{(-\lambda\chi)}$ nell'integrale funzionale pesa le differenti topologie nella somma sui world-sheet. Quindi il termine di Eulero nell'azione genera le potenze della costante di accoppiamento della teoria di stringa che possiamo definire come
\begin{equation}
g_o^2 \sim g_s \sim e^{\lambda} \ .
\end{equation} 
La costante $\lambda$ non \`e un parametro libero della teoria, ma \`e legata al valore di vuoto di un campo detto dilatone. Differenti valori di $\lambda$ non corrispondono quindi a diverse teorie ma a differenti \emph{background} di una sola teoria.

\section{Metodo di Faddeev-Popov}

L'integrale funzionale (\ref{intfunzionale}) \`e mal definito, dal momento che l'integrazione avviene su infinite copie delle configurazioni fisiche collegate tra loro da trasformazioni di simmetria del tipo $diff \times Weyl$. Occorre dividere l'integrale funzionale per il volume del gruppo di simmetria locale,
\begin{equation}
Z \equiv \int\frac{[dX][dg]}{V_{diff \times Weyl}}e^{-S} \ .
\end{equation}
Questo equivale ad integrare sulle sole configurazioni fisiche inequivalenti. Per farlo si deve fissare opportunamente un gauge, restringendo l'integrazione su una sezione dello spazio delle configurazioni $(X, g)$ che intersechi una sola volta ciascuna classe di equivalenza di gauge.

Nel capitolo precedente si \`e visto come sia possibile fissare localmente la metrica sul world-sheet nella forma diagonale
\be
\hat g_{ab}(\sigma)=\delta_{ab} \ ,
\ee
passando quindi da una metrica curva ad una piatta. Questo dal momento che in due dimensioni il tensore di Weyl si annulla identicamente. La trasformazione di Weyl dello scalare di Ricci \`e
\be
g'^{1/2}R'=g^{1/2}(R-2\nabla^2 \omega) \ .
\ee 
Risolvendo l'equazione $2\nabla^2 \omega=2$ per $\omega$ si fissa $R'=0$. Dal momento che in due dimensioni il tensore di Riemann \`e proporzionale allo scalare di Ricci
\be
R_{abcd}=\frac{1}{2}(g_{ac}g_{bd}-g_{ad}g_{bc})R \ ,
\ee
si fissa a zero anche il tensore di Riemann, che equivale a scegliere una metrica piatta. Come si \`e visto la teoria con la metrica fissata conserva una invarianza conforme residua. Rivediamolo brevente per la teoria euclidea.  A tal scopo introduciamo coordinate euclidee complesse $z=\sigma^1 + i\sigma^2$, con metrica $ds^2=dzd \bar z$. Una trasformazione olomorfa di $z$,
\be
z'\equiv \sigma'^1+ i \sigma'^2 = f(z) \ ,
\ee
combinata con una trasformazione di Weyl, manda la metrica in
\be
ds'^2=e^{2\omega}|\partial_z f|^{-2}dz'd\bar z' \ ,
\ee
e quindi scegliendo $\omega= ln|\partial_z f|$, la metrica \`e invariante. Si \`e trovato che il gruppo di simmetria conforme \`e il sottogruppo delle trasformazioni $diff \times Weyl$, che lasciano invariata la metrica unitaria. \`E un risultato di grande valore, dal momento che indica la possibilit\`a di utilizzare in Teoria delle Stringhe gli strumenti tipici delle Teorie di Campo conformi bidimensionali.

Dopo aver fissato la metrica l'integrale funzionale corre lungo una sezione parametrizzata da $X^\mu$ soltanto. Per ottenere la misura corretta occorre seguire, come nelle usuali teorie di campo, la procedura di Faddeev-Popov. L'idea \`e separare l'integrale funzionale in un integrale sul gruppo di gauge moltiplicato per un integrale lungo la sezione di gauge, in modo da poter normalizzare il primo termine. Il determinante di Faddeev e Popov \`e lo Jacobiano di questo cambio di variabili.

Consideriamo la combinazione di una trasformazione di coordinate e di una trasformazione di Weyl, che indichiamo come $\zeta$,
\be
\label{zeta}
\zeta : g_{ab}(\sigma) \to g^\zeta_{ab}(\sigma')=e^{2\omega(\sigma)}\frac{\partial \sigma^c}{\partial \sigma'^a}\frac{\partial \sigma^d}{\partial \sigma'^b}g_{cd}(\sigma) \ .
\ee

La definizione consueta della misura di Faddeev-Popov $\Delta_{FP}$ \`e
\be
\label{FP}
1=\Delta_{FP}(g)\int [d\zeta]\delta(g-\hat g^\zeta) \ ,
\ee
dove $\hat g_{ab}$ indica la metrica in cui ci si vuole portare e si \`e indicata con $[d\zeta]$ la misura di integrazione gauge invariante sul gruppo $diff \times Weyl$. Non \`e difficile dimostrare che il determinante di Faddeev-Popov \`e gauge invariante:
\ba
\Delta_{FP}(g^\zeta)^{-1}&=& =\int [d\zeta']\delta(g^\zeta-\hat g^{\zeta'})=\int [d\zeta']\delta(g-\hat g^{\zeta^{-1}\cdot \zeta'})\nonumber \\
&=& \int [d\zeta'']\delta(g-\hat g^{\zeta''})=\Delta_{FP}(g)^{-1} \ ,
\ea
dove $\zeta''=\zeta^{-1}\cdot \zeta'$, nella seconda uguaglianza si \`e fatto uso dell'invarianza di gauge della funzione delta, e nella terza dell'invarianza della misura. 

Inserendo l'identit\`a (\ref{FP}) nell'integrale funzionale, si ottiene
\begin{equation}
Z [\hat g]= \int\frac{[d\zeta][dX][dg]}{V_{diff \times Weyl}}\Delta_{FP}(g)\delta(g-\hat g^\zeta)e^{-S[X, g]} \ .
\end{equation}
Integrando su $g_{ab}$ e rinominando la variabile di integrazione $X \to X^\zeta$, si ottiene
\be
Z [\hat g]= \int\frac{[d\zeta][dX^\zeta]}{V_{diff \times Weyl}}\Delta_{FP}(\hat g^\zeta)e^{-S[X^\zeta, \hat g^\zeta]} \ .
\ee
Ricordando l'invarianza di gauge di $\Delta_{FP}$, dell'azione e della misura dell'integrale funzionale si scrive
\be
Z [\hat g]= \int\frac{[d\zeta][dX]}{V_{diff \times Weyl}}\Delta_{FP}(\hat g)e^{-S[X, \hat g]} \ ,
\ee
dove l'integrando non dipende da $\zeta$. L'integrazione su $\zeta$ porta il volume del gruppo che si cancella con il fattore al denominatore lasciando
\be
Z [\hat g]= \int [dX] \Delta_{FP}(\hat g)e^{-S[X, \hat g]} \ .
\ee

Per calcolare esplicitamente il determinante di Faddeev-Popov consideriamo la gauge completamente fissata dalla procedura di gauge fixing, in modo tale che per un solo valore di $\zeta$ la funzione $\delta(g-\hat g^\zeta)$ sia diversa da zero (in questo modo non si avranno i problemi legati alle \emph{copie di Gribov}). Questo \`e possibile dal momento che la simmetria globale residua non modifica le propriet\`a locali del world-sheet. Consideriamo trasformazioni infinitesime di gauge: la variazione infinitesima della (\ref{zeta}) risulta essere
\be
\label{varmet}
\delta g_{ab}=2\delta \omega g_{ab}- \nabla_a \delta \sigma_b - \nabla_b \delta \sigma_a \ .
\ee
Definendo un operatore differenziale $P_1$ che mandi vettori in tensori a due indici a traccia nulla
\be
\label{P1}
(P_1\delta \sigma)_{ab}=\frac{1}{2}(\nabla_a \delta \sigma_b + \nabla_b \delta \sigma_a - g_{ab}\nabla_c\delta \sigma^c) \ ,
\ee
la \ref{varmet} pu\`o essere riscritta nella forma
\be
\delta g_{ab}=(2 \delta \omega - \nabla_c \delta \sigma^c)g_{ab}- 2(P_1 \delta \sigma)_{ab} \ .
\ee
Per trasformazioni $\zeta$ vicine all'identit\`a, il determinante di Faddeev e Popov si scrive quindi
\be
\Delta_{FP}(\hat g)^{-1}=\int [d\delta \omega d\delta \sigma]\delta \left[-(2 \delta \omega - \nabla_c \delta \sigma^c)g_{ab}+2(P_1 \delta \sigma)_{ab} \right] \ ,
\ee
e introducendo la rappresentazione esponenziale della funzione $\delta$ si ottiene
\be
\Delta_{FP}(\hat g)^{-1}=\int [d\delta \omega d\delta \sigma d \beta] e^{2\pi i \int d^2\sigma \hat g^{1/2}\beta^{ab}\left[-(2 \delta \omega - \nabla_c \delta \sigma^c)g_{ab}+2(P_1 \delta \sigma)_{ab} \right]} \ ,
\ee
dove $\beta^{ab}$ \`e un campo tensoriale simmetrico. Integrando su $\delta \omega$ si ottiene una funzione $\delta$ che forza $\beta^{ab}$ ad avere traccia nulla. Denotando il tensore a traccia nulla con $\beta'^{ab}$ possiamo scrivere 
\be
\Delta_{FP}(\hat g)^{-1}= \int [d \beta'd\delta \omega]e^{4\pi i \int d^2\sigma \hat g^{1/2}\beta'^{ab}(\hat P_1 \delta \sigma)_{ab}} \ ,
\ee
dove il cappuccio sull'operatore differenziale indica la presenza della metrica $\hat g_{ab}$. \`E possibile, ricordando le propriet\`a delle variabili anticommutanti, invertire l'integrale funzionale sostituendo a ciascun campo bosonico da integrare un corrispondente campo anticommutante. Questi campi anticommuntanti prendono il nome di campi \emph{ghost}. Quindi definendo i campi anticommutanti
\ba
\delta \sigma^a \to c^a \ , \nonumber \\
\beta'_{ab} \to b_{ab} \ ,
\ea
dove $b_{ab}$ \`e a traccia nulla, possiamo scrivere
\be
\Delta_{FP}(\hat g)=\int [db\, dc]\,  e^{-S_g} \ .
\ee
L'azione $S_g$ dei campi ghost, scegliendo una normalizzazione conveniente, \`e
\be
S_g=\frac{1}{2\pi}\int d^2\sigma \hat g^{1/2}b_{ab} \hat \nabla^a c^b = \frac{1}{2\pi}\int d^2\sigma \hat g^{1/2}b_{ab}(\hat P_1 c)^{ab} \ .
\ee
La procedura di gauge fixing introduce nell'integrale di Polyakov l'azione di campi ghost `non fisici'. Localmente l'integrale funzionale \`e
\be
\label{polyakovghost}
Z [\hat g]= \int [dX\, db\, dc] e^{-S_X - S_g} \ .
\ee
L'azione \`e quadratica nei campi e l'integrale pu\`o essere calcolato esattamente con le regole degli integrali gaussiani. Il calcolo dell'integrale porta a scrivere l'integrale di Polyakov come 
\be
Z [\hat g] \sim (\det{\hat \nabla^2})^{-D/2}\det{\hat P_1} \ ,
\ee
dove il primo determinante ha origine dall'integrazione sui campi $X$ e il secondo dall'integrazione sui campi ghost.

Nella gauge conforme, in coordinate olomorfe $z, \bar z$, l'azione dei campi ghost diventa
\ba
S_g &=& \frac{1}{2\pi}\int d^2z(b_{zz}\nabla_{\bar z}c^z+b_{\bar z \bar z}\nabla_z c^{\bar z}) \nonumber \\
&=& \frac{1}{2\pi}\int d^2z(b_{zz}\partial_{\bar z}c^z+b_{\bar z \bar z}\partial_z c^{\bar z}) \ .
\ea
Le derivate covarianti in $\bar z$ di tensori con soli indici $z$ si riducono a derivate ordinarie, e vice versa, per via delle propriet\`a della connessione.

Nel caso di stringa aperta si deve anche considerare la presenza di bordi sul world sheet. \`E conveniente considerare l'integrale funzionale definito su campi presenti in regioni fissate. In questo modo l'invarianza sotto diffeomorfismi del world sheet deve essere limitata a cambi di coordinate che lascino i bordi invariati. La variazione di coordinate sulla superficie di universo $\delta \sigma^a$ deve avere una componente normale nulla,
\be
n_a\delta \sigma^a=0 \ .
\ee
Questa propriet\`a viene naturalmente trasmessa al corrispondente campo anticommutante, per cui si ha
\be
n_a c^a=0 \ .
\ee
Il campo $c^a$ \`e quindi propozionale al vettore tangente $t^a$. Le equazioni del moto impongono una condizione di bordo su $b_{ab}$
\be
\int_{\partial M} ds n^ab_{ab}\delta c^b=0 \ ,
\ee
che insieme alla condizione di bordo su $c^a$ implica
\be
n_a t_b b^{ab}=0 \ .
\ee

L'integrale funzionale nella forma (\ref{polyakovghost}) \`e ben definito dal momento che non presenta problemi di multiplo conteggio e pu\`o essere calcolato esplicitamente. La simmetria di gauge \`e stata fissata, ma l'integrale funzionale contiene una simmetria globale residua, \emph{l'invarianza BRST}, come si vedr\`a in dettaglio nei prossimi paragrafi.

\section{Quantizzazione BRST}

Prima di discutere la quantizzazion BRST nel caso della stringa, passeremo brevemente in rassegna le sue propriet\`a in generale. Consideriamo una teoria con campi $\phi_i$, che abbia una qualche simmetria di gauge, in modo tale che le trasformazioni di gauge soddisfino l'algebra\footnote{Questa non \`e l'algebra pi\`u generale possibile, ma \`e sufficiente per una discussione introduttiva.}
\be
\label{gauge}
[\delta_\alpha, \delta_\beta]=f_{\alpha \beta}^{\, \, \, \, \gamma} \delta_\gamma \ .
\ee
Il gauge pu\`o essere fissato imponendo una condizione appropiata,
\be
F^A(\phi_i)=0 \ .
\ee
Usando il metodo di Faddeev-Popov, l'integrale funzionale si scrive come
\ba
\label{azioneFP}
\int \frac{[d\phi]}{V_{gauge}} e^{-S_0} &\sim & \int [d\phi][db_A][dc^\alpha] \delta(F^A(\phi))e^{-S_0-\int b_A(\delta_\alpha F^A)c^\alpha} \nonumber \\
&\sim & \int [d\phi][dB_A][db_A][dc^\alpha]e^{-S_0-\int B_AF^A(\phi) -\int b_A(\delta_\alpha F^A)c^\alpha} \nonumber \\
&=& \int [d\phi][dB_A][db_A][dc^\alpha]e^{-S} \ ,
\ea
dove 
\be
S= S_0+ S_1+ S_2 \ , \qquad S_1=i\int B_AF^A(\phi) \ , \qquad S_2= \int b_A(\delta_\alpha F^A)c^\alpha \ .
\ee
\`E utile osservare come l'indice $\alpha$ associato al campo ghost $c_\alpha$ \`e in corrispondenza uno ad uno con i parametri delle trasformazioni di gauge (\ref{gauge}), mentre l'indice $A$ del campo di ghost $b_A$ \`e legato alle condizioni di gauge.

L'azione complessiva S \`e invariante sotto le trasformazioni di Becchi-Rouet-Stora-Tyutin (BRST) \cite{Becchi:1975nq, Tyutin},
\ba
\delta_{BRST}\phi_i &=& -i\epsilon c^\alpha \delta_\alpha \phi_i \ ,\nonumber \\
\delta_{BRST}b_A &=& -\epsilon B_A \ ,\nonumber \\
\delta_{BRST}c^\alpha &=& -\frac{1}{2}\epsilon c^\beta c^\gamma f_{\beta \gamma}^{\, \, \, \,\alpha} \ ,\nonumber \\
\delta_{BRST}B_A &=& 0 \ ,\nonumber \\
\ea
dove il parametro $\epsilon$ deve essere una costante anticommutante. La prima trasformazione \`e semplicemente la trasformazione di gauge originaria su $\phi_i$ con il parametro di gauge sostituito con il campo ghost $c_\alpha$. C'\`e un numero di ghost conservato che \`e $+1$ per $c_\alpha$, $-1$ per $b_A$ e $\epsilon$, e $0$ per gli altri campi. Non \`e difficile accorgersi che i termini aggiuntivi dell'azione dovuti al gauge fixing in (\ref{azioneFP}) possono essere scritti in termini di una trasformazione BRST
\be
\delta_{BRST}(b_AF^a)=\epsilon[B_AF^A(\phi)+b_Ac^\alpha \delta_\alpha F^A(\phi)] \ .
\ee
La simmetria BRST \`e un'estensione della simmetria di gauge che sopravvive alla procedura di scelta della gauge. Consideriamo ora l'effetto di una piccola variazione della condizione di gauge $\delta F^A$ su un ampiezza fisica
\be
\label{fisivarianza}
\epsilon \delta_F \langle \psi | \psi' \rangle = - i \langle \psi | \delta_{BRST}(b_A\delta F^A)|\psi' \rangle =
\langle \psi | \{Q_{BRST}, b_A\delta F^A \}|\psi' \rangle \ ,
\ee
dove $Q_{BRST}$ \`e la carica conservata legata alla simmetria BRST. L'ampiezza corrispondente ad una osservabile non deve cambiare per variazioni della condizione di gauge. Dal momento che i campi $c^\alpha$ e $b_A$ sono legati rispettivamente al parametro di gauge $\epsilon^\alpha$ e al moltiplicatore di Lagrange $B_A$, \`e naturale supporre che essi siano reali. Questo porta ad assumere che $Q_{BRST}^\dag=Q_{BRST}$, e insieme alla (\ref{fisivarianza}) porta direttamente al risultato
\be
\label{esatti}
Q_{BRST}|\phi \rangle = 0 \ ,
\ee
Ovvero tutti gli stati fisici $|\phi \rangle$ devono essere BRST invarianti.

Per verificare se la carica BRST \`e effettivamente conservata, occorre verificare che essa commuti con la variazione dell'Hamiltoniana conseguente alla variazione della condizione di gauge,
\ba
0&=&[Q_{BRST}, \delta H] = [Q_{BRST}, \delta_B(b_A\delta F^A)] \nonumber \\
&=& [Q_{BRST}, \{Q_{BRST}, b_A\delta F^A\}]=[Q^2_{BRST}, b_A\delta F^A] \ .
\ea
Dal momento che deve essere verificata per variazioni arbitrarie della condizione di gauge, si deve avere
\be
Q^2_{BRST}=0 \ ,
\ee
ovvero la carica deve essere nilpotente per avere una descrizione quantistica consistente. Il caso $Q^2_{BRST}=cost.$ \`e escluso dal momento che $Q^2_{BRST}$ ha numero di ghost $+2$. Non \`e difficile verificare direttamente che, agendo due volte sui campi con le trasformazioni BRST, tutti i campi sono invarianti.

Vediamo le conseguenze fisiche delle propriet\`a di $Q_{BRST}$. Uno stato $Q_{BRST}|\chi \rangle$ sar\`a annichilato da $Q_{BRST}$ per qualsiasi $|\chi \rangle$ ed \`e quindi fisico. Eppure questo stato \`e ortogonale a tutti gli stati fisici, compreso se stesso. Tuttte le ampiezze fisiche che conivolgono questi stati sono quindi nulle, e stati di questo tipo sono per questo detti \emph{stati nulli}. Due stati fisici che differiscano per uno stato nullo,
\be
|\psi' \rangle = |\psi \rangle + Q_{BRST}|\chi \rangle
\ee
avranno quindi gli stessi prodotti interni con tutti gli stati fisici, e saranno pertanto fisicamente indistinguibili. Come nel caso della quantizzazione covariante tradizionale, i veri stati fisici corrispondono all'insieme delle classi di equivalenza composte da stati che differiscono fra loro per stati nulli. Questa \`e una costruzione naturale per operatori nilpotenti, ed \`e conosciuta come la coomologia di $Q_{BRST}$. In coomologia ci si riferisce agli stati annichilati da $Q_{BRST}$ come stati \emph{chiusi}, mentre a quelli della forma $Q_{BRST}|\chi \rangle$ come stati \emph{esatti}. Lo spazio fisico degli stati \`e quindi
\be
{\cal H}_{BRST}=\frac{{\cal H}_{chiusi}}{{\cal H}_{esatti}} \ .
\ee

\subsection{Quantizzazione BRST in Teoria delle Stringhe}

Vediamo le trasformazioni BRST in teoria delle stringhe. Consideriamo oltre all'azione di Polyakov e dei campi ghost il termine di gauge fixing
\be
\frac{i}{4\pi}\int d^2\sigma g^{1/2}B^{ab}(\delta_{ab}-g_{ab}) \ .
\ee
Integrando su $B_{ab}$ e utilizzando le equazioni del moto di $g_{ab}$ per eliminare $B_{ab}$ dalle trasformazioni, si ottengono le trasformazioni BRST nella forma
\ba
\delta_{BRST}X^\mu &=& i\epsilon (c\partial + \tilde c \bar \partial)X^\mu \ , \nonumber \\
\delta_Bb &=&  i\epsilon (T^X+T^g) \ ,  \qquad \delta_B\tilde b =  i\epsilon (\tilde T^X+\tilde T^g) \ , \nonumber \\
\delta_Bc &=&  i\epsilon (c\partial + \tilde c \bar \partial)c \ ,  \qquad \delta_B\tilde c =  i\epsilon (c\partial + \tilde c \bar \partial)\tilde c \ .
\ea
Come si \`e visto, il ghost di Weyl forza $b_{ab}$ ad essere a traccia nulla. Dal teorema di Noether si trova la corrente della simmetria
\ba
j_{BRST}&=&cT^X+\frac{1}{2}:cT^g:+\frac{3}{2}\partial^2 c \ , \nonumber \\
&=&cT^X+:bc\partial c: + \frac{3}{2}\partial^2c: \ , 
\ea
e la corrispondente corrente $\tilde j_B$. L'ultimo termine \`e una derivata totale e non contribuisce per questo alla carica BRST, ma viene aggiunto per garantire che $j_{BRST}$ trasformi come una corrente (tensore di rango 1). Le espansioni OPE per le correnti BRST con i campi dei ghost e con un generico tensore di materia sono
\ba
j_{BRST}\,b(0) &\sim & \frac{3}{z^3}+\frac{1}{z^2}j^g(0)+\frac{1}{z}T^{X+g}(0) \ , \nonumber \\
j_{BRST}\,c(0) &\sim & \frac{1}{z}c\partial c(0) \ , \nonumber \\
j_{BRST}\,{\cal O}^m(0) &\sim & \frac{h}{z^2}c{\cal O}^m(0)+\frac{1}{z}[h(\partial c){\cal O}^m(0)+c\partial {\cal O}^m(0)] \ .
\ea
L'operatore di carica \`e
\be
Q_{BRST}=\frac{1}{2\pi i} \oint (dz j_{BRST}+d\bar z\tilde j_{BRST}) \ .
\ee
Dall'azione di ghost si possono ricavare le equazioni del moto dei campi
\be
\partial_{\bar z}b_{zz}=\partial_{z}b_{\bar z \bar z}=\partial_{\bar z}c^z=\partial_{z}c^{\bar z}=0 \ .
\ee
Imponendo le condizioni di periodicit\`a nel caso di stringa chiusa o le condizioni ai bordi nel caso di stringa aperta, si possono espandere i campi ghost in termini di modi di Fourier.

L'espansione OPE porta a trovare
\be
\{Q_{BRST}, b_m \}=L_m^X+L_m^g \ .
\ee
In termini di modi dei ghost si trova
\ba
\label{caricaBRST}
Q_{BRST} &=& \sum_{n=-\infty}^{\infty}(c_nL_{-n}^m+ \tilde c_n\tilde L_{-n}^m) \nonumber \\
&+& \sum_{m, n=-\infty}^\infty \frac{(m-n)}{2}:(c_mc_nb_{-m-n}+\tilde c_m \tilde c_n \tilde b_{-m-n}):
+a^B(c_0+\tilde c_0) \ .
\ea
La costante $a^B=a^g$ \`e fissata a $-1$ calcolando l'anticommutatore
\be
\{Q_{BRST}, b_0 \}=L_0^X+L_0^g \ .
\ee

Per verificare la propriet\`a di nilpotenza della carica BRST occorre calcolare l'anticommutatore
\be
\{Q_{BRST}, Q_{BRST} \} \ ,
\ee
che risulta essere uguale  a $0$ solo per la dimesione critica $D=26$. Si tratta di un risultato atteso. La quantizzazione della stringa mostra infatti che c'\`e un anomalia nella simmetria di gauge (anomalia di Weyl) che si cancella solo per $D=26$.

In generale, in presenza di un gruppo di simmetria residuo, che non sia stato fissato dalle condizioni di gauge (in teoria delle stringhe si tratta del gruppo conforme i cui generatori sono $L_m$ e $\tilde L_m$), si deve richiedere che i suoi generatori si annullino negli elementi di matrice fisici. Dato un gruppo di simmetria residuo di generatori $G_I$, questi formeranno un'algebra
\be
\label{galgebra}
[G_I, G_J]=ig^K_{\phantom{K}IJ}G_K \ .
\ee
Associata ad ogni generatore c'\`e una coppia di ghost, $b_I$ e $c^I$, con le relazioni di anticommutazione
\be
\label{cbalgebra}
\{c^I, b_J\}=\delta_I^J \ , \qquad \{c^I, c^J\}=\{b_I, b_J\}=0 \ ,
\ee
La forma generale della carica BRST, come pu\`o vedere dalla (\ref{caricaBRST}), \`e
\be
Q_{BRST}=c^IG_I^m-\frac{i}{2}g^K_{\phantom{K}IJ}c^Ic^Jb_K=c^I\left(G_I^m+\frac{1}{2}G_I^g\right) \ ,
\ee
dove $G_I^m$ \`e la parte di materia di $G_I$ e
\be
G_I^g=-ig^K_{\phantom{K}IJ}c^Jb_k \ ,
\ee
\`e la parte dovuta ai campi ghost. Sia $G_I^m$ che $G_I^g$ soddisfano l'algebra (\ref{galgebra}). Usando i commutatori (\ref{galgebra}) e (\ref{cbalgebra}), si trova
\be
Q_{BRST}^2=\frac{1}{2}\{Q_{BRST}, Q_{BRST} \}=-\frac{1}{2}g^K_{\phantom{K}IJ}g^M_{\phantom{M}KL}c^Ic^Jc^Lb_M =0 \ .
\ee
L'uguaglianza a $0$ \`e dovuta all'identit\`a di Jacobi dell'algebra dei $G_I$ che richiede che $g^K_{\phantom{K}IJ}g^M_{\phantom{M}KL}$ si annulli antisimmetrizzando gli indici $IJK$. Il termine di carica centrale che si \`e ignorato deve essere introdotto opportunamente. Questa generalizzazione permetter\`a di trattare il caso di superstringa in maniera diretta.

\subsection{Spettro di stringa in quantizzazione BRST}

Come visto nella discussione precedente, gli stati fisici sono quelli annichilati dall'operatore $Q_{BRST}$ che non siano esatti, ovvero non della forma $Q_{BRST}|\chi \rangle$. C'\`e una condizione addizionale che gli stati fisici devono rispettare,
\be
\label{siegel}
b_0|\phi \rangle=0 \ .
\ee
Questa condizione, nota come \emph{gauge di Siegel},  \`e dovuta al fatto che quando si calcolano le ampiezze d'interazione degli stati fisici, i propagatori contengono sempre fattori $b_0$, che effettivamente proiettano gli stati fisici sul sottospazio che soddisfa la (\ref{siegel}) dal momento che $b_0^2=0$.

Lo spazio di Hilbert degli stati sar\`a dato dal prodotto tensoriale fra lo spazio di Hilbert degli stati creati dagli oscillatori di $X^\mu$ e quello creato dagli oscillatori dei campi ghost. Il vuoto degli stati di ghost \`e definito come lo stato annichilato dai modi di oscillazione dei ghost positivi
\be
b_{n > 0}|ghost \, vacuum \rangle =0 \ , \qquad c_{n > 0}|ghost \, vacuum \rangle=0 \ .
\ee
C'\`e una sottigliezza dovuta alla presenza di zero modi $b_0$ e $c_0$ che soddisfano le condizioni $b_0^2=0$  $c_0^2=0$ e $\{b_0, c_0\}=1$. Queste relazioni definiscono l'algebra delle matrici $\gamma$ in due dimensioni. Questa rappresentazione contiene due stati: uno stato ``spin alto'' e uno stato ``spin basso'' tali che 
\ba
b_0|\downarrow \rangle &=&0 \ ,\qquad b_0|\uparrow \rangle = |\downarrow \rangle \ ,\nonumber \\
c_0|\uparrow \rangle &=& 0 \ ,\qquad c_0|\downarrow \rangle = |\uparrow \rangle \ .
\ea
Imporre anche la condizione (\ref{siegel}) implica che il corretto vuoto di ghost sia lo stato $|\downarrow \rangle$. Gli stati vengono creati a partire da questo vuoto agendo con i modi negativi dei ghost $b_m$, $c_n$. Non si pu\`o per\`o agire con $c_0$ dal momento che lo stato ottenuto non rispetterebbe la condizione (\ref{siegel}). 

Vediamo ora i primi livelli dello spettro di stringa di aperta. Al livello zero si ha un solo stato, il prodotto scalare dei vuoti dei due spazi di Hilbert che indichiamo come $|\downarrow , k^\mu \rangle$. Si deve avere
\be
Q_{BRST}|\downarrow , k^\mu \rangle = 0 \ ,
\ee
dal momento che tutti i termini di $Q_{BRST}$ contengono $L_0$ e operatori di annichilazione. Questo mostra anche che non esistono a questo livello stati esatti, e quindi ogni stato invariante corrisponde a una classe di coomologia. Si tratta degli stati del tachione.

Al primo livello i possibili operatori sono $\alpha_{-1}^\mu$, $b_{-1}$ e $c_{-1}$. Lo stato pi\`u generale che si pu\`o costruire \`e
\be
|\psi_1 \rangle=(\zeta \cdot \alpha_{-1}+\xi_1c_{-1}+\xi_2c_{-1})|\downarrow , k^\mu \rangle \ ,
\ee
che contiene 28 parametri: i 26 parametri del vettore di polarizzazione $\zeta_\mu$ e le due costanti $\xi_1$ e $\xi_2$. La condizione BRST richiede
\ba
Q_{BRST}|\psi_1 \rangle &=& (c_{-1}k \cdot \alpha_1+c_{1}k \cdot \alpha_{-1}) |\psi_1 \rangle \nonumber \\
&=& 2((k\cdot \zeta)c_{-1}+\xi_1 k \cdot \alpha_{-1}) |\downarrow , k^\mu \rangle = 0 \ .
\ea
I termini proporzionali a $c_0$ sommano a zero per la condizione di mass-shell ($k^2=0$), e sono stati per questo omessi. L'uguaglianza a zero vale solo per $k\cdot \zeta=\xi_1=0$, e quindi rimangono solo $26$ stati linearmente indipendenti. Occorre controllare che questi stati non siano Q-esatti. Gli stati invarianti con norma nulla vengono creati da $c_{-1}$ e $k \cdot \alpha_{-1}$. Il pi\`u generale stato Q-esatto a questo livello, per $k^2=0$, \`e
\ba
Q_{BRST}|\chi \rangle = 2((k \cdot {\zeta}')c_{-1} + {\xi_1}' k \cdot \alpha_{-1}) |\downarrow , k^{\mu} \rangle \ .
\ea
Quindi lo stato $c_{-1}|0; k \rangle$ \`e Q-esatto, mentre il vettore di polarizzazione  \`e trasverso con la relazione $\zeta_\mu \sim \zeta_\mu+\xi_1'k_\mu$. Questo lascia i 24 stati di polarizzazione di una particella vettoriale a massa nulla, in accordo con quanto trovato con gli altri metodi di quantizzazione. Non \`e difficile verificare che gli stati fisici trovati sono a norma positiva.

La stessa procedura pu\`o essere seguita per i livelli pi\`u alti. Nel caso di stringhe chiuse occorre tenere conto degli operatori dei modi destri e sinistri.

\section{Ampiezze d'urto}

L'idea di considerare come proprie della teoria le sole interazioni implicite nella somma sui world-sheet che abbiano come bordi gli stati iniziali e finali di stringa \`e estremamente semplice e naturale. In generale \`e per\`o molto complicato definire stati iniziali e finali consistenti con le simmetrie locali della superficie d'universo. Nel limite in cui si considerano sorgenti di stringa infinitamente lontane dalla regione di interazione, la definizione delle ampiezze di interazione \`e notevolmente semplificata. Per assegnati stati iniziali e finali, le ampiezze d'interazione di questi processi asintotici definiscono elementi della matrice S.

Non \`e difficile dare alcuni argomenti di tipo euristico sul tipo di superfici di universo legate ai processi asintotici. Si pu\`o immaginare che lontano dalla zona d'interazione le stringhe si propaghino liberamente. Come \`e facile immaginare, una striga chiusa che si propaghi liberamente descrive un cilindo, che pu\`o essere rappresentato in coordinate complesse $w$ come
\be
-2\pi t \leq Im(w) \leq 0 \ ,\qquad w \sim w + 2\pi \ ,
\ee
dove l'estremo inferiore del cilindro $Im(w) = - 2 \pi t$ \`e dato dalla sorgente esterna, mentre l'estremo superiore si congiunge al resto della superficie d'universo nella regione di interazione. Il limite in cui si hanno processi asintotici si ha, naturalmente, per $t \to \infty$. Il world-sheet complessivo di un processo di interazione di stringhe chiuse sar\`a quindi una superficie chiusa (la regione di interazione) su cui si inseriscono i tubi di propagazione delle stringhe entranti e uscenti. Nel limite $t \to \infty$, tenendo conto della simmetria conforme della teoria, i tubi di propagazione possono essere immaginati infinitamente lunghi e infinitamente stretti, e il world sheet d'interazione si riduce quindi ad una superficie chiusa con una ``puntura'' per ciscun stato esterno. Nel caso pi\`u semplice la superficie d'interazione \`e una sfera. In temini pi\`u precisi si pu\`o considerare la descrizione conformemente equivalente del tubo di propagazione che si ha in termini della coordinata $z$,
\be
\label{punture}
z=e^{-iw} \ , \qquad e^{-2\pi t} \leq |z| \leq 1 \ ,
\ee
che mappa i punti del cilindro nel disco unitario. In questa descrizione lo stato iniziale \`e il cerchio piccolo di raggio $e^{-2 \pi t}$, nel limite $t \to \infty$ il cerchio piccolo si riduce ad un punto.

Le considerazioni fatte valgono anche nel caso di stringhe aperte. Una stringa aperta propagandosi liberamente descrive un `nastro', ovvero una superficie bidimensionale con bordi, che pu\`o essere rappresentato come
\be
-2\pi t \leq Im(w) \leq 0 \	, \qquad 0 \leq Re(w) \leq \pi \ ,
\ee
dove $Im (w)=-2 \pi t$ \`e la sorgente e $Re(w)=0, \pi$ sono i bordi descritti dagli estremi della stringa. La mappa $z=-e^{-i w}$, 
\be
e^{-2\pi t}\leq |z| \leq 1 \ , \qquad Im(z) \geq 0 \ ,
\ee
manda il nastro nel semicerchio di raggio unitario posizionato all'origine nel semipiano superiore, a meno di un piccolo semicerchio di raggio $e^{-2\pi t}$ dato dallo stato iniziale. Nel limite $t \to \infty$ le sorgenti si riducono ad un punto sul bordo del semipiano.

Come si \`e visto, ogni sorgente di stringa diventa una ``perturbazione'' locale sul world-sheet. Ad ogni stringa entrante o uscente di momento D-dimensionale $k^\mu$ e stato interno $j$, si pu\`o far corrispondere un \emph{operatore di vertice} locale ${\cal V}_j(k)$. Il segno di $k^0$ distingue gli stati entranti da quelli uscenti, $k^\mu = \pm (E, k)$ dove il segno meno indica gli stati entranti.

Si \`e visto che l'integrale di Polyakov \`e una somma su tutte le superfici che collegano gli stati iniziali a quelli finali. Considerando solo superfici d'universo compatte con punture che rappresentino gli stati esterni, l'integrale funzionale porta naturalmente a definire gli elementi della matrice di interazione come
\be
\label{ampiezze1}
S_{j_1, \dots, j_n}(k_1, \dots, k_n)= \sum_{\stackrel{topologie}{compatte}}\int \frac{[dX \, dg]}{V_{diff\times Weyl}}e^{-S_X-\lambda \chi}\prod_{i=1}^n \int d^2\sigma_i g(\sigma_i)^{1/2}{\cal V}_{j_i}(k_i, \sigma_i) \ .
\ee
Gli operatori di vertice, che saranno discussi nei prossimi capitoli, sono integrati sul world-sheet per rendere i loro contributi invarianti sotto diffeomorfismi.

\begin{figure}
\begin{center}
\includegraphics{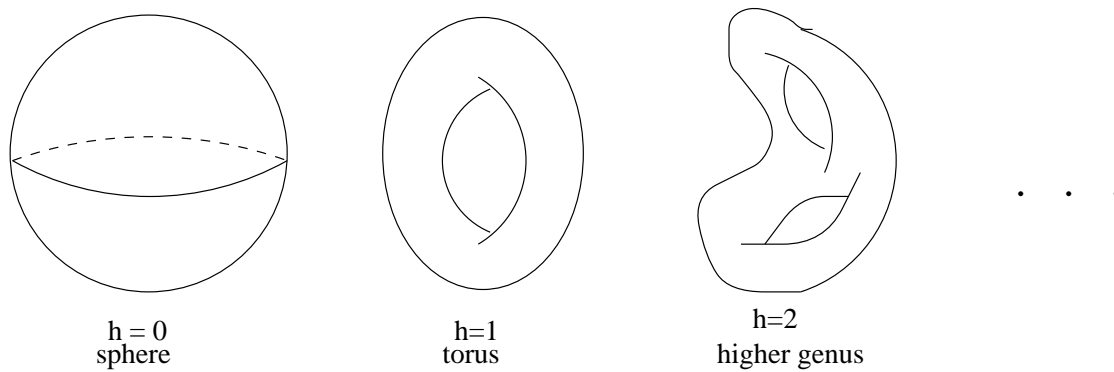}
\end{center}
\caption{Superfici orientate chiuse compatte connesse di genere 0, 1, 2}
\label{tori}
\end{figure}
I processi di interazione possono naturalmente coinvolgere anche topologie non connesse, ma questo tipo di ampiezze possono essere pensate come fattorizzazione di processi con topologie connesse: sar\`a quindi sufficiente limitarsi a questi ultimi. In due dimensioni la classificazione delle topologie compatte e connesse \`e ben nota. Ogni superficie compatta connessa orientata senza bordi \`e equivalente ad una sfera con $h$ manici (vedi figura \ref{tori}). Si \`e visto che il numero di Eulero $\chi$ determina il peso delle superfici di world-sheet nell'espansione di Polyakov. I modelli di sole stringhe chiuse hanno la particolarit\`a di avere un solo tipo di contributo ad ogni ordine della teoria perturbativa, con 
\begin{equation}
\label{Eulero1}
\chi=2 - 2h \ ,
\end{equation}
e il loro sviluppo perturbativo \`e una somma su superfici chiuse di Riemann con un numero crescente di manici \emph{h}.

Come si \`e detto, nelle teorie di stringhe aperte e chiuse non orientate lo sviluppo coinvolge anche superfici di Riemann non orientabili e con bordi che, quindi, possono contenere un numero variabile di due nuove strutture, bordi, \emph{b}, e crosscap, \emph{c}. In questo caso
\begin{equation}
\label{Eulero2}
\chi=2 - 2h - b - c \ ,
\end{equation}
e la serie perturbativa ora include potenze sia pari che dispari di $g_s$. Il genere \emph{g} di una superficie viene definito come:
\begin{equation}
g=h+\frac{b}{2}+\frac{c}{2} \ .
\end{equation}

\begin{figure}[ht!]
\begin{center}
\includegraphics{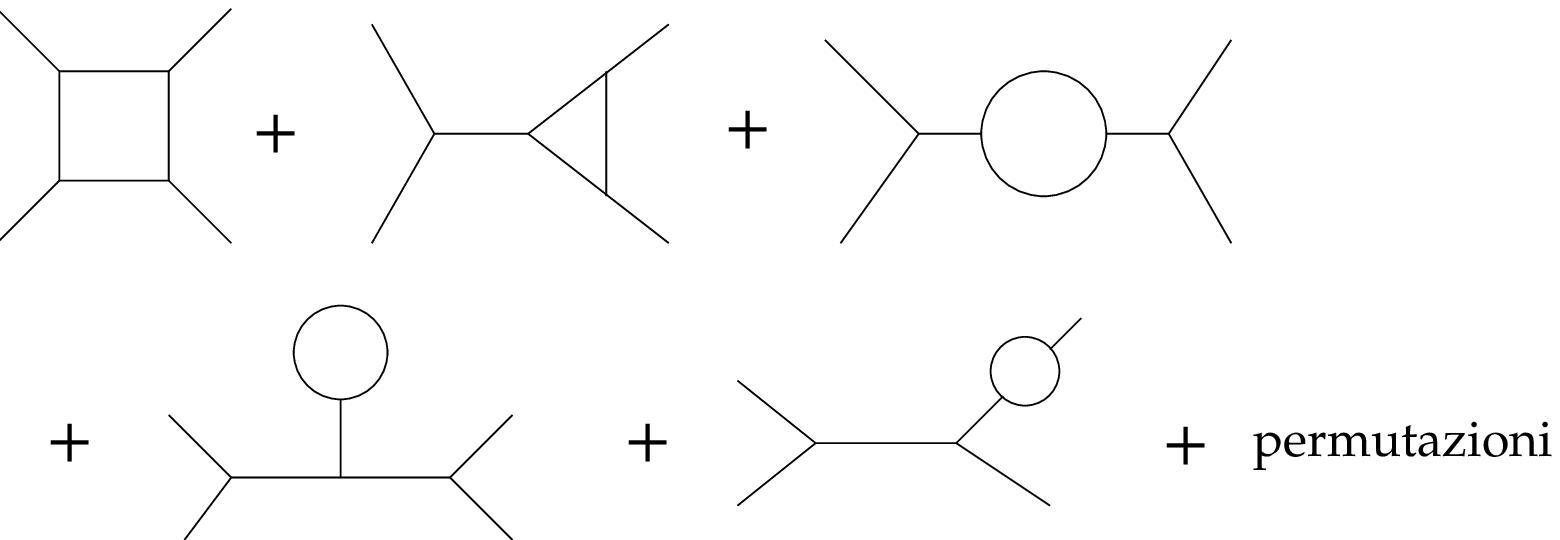}
\end{center}
\begin{center}
\includegraphics{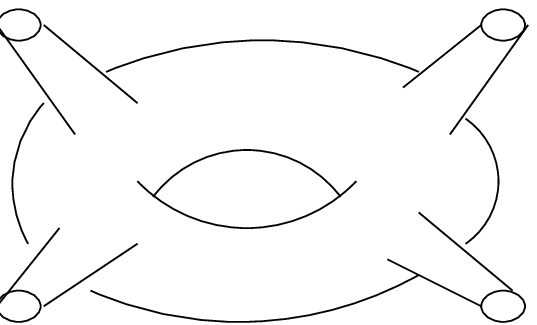}
\end{center}
\caption{(a) Grafici di Feynman all'ordine ad un loop in una teoria $\lambda \phi^3$ per interazioni a quattro particelle. (b) Grafico di interazione di quattro stringhe all'ordine ad un loop in una teoria di stringhe chiuse orientate. }
\label{diagrammi}
\end{figure}
Alla luce di queste osservazioni, emerge immediatamente la relativa semplicit\`a dell'espansione perturbativa della Teoria delle Stringhe. Il numero di distinte topologie \`e molto piccolo comparato al numero di grafici di Feynman distinti in Teoria dei Campi: un singolo grafico di di stringa contiene molti grafici di campo dell'ordine perturbativo corrispondente (si vedano ad esempio all'ordine ad un loop i grafici di campo e di stringa, vedi figura \ref{diagrammi}). Questa caratteristica \`e stata usata anche in tecniche di calcolo dei diagrammi di Teoria dei Campi.

\section{Spazio dei moduli}

Si \`e visto che l'ampiezze di stringa (\ref{ampiezze1}) sono date, in generale, da un integrale funzionale sullo \emph{spazio delle metriche}, ${\mathcal G}_r$, per una topologia $r$ del world-sheet. In una teoria di stringhe chiuse orientate, come si \`e visto, $r$ pu\`o essere pensato come il numero di manici $h$. Quozientando lo spazio delle metriche per il gruppo delle simmetrie di gauge, si ottiene lo \emph{spazio dei moduli},
\be
{\mathcal M}_r=\frac{{\mathcal G}_r}{(diff\times Weyl)_r} \ .
\ee
In generale lo spazio dei moduli \`e parametrizzato da un numero finito di \emph{moduli}. 

La scelta di una metrica non fissa completamente la gauge. Infatti pu\`o esistere un sottogruppo di $diff\times Weyl$ che lasci la metrica invariante, detto \emph{Gruppo Conforme di Killing} $(CKG)$. 

In presenza di operatori di vertice nell'integrale funzione, \`e utile considerare la loro posizione sulla superficie d'interazione come moduli della superficie, alla stessa stregua dei \emph{moduli della metrica}. In questo modo lo spazio dei moduli su cui si definisce l'integrale di Polyakov sar\`a dato da
\be
{\mathcal M}_{r, n}=\frac{{\mathcal G}_r \times {\mathcal M}^n}{(diff\times Weyl)_r} \ ,
\ee
dove ${\mathcal M}^n$ \`e lo spazio dei moduli che definscono le posizioni dei vertici di interazione. Il gruppo di simmetria residuo $CKG$, pu\`o essere fissato, assegnando ad un numero opportuno di vertici la posizione sul world-sheet. 

In generale, $diff_r$ non \`e connesso, il quoziente  del gruppo dei diffeomorfismi rispetto alla componente connessa che contiene l'identit\`a, 
\be
\frac{diff_r}{diff_{r0}} \ ,
\ee
\`e detto \emph{gruppo modulare}.

Riassumendo, per studiare in dettaglio lo spazio delle metriche in seguito al gauge fixing, occorre tener conto da un lato dei parametri della metrica che non possono essere rimossi dalle simmetrie (\emph{moduli}), e dall'altro delle simmetrie residue codificate nel gruppo di Killing conforme. 

Bisogna cercare le variazioni infinitesime della metrica che non siano equivalenti a trasformazioni di gauge $diff\times Weyl$ e quindi corrispondono a variazioni dei moduli. Occorre anche cercare le trasformazioni infinitesime di $diff\times Weyl$ che non cambino la metrica, che definiscono i vettori conformi di Killing $CKV$, elementi di $CKG$.

Una trasformazione infinitesima del gruppo di gauge si \`e vista essere
\be
\label{infgauge}
\delta g_{ab}=-2(P_1\delta \sigma)_{ab}+(2 \delta \omega - \nabla \cdot \delta \sigma)g_{ab} \ ,
\ee
dove l'operatore simmetrico a traccia nulla $P_1$ \`e stato definito nella (\ref{P1}). I moduli corrispondono a variazioni della metrica $\delta' g_{ab}$ che siano ortogonali a tutte le variazioni della forma (\ref{infgauge}),
\ba
\label{overlap}
0 &=& \int d^2 \sigma g^{1/2}\delta'g_{ab}\left[-2(P_1\delta \sigma)^{ab}+ (2 \delta \omega - \nabla \cdot \delta \sigma)g^{ab} \right]\nonumber \\
&=& \int d^2 \sigma g^{1/2} \left[-2(P_1^T\delta'g)_a\delta \sigma^a+ \delta'g_{ab}g^{ab}(2 \delta \omega - \nabla \cdot \delta \sigma)g^{ab} \right] \ .
\ea
L'operatore trasposto $P_1^T=-\nabla^bu_{ab}$, che manda tensori a due indici a traccia nulla in vettori, \`e definito come
\be
\label{prodscal}
(u, P_1 v)=\int d^2 \sigma g^{1/2} u \cdot P_1 v = (P_1^Tu, v)=  \int d^2 \sigma g^{1/2} P_1^T u \cdot v \ .
\ee
Perch\`e l'integrale (\ref{overlap}) sia nullo per $\delta \omega$ e $\delta \sigma$ arbitrari si deve avere
\ba
\label{modeq}
g^{ab}\delta' g_{ab}&=&0 \ ,\nonumber \\
(P_1^T \delta' g)_a &=& 0 \ .
\ea
La prima condizione impone che $g'_{ab}$ sia a traccia nulla. Per ogni soluzione distinta di queste equazioni si avr\`a un modulo della metrica.

I CKV sono trasformazioni (\ref{infgauge}) tali che $\delta g_{ab}=0$. Prendendo la traccia di questa equazione si fissa $\delta \omega$. L'equazione per i vettori conformi di Killing si riduce a
\be
\label{CKV}
(P_1\delta \sigma)_{ab}=0 \ .
\ee
Ponendosi in gauge conforme le equazioni (\ref{modeq})  e (\ref{CKV}) prendono forma semplice
\ba
\label{modeqCKV}
\partial_{\bar z}\delta' g_{zz} &=& \partial_{z}\delta' g_{\bar z \bar z}= 0 \ , \nonumber \\
\partial_{\bar z} \delta z &=& \partial_{z} \delta \bar z = 0 \ ,
\ea
le variazioni dei moduli corrispondono a differenziali olomorfi quadratici e i $CKV$ a campi vettoriali olomorfi. 

I moduli metrici corrispondono al kernel di $P_1^T$, e i CKV al kernel di $P_1$. Il teorema di Riemann-Roch collega il numero di moduli (reali) della metrica $\mu = \dim ker P_1^T$, il numero di vettori di Killing conformi $\kappa = \dim ker P_1$, e il numero di Eulero $\chi$ (\ref{Eulero1}, \ref{Eulero2}),
\be
\label{RimRoch}
\mu - \kappa = -3 \chi \ .
\ee 
Con una trasformazione di Weyl \`e sempre possibile porsi in una metrica in cui $R$ sia costante. In questo modo il segno di $R$ determina il segno di $\chi$ (\ref{chi}). Dal momento che $P_1^TP_1=-\frac{1}{2}\nabla^2-\frac{1}{4}R$, si ha 
\ba
\int d^2 \sigma g^{1/2} (P_1\delta \sigma)_{ab}(P_1\delta \sigma)^{ab} &=&  \int d^2 \sigma g^{1/2} \delta \sigma_a (P_1^TP_1 \delta \sigma)^a \nonumber \\
&=& \int d^2 \sigma g^{1/2} \left(\frac{1}{2}\nabla_a \delta \sigma_b \nabla^a\delta \sigma^b-\frac{R}{4}\delta \sigma_a\delta \sigma^a \right) \ .
\ea
Per $\chi$ negativo, il termine di destra dell'equazione \`e strettamente positivo, quindi $P_1\delta \sigma$ non pu\`o essere mai nullo. In maniera simile si pu\`o dimostrare che $P_1^T\delta' \sigma$ non pu\`o essere mai nullo per superfici con $\chi$ positivo. Riassumendo si \`e trovato che
\ba
\chi > 0 &:& \qquad \kappa = 3 \chi \ , \qquad \mu =0 \ , \nonumber \\
\chi < 0 &:& \qquad \kappa = 0  \ , \qquad \mu =-3 \chi\ .
\ea

\section{Superfici di Riemann}

\subsection{Superfici di Riemann e variet\`a Riemanniane}

Data una variet\`a \`e possibile definirne un ricoprimento con un insieme di aperti che si sovrappongano. All'interno degli aperti definiamo delle coordinate $\sigma_m^a$, dove $m$ indicizza gli aperti e $a$ le dimensioni della variet\`a. Quando due aperti $m$ e $n$ si sovrappongono, le rispettive coordinate saranno legate da \emph{funzioni di trasizione}, $f_{mn}$, differenziabili
\be
\sigma_m^a=f_{mn}^a(\sigma_n) \ .
\ee
Per una variet\`a Riemanniana saranno definite metriche $g_{m, ab}(\sigma_m)$ in ogni aperto, collegate, nelle zone di sovrapposizione degli aperti, dalla legge di trasformazione dei tensori.

Per una variet\`a complessa, si avranno coordinate complesse $z_m^a$ in ogni aperto. In questo caso l'indice $a$ prende i valori da uno a $d/2$, dove $d$ \`e la dimensione della variet\`a. Le funzioni di transizione dovranno essere funzioni olomorfe,
\be
z_m^a=f_{mn}^a(z_n) \ .
\ee
\`E possibile definire funzioni olomorfe sulla variet\`a dal momento che questa propriet\`a non dipende dalle coordinate $z_m^a$ usate. Due variet\`a complesse sono equivalenti se esiste una mappa olomofa biunivoca fra loro. Un cambio di coordinate olomorfo in ogni aperto definisce per questo una superficie equivalente. 

Nel caso di una dimensione complessa (due dimensioni reali), le variet\`a differenziali sono dette \emph{superfici di Riemann}, e si ha la corrispondenza
\be
superfici\, di\, Riemann \longleftrightarrow varieta'\, Riemanniane\, mod\, Weyl \ .
\ee
Per verificare questo isomofismo, si pu\`o partire da una variet\`a Rimanniana e porsi, come gi\`a visto per la gauge conforme, in ogni aperto, in sistemi di coordinate $z_m$ tali che
\be
\label{metrolo}
ds^2 \propto dz_md \bar z_m \ .
\ee
Le metriche di aperti vicini non devono essere necessariamente uguali, ma le funzioni di transizione devono essere olomorfe per preservare la forma delle metriche (\ref{metrolo}). Questo prova la prima parte dell'isomorfismo. Per la mappa inversa, si pu\`o scegliere una metrica $dz_md \bar z_m$ nell'aperto $m$-esimo congiungendola con continuit\`a alle metriche degli aperti sovrapposti per ottenere una variet\`a Riemanniana.

\subsection{Alcune superfici di Riemann}

Vediamo in dettaglio le superfici di genere pi\`u basso che, come si vedr\`a sono cruciali nella determinazione delle ampiezza ai primi ordini della teoria perturbativa di stringa. 

La sfera $\chi=2$, $g=0$, il disco e il piano proiettivo $(\chi=1,g=1/2)$ sono le uniche superfici di Riemann con numero di Eulero positivo. 
 
\subsubsection{La sfera}

La Sfera $S_2$ pu\`o essere ricoperta con due aperti coordinati che si sovrappongano. Consideriamo due dischi $|z|<\rho$ e $|u|<\rho$ con $\rho>1$ uniti tramite l'identificazione $u=1/z$. Nel limite $\rho \to \infty$, la coordinata $z$ \`e ben definita ovunque tranne che per $u=0$. Si pu\`o quindi lavorare nel disco $z$, ricordandosi di verificare che nel punto di singolarit\`a tutto vada come deve.

La sfera quindi pu\`o essere studiata come una superficie piatta, dotandola di metriche piatte nei due aperti connesse da trasformazioni conformi (trasformazioni di gauge della simmetria residua). Una metrica conforme, come si \`e visto, \`e della forma
\be
ds^2=e^{2\omega (z, \bar z)}dz d\bar z \ ,
\ee
Differenziando la relazione di identificazione, si ha $dzd\bar z=|z|^4 dud\bar u$, la condizione di nonsingolarit\`a della metrica in $u=0$ \`e che il fattore esponenziale sia almeno dell'ordine $|z|^{-4}$ nel limite $z \to \infty$. 

Ricordando i risultati precedentemente derivati, la sfera non ha moduli ma ha 6 vettori di Killing conformi. Le metriche saranno quindi tutte equivalenti a meno di diffeomorfismi e di trasformazioni di Weyl. Le equazioni infinitesime (\ref{modeqCKV}), nel caso della caso della sfera, devono valere in entrambi gli aperti coordinati. Dal momento che 
\ba
\delta u &=& \frac{\partial u}{\partial z}\delta z = - z^{-2}\delta z \ , \nonumber \\
\delta g_{uu} &=& \left(\frac{\partial u}{\partial z}\right)^{-2}\delta g_{zz}=z^4 \delta g_{zz} \ .
\ea
I differenziali quadratici olomorfi $\delta g_{zz}$, devono essere olomofi in $z$ per le (\ref{modeqCKV}), e andare come $|z|^{-4}$ all'infinito, in modo da verificare le equazioni anche nell'aperto $u$. Allo stesso modo i CKV devono crescere per $z \to \infty$ al pi\`u come $z^2$. La soluzione per $\delta z$ \`e della forma
\ba
\delta z &=& a_0+a_1z+a_2z^2 \ , \nonumber \\
\delta \bar z &=& a^\ast_0+a^\ast_1\bar z+a^\ast_2\bar z^2 \ ,
\ea
dove gli $a_i$, sono i sei parametri reali (tre complessi), previsti dal teorema di Rimann-Roch. In queste trasformazioni si pu\`o riconoscere la forma infinitesima del gruppo $PSL(2, \mathbb C)$. Una trasformazione generale di questo gruppo \`e della forma
\be
\label{MoebiusGroup}
z \to z'=\frac{\alpha z+ \beta}{\gamma z + \delta} \ ,
\ee
con $\alpha, \beta, \delta, \gamma$ complessi, tali che $\alpha \delta - \beta \gamma = 1$. Naturalmente cambiando segno contemporaneamente a tutti i parametri complessi, la trasformazione \`e invariata.

\subsubsection{Il disco}

Ai fini della derivazione delle ampiezze \`e utile costruire il disco $D_2$ identificando i punti della sfera sotto riflessione. La relazione di riflessione pi\`u semplice \`e
\be
z'=\bar z \ ,
\ee
che permette di considerare come regione fondamentale il semipiano superiore. L'asse reale \`e il luogo dei punti fissi rispetto alla relazione di riflessione e diventa quindi il bordo del disco. Una relazione di riflessione pi\`u complessa \`e 
\be
z'=\frac{1}{\bar z} \ .
\ee
Considerando coordinate polari $z=re^{i\phi}$, si comprende che questa riflessione inverte il raggio lasciano l'angolo fissato. La regione fondamentale \`e il disco unitario.

Anche il disco non ha moduli. I vettori di Killing Conformi saranno il sottogruppo delle trasformazioni infinitesime di $PSL(2, \mathbb C)$, che lascino il bordo invariato. Queste trasformazioni, ricordando la relazione di riflessione, saranno della forma (\ref{MoebiusGroup}) ma con parametri reali. 

\subsubsection{Il Piano Proiettivo}

Il piano proiettivo $RP_2$ pu\`o essere pensato come un'identificazione della sfera sotto $\mathbb Z_2$. L'identificazione antipodale
\be
\label{antipod}
z'=-\frac{1}{\bar z} \ ,
\ee
non ha punti fissi e quindi non definisce bordi. Lo spazio risulta non orientato. La regione fondamentale pu\`o essere fissata sia nel semipiano superiore, sia nel disco unitario $|z|\leq 1$. La linea definita dall'equatore \`e detta crosscap ed \`e responsabile della non orientabilit\`a della superficie. Il piano proiettivo reale non ha  moduli, mentre i CKV sono il sottogruppo di $PSL(2, \mathbb C)$ che rispetta l'identificazione (\ref{antipod}), e corrisponde al gruppo delle rotazioni $SO(3)$.

Studiamo ora le superfici con $\chi=0$, $g=1$ che, come si vedr\`a, sono legate alle ampiezze di vuoto in teoria delle stringhe. Le possibili superfici sono il \emph{toro} ($h=1, c=0, b=0$), la \emph{bottiglia di Klein} ($h=0, c=0, b=2$), l'anello ($h=0, c=0, b=2$) e il \emph{nastro di M$\ddot{o}$bius} ($h=0, c=1, b=1$). In generale queste superfici possono essere aperte su un piano con un numero opportuno di tagli. In particolare, per superfici con numero di Eulero nullo \`e possibile definire una metrica euclidea sul piano. 

\subsubsection{Il toro}

Il toro $T^2$ \`e una superficie chiusa orientabile e pu\`o essere immaginato come un cilindo i cui estremi vengano sovrapposti. Nel piano complesso il toro \`e definito dalle identificazioni
\begin{equation}
z\sim z +n\lambda_1 + m\lambda_2,
\end{equation}
Dove $\lambda_1$ e $\lambda_2$ sono vettori nel piano complesso e $m$ e $n$ son due numeri interi. Queste identificazioni definiscono un reticolo la cui cella fondamentale \`e un parallelogramma con i lati opposti identificati, e con un opportuno riscalamento si pu\`o sempre scegliere il lato orizzontale di lunghezza unitaria. Il toro \`e quindi definito dall'assegnazione di un unico parametro complesso $\tau=\tau_1 + i\tau_2$ con parte immaginaria positiva $\tau_2$, uguale al rapporto fra il lato obliquo e quello orizzontale della cella fondamentale, che viene detto \emph{parametro di Teichm\"uller}, o modulo del toro. Non tutti i moduli $\tau$ definiscono tori inequivalenti. \`E infatti possibile ridefinire la cella fondamentale, ottendo tori equivalenti, traslando il lato orizzontale superiore di multipli della lunghezza orizzontale, o scambiando il lato orizzontale e quello obliquo della cella. Queste due operazioni sono rispettivamente generate dalle trasformazioni 
\be
T \ : \ \tau \rightarrow \tau+1 \ , \qquad \qquad S \ : \ \tau \rightarrow -\frac{1}{\tau} \ .
\ee
T e S sono i generatori del gruppo modulare $PSL(2,\mathbb{Z})=SL(2,\mathbb{Z})/\mathbb{Z}_{2}$, la cui azione su $\tau$ \`e 
\begin{equation} 
\label{mod}
\tau \rightarrow \tau' = \frac{a\tau + b}{c\tau + d}
\qquad con \qquad ad-bc=1 \qquad
e \qquad a, b, c, d \in \mathbb{Z} \ ,
\end{equation}
dove il quoziente $\mathbb{Z}_2$ \`e dovuto al fatto che invertendo i segni di $a, b, c$ e $d$ $\tau$ \`e invariante. Queste trasformazioni possono essere viste come `grandi' diffeomorfismi sul toro, scrivendole nella forma
\ba
\left[\begin{array}{rc}
\sigma^1 \\
\sigma^2 \\
\end{array} \right] \,
= \ \left[ \begin{array}{rr}
d & b  \\
c & a  \\
\end{array} \right] 
\left[\begin{array}{rc}
\sigma'^1 \\
\sigma'^2 \\
\end{array} \right]
\, .
\ea

\begin{figure}
\begin{center}
\epsfbox{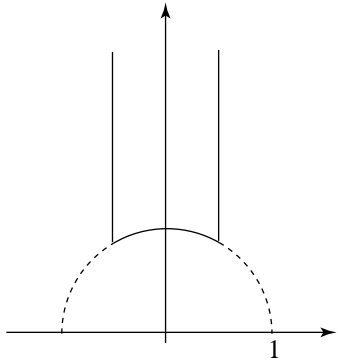}
\end{center}
\caption{Regione fondamentale del toro.}
\label{fig3}
\end{figure}

Usando le trasformazioni modulari (\ref{mod}), si pu\`o mostrare che ogni $\tau$ \`e equivalente ad un solo punto nella regione 
\begin{equation}
\mathcal{F}=\{-\frac{1}{2}<\tau_1\leq\frac{1}{2} , \
\mid\tau\mid \ \geq 1\}
\end{equation} 
i cui bordi sono identificati come in figura. La regione del $\tau$-piano $\mathcal{F}$ \`e detta \emph{regione fondamentale} dello spazio dei moduli per il toro.

Per concludere questa breve presentazione del toro occorre ricordare la presenza di due vettori di Killing conformi che corrispondono a traslazioni rigide sulla superficie bidimensionale 
\be
\sigma^a \to \sigma^a + v^a \ .
\ee
Oltre a questo sottogruppo del gruppo $diff\times Weyl$, anche le trasformazioni discrete 
\be
\sigma^a \to - \sigma^a
\ee
lasciano la metrica sul toro invariata.

\begin{figure}
\begin{center}
\includegraphics[width=10cm,height=9cm]{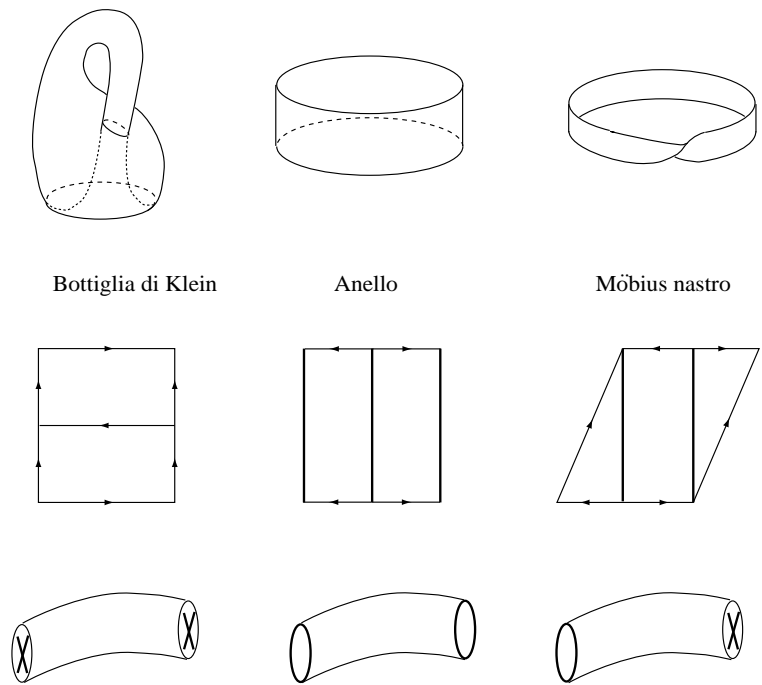}
\end{center}
\caption{bottiglia di Klein, anello e nastro di M\"{o}bius}
\label{figsurfs}
\end{figure}

\subsubsection{La bottiglia di Klein}

La bottiglia di Klein $K_2$ \`e una superficie chiusa non orientabile, e pu\`o essere vista come un cilindro i cui estremi siano congiunti dopo una trasformazione di parit\`a $\Omega$. Corrisponde al piano complesso con le identificazioni
\begin{equation}
z\sim z + n \sim -\overline{z} + it,
\end{equation}
dove l'unico modulo $t$ \`e definito sull'intevallo $[0, \infty]$. $K_2$ non ha alcun ricoprimento nello spazio tridimensionale Euclideo privo di auto-intersezioni, ma \`e ottenibile dal toro doppiamente ricoprente di modulo puramente immaginario $\tau= 2it$ con l'identificazione
\begin{equation}
z'=-\bar{z} + it \ .
\end{equation}
C'\`e una seconda possibilit\`a nella scelta del poligono fondamentale, ottenuta dimezzando il lato orizzontale e raddoppiando quello verticale, lasciando quindi l'area della cella inalterata. Il risultato \`e una rappresentazione equivalente di $K_2$ come un tubo che termina su due crosscap. Il tubo \`e dato dalla regione intena al poligono i cui lati orizzontali hanno ora la stessa orientazione, mentre i crosscap sono i due lati verticali, dove i punti che differiscono per traslazioni di met\`a della lunghezza dei lati sono identificati a coppie.

\subsubsection{L'anello} 

L'anello $C_2$ \`e una superficie orientabile con due bordi. Nel piano complesso \`e la regione
\begin{equation}
0 \leq Re (z) \leq 1 \ , \qquad z\sim z+it \ ,
\end{equation}
una striscia di larghezza $1$ e lunghezza $it$ con i lati orizzontali identificati. Il modulo $t$ \`e definito sull'intevallo $[0, \infty]$, e anche in questo caso non c'\`e invarianza modulare. Il ricoprimento doppio del cilindro \`e il toro di modulo puramente immaginario $\tau=it$, da cui esso pu\`o essere ottenuto con l'identificazione
\begin{equation}
z'=-\bar{z} \ ,
\end{equation}
una riflessione rispetto all'asse immaginario. Le linee $\sigma^1=0,1$ sono fisse rispetto all'involuzione e corrispondono ai bordi.

\subsubsection{Il nastro di M\"obius}
L'ultima superficie con $g=1$ \`e il nastro di M\"{o}bius $M_2$, una superficie non orientabile con un bordo che pu\`o essere vista come una striscia chiusa con un rivolgimento indotto da una parit\`a $\Omega$,
\begin{equation}
\label{mob}
0 \leq Re (z) \leq 1, \qquad z\sim -\bar{z} +1 + it \ .
\end{equation}
Si noti che in questo caso i due lati verticali della cella fondamentale descrivono differenti porzioni di un unico bordo. Anche per questa superficie \`e possibile ottenere una rappresentazione diversa, molto importante, con una ridefinizione della cella fondamentale. Raddoppiando il lato verticale e dimezzando quello orizzontale, $M_2$ \`e infatti rappresentabile come un tubo fra un bordo e un crosscap. Infatti nel nuovo poligono fondamentale uno dei lati verticali corrisponde all'unico bordo del nastro di M\"{o}bius, mentre nell'altro i punti sono identificati a coppie dopo la traslazione verticale tramite l'involuzione (\ref{mob}), ed \`e quindi un crosscap. \`E importante osservare che in questo caso, a differenza dei precedenti, si ha un toro doppiamente ricoprente ma con un modulo non puramente immaginario
\begin{equation}
\tau = \frac{1}{2}+\frac{1}{2}i\tau_2
\end{equation}

Per superfici di genere pi\`u alto vale una relazione di equivalenza: tre crosscap possono essere sostituiti con un manico e un crosscap. Questo limita l'espansione di Polyakov a superfici con numero arbitrario di manici $h$ e di bordi $b$, ma solo con 0,1 o 2 crosscap $c$.

\subsection{Descrizioni equivalenti dello spazio dei moduli}

La descrizione pi\`u naturale dello spazio dei moduli si ha scegliendo una sezione dello spazio delle metriche in modo da avere una sola metrica per ogni classe di equivalenza. Si definisce in questo modo una famiglia di metriche $\hat g_{ab}(t, \sigma)$ parametrizzate dai moduli $t^k$. Una descrizione equivalente  si ha considerando la metrica fissata e codificando i moduli nalla regione delle coordinate. 

Nel caso del toro si pu\`o definire la regione di coordinate
\be
0 \leq \sigma^1 \leq 1 \ , \qquad 0 \leq \sigma^2 \leq 1 \ , 
\ee
e quindi
\be
\label{periodic}
(\sigma^1,\sigma^2) \sim (\sigma^1,\sigma^2) + (m, n) \ .
\ee
In questo sitema di coordinate non \`e possibile fissare una metrica unitaria preservando le condizioni di periodicit\`a (\ref{periodic}), ma \`e possibile sceglire la metrica della forma
\be
ds^2=|d\sigma^1+ \tau d\sigma^2|^2 \ ,
\ee
dove $\tau$ \`e una costante complessa. 

Possiamo passare ad una descrizione equivalente in cui la metrica sia unitaria definendo nuove coordinate
\be
\tilde \sigma^a \sim \tilde \sigma^a + (m u^a + n v^a) \ ,
\ee
con periodicit\`a definita dai due vettori arbitrari $u^a$ e $v^a$.  Possiamo fissare $u=(1, 0)$ ruotando e riscalando il sistema di coordinate. Rimangono non fissati i due parametri di $v$. Definendo $w=\tilde \sigma^1 + i\tilde \sigma^2$, la metrica \`e $ds^2 = dw d\bar w$, e la periodicit\`a \`e
\be
w \sim w + (m+n\tau) \ ,
\ee
con $\tau= v^1+iv^2$. Naturalmente per $w=\sigma^1+\tau \sigma^2$ si ritorna alla descrizione (\ref{periodic}). La descrizione del toro in termini di coordinate complesse $w$ illustra l'idea di una variet\`a complessa. Si pensi ad esempio di segliere un solo aperto un po' pi\`u largo del poligono fondamentale del toro. In questo modo le condizioni di periodicit\`a sono le funzioni di transizione nelle zone di sovrapposizioni fra i bordi opposti dell'aperto. 

Nel definire l'integrale funzionale sulle metriche \`e pi\`u semplice considerare la metrica come una funzione di coordinate fissate come nel caso (\ref{periodic}). Al contrario per studiare una teoria quantistica di campo su una data superficie \`e pi\`u semplice adottare una descrizione in cui la metrica sia unitaria con funzioni di transizione dipendenti dai moduli.


\chapter{Ampiezze di stringa bosonica}

Partendo dall'idea della somma sulle ``storie'' di stringa, si \`e giunti a definire la formula (\ref{ampiezze1}) per le ampiezze di interazione. Per studiare in dettaglio le ampiezze di stringa per una data topologia, e quindi per un dato ordine perturbativo, la strategia \`e quella di ridurre l'integrale funzionale sulle metriche ad un integrale sui moduli che descrivono lo spazio delle metriche inequivalenti e sulle orbite di del gruppo di gauge $diff\times Weyl$. Questo si ottiene scegliendo un gauge e utilizzando la procedura di Faddeev-Popov.

Si pu\`o dimostrare che per uno spazio tempo piatto euclideo, in dimensione critica, le ampiezze si riducono a integrali ben definiti sullo spazio dei moduli. In particolare, in questo capitolo si svilupper\`a il formalismo funzionale per il calcolo della ampiezze e si calcoleranno esplicitamente le ampiezze all'ordine ad albero, e ad un loop per la stringa bosonica.

\section{Ampiezze come integrali sullo spazio dei moduli}

Data una teoria conforme con generici campi di materia $\phi_i$ (per i quali $c=\tilde c=26$), l'integrale di Polyakov euclideo per la matrice S \`e
\be
\label{ampiezze2}
S_{j_1, \dots, j_n}(k_1, \dots, k_n)= \sum_{\chi}\int \frac{[d\phi \, dg]}{V_{diff\times Weyl}}e^{-S_m-\lambda \chi}\prod_{i=1}^n \int d^2\sigma_i g(\sigma_i)^{1/2}{\cal V}_{j_i}(k_i, \sigma_i) \ ,
\ee
Dove la somma \`e sulle diverse caratteristiche di Eulero delle superfici. Vogliamo ora mostrare come, con un'opportuna scelta del gauge, l'integrale funzionale sulle metriche non viene completamente eliminato, ma si riduce ad un integrale finito-dimensionale sullo spazio dei moduli della superficie. In particolare, vogliamo mostrare  che l'integrale sulle metriche e sulle posizioni dei vertici diventa un integrale sul gruppo di gauge, sui moduli e sulle posizioni (di una parte) dei vertici,
\be
[dg]d^{2n}\sigma \to [d\zeta]d^\mu t d^{2n-\kappa}\sigma \ .
\ee
Dopo aver fattorizzato il volume del gruppo di gauge, lo Jacobiano per questa trasformazione fornisce precisamente la misura sullo spazio dei moduli. 

Il determinante di Faddeev-Popov pu\`o essere definito pi\`u precisamente come
\be
\label{FP2}
1= \Delta_{FP}(g, \sigma)\int_F d^\mu t\int_{diff\times Weyl}[d\zeta]\delta(g-\hat g(t)^\zeta)\prod_{(a, i)\in f}\delta \left(\sigma_i^a-\hat \sigma_i^{\zeta a} \right) \ .
\ee
Ogni metrica \`e genericamente equivalente, a meno di una trasformazione $Diff\times Weyl$, alla metrica di riferimento $\hat g(t)$, per un solo valore di $t$ e $\zeta$, o, nel caso in cui siano presenti simmetrie residue discrete, per un numero finito $n_R$ di valori distinti.

Inserendo l'espressione (\ref{FP2}) in (\ref{ampiezze2}) e invertendo l'ordine delle integrazioni si ottiene
\ba
S_{j_1, \dots, j_n}(k_1, \dots, k_n)= \sum_{\stackrel{topologie}{compatte}}\int_F d^\mu t \Delta_{FP}(\hat g(t), \hat \sigma)\int [d\phi]\int \prod_{(a, i)\notin f} d\sigma_i^a \nonumber \\
\times e^{-S_m[\phi, \hat g(t)]-\lambda \chi} \prod_{i=1}^n \left[ \hat g(\sigma_i)^{1/2}{\mathcal V}_{j_i}(k_i; \sigma_i) \right] \ .
\ea
L'integrale \`e ora sullo spazio dei moduli $F$ e sulle coordinate dei vertici rimaste non fissate. Le simmetrie di gauge residue, legate al $CKG$, sono state utilizzate per fissare $\kappa$ posizioni dei vertici di interazione. 

Valutiamo ora il determinante $\Delta_{FP}$. Le funzioni $\delta$, come si \`e osservato possono essere diverse da zero in un numero finito $n_R$ di punti legati da simmetrie discrete residue. Per calcolare la misura di Faddeev-Popov, si pu\`o espandere nell'intorno di uno di questi punti, tenendo conto eventualmente della simmetria discreta e dividendo per $n_R$. La variazione generale della metrica \`e una variazione di gauge pi\`u una variazione nel modulo $t^\mu$,
\be
\delta g_{ab}= \sum_{k=1}^{\mu}\delta t^k\partial_{t^k}\hat g_{ab}-2\hat g_{ab}-2(\hat P_1 \delta \sigma)_{ab}+(2\delta \omega-\hat \nabla \cdot \delta \sigma)\hat g_{ab} \ .
\ee
Scrivendo le funzioni e i funzionali $\delta$ in forma esponenziale come integrali su $\sigma^a$ e su $\beta_{ab}$, e  integrando quindi su $\delta \omega$ per ottenere il tensore $\beta'_{ab}$ a traccia nulla per il determinate di Faddeev-Popov si ottiene l'espressione
\ba
\Delta_{FP}(\hat g,\hat \sigma)^{-1} = &n_R&\int d^\mu \delta t [d \delta \sigma \, d \delta \omega]\prod_{(a, i)\in f}\delta \left(\delta \sigma^a(\hat \sigma_i) \right) \nonumber \\
=&n_R& \int d^\mu \delta t d^\kappa x [d \beta' \, d\delta \sigma] \nonumber \\
&\times & \exp{\left[ 2\pi i (\beta', 2\hat P_1 \delta \sigma-\delta t^k \partial_k \hat g)+ 2\pi i \sum_{(a, i)\in f}x_{ai}\delta \sigma^a(\hat \sigma_i)\right]} \ .
\ea
Si possono ora sostituire le variabili bosoniche con variabili grassmaniane per invertire il determinante,
\ba
\delta \sigma &\to & c^a \, \nonumber \\
\beta'^{ab} &\to & b_{ab} \, \nonumber \\
x_{ai} &\to & \eta_{ai} \, \nonumber \\
\delta t^k &\to & \xi^k \ . 
\ea
Scegliendo una normalizzazione opportuna per i nuovi campi, ottiene infine
\ba
\Delta_{FP}(\hat g,\hat \sigma)&=& \frac{1}{n_R} \int [db\, dc]d^\mu \xi d^\kappa \eta \nonumber \\
&\,& \times exp{\left[-\frac{1}{4\pi}(b, 2\hat P_1c-\xi^k \partial_k \hat g)+\sum_{(a, i)\in f}\eta_{ai}c^a(\hat \sigma_i)\right]} \nonumber \\
&=& \frac{1}{n_R}\int [db\, dc] e^{-S_g}\prod_{k=1}^\mu \frac{1}{4\pi}(b,\partial_k \hat g)\prod_{(a,i)\in f}c^a(\hat \sigma_i) \ .
\ea
dopo aver effettuato l'integrazione sulle variabili grassmaniane $\eta_{ai}$ e $\xi^k$. 

Nel sostituire l'espressione trovata nell'integrale funzionale si pu\`o fissare il segno complessivo in modo tale da ottenere un risultato positivo. L'espressione finale per la matrice S \`e pertanto
\ba
\label{ampiezze3}
S_{j_1, \dots, j_n}(k_1, \dots, k_n) &=& \sum_{\stackrel{topologie}{compatte}}\int_F \frac{d^\mu t}{n_R}\int [d\phi \, db \, dc]e^{-S_m-S_g-\lambda \chi} \nonumber \\
&\times & \prod_{(a, i)\notin f}\int d\sigma_i^a \prod_{k=1}^\mu \frac{1}{4\pi}(b, \partial_k \hat g)\prod_{(a, i)\in f}c^a(\hat \sigma_i)\prod_{i=1}^n \hat g(\sigma_a)^{1/2}{\mathcal V}_{j_i}(k_i, \sigma_i) \ . \nonumber\\ 
\,
\ea
Questo risultato pu\`o essere esteso a tutte le costruzioni di stringa bosonica, estendendo eventualmente la classe delle superfici incluse e variando il tipo di vertici permessi. \`E importante notare che ogni coodinata fissata utilizzando le simmetrie di gauge residue associate ai $CKV$, l'integrazione sulla posizione di un vertice venga sostituita con un fattore $c_i^a$, mentre ogni modulo metrico da luogo ad un inserzione di $b$.

\subsection{Calcolo della misura di Faddeev-Popov}

L'espressione (\ref{ampiezze3}) per le ampiezze di interazione pu\`o essere ulteriormente ridotta, calcolando direttamente il termine di Faddeev-Popov nell'integrale funzionale. 

Definiamo una base completa di funzioni su cui espandere i campi di ghost. Ricordando le definizioni del prodotto scalare (\ref{prodscal}), l'azione di ghost pu\`o essere scritta in forme equivalenti,
\be
S_g=\frac{1}{2\pi}\left(b, P_1c\right)=\frac{1}{2\pi}\left(P_1^Tb, c\right) \ .
\ee
Dal momento che l'operatore $P_1$ manda vettori in tensori a due indici, non \`e possibile diagonalizzarlo. Si possono per\`o diagonalizzare $P_1^TP_1$ e $P_1P_1^T$,
\be
P_1^T P_1 C_J^a={\nu'}_J^2 C_J^a \ , \qquad P_1 P_1^T B_{Kab}=\nu_K^2B_{Kab} \ .
\ee
Le autofunzioni corrispondenti possono essere scelte reali e normalizzate in relazione ai rispettivi prodotti scalari,
\ba
(C_J, C_J') &=& \int d^2 \sigma g^{1/2}C_J^aC_{J'a}= \delta_{JJ'} \ , \nonumber \\
(B_K, B_K') &=& \int d^2 \sigma g^{1/2}B_{Kab}B_{K'}^{ab} = \delta_{KK'} \ .
\ea
Si pu\`o inoltre osservare che
\be
(P_1 P_1^T)P_1 C_J= P_1(P_1^T P_1)C_J = {\nu'}_J^2 P_1 C_J \ ,
\ee
e questo comporta che $P_1C_J$ \`e una autofunzione di $P_1P_1^T$. Nello stesso modo si dimostra che $P_1^TB_K$ \`e una autofunzione di $P_1^TP_1$. Esiste quindi una corrispondenza biunivoca fra le autofunzioni dei due operatori, eccetto nei casi $P_1C_J=0$ e $P_1^TB_K=0$, ovvero per gli autovalori nulli degli operatori $P_1^TP_1$ e $P_1P_1^T$. Questi autovalori sono i $\kappa$ vettori di Killing conformi e i $\mu$ differenziali quadratici olomorfi. Indichiamo le autofunzioni di autovalore nullo come $C_{0j}$ e $B_{0k}$, e le restanti autofunzioni con $C_J$, $B_K$ per $J, K =1, 2, \dots \,$. Le autofunzioni corrispondenti ad autovalori non nulli sono quindi legate dalle relazioni
\be
B_{Jab}=\frac{1}{\nu_J}(P_1C_J)_{ab} \ , \qquad \nu_J= \nu_J' \neq 0 \ .
\ee 

I campi di ghost possono essere sviluppati nelle basi di autofunzioni appena definite,
\be
c^a(\sigma)= \sum_J c_JC^a_J(\sigma) \ , \qquad b_{ab}= \sum_{K}b_KB_{Kab}(\sigma) \ ,
\ee
e in termini dei modi l'integrale funzionale $\Delta_{FP}$ sui ghost diventa
\be
\int \prod_{k=1}^\mu db_{0k}\prod_{j=1}^\kappa d c_{0j}\prod_{J}db_Jdc_Je^{-\frac{\nu_J b_Jc_J}{2\pi}}\prod_{k=1}^\mu \frac{1}{4\pi}(b, \partial_k\hat g)\prod_{(a,i)\in f}c^a(\sigma_i) \ .
\ee
Gli integrali sulle variabili grassmaniane sono diversi da zero solo quando la variabile compare nell'integrando. I modi nulli $c_{0j}$ e $b_{0k}$ non compaiono nell'azione, ma solo nelle inserzioni, e correttamente il  numero di inserzioni di ciascun tipo di ghost \`e uguale rispettivamente a $\kappa$ e $\mu$, esattamente i numeri di modi zero.  L'integrale funzionale risultante fattorizza nella forma
\ba
\Delta_{FP}=\int \prod_{k=1}^\mu db_{0k}&\,& \prod_{k'=1}^\mu \left[\sum_{k''=1}^\mu \frac{b_{0k''}}{4\pi}(B_{0k''},\partial_{k'}\hat g)\right] \nonumber \\
&\,& \times \int \prod_{j=1}^\kappa dc_{0j}\prod_{(a,i)\in f} \left[\sum_{j'=1}^\kappa c_{0j'}C_{0j'}^a(\sigma_i) \right] \nonumber \\
&\,& \times \int \prod_J db_J \, dc_J e^{-\frac{\nu_J b_J c_J}{2\pi}} \ .
\ea
La saturazione degli integrali sui modi zero dei due campi ghost con le variabili grassmaniane da luogo a due determinanti finito-dimensionali. I modi restanti danno invece luogo ad un prodotto infinito che ricostruisce un determinante funzionale. Si ha quindi
\be
\label{detFP}
\Delta_{FP}= \det{\frac{(B_{0k}, \partial_{k'}\hat g)}{4\pi}}\det{C_{0j}^a(\sigma_i){\det}'{\left(\frac{P_1^TP_1}{4\pi^2}\right)^{1/2}}} \ ,
\ee
dove il determinante ``primato'' indica l'assenza degli zero modi e $C_{0j}^a$ \`e correttamente una matrice quadrata dal momento che sia gli indici $(a, i) \in f$ che gli indici $j$ corrono sui $\kappa$ valori.

\subsection{Differenziali di Beltrami e misura di integrazione}

Per derivare la misura di Faddeev-Popov anche nella rappresentazione in cui i moduli siano codificati nelle funzioni di transizione introduciamo il concetto di \emph{differenziali di Beltrami}. I differenziali di Beltrami sono una base nello spazio duale a quello dei differenziali olomorfi quadratici, rispetto ai quali possono essere integrati. 

Data una superficie di Riemann, consideriamo un suo ricoprimento di aperti con coordinate complesse $z_m$, dove l'indice $m$ identifica gli aperti, e funzioni di transizioni olomorfe. Dato un punto $t_0$ dello spazio dei moduli, sia definita in queste coordinate una metrica $\hat g(t_0)$, equivalente a meno di trasformazioni di Weyl a $dz_md \bar z_m$.

Vogliamo descrivere l'effetto della variazione dei moduli prima nella rappresentazione in cui la metrica dipenda esplicitamente da essi per passare quindi nella rappresentazione in cui la metrica \`e fissata e le funzioni di transizione dipendono esplicitamente dai moduli.

Definiamo il differenziale di Beltrami nella prima descrizione come
\be
\mu_{ka}^b=\frac{1}{2}\hat g^{bc}\partial_k \hat g_{ac} \ ,
\ee
e le inserzione del campo $b$ sono
\be
\label{binserzione}
\frac{1}{2\pi}(b, \mu_k)=\frac{1}{2\pi}\int d^2z \left(b_{zz}\mu_{k \bar z}^z+b_{\bar z \bar z}\mu_{kz}^{\bar z} \right) \ .
\ee
Nella seconda descrizione dopo una variazione dei moduli $\delta t^k$, si avranno le variazioni di coordinate
\be
\label{ax}
z'_m=z_m+\delta t^k v_{km}^{z_m}(z_m, \bar z_m) \ .
\ee
Le metriche risultanti devono essere equivalenti, a meno di trasformazioni di Weyl, per i punti dello spazio dei moduli $t_0^k$ e $t_0^k+\delta t^k$,
\be
dz_md \bar z_m \propto dz_md \bar z_m + \delta t^k\left(\mu_{kz_m}^{\bar z_m}dz_mdz_m + \mu_{k\bar z_m}^{z_m} d\bar z_md \bar z_m \right) \ . 
\ee
La variazione delle coordinate pu\`o quindi essere espressa in termini del differenziale di Beltrami, 
\be
\mu_{kz_m}^{\bar z_m}=\partial_{z_m}v_{kz_m}^{\bar z_m} \ , \qquad \mu_{k\bar z_m}^{z_m}=\partial_{\bar z_m}v_{kz_m}^{z_m} \ .
\ee
In questa forma si risconosce la versione infinitesimale dell'\emph{equazione di Beltrami}. In realt\`a, queste equazioni non determinano completamente $v_{kz_m}^{\bar z_m}$ e $v_{kz_m}^{z_m}$. La parte olomorfa di $v_{kz_m}^{\bar z_m}$ e la parte antiolomorfa di $v_{kz_m}^{z_m}$, che restano non fissate, corrispondono alla possibilit\`a di operare riparametrizzazioni olomorfe. Sostituendo nella (\ref{binserzione}) e integrando per parti si trova
\be
\frac{1}{2\pi}(b, \mu_k)=\frac{1}{2\pi i}\sum_m \oint_{C_m} \left(dz_m v_{km}^{\bar z_m}b_{z_mz_m}-d\bar z_mv_{km}^{\bar z_m}b_{\bar z_m\bar z_m} \right) \ ,
\ee
dove l'integrazione \`e in senso antiorario lungo i contorni $C_m$ che circondano gli aperti. 

Ricordando la (\ref{ax}), la derivata delle funzioni di trasizione rispetto ai moduli risulta essere 
\ba
\left. \frac{\partial z_m}{\partial t^k}\right|_{z_n}&=& \frac{d z_m}{d t^k}-\left. \frac{\partial z_m}{\partial z_n}\right|_t\frac{d z_n}{d t^k}=\nonumber \\
&=&v_{km}^{z_m}-\left. \frac{\partial z_m}{\partial z_n}\right|_tv_{kn}^{z_n}=v_{km}^{z_m}-v_{kn}^{z_n} \ .
\ea 
I contorni di integrazione degli aperti vicini si posso combinare in modo da scrivere
\be
\frac{1}{2\pi}(b, \mu_k)=\frac{1}{2\pi i}\sum_{mn}\int_{C_{mn}}\left(dz_m \left. \frac{\partial z_m}{\partial t^k}\right|_{z_n}b_{z_mz_m}-d\bar z_m\left. \frac{\partial \bar z_m}{\partial t^k}\right|_{z_n} b_{\bar z_m \bar z_m}\right) \ .
\ee
La somma corre su tutte le coppie di aperti che si sovrappongano. I contorni $C_{mn}$ sono definiti intorno alle sovrapposzioni degli aperti $m$ e $n$, in senso antiorario dal punto di vista di $m$. Le inserzioni sono ora espresse in termini delle funzioni di transizione.  

\section{Operatori di vertice}

Come si \`e visto, la costruzione dell'ampiezze di stringa richiede che gli stati esterni vengano mappati in un numero finito di punti. In ognuno di questi punti in cui una stringa entra o esca devono comparire operatori locali con i giusti numeri quantici degli stati di interesse. Si \`e quindi portati all'idea di associare ad ogni stato della stringa un qualche operatore locale della teoria di campo conforme bidimensionale. Questi operatori sono detti \emph{operatori di vertice}, e una richiesta importante \`e che gli operatori di vertice abbiano dimensione conforme $(1, 1)$ per le stringhe chiuse e $1$ per le stringhe aperte. Questo garantisce infatti la loro integrabilit\`a sul world-sheet.

Un operatore di vertice di uno stato fisico di momento fissato $k$ deve obbedire ad alcune condizioni di covarianza. Essenzialmente, esso deve essere consistente con tutte le simmetrie della corrispondente teoria. Per teorie di stringa bosonica gli operatori di vertice devono quindi avere seguenti propriet\`a:
\begin{enumerate}
\item invarianza per traslazioni spazio-temporali. La dipendenza dal campo $X^\mu$ pu\`o quindi essere indotta solo da fattori della forma $\exp{(ik\cdot X)}$ o da derivate di $X^\mu$; 
\item invarianza per trasformazioni di Lorentz. Gli indici spazio-temporali $\mu$ devono essere tutti saturati;
\item invarianza per riparametrizzazioni della superficie di universo. Gli indici bidimensionali devono essere tutti contratti e deve comparire un fattore di volume $g^{1/2}$;
\item invarianza per trasformazioni di Weyl.
\end{enumerate}

L'operatore di vertice pi\`u generale \`e quindi della forma della forma
\be
{\mathcal V}(\epsilon, k)= \int_M d^2\sigma g^{1/2}W(\epsilon, \nabla X, R)e^{ik\cdot X} \ ,
\ee
dove $W$ \`e un polinomio scalare e $\epsilon$ \`e un tensore di polarizzazione. Ad esempio l'operatore di vertice per un tachione di stringa chiusa \`e
\ba
V_0&=&2g_c\int d^2\sigma g^{1/2}e^{ik\cdot X} \nonumber \\
&=& g_c\int d^2z:e^{ik\cdot X}:\ ,
\ea
dove abbiamo anche incluso la costante di accoppiamento di stringa chiusa $g_c$. In maniera analoga, per il primo stato eccitato si ottiene
\be
V_1= \int d^2z:\partial X^\mu \bar \partial X^\nu e^{ik\cdot X}: \ .
\ee
La richiesta dell'invarianza di Weyl per i vertici porta alla condizione di mass-shell sugli impulsi. In Teorie delle Stringhe \`e quindi possibile soltanto definire operatori di vertice \emph{on-shell}. 

L'estensione al caso di stringa aperta coinvolge integrali sui bordi, ad esempi il vertice del tachione \`e
\be
V_0=g_o\int_{\partial M} ds \left[e^{ik\cdot X}\right]_r \ ,
\ee
che risulta invariante per trasformazioni di Weyl se la condizione di mass-shell $k^2=1/\alpha'$ \`e verificata. Per il vertice del fotone si trova invece
\be
V_1=-i\frac{g_o}{(2\alpha')^{1/2}}e_\mu \int_{\partial M} ds \left[\dot X e^{ik\cdot X}\right]_r \ ,
\ee
che risulta invariante per trasformazioni di Weyl per $k^2=0$.

\section{Ampiezze al livello ad albero}

All'ordine pi\`u basso le ampiezze di interazione coinvolgono superfici con caratteristica di Eulero positiva e numeri variabili di vertici di interazione. A questo livello le ampiezze, dette \emph{ampiezze ad albero}, corrispondono alla teoria classica. Le correzioni quantistiche si ottengono tenendo conto, nello sviluppo perturbativo, delle correzioni a loop.

\subsection{Funzioni di correlazione}

Prima di specializzare i calcoli dei valori di aspettazione al caso delle tre superfici che compaiono nelle ampiezze ad albero, facciamo alcune considerazioni generali per una superficie compatta arbitraria $M$. Il punto di partenza, come in Teoria dei Campi, \`e un funzionale generatore, che indichiamo in generale come
\be
Z[J]= \langle{\, e^{i\int d^2 \sigma J(\sigma)\cdot X(\sigma)}\,}\rangle \ ,
\ee
dove la notazione $\langle{...}\rangle $ indica l'integrale funzionale che definisce la teoria, con l'inserzione di opportuni campi. Consideramo un set ${\mathcal X}_I(\sigma)$ di autosoluzioni dell'operatore cinetico $\nabla^2$, opportunamente normalizzate,
\ba
\nabla^2{\mathcal X}_I &=& -\omega^2_I{\mathcal X}_I \ , \nonumber \\
\int d^\sigma g^{1/2} {\mathcal X}_I{\mathcal X}_{I'} &=& \delta_{II'} \ .
\ea
Espandendo i campi $X^\mu(\sigma)$ e $J^\mu(\sigma)$ in termini delle autosoluzioni 
\ba
X^\mu(\sigma) &=& \sum_I x_I{\mathcal X}_I(\sigma) \ , \nonumber \\
J^\mu(\sigma) &=& \sum_I J\mu_I{\mathcal X}_I(\sigma) \ ,
\ea
l'integrale funzionale diventa
\be
Z[J]=\prod_{I, \mu}\int dx_I^\mu e^{-\frac{\omega_I^2x_I^\mu x_{I\mu}}{4\pi \alpha'}+ix_I^\mu J_{I\mu}} \ .
\ee
Si ottiene quindi un prodotto di integrali gaussiani, che si possono esprimere in termini degli autovalori, eccetto che per il modo nullo ${\mathcal X}_0$, per il quale l'azione \`e nulla, che da luogo ad una funzione delta, 
\ba
\label{ampiezze4}
Z[J]&=& i(2\pi)^d \delta^d(J_0)\prod_{I\neq 0}\left(\frac{4\pi^2 \alpha'}{\omega_I^2}\right)^{d/2}e^{-\frac{\pi \alpha'J_I\cdot J_I}{\omega^2_I}} \nonumber \\
&=& i(2\pi)^d \delta^d(J_0) \left( {\det}'{\frac{-\nabla^2}{4\pi^2 \alpha'}} \right)^{-d/2}e^{-\frac{1}{2}\int d^2\sigma d^2\sigma' J(\sigma)\cdot J(\sigma')G'(\sigma, \sigma')} \ .
\ea
Nell'espressione ottenuta, per riscrivere l'esponenziale, \`e stata utilizzata l'espressione dei coefficienti di espansione della corrente $J^\mu$
\be
J_I^\mu = \int d^2\sigma J^\mu (\sigma){\mathcal X}_I(\sigma) \ .
\ee
Per ottenere degli integrali gaussiani con segno corretto, si sono ruotati i modi di tipo tempo, effettuando la consueta scelta di un tempo immaginario $x_I^0 \to -ix_I^d$ per $I \neq 0$. Per mantenere la corretta funzione $\delta$, si \`e poi contro-ruotato il modo zero di tipo tempo, $x_0^0$, producendo il fattore moltiplicativo $i$. La funzione primata
\be
G'(\sigma, \sigma')=\sum_{I \neq 0}\frac{2\pi \alpha'}{\omega_I^2}{\mathcal X}_I(\sigma){\mathcal X}_I(\sigma') \ ,
\ee
dove la somma \`e su tutti i modi eccetto il modo nullo, \`e la funzione di Green, e soddisfa l'equazione differenziale
\ba
\label{Greeneq}
-\frac{1}{2\pi \alpha'}\nabla^2G(\sigma, \sigma') &=& \sum_{I\neq 0}{\mathcal X}_I(\sigma){\mathcal X}_I(\sigma') \nonumber \\
&=& g^{-1/2}\delta^2(\sigma - \sigma')-{\mathcal X}_0^2 \ ,
\ea
in virt\`u della propriet\`a di completezza delle funzioni ${\mathcal X}_I$. La funzione di Green ordinaria per una sorgente puntiforme, in cui compare solo la funzione delta, nel caso di stringa \`e corretta dalla presenza del termine ${\mathcal X}_0^2$. Questo pu\`o essere interpretato come un contributo di carica di background che neutralizza una carica localizzata, che come tale non pu\`o essere presente su una superficie compatta.

\subsection{Ampiezze sulla sfera}

Specializziamo la soluzione trovata alla sfera. La soluzione dell'equazione differenziale (\ref{Greeneq}) sulla sfera \`e
\be
G'(\sigma_1, \sigma_2)=-\frac{\alpha'}{2}\ln{|z_1-z_2|^2}+f(z_1, \bar z_1)+ f(z_2, \bar z_2) \ ,
\ee
dove la funzione $f(z, \bar z)$ che, come si vedr\`a, non compare nelle ampiezze, \`e della forma
\be
f(z, \bar z) = \frac{\alpha' {\mathcal X}_0^2}{4}\int d^2z'e^{2\omega(z', \bar z')}\ln{|z-z'|^2}+ k \ ,
\ee
dove $k$ \`e una costante.

Consideriamo l'ampiezza d'urto per $n$ tachioni, inserendo i vertici nell'integrale funzionale,
\be
A_{S_2}^n = \Big{\langle}{\left[e^{ik_1\cdot X(\sigma_1)}\right]_r\left[e^{ik_2\cdot X(\sigma_2)}\right]_r \dots \left[e^{ik_n\cdot X(\sigma_n)}\right]_r}\Big{\rangle}_{S_2} \ .
\ee
Questo corrisponde a 
\be
J(\sigma)=\sum_{i=1}^n k_i \delta^2(\sigma-\sigma_i) \ ,
\ee
e sostituendo nell'ampiezza (\ref{ampiezze4}) si ottiene 
\ba
A_{S_2}^n(k, \sigma) &=& i{\mathcal X}_0^{-d}\left({\det}'{\frac{-\nabla^2}{4\pi^2\alpha'}} \right)_{S_2}^{-d/2}(2\pi)^d \delta^d(\sum_i k_i) \nonumber \\
&\times &\exp{\left(-\sum_{\stackrel{i,j=1}{i<j}}^n k_i\cdot k_j G'(\sigma_i, \sigma_j)-\frac{1}{2}\sum_{i=1}^n k_i^2G'_r(\sigma_i, \sigma_i)\right)} \ .
\ea
\`E importante notare come la funzione $\delta$ ora garantisca opportunamente la conservazione dell'impulso. Il determinante pu\`o essere regolarizzato, e la funzione di Green \`e stata rinormalizzata, introducendo un contro-termine
\be
G'_r(\sigma, \sigma')=G'(\sigma, \sigma')-\frac{\alpha'}{2}\ln{d^2(\sigma, \sigma')} \ ,
\ee
dove $d^2(\sigma, \sigma')$ \`e la distanza fra punti lungo le geodetiche. Per brevi distanze, in metrica conforme, si ha $d^2(\sigma, \sigma') \approx (\sigma - \sigma')^2e^{2\omega(\sigma)}$, quindi si ha per la funzione rinormalizzata,
\be
G'_r(\sigma, \sigma)=2f(z, \bar z)+\alpha' \omega(z, \bar z) \ .
\ee
Sostituendo nell'ampiezza sulla sfera si ha
\be
A_{S_2}^n(k, \sigma) = iC_{S_2}^X(2\pi)^d \delta^d(\sum_{i}k_i)\exp{\left(-\frac{\alpha'}{2}\sum_i k_i^2 \omega(\sigma_i)\right)}\prod_{\stackrel{i,j=1}{i<j}}^n |z_{ij}|^{\alpha' k_i \cdot k_j} \ , 
\ee
in cui, come atteso, non compare la funzione $f$ e la costante \`e
\be
C_{S_2}^X={\mathcal X}_0^{-d}\left({\det}'{\frac{-\nabla^2}{4\pi^2\alpha'}} \right)_{S_2}^{-d/2} \ .
\ee

L'ampiezza d'urto di stati eccitati di stringa sar\`a del tipo
\be
\Big{\langle} \prod_{i=1}^n \left[e^{ik_i\cdot X(z_i, \bar z_i)}\right]_r \prod_{j=1}^p \partial X^\mu_j(z'_j) \prod_{k=1}^q \bar \partial X^{\nu_k}(\bar z''_k) \Big{\rangle}_{S_2} \ ,
\ee
dal momento che i vertici sono esponenziali per derivate di $X^\mu$. L'ampiezza si calcola sommando su tutte le contrazioni, dove ogni $\partial X$ e $\bar \partial X$ deve essere contratta sia con gli esponenziali che con le altre derivate di $X$. Il risultato finale \`e
\ba
iC_{S_2}^X(2\pi)^d \delta^d(\sum_i k_i)\exp{\left(-\frac{\alpha'}{2}\sum_i k_i^2 \omega(\sigma_i)\right)}\prod_{\stackrel{i,j=1}{i<j}}^n |z_{ij}|^{\alpha' k_i \cdot k_j} \nonumber \\
\times \Big{\langle}\prod_{j=1}^p \left[v^{\mu_j}(z'_j)+q^{\mu_j}(z'_j) \right] \prod_{k=1}^q \left[\tilde v^{\nu_k}(\bar z''_k)+\tilde q^{\nu_k}(\bar z''_k) \right]  \Big{\rangle}_{S_2} \ ,
\ea
dove 
\be
v^\mu = - i \frac{\alpha'}{2}\sum_{i=1}^n \frac{k_i^\mu}{z-z_i} \ , \qquad \tilde v^\mu(\bar z)=- i \frac{\alpha'}{2}\sum_{i=1}^n \frac{k_i^\mu}{\bar z-\bar z_i} \ ,
\ee
provengono dalle contrazioni con gli esponenziali.

Il contributo dei ghost delle ampiezze, come abbiamo visto, \`e esprimibile in termini del determinante (\ref{detFP}). Ricordando la struttura delle ampiezze, il numero di CKV della sfera impone che l'unico valore di aspettazione indipendente dei campi ghost, che compare nelle ampiezze ad albero, sia  
\be
\langle c(z_1)c(z_2)c(z_3)\tilde c(\bar z_4)\tilde c(\bar z_5)\tilde c(\bar z_6)\rangle_{S_2} \ ,
\ee
I sei vettori di Killing della sfera compongono una base complessa non ortonormale
\ba
(1,0)\ , \qquad (z, 0)\ , \qquad (z^2, 0) \ , \qquad
(0,1)\ , \qquad (0, \bar z)\ , \qquad (0, \bar z^2) \ . \qquad
\ea
In questa base il determinante degli zero modi del campi $c$ fattorizza in due blocchi
\be
\det{C_{0j}^a} = C_{S_2}^g
\det{
\left| \begin{array}{ccc}
1 & 1 & 1 \\
z_1 & z_2 & z_3 \\
z_1^2 & z_2^2 & z_3^2 \\
\end{array} \right|}
\det{
\left| \begin{array}{ccc}
1 & 1 & 1 \\
\bar z_1 & \bar z_2 & \bar z_3 \\
\bar z_1^2 & \bar z_2^2 & \bar z_3^2 \\
\end{array} \right|}=
C_{S_2}^g z_{12}z_{13}z_{23}\bar z_{45}\bar z_{46}\bar z_{56} \ ,
\ee
dove la notazione $z_{ij}$ sta per $z_i - z_j$ e la costante $C_{S_2}^g$ \`e uno Jacobiano finito dimensionale, dovuto alla scelta di una base non ortonormale.

\subsection{Ampiezze sul disco}

Per scrivere l'ampiezza sul disco, partendo dal caso della sfera, \`e necessario introdurre una riflessione. L'identificazione $z'=\bar z$ porta a restringere $z$ al semipiano superiore del piano complesso, e introducendo al contempo nella (\ref{Greeneq}) una carica immagine, che porta a una funzione di Green del tipo
\be
G'(\sigma_1, \sigma_2)=-\frac{\alpha'}{2}\ln{|z_1-z_2|^2}-\frac{\alpha'}{2}\ln{|z_1-\bar z_2|^2} \ .
\ee
L'ampiezza d'urto fra tachioni diventa
\ba
\Big{\langle} \prod_{i=1}^n \left[e^{ik_i\cdot X(z_i, \bar z_i)}\right]_r \Big{\rangle}_{D_2} &=& iC_{D_2}^X(2\pi)^d \delta^d(\sum_i k_i)\prod_{i=1}^n|z_i - \bar z_i|^{\alpha'k_i^2/2} \nonumber \\
&\times &\prod_{\stackrel{i, j=1}{i<j}}^n |z_i - z_j|^{\alpha' k_i\cdot k_j}|z_i - \bar z_j|^{\alpha' k_i \cdot k_j} \ ,
\ea
mentre per ampiezze d'urto fra stati eccitati di stringa occorre nuovamente sommare su tutte le contrazioni.

Per vertici sul bordo del disco, i due termini della funzione di Green a punti uguali sono entrambi divergenti, e sottraendo il consueto controtermine del primo termine, la funzione resta divergente. Occorre pertanto definire un ordinamento normale di bordo raddoppiando il controtermine,
\be
\stackrel{\star}{\star}X^\mu(y_1)X^\nu(y_2)\stackrel{\star}{\star}=X^\mu(y_1)X^\nu(y_2)+ 2 \alpha' \eta^{\mu \nu}\ln{|y_1 - y_2|} \ ,
\ee
dove i punti $y$ sono sull'asse reale.

Per calcolare le ampiezze dei ghost nel caso del disco, il modo pi\`u semplice \`e utilizzare nuovamente la ``duplicazione''. In analogia con il caso precedente, si considera pertanto il caso della sfera, con le opportune restrizioni sui campi
\be 
\tilde b(\bar z)=b(z')\ , \qquad \tilde c(\bar z)=c(z')\ , \qquad z'=\bar z \ , Im(z)>0 \ .
\ee
e il risultato \`e
\be
\langle c(z_1)c(z_2)c(z_2)\rangle_{D_2}=C_{D_2}^gz_{12}z_{13}z_{23} \ .
\ee

\subsection{Ampiezze sul piano proiettivo}

Con il metodo delle cariche immagine, la funzione di Green risulta essere in questo caso
\be
G'(\sigma_1, \sigma_2)=-\frac{\alpha'}{2}\ln{|z_1-z_2|^2}-\frac{\alpha'}{2}\ln{|1+z_1\bar z_2|^2} \ ,
\ee
e l'ampiezza tachionica \`e quindi 
\ba
\Big{\langle} \prod_{i=1}^n \left[e^{ik_i\cdot X(z_i, \bar z_i)}\right]_r \Big{\rangle}_{D_2} &=&
iC_{RP_2}^X(2\pi)^d \delta^d(\sum_i k_i)\prod_{i=1}^n|1+z_i\bar z_i|^{\alpha'k_i^2/2} \nonumber \\
&\times &\prod_{\stackrel{i, j=1}{i<j}}^n |z_i - z_j|^{\alpha' k_i\cdot k_j}|1 + z_i \bar z_j|^{\alpha' k_i \cdot k_j} \ .
\ea 

Anche le caso del piano proiettivo si pu\`o utilizzare la stessa tecnica di ``duplicazione'' per calcolare la parte di ampiezza dovuta ai campi di ghost. L'involuzione $z'=-\bar z^{-1}$ porta a richiedere
\be
\tilde b(\bar z)=\left(\frac{\partial z'}{\partial \bar z}\right)^2b(z')=z'^4b(z')\ , \qquad \tilde c(\bar z)=\left(\frac{\partial z'}{\partial \bar z}\right)^{-1}c(z')=z'^{-2}c(z')\ ,
\ee
e l'ampiezza si trova essere
\be
\langle c(z_1)c(z_2)c(z_2)\rangle_{RP_2}=C_{RP_2}^gz_{12}z_{13}z_{23} \ .
\ee

\section{Ampiezze per stringhe chiuse e aperte}

\subsection{Ampiezza di Veneziano}

Studiamo, a titolo di esempio, l'ampiezza di interazione di quattro tachioni di stringhe aperte. L'ampiezza all'ordine pi\`u basso ha la topologia del disco, che possiamo rappresentare come il semipiano superiore, indicando le coordinare reali del bordo con $y$. Dopo aver fissato la metrica si possono utilizzare, come si \`e visto, tre dei CKV per fissare tre dei quattro vertici di interazione in posizioni $y_1$, $y_2$, $y_3$. Ogni coordinata di vertice di interazione fissata viene sostituita dall'inserzione di un campo di ghost $c$. Ogni vertice di interazione porta un fattore di accoppiamento di stringa $g_o$. Tenendo conto dei due possibili differenti ordinamenti dei vertici fissati, e prendendo il modulo del contributo dei ghost, l'ampiezza d'urto si scrive
\ba
\label{Venez1}
S_{D_2}(k_1, k_2, k_3, k_4) &=& \nonumber \\
&=&g_o^4e^{-\lambda}\int_{-\infty}^{\infty}dy_4 \Big{\langle}\prod_{i=1}^3\stackrel{\star}{\star}c^1(y_i)e^{ik_i\cdot X(y_i)}\stackrel{\star}{\star} \, \stackrel{\star}{\star}e^{ik_4\cdot X(y_4)}\stackrel{\star}{\star}\Big{\rangle}+ (k_2 \leftrightarrow k_3) \nonumber \\
&=&ig_o^4C_{D_2}(2\pi)^{26}\delta^{26}(\sum_i k_i)|y_{12}y_{13}y_{23}|\int_{-\infty}^{\infty}dy_4 \prod_{i<j}|y_{ij}|^{2\alpha' k_i\cdot k_j} \nonumber\\
&\,& + (k_2 \leftrightarrow k_3) \ .
\ea
Convenzionalmente si sceglie $y_1=0$, $y_2=1$ e $y_3\to \infty$  e si adottano le variabili di Mandelstam
\be
s=-(k_1+k_2)^2 \ ,\qquad t=-(k_1+k_3)^2 \ ,\qquad u=-(k_1+k_4)^2 \ ,
\ee 
che sono legate dalla conservazione del momento e dalla condizione di mass-shell
\be
s+t+u=\sum_i m_i = -\frac{4}{\alpha'} \ .
\ee
Utilizzando la relazione $2\alpha' k_i\cdot k_j=-2+\alpha'(k_i+k_j)^2$, e introducendo le variabili di Mandelstam, l'ampiezza (\ref{Venez1}) diventa
\ba
S_{D_2}(k_1, k_2, k_3, k_4) &=& ig_o^4C_{D_2}(2\pi)^{26}\delta^{26}(\sum_i k_i)\nonumber \\
&\times &\left[\int_{-\infty}^{\infty}dy_4|y_4|^{-\alpha'u-2}|1-y_4|^{-\alpha't-2}+(t\to s)\right] \ .\nonumber \\ 
\ea
L'espressione trovata pu\`o essere riscritta in forma pi\`u elegante, mettendo in evidenza la simmetria nei diversi canali. L'integrale si divide in tre regioni: $-\infty <y_4<0$, $0 <y_4< 1$, e $0 <y_4< \infty$, in cui si hanno diversi ordinamenti per i  vertici. Si possono utilizzare le trasformazioni (\ref{MoebiusGroup}) a parametri reali in modo tale da avere contributi solo nella seconda regione d'integrazione, in questo modo l'ampiezza diventa
\be
\label{Venez2}
S_{D_2}(k_1, k_2, k_3, k_4) = ig_o^4C_{D_2}(2\pi)^{26}\delta^{26}(\sum_i k_i)\left[I(s, t)+ I(t, u)+I(u, s) \right] \ ,
\ee
con
\be
I(s, t)=\int_0^1 dy y^{-\alpha's-2}(1-y)^{-\alpha't-2} \ .
\ee

L'ampiezza (\ref{Venez2}) pu\`o essere riscritta utilizzando la funzione beta di Eulero
\be
B(a, b)=\int_0^1 dy y^{a-1}(1-y)^{b-1} \ ,
\ee
che pu\`o essere espressa in termini di funzioni $\Gamma$. Definendo infatti una nuova variabile $y=v/w$, a $w$ fissato, si ha
\be
w^{a+b-1}B(a, b)= \int_0^w dv v^{a-1}(w-v)^{b-1} \ ,
\ee
e moltiplicando entrambi i lati dell'equazione per $\int_0^\infty dw e^{-w}$, si trova
\be
\Gamma(a+b)B(a, b)=\int_0^\infty dv v^{a-1}e^{-v}\int_0^\infty d(w-v)(w-v)^{b-1}e^{-(w-v)}= \Gamma(a)\Gamma(b) \ ,
\ee
da cui segue immediatamente che
\be
B(a, b)=\frac{\Gamma(a)\Gamma(b)}{\Gamma(a+b)} \ .
\ee
Osservando che
\be
I(s,t)=B(- \alpha's -1, - \alpha't -1) \ ,
\ee
e fissando la costante richiedendo l'unitariet\`a della matrice S,  l'ampiezza pu\`o essere scritta nella forma
\ba
\label{Venez3}
S_{D_2}(k_1, k_2, k_3, k_4) &=& ig_o^2C_{D_2}(2\pi)^{26}\delta^{26}(\sum_i k_i) \nonumber \\
&\,& \times \left[\frac{\Gamma(- \alpha's -1)\Gamma(- \alpha't -1)}{\Gamma(- \alpha's - \alpha't -2)}
+ (t \to u)+(s \to u) \right] \ ,
\ea
dove 
\be
\label{Venez4}
A(s, t)=\frac{\Gamma(- \alpha's -1)\Gamma(- \alpha't -1)}{\Gamma(- \alpha's - \alpha't -2)}
\ee
\`e la celebre \emph{Ampiezza di Veneziano}, derivata originariamente per descrivere le interazioni forti \cite{veneziano}.

L'ampiezza di Veneziano ha due propriet\`a notevoli: la simmetria manifesta sotto lo scambio di $s$ con $t$, detta \emph{dualit\`a planare}, e la propriet\`a che i residui nei poli in $s$ siano polinomi in $t$ e viceversa. Entrambe le propriet\`a sono caratteristiche importanti della Teoria delle Stringhe. In Teoria dei Campi, infatti, le ampiezze ad albero associate ai diagrammi di Feynman hanno individualmente solo un polo in uno dei canali.

\begin{figure}
\begin{center}
\kern.75true cm\vbox{\epsfxsize=5.00in\epsfbox{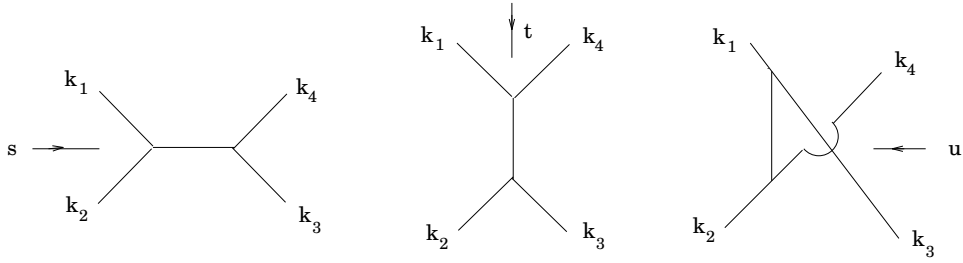}}
\end{center}
\caption{Processi nei canali $s$, $t$ e $u$.}
\label{fig8}
\end{figure}

La dualit\`a planare dell'ampiezza di Veneziano riflette la deformabilit\`a del disco, che consente di avvicinare a coppie i vertici di interazione, modificando il processo. Questo comporta che l'ampiezza completa (\ref{Venez3}) condensa, nel caso di stringa, diversi diagrammi d'urto di Feynman, che corrispondono ai diversi canali $s$, $t$ e $u$ in cui si pu\`o avere il processo d'interazione (vedi figura \ref{fig8}). 

La funzione $\Gamma(x)$ ha la propriet\`a di avere infiniti poli sul semiasse reale negativo, in corrispondenza degli interi. Nell'intorno di questi poli, per $x \approx -n$, si ha
\be
\Gamma(x) \stackrel{x \to -n}{\sim}\frac{(-)^n}{n!}\frac{1}{x+n} \ ,
\ee
mentre dal momento che $\Gamma(x+1)=x\Gamma(x)$, si trova
\be
\frac{\Gamma(y)}{\Gamma(y-n)}=(y-1)(y-2)(\dots)(y-n) \ .
\ee
La struttura dei residui dell'ampiezza sar\`a quindi del tipo
\be
A(s, t) \stackrel{x \to -n}{\sim}\frac{(-)^n}{n!}\frac{(y-1)(y-2)(\dots)(y-n)}{x+n} \ .
\ee
L'ampiezza di Veneziano, considerando i giusti argomenti delle funzioni gamma, ha quindi infiniti poli e, ad per esempio, per il canale $s$, in corrispondenza di $\alpha' s = -1, 0, 1, 2, \dots$. L'ampiezza di Veneziano descrive quindi, un processo d'urto fra particelle scalari che scambiano tra loro infinite particelle con spin e massa crescente.

\subsection{Ampiezza di Shapiro-Virasoro}

Il calcolo dell'ampiezza d'urto di stringhe chiuse 	procede in maniera simile al caso di stringhe aperte. Per quattro stringhe tachioniche si trova,
\be
S_{S_2}(k_1, k_2, k_3, k_4)=g_c^4e^{-2\lambda}\int_{\mathbb C} d^2z_4 \Big{\langle}\prod_{i=1}^3 :\tilde cce^{ik_i\cdot X}(z_i, \bar z_i): \, :e^{ik_4\cdot X}(z_4, \bar z_4):\Big{\rangle}_{S_2} \ ,
\ee
dove l'integrazione avviene su tutto il piano complesso ${\mathbb C}$. Calcolando i valori di aspettazione, e fissando i vertici di interazione nei punti $z_1=0$, $z_2=1$ e $z_3=\infty$, l'ampiezza si riduce a
\ba
S_{S_2}(k_1, k_2, k_3, k_4)&=& g_c^4C_{S_2}(2\pi)^{26}\delta^{26}(\sum_i k_i) \nonumber \\
&\,& \times \int_{\mathbb C}d^2z_4|z_4|^{-\alpha'u/2-4}|1-z_4|^{-\alpha't/2-4} \ .
\ea
In questo caso le variabili di Mandelstam sono soggette al vincolo $s+t+u=-16/\alpha'$. Questa ampiezza ha poli nella variabile $u$ per $-\infty < z_4 < 0$, nella variabile $t$ per $0 < z_4 < 1$ e in $s$ per $1 < z_4 < \infty$. I poli si hanno per i valori
\be
\alpha' s, \alpha' t, \alpha' u = -4, 0, 4, 8, \dots \ ,
\ee
che sono correttamente i valori dei quadrati delle masse degli stati di stringa chiusa.

Anche l'ampiezza di striga chiusa pu\`o essere espressa in temini di funzioni gamma,
\be
S_{S_2}(k_1, k_2, k_3, k_4)=\frac{8\pi i g_c^2}{\alpha'}(2\pi)^{26}\delta^{26}(\sum_i k_i)C(-1-\alpha't/4, -1-\alpha'u/4) \ ,
\ee
dove
\be
C(a,b)=2\pi \frac{\Gamma(a)\Gamma(b)\Gamma(c)}{\Gamma(a+b)\Gamma(a+c)\Gamma(b+c)} \ , \qquad a+b+c=1 \ .
\ee
e utilizzando gli argomenti corretti si ha
\be
C \sim
\frac{\Gamma(-1-\alpha't/4)\Gamma(-1-\alpha'u/4)\Gamma(-1-\alpha's/4)}{\Gamma(-2-\alpha't/4-\alpha'u/4)\Gamma(-2-\alpha't/4-\alpha's/4)\Gamma(-2-\alpha'u/4-\alpha's/4)} \ .
\ee
Questa \`e l'\emph{ampiezza di Shapiro-Virasoro}. In questa ampiezza la simmetria fra i canali $s$, $t$ e $u$ \`e completa.

\subsection{Fattori di Chan-Paton e interazioni di gauge}

\`E possibile introdurre una generalizzazione particolarmente interessante delle teorie di stringa aperta, che ne rispetta le simmetrie, ammettendo la presenza agli estremi della stringa di gradi di libert\`a aggiuntivi non dinamici, noti come \emph{cariche di Chan-Paton} \cite{cp}. Uno stato di stringa aperto pu\`o cos\`i essere scritto nella forma
\be
|N; k; ij\rangle \ ,
\ee
dove gli indici $i$ e $j$ associati agli estremi della stringa vanno da $1$ a $n$. Dal momento che le cariche di Chan-Paton vengono introdotte senza termini dinamici nella Lagrangiana, tutte le simmetrie della teoria sono preservate. \`E possibile definire una base degli stati del tipo
\be
|N; k; a\rangle = \sum_{i,j=1}^n \lambda_{ij}^a|N; k; ij\rangle 
\ee
con $\lambda_{ij}^a$ una opportuna rappresentazione di un gruppo di simmetria. Questo \`e un modo molto naturale di introdurre gruppi di simmetria, anche non abeliani, in Teoria delle Stringhe. 

Le ampiezze di stringa aperta per uno stesso tipo bosoni esterni possono essere generalizzate definendo ampiezze ``rivestite''
\be
A(1, \dots, n)\tr(\lambda^a_1\dots \lambda^a_2) \ ,
\ee
dove $A(1, \dots, n)$ indica l'ampiezza calcolata senza cosiderare i gradi di libert\`a aggiuntivi. Questa \`e una scelta abbastanza naturale dal momento che le cariche di Chan-Paton non evolvono, e quindi l'interazione fra gli estremi delle stringhe deve avvenire fra estremi che si trovino nello stesso stato di carica. A livello ad albero, nell'ampiezza di disco, ciscun estremo destro di stringa deve avere alla sua sinistra l'estremo di un'altra stringa nello stesso stato.

Le ampiezze modificate devono inoltre rispettare i vincoli di unitariet\`a della teoria. La risoluzione di questi vincoli \cite{Marcus_Sagnotti} ha dimostrato la possibilit\`a di introdurre in teoria di stringhe tutti i gruppi classici, $SO(N)$, $Sp(N)$, $U(N)$. I due estremi delle stringhe aperte devono avere valori nelle rappresentazioni fondamentali dei gruppi classici. 

\section{Ampiezze ad un loop}

I primi diagrammi a loop che entrano nell'espansione perturbativa della Teoria delle Stringhe sono associati a superfici di Riemann con numero di Eulero nullo. In particolare in questa sezione discutiamo il diagramma di toro che \`e associato, nelle teorie di stringa chiusa, al primo ordine perturbativo. Altre ampiezze, presenti in teorie aperte e non orientate, saranno studiate in dettaglio, nel caso in cui non si abbiano inserzioni di vertici, nel capitolo 4.

Nel caso del toro la funzione di Green (\ref{Greeneq}) deve essere periodica in entrambe le direzioni definite sulla superficie. Ci si aspetta anche che sia la somma di un contributo olomorfo e di uno antiolomorfo per analogia con il caso di genere zero e per le propriet\`a del laplaciano. Le funzioni theta di Jacobi (si veda l'Appendice A) rispondono bene a queste propriet\`a, e un buon candidato \`e 
\be
G'(w, \bar w; w', \bar w') \sim -\frac{\alpha'}{2}\ln{\Big| \vartheta_1\left(\frac{w-w'}{2\pi}\Big| \tau\right)\Big|^2} \ ,
\ee
dove $\vartheta_1$ \`e una funzione theta di Jacobi  che tende a zero linearmente quando il suo primo argomento tende all'origine. Il logaritmo \`e somma di un contributo olomorfo e antiolomorfo, \`e quindi annullato da l'azione di $\partial \bar \partial$. La funzione proposta non \`e per\`o doppiamente periodica, a causa delle proprieta di trasformazione delle funzioni $\vartheta$: sotto una trasformazione $w \to w + 2\pi \tau$, si genera infatti un termine $-\alpha'[Im(w-w')+\pi \tau_2 ]$. Inoltre occorre introdurre, come si \`e gi\`a avuto modo di notare, anche un termine di carica di background. Tendendo conto di queste due osservazioni, la funzione di Green si trova essere
\be
G'(w, \bar w; w', \bar w') = -\frac{\alpha'}{2}\ln{\Big| \vartheta_1\left(\frac{w-w'}{2\pi}\Big| \tau\right)\Big|^2} + \alpha'\frac{[Im(w-w')]^2}{4\pi \tau_2}+k(\tau, \tau) \ ,
\ee
dove la funzione $k$ si determina per ortogonalit\`a rispetto a ${\mathcal X}_0$, ma come nel caso della sfera non contribuisce nelle ampiezze. Il valore di aspettazione per un prodotto di operatori di vertici si trova come nel caso della sfera, e risulta essere
\ba
\Big\langle \prod_{i=1}^n : e^{ik_i\cdot X(z_i, \bar z_i)}: \Big\rangle_{T_2} \, &=& \, iC_{T_2}^X(\tau)(2\pi)^d \delta^d(\sum_i k_i) \nonumber \\
&\times & \prod_{i<j} \Big| \frac{2\pi}{\partial_\nu \vartheta_1(0|\tau)}\vartheta_1\left(\frac{w_{ij}}{2\pi}\Big| \tau\right)\exp{\left[-\frac{(Im w_{ij})^2}{4\pi\tau_2}\right]}\Big|^{\alpha'k_i \cdot k_j} \ .
\ea
L'ampiezza di Toro pu\`o essere ricavata dall'espressione (\ref{ampiezze4}), tenendo conto della presenza di due moduli reali e di due vettori di Killing conformi. La sua forma generale \`e
\be
S_{T^2}(1; 2; \dots; n) 
= \frac{1}{2}\int_{\mathcal F} d\tau d\bar \tau \Big\langle B\tilde B c \tilde c {\mathcal V}_1(w_1, \bar w_2)\prod_{i=2}^n \int dw_id\bar w_i {\mathcal V}_i (w_i, \bar w_i) \Big\rangle_{T_2} \ ,
\ee
dove si sono utilizzati i CKV per fissare uno dei vertici di interazione. L'integrazione avviene sulla regione fondamentale dello spazio dei moduli, mentre il fattore moltiplicativo $1/2$ tiene conto dell'ulteriore simmetria discreta ${\mathbb Z}_2$ che, come si \`e visto, caratterizza il toro. 

L'inserzione dei campi di ghost \`e in questo caso 
\be
B=\frac{1}{4\pi}(b, \partial_\tau g) = \frac{1}{2\pi}\int d^2w b_{ww}(w)\partial_\tau g_{\bar w \bar w} \ .
\ee
Se consideriamo una variazione della metrica $\delta g_{ww}=\epsilon^\ast$, la nuova metrica \`e
\ba
ds^2 &=& dwd\bar w + \epsilon^\ast dw^2 + \epsilon d\bar w^2 \nonumber \\
&=& (1+\epsilon^\ast+\epsilon)d[w+\epsilon(\bar w-w)]d[\bar w + \epsilon^\ast(w-\bar w)]+O(\epsilon^2) \ ,
\ea
che risulta equivalente (a meno di una trasformazione di Weyl) ad una metrica della forma $dw'd\bar w'$ con $w'=w+\epsilon(\bar w - w)$, di periodicit\`a
\be
w'\cong w'+ 2\pi \cong w' + 2\pi (\tau-2i\tau_2 \epsilon) \ .
\ee
Quindi la variazione della metrica equivale ad una trasformazione del modulo
\be
\label{varmod}
\delta \tau = -2i\tau_2 \epsilon \ .
\ee
Tornando al calcolo delle inserzioni, il risultato (\ref{varmod}) permette di calcolare esplicitamente la derivata della metrica rispetto a $\tau$, e quindi
\be
B= \frac{i}{4\pi \tau_2}\int d^2w b_{ww}(w) = 2\pi i b_{ww}(0) \ .
\ee
L'indipendenza dalla posizione del campo di ghost \`e dovuta al fatto che sul toro i differenziali quadratici sono costanti, mentre l'integrale d'area \`e semplicemente $\int 2d \sigma^1 d \sigma^2 = 2(2\pi)^2\tau_2$. 

Si pu\`o ulteriormente manipolare l'espressione delle ampiezze per porre tutti i vertici sullo stesso piano. Dal momento che i $CKV$ sono costanti, il valore di aspettazione dei campi $c$, \`e indipendente dalla posizione. Si pu\`o quindi fissare la posizione dei campi di ghost in maniera arbitraria, e l'operatore di vertice pu\`o essere reso libero di muoversi sulla superficie del toro, ricordandosi per\`o di mediare su tutte le sue possibili traslazioni. Per farlo si introduce nell'integrale funzionale l'integrale
\be
\int \frac{dwd\bar w}{2(2\pi)^2\tau_2} \ ,
\ee
dove il denominatore \`e opportunamente l'area del toro che corrisponde al volume dei CKV. Raccogliendo i risultati trovati, l'ampiezza diventa
\be
S_{T^2}(1; 2; \dots; n) 
= \int_{\mathcal F} \frac{d\tau d\bar \tau}{4\tau_2} \Big\langle b(0)\tilde b(0) c(0) \tilde c(0) \prod_{i=1}^n \int dw_id\bar w_i {\mathcal V}_i (w_i, \bar w_i) \Big\rangle_{T_2} \ ,
\ee
e in particolare l'ampiezza di vuoto \`e 
\be
Z_{T^2}=\int_{\mathcal F} \frac{d\tau d\bar \tau}{4\tau_2}\Big\langle b(0)\tilde b(0) c(0) \tilde c(0)\Big\rangle_{T_2}  \ ,
\ee
che calcoleremo esplicitamente nel formalismo del cono di luce e che, come si vedr\`a, contiene molte informazioni importanti sulla teoria.


\chapter{Funzioni di partizione}

\section{Ampiezze di vuoto ad un loop}

In Teoria dei campi le ampiezze di vuoto ad un loop sono completamente determinate dallo spettro della teoria e non contengono molte informazioni a parte la loro relazione con la costante cosmologica. Al contrario in Teoria delle Stringhe le ampiezze di vuoto permettono di studiare lo spettro perturbativo della teoria libera e di estrarre condizioni di consistenza che la rendano priva di divergenze e di anomalie.

Per calcolare le ampiezze ad un loop per stringhe chiuse e aperte \`e utile partire dalla Teoria dei Campi. Iniziamo dal caso pi\`u semplice, calcolando l'energia di vuoto per un campo scalare massivo in D-dimensioni
\be
\label{se}
S_{(E)} \ = \ \int \ d^Dx \ \frac{1}{2} \ \left(\dd_\mu\phi \ \dd^\mu\phi+M^2\phi^2 \right) \ .
\ee
L'integrale funzionale della teoria euclidea, ottenuto dopo una rotazione di Wick, \`e
\begin{equation}
\label{funzionale}
Z[J] = \ \int [\mathcal{D}\phi] \ e^{-S_{(E)}-J\phi}=e^{-W[J]} ,
\end{equation}
dove W[J] \`e il funzionale generatore delle funzioni di Green connesse. L'\emph{azione effettiva} \`e definita da una trasformazione di Legendre:
\be
\label{gamma}
\Gamma [\bar{\phi}]= W[J] + \int\ d^Dx \ J\bar{\phi},
\ee
e $\Gamma[\bar{\phi}]$ \`e il funzionale generatore delle funzioni irriducibili ad una particella. Nel limite classico $\hbar \to 0$ si trova, sostituendo l'espressione (\ref{gamma}) in (\ref{funzionale}),
\be
\Gamma [\bar{\phi}]= S[\bar{\phi}] + O(\hbar).
\ee
Dalla (\ref{gamma}) si pu\`o calcolare la derivata di $\Gamma[\bar{\phi}]$ a J fissato
\be
\frac{\delta \Gamma [\bar{\phi}]}{\delta \bar{\phi}}= -J \ ,
\ee
e quindi per $J=0$ si ottengono i punti estremali dell'azione effettiva che possono essere interpretati come 
energie dei vuoti corrispondenti. Nel caso della teoria libera (\ref{se}) si ha quindi per l'energia di vuoto l'espressione
\be
e^{-\Gamma} \ = \ \int [\mathcal{D}\phi] \ e^{-S_{(E)}}\sim \left[\mathrm{det}(-\Box+M^2) \right]^{-\frac{1}{2}} \ ,
\ee
che si ottiene dall'integrale gaussiano generalizzato, e quindi
\be
\Gamma \ = \ \frac{1}{2} \mathrm{tr}\left[\ln(-\Box+M^2)\right] \ .
\ee
Per estrarre la dipendenza dalla massa M \`e utile l'identit\`a
\be
\mathrm{tr}(\ln{A}) \ = \ -\int_{\epsilon}^{\infty}\frac{dt}{t} \ 
\mathrm{tr}\left(e^{-tA}\right) \ ,
\ee
dove $\epsilon$ \`e un cutoff ultravioletto e $t$ \`e un parametro di Schwinger. La traccia \`e facilmente calcolabile usando una base che diagonalizzi l'operatore cinetico, ovvero un set completo di autostati dell'impulso
\be
\label{imp}
\Gamma \ = \ -\frac{V}{2}\int_{\epsilon}^{\infty} \frac{dt}{t} \
\int\frac{d^Dp}{(2\pi)^D} \ e^{-tp^2} \ e^{-tM^2} \ ,
\ee
dove V \`e il volume spazio-temporale. Integrando sugli impulsi si trova quindi l'energia di vuoto ad un loop per un grado di libert\`a bosonico:
\be
\label{Gamma_bose}
\Gamma \ = \ -\frac{V}{2(4\pi)^{\frac{D}{2}}} \
\int_{\epsilon}^{\infty} \frac{dt}{t^{\frac{D}{2}+1}} \ e^{-tM^2} \ .
\ee
Ripetendo il calcolo per un fermione di Dirac, le regole per l'integrazione di variabili anticommutanti introducono un segno $(-)$, e ricordando che in $D$ dimensioni gli spinori di Dirac hanno $2^{\frac{D}{2}}$ gradi di libert\`a si ottiene
\be
\label{Gamma_fermi}
\Gamma \ = \ \frac{V2^{\frac{D}{2}}}{2(4\pi)^{\frac{D}{2}}} \
\int_{\epsilon}^{\infty} \frac{dt}{t^{\frac{D}{2}+1}} \ e^{-tM^2} \ .
\ee

Per poter usare le formule trovate per calcolare l'energia di vuoto ad un loop in teoria delle stringhe occorre un'opportuna generalizzazione. Si \`e visto che $\Gamma$ dipende solo dalle masse dei modi fisici che sono presenti nel loop di vuoto, e quindi nel caso in cui siano presenti pi\`u particelle (o pi\`u eccitazioni di stringa), si avr\`a una traccia sulle masse. Si deve infine tenere conto del segno e della molteplicit\`a dei contributi fermionici. Queste osservazioni possono essere raccolte nell'espressione generale
\be
\label{Gamma}
\Gamma \ = \ -\frac{V}{2(4\pi)^{\frac{D}{2}}} \
\int_{\epsilon}^{\infty} \frac{dt}{t^{\frac{D}{2}+1}} \ \mathrm{Str}\left(e^{-tM^2}\right) \ ,
\ee
dove la supertraccia $\mathrm{Str} \ $ \`e 
\be
\mathrm{Str} \ = \ \sum_{bosoni} \ - \ \sum_{fermioni} \ .
\ee

Prima di applicare questa formula al calcolo delle ampiezze di vuoto di stringa torniamo per un momento alle superfici 
di Riemann di genere $g=1$.  Il toro rappresenta una stringa chiusa orientata che si propaga in un loop, e il suo modulo $\tau$ pu\`o essere interpretato fisicamente come il tempo proprio della stringa nel loop. L'invarianza modulare indica che c'\`e  un'infinit\`a di scelte equivalenti nella scelta del tempo sul world-sheet. Nella teoria questa invarianza \`e  peculiare, dal momento che introduce un cut-off naturale ultravioletto, come si \`e  visto definendo la regione fondamentale nel $\tau$-piano. Al contrario, nelle altre superfici di genere $g=1$ sono possibili due scelte del tempo proprio legate sostanzialmente dalle trasformazioni $S$, che danno luogo a diverse interpretazioni dei diagrammi. Non c'\`e quindi alcuna simmetria che protegga dalle divergenze, che vanno cancellate imponendo condizioni di cancellazione.

La bottiglia Klein, come si \`e visto, pu\`o essere interpretata alternativamente come un diagramma di vuoto di stringa chiusa non orientata o come un diagramma di propagazione ad albero di stringa chiusa fra due crosscap. Le due rappresentazioni corrispondono a scelte differenti del tempo proprio, rispettivamente il lato verticale e quello orizzontale dei due poligoni fondamentali.
Anche nell'anello sono possibili due differenti scelte del tempo sul world-sheet. Scegliendo la direzione verticale si interpreta l'anello come un diagramma di vuoto di una stringa aperta, mentre scegliendo il tempo sull'asse reale, si ha un diagramma di propagazione ad albero di una stringa chiusa fra due bordi.\\
La striscia di M\"{o}bius, scegliendo il tempo ``verticale'', descrive una stringa aperta non orientata in un diagramma di vuoto. Al contrario, ridefinendo il poligono fondamentale, come si \`e visto, e, scegliendo un tempo proprio ``orizzontale'', si ha la propagazione ad albero di una stringa chiusa fra un bordo e un crosscap. 
Solitamente si fa riferimento alle ampiezze di vuoto come \emph{ampiezze nel canale diretto}, e alle stesse nella rappresentazione ad albero come \emph{ampiezze nel canale trasverso}.

\section{Funzioni di partizione della stringa bosonica}

\subsection{Teoria chiusa orientata nel formalismo del cono di luce}

Applichiamo l'espressione (\ref{Gamma}) per una teoria di sola stringa bosonica chiusa in dimensione critica $D=26$, il cui spettro di massa \`e dato da
\be
\label{Mass_close}
M^2 \ = \ \frac{2}{\alpha'} \ \left[N^\bot +\bar N^\bot - 2\right] \ ,
\ee
con la ``level matching condition''
\be
\qquad N^\bot - \bar N^\bot = 0 \ . 
\ee
Sostituendo nella (\ref{Gamma}) con $D=26$  la (\ref{Mass_close}) e imponendo il vincolo di level-matching introducendo una $\delta$-function, si ottiene
\be
\Gamma \ = \ -\frac{V}{2(4\pi)^{13}} \ \int_{-\frac{1}{2}}^{\frac{1}{2}}d s
\int_{\epsilon}^{\infty} \frac{dt}{t^{14}} \ \mathrm{tr}
\left(e^{ -\frac{2}{\alpha'}  \left[N^\bot +\bar N^\bot - 2\right]t} \ e^{ 2\pi i \left[ N^\bot -\bar N^\bot \right]s} \right) \ ,
\ee
che, definendo un parametro di Schwinger complesso  $\tau=s+i\frac{t}{\pi\alpha'} \ $ e definendo
$q \ = \ e^{2\pi i \tau} \ $, $ \ \bar q \ = \ e^{-2\pi i \bar\tau} \ $, pu\`o essere scritta nella forma elegante
\be
\Gamma \ = \ -\frac{V}{2(4\alpha'\pi^2)^{13}} \ \int_{-\frac{1}{2}}^{\frac{1}{2}}d \tau_1
\int_{\epsilon}^{\infty} \frac{d\tau_2}{\tau_2^{14}} \ \mathrm{tr}
\left(\ q^{N^\bot - 1} \ \bar q^{ \ \bar N^\bot -1} \right) \ .
\ee
Questa \`e l'ampiezza per una stringa chiusa che si propaghi in un loop, e l'integrazione sulla varibile complessa $\tau$ dovrebbe equivalere ad integrare sullo spazio dei moduli del toro, dal momento che il calcolo dell'integrale sui cammini, richiede che si sommi su tutte le metriche rappresentanti tutte le superfici topologicamente inequivalenti, e quindi nel nostro caso su tutti i possibili tori. Si \`e visto infatti che il toro \`e univocamente determinato, assegnata una metrica piatta, dal modulo complesso $\tau$ con parte immaginaria positiva. 
In realt\`a il calcolo seguito fin qui porta un problema di multiplo conteggio, dal momento che non tutti i punti del $\tau$-piano rappresentano tori inequivalenti. Occorre, come si \`e detto, restringersi nell'integrazione alla regione fondamentale $\mathcal{F}$, e a meno di una costante di normalizzazione, l'ampiezza del toro \`e quindi
\be
\label{torus}
\mathcal{T} \ = \ \int_{\mathcal{F}}\frac{d^2\tau}{\tau_2^2} \
\frac{1}{\tau_2^{12}} \ \mathrm{tr}\left( q^{ \ N^\bot - 1} \
\bar q^{ \ \bar N^\bot - 1}\right) \ .
\ee

Possiamo ora riprodurre il calcolo dell'ampiezza di vuoto per una stringa chiusa. Cosideriamo la cella fondamentale di un toro di modulo $\tau$ nel piano complesso e interpretiamo l'asse verticale come dimensione temporale e quello orizziontale come dimensione spaziale. Immaginiamo una stringa di lunghezza unitaria che al tempo $t=0$ giaccia sull'asse orizzontale, e dopo un tempo $t=\tau_2$ sia propagata raggiungendo il lato superiore della cella, traslando di $x= Re ({\tau})=\tau_1$. Gli operatori per le traslazioni temporali e spaziali sono rispettivamente l'Hamiltoniana $H = L_0^\bot+\bar L_0^\bot - 2$ \ e l'impulso 
$P = L_0^\bot -  \bar L_0^\bot$. Si ritrova cos\`i il risultato precedente
\be
Z = \mathrm{Tr} \left\{ e^{-2\pi \tau_2 H} e^{-2\pi i \tau_1 P} \right\}
\ = \mathrm{Tr} \left\{ q^{ \ L_0^\bot -1}  \bar q^{ \ \bar L_0^\bot -1} \right\} \ ,
\ee
dove Tr \`e una somma sulle variabile discrete e un'integrazione su quelle continue. 

A questo punto calcoliamo esplicitamente l'ampiezza di toro (\ref{torus}), partendo dalla traccia (la somma sull'indice $i$ delle dimensioni \`e sottintesa)
\be
\mathrm{tr} \ q^{N^\bot -1} = \ \frac{1}{q} \ \mathrm{tr}\left(q^{ \ \sum_{n=1}^\infty \alpha_{-n}^i\alpha_n^i}\right) \ ,
\ee
che possiamo scrivere esplicitamente in una base di tutti gli stati di stringa della forma $\alpha_{-n} |0\rangle$, $(\alpha_{-n})^2 |0\rangle$, ecc. in cui si ha:
\be
q^{ \ \alpha_{-n}\alpha_n}\ 
=\ \left( \begin{array}{ccccc}
1 & \, & \, & \, & \, \\
\, & q^n & \, & \, & \, \\
\, & \, & q^{2n} & \, & \, \\
\, & \, & \, & q^{3n} & \, \\ 
\, & \, & \, & \, & \ddots \\
\end{array} \right)
\ee
e quindi si ha
\be
\label{tr-bose}
\mathrm{tr} \left( q^{ \ \sum_{n=1}^\infty \alpha_{-n}^i\alpha_n^i} \right) = 
\prod_{i=1}^{24}\prod_{n=1}^\infty \sum_k q^{nk} = 
\frac{1}{\prod_{n=1}(1-q^n)^{24}} \ .
\ee
Usando la funzione $\eta$ di Dedekind \cite{ww}, definita come:
\be
\label{eta}
\eta(\tau) \ = \ q^{\frac{1}{24}} \prod_{n=1}^\infty \ (1-q^n) \ ,
\ee
possiamo riscrivere l'ampiezza di toro con i risultati ottenuti nella forma
\be
\label{torus_bose}
\mathcal{T}_{bosonica}\ = \ \int \frac{d^2\tau}{\tau_2^2} \
\frac{1}{\tau_2^{12} \ (\eta \ \bar\eta)^{24}} \ .
\ee
$\mathcal{T}$ \`e invariante modulare, dal momento che la misura d'integrazione \`e invariante modulare e, usando le propriet\`a di trasformazione della funzione $\eta$,
\be
T \ : \ \eta(\tau)\rightarrow\eta(\tau+1) \ = \ e^{\frac{i\pi}{12}} \
\eta(\tau) \ , \qquad
S \ : \ \eta(\tau)\rightarrow \eta\left(-\frac{1}{\tau}\right) \ = \
\sqrt{-i\tau} \ \eta(\tau) \ ,
\ee
si verifica che anche il fattore $\tau_2^{1/2}(\eta\bar\eta)$ \`e invariante.

Si \`e detto che le ampiezze di vuoto scritte meritano il nome di funzioni di partizione, la ragione \`e che, espandendo l'integrando in potenze di $q$ e $\bar q$ e moltiplicando per un fattore $4/ \alpha'$ si ottengono i livelli delle eccitazioni in $M^2$, moltiplicati per coefficienti che forniscono le loro degenerazioni. La degenerazione \`e in questo caso il numero di partizioni del livello in numeri interi. Ad esempio, al livello tre si avrebbe degenerazione tre dal momento che si hanno le tre possibilit\`a $\alpha_3$, $\alpha_1\alpha_2$ e $\alpha_1\alpha_1\alpha_1$. 
Per il toro, in $D=26$ nei livelli pi\`u bassi si ha
\be
\label{espansione_toro}
{\cal T}: \qquad
\frac{1}{(\eta \ \bar\eta)^{24}}\ \simeq \ \frac{1}{q \bar q}[1 + 24(q + \bar q) + (24)^2q \bar q + \dots] \ ,
\ee
dove il termine $(q \bar q)^{-1}$ \`e la base della torre degli stati e indica che a livello zero si ha un tachione. Eliminando gli stati che non rispettano la condizione di level-matching, restano infiniti termini del tipo $q^{w_1} \bar q^{w_2}$ che indicano la presenza di campi massivi di ``peso'' $(w_1,w_2)$, con $w_1=w_2$. Ad esempio a livello uno gli stati di massa nulla hanno molteplicit\`a $24^2$, e sono, come atteso, gli stati del multipletto ($G_{\mu \nu}$, $B_{\mu \nu}$, $\Phi$).

\subsection{Discendenti aperti}

La costruzione di teorie consistenti di stringhe chiuse e aperte non orientate a partire da teorie di sole stringhe chiuse \`e basata sulla proiezione di orientifold \cite{cargese, open_string}. L'idea \`e costruire dalle ampiezze di vuoto ad un loop di stringhe orientate, le ampiezze di stringhe non orientate per mezzo dell'operatore $\Omega$ di world-sheet. Per una stringa chiusa si vede che lo spettro \`e invariante sotto parit\`a del world-sheet, $\sigma \to -\sigma$, dove l'azione naturale di $\Omega$ scambia i modi sinistri e destri
\begin{equation}
\Omega: \ \ \alpha_n^{\mu} \leftrightarrow \tilde{\alpha}_n^{\mu} \ .
\end{equation} 
Per ottenere stringhe non orientate si devono identificare i modi destri e sinistri della stringa. Questo corrisponde a proiettare lo spettro degli stati su uno dei due autospazi associati ai due autovalori di $\Omega$, $\pm 1$. La proiezione deve essere consistente con le interazioni di stringa. Dal momento che l'urto di due stringhe chiuse antisimmetriche sotto lo scambio dei loro modi sinistri e destri darebbe luogo ad uno stato simmetrico, la sola opzione in questo caso \`e proiettare sugli stati simmetrici. Dal punto di vista del calcolo delle ampiezze di vuoto la proiezione si ottiene inserendo nella traccia sugli stati in (\ref{torus_bose}) l'operatore 
\be
P=\frac{1+\Omega}{2} \ .
\ee
L'azione del proiettore equivale a $\mathcal{T}\rightarrow \mathcal{T}/2+\mathcal{K}$, con $\mathcal{K}$ la funzione di partizione per la bottiglia di Klein
\be
{\cal K} \ = \ \frac{1}{2} \int_{\cal F_{\cal K}} \frac{d^2\tau}{\tau_2^{14}} \ \mathrm{tr} \ \left( q^{N^\bot- 1}
\bar q^{\bar N^\bot- 1} \Omega\right) \ ,
\ee
dove
\be
\sum_{L,R}\langle L,R\mid q^{N^\bot-1}\bar
q^{\bar N^\bot-1}\Omega\mid L,R\rangle
\ = \ \sum_{L,R}\langle L,R\mid q^{N^\bot-1}\bar
q^{\bar N^\bot-1}\mid R,L\rangle \ ,
\ee
e l'ortogonalit\`a degli stati riduce la somma al solo sottospazio diagonale, identificando $N^\bot$ con
$\bar N^\bot$,
\be
\sum_{L}\langle LR\mid (q\bar q)^{N^\bot-1}\mid LR\rangle \ = \frac{1}{\eta (2i\tau_2)^{24}} \ , 
\ee
dal momento che $q\bar q=e^{-4\pi \tau_2}= e^{2 \pi i (2i\tau_2)}$. Si \`e trovata un'espressione che dipende naturalmente dal modulo del toro doppiamente ricoprente della bottiglia di Klein. Per fissare il dominio di integrazione $\cal F_{\cal K}$ occorre ricordare che, come si \`e osservato, non c'\`e invarianza modulare e, quindi, occorre integrare $\tau_2$ su tutto il dominio di definizione, che coincide con il semiasse immaginario positivo. In conclusione si ottiene
\be
\label{Klein}
{\cal K} \ = \ \frac{1}{2} \int_0^{\infty} \frac{d^2\tau}{\tau_2^{14}} \frac{1}{\eta^{24}(2i\tau_2)} \ .
\ee

L'espansione dell'ampiezza di vuoto della bottiglia di Klein (\ref{Klein}) in funzione di $q$ e $\bar q$  \`e
\be
\label{espansione_klein}
{\cal K}: \qquad \frac{1}{(\eta^{24})(2i\tau_2)}\ \simeq \ \frac{1}{q \bar q}[1 + 24(q \bar q) + \dots] \ ,
\ee
e per conoscere lo spettro della teoria proiettata occorre sommare il contributo della bottiglia di Klein (\ref{espansione_klein}) a quello del toro (\ref{espansione_toro}) opportunamente dimezzato e ridotto ai termi che soddisfano la condizione di ``level matching'':
\be
\label{espansione_k_t}
{\frac{1}{2}\cal T + \cal K}: \qquad \frac{1}{q \bar q}[1 + \frac{24(24+1)}{2}(q \bar q) + \dots] \ .
\ee
Si \`e ottenuto il risultato atteso: sono scomparsi tra gli stati di massa nulla i gradi di libert\`a del tensore antisimmetrico $B_{\mu \nu}$. 

Si \`e visto che sono possibili due diverse scelte naturali per il ``tempo'' sul world-sheet e che a queste corrispondono due diverse rappresentazioni diagrammatiche dell'ampiezza di bottiglia di Klein. Le due rappresentazioni sono legate da una trasformazione $S$, e comportano una ridefinzione della cella fondamentale. Per passare nel canale trasverso \`e sufficiente raddoppiare il modulo su cui si integra $t = 2\tau_2$ generando un fattore moltipicativo $2^{13}$, ed effettuare quindi una trasformazione $S$ : $t_2\rightarrow 1/\ell$, ottendo
\be
\label{tildeKI}
\tilde{\cal K} = \frac{2^{13}}{2} \int_0^{\infty} \ d \ell  \
\frac{1}{\eta^{24}(i \ell)} \, .
\ee
L'ampiezza $\tilde{\cal K}$ \`e interpretata come l'ampiezza di propagazione di una stringa chiusa fra due crossacap in un tempo ``orizzontale'' $\ell$, ed \`e pertanto un diagramma ``ad albero''. Come si \`e gi\`a detto, non c'\`e per la bottiglia di Klein invarianza sotto l'azione del gruppo modulare. Per ottenere una teoria che sia priva di divergenze, occorre per questo aggiungere alla teoria i settori di stringa aperta per poter poi imporre condizioni di cancellazione delle divergenze fissando i gradi di libert\`a delle cariche di Chan-Paton.

Per calcolare l'ampiezza di Anello nel canale diretto sostituiamo l'operatore di massa di stringa aperta $M= \  \alpha'(N^\bot- 1)$ nella (\ref{Gamma}) 
\be
\label{open}
{\cal A} = \frac{1}{2} \int_0^\infty \ \frac{d
\tau_2}{\tau_2^{14}} \ {\rm tr} \ q^{\frac{1}{2}(N^\bot - 1)}  \ ,
\ee
dove $q=e^{-2 \pi \tau_2}$ e, la traccia \`e su stati del tipo $\lambda_{ij}\mid k;  ij\rangle$. Gli indici $i$ e $j$ sono riferiti alle cariche di Chan-Paton di moltiplicit\`a N agli estremi della stringa aperta. Calcolando la traccia e tenendo conto della presenza di gradi di libert\`a aggiuntivi, si ottiene
\be
{\cal A} = \frac{N^2}{2} \int_0^{\infty} \ \frac{ d \tau_2}{\tau_2^{14}} \
\frac{1}{\eta^{24}( {\textstyle{1\over 2}} i \tau_2)} \ .
\ee
Anche in questo caso c'\`e una dipendenza naturale dal modulo del toro doppiamente ricoprente $\tau = \frac{1}{2} i \tau_2$. L'ampiezza del canale trasverso si ottiene utilizzando come varibile di integrazione il modulo del toro doppiamente ricoprente $t = \tau_2/2$, che porta un fattore $2^{-13}$, e quindi effettuando una trasformazione $S$ : $t_2\rightarrow 1/\ell$, ottenendo infine
\be
\label{tildeA}
\tilde{\cal A} = \frac{2^{-13} \ N^2}{2} \int_0^{\infty} \
d \ell \ \frac{1}{\eta^{24}( i \ell )} \ .
\ee
L'interpretazione di questa ampiezza al livello ad albero come propagazione di una stringa chiusa fra due bordi, porta ad associare N al coefficiente di riflessione sui bordi.

L'ultimo settore da studiare \`e quello di stringa aperta non orientata. Come nel caso di stringa chiusa, le ampiezze non orientate si ottengono inserendo nella traccia di anello il proiettore
\be
\label{proiezione2}
P=\frac{(1+\epsilon \Omega)}{2} \ ,
\ee
e tenendo conto che nel caso di stringhe aperte la parit\`a sul world-sheet ha un'azione leggermente differente dal caso chiuso, $\Omega : \sigma \to \pi -\sigma$, e quindi sugli oscillatori 
\begin{equation}
\label{azione_oscillatori}
\Omega: \ \ \alpha_n^{\mu} \leftrightarrow (-1)^n\alpha_n^{\mu} \ .
\end{equation} 
L'azione su uno stato generico deve essere scritta tenendo conto della struttura aggiuntiva introdotta con le cariche di Chan Paton. In generale ci si aspetta che l'azione di $\Omega$ inverta le cariche agli estremi, ma si deve anche tenere conto della sua azione sui fattori di Chan Paton. Questa ulteriore libert\`a permette di lasciare indeterminato il segno $\epsilon$ nella proiezione (\ref{proiezione2}), che come si vedr\`a \`e fissato dalla simmetria delle matrici di Chan Paton e quindi dalla scelta del gruppo di gauge.

Consideriamo ora l'ampiezza del nastro di M\"obius, che definiamo come
\be
{\cal M} = \ \frac{\epsilon}{2} \int_0^\infty \ \frac{d
\tau_2}{\tau_2^{14}} \ {\rm tr} (\ q^{\frac{1}{2}(N^\bot - 1)}\Omega)  \ .
\ee
Nel calcolare la traccia si deve tener conto dell'operatore $\Omega$, che restringe la somma sugli indici $i$, $j$ agli stati diagonali e quindi porta un fattore N, mentre sugli oscillatori, come visto nella (\ref{azione_oscillatori}), genera un segno $(-1)^n$, 
\be
\mathrm{tr} \left( q^{ \ \ \frac{1}{2}\sum_{n=1}^\infty \alpha_{-n}^i\alpha_n^i} \Omega \right) = 
N \prod_{i=1}^{24}\prod_{n=1}^\infty \sum_k (-)^{nk}q^\frac{nk}{2} = 
\frac{N}{\prod_{n=1}(1-(-)^nq^\frac{n}{2})^{24}} \ .
\ee
Nell'espressione trovata si riconosce la dipendenza dal modulo del toro doppiamente ricoprente, osservando che $q^{\frac{n}{2}}(-)^n = e^{2\pi i n(1/2+i\tau_2/2)}$. Definendo un funzione $\hat{\eta}$ come
\be
\hat \eta({\textstyle{1\over 2}} i \tau_2 + {\textstyle{1\over 2}}) \ = 
\ (\sqrt {q})^{\frac{1}{24}} \prod_{n=1}^\infty \ (1-(-)^nq^{\frac{n}{2}}) \ ,
\ee
che differisce dalla ${\eta}$ di Dedekind per una fase, l'ampiezza di nastro di M\"obius \`e quindi 
\be
{\cal M} = \frac{\epsilon N}{2} \int_0^{\infty} \
\frac{ d \tau_2}{\tau_2^{14}} \
\frac{1}{\hat{\eta}^{24}({\textstyle{1\over 2}} i \tau_2 + {\textstyle{1\over 2}})} \ .
\ee
In generale, come si vedr\`a anche in altri casi, il passaggio al canale trasverso per il nastro di M\"obius richiede qualche accortezza, a causa del modulo complesso del toro doppiamente ricoprente. La trasformazione su $\tau$ che definisce il passaggio all'ampiezza ad albero \`e
\be
P:\ \frac{1}{2}+i\frac{\tau_2}{2} \to \frac{1}{2}+i\frac{1}{2\tau_2} \ ,
\ee
che pu\`o essere ottenuta da una sequenza di trasformazioni S e T, 
\be
P=TST^2S \ ,
\ee
Per $\hat \eta$ la ridefinizione di fase introduce un'operazione di coniugio, e si ha
\be
P=T^{1/2}ST^2ST^{1/2} \ .
\ee
Nel nostro caso per passare al canale trasverso ridefiniamo $\tau_2 \to 1/t$, che corrisponde ad effetturare una trasformazione P. \`E semplice mostrare che
\be
\hat \eta({\textstyle{i\over {2t}}} + {\textstyle{1\over 2}}) \ = \sqrt t \hat \eta({\textstyle{it\over {2}}} + {\textstyle{1\over 2}})
\ee
e quindi
\be
\tilde{\cal M} = \ \frac{\epsilon N}{2} \int_0^{\infty} \ d t  \
\frac{1}
{\hat{\eta}^{24}(\frac{1}{2}it + {\textstyle{1\over 2}})} \ ,
\ee
e infine, con $\ell=t/2$, si ha 
\be
\label{MtildeI}
\tilde{\cal M} = 2 \ \frac{\epsilon N}{2} \int_0^{\infty} \ d \ell  \
\frac{1}
{\hat{\eta}^{24}(i \ell + {\textstyle{1\over 2}})} \ .
\ee
\\
A questo punto vediamo per il canale diretto, dove si hanno stringhe aperte, l'effetto dell'operatore di proiezione sviluppando in potenze di di $\sqrt q$ le ampiezze di anello e di M\"obius
\be
{\cal A + \cal M}: \qquad \frac{N^2-\epsilon N}{2\sqrt q} + 24\frac{N^2-\epsilon N}{2} + \dots \ .
\ee
Quindi per $\epsilon = +1$ si hanno $N(N-1)/2$ vettori di massa nulla che completano la rappresentazione aggiunta di un gruppo $SO(N)$, mentre per $\epsilon = -1$ si hanno gli $N(N+1)/2$ vettori di massa nulla dell'aggiunta di un gruppo $USp(N)$.

La teoria costruita fin qui presenta divergenze ultraviolette nel canale diretto nel limite $\tau_2 \to 0$: infatti fatta eccezione per il toro, nelle altre ampiezze di genere $g=1$ l'integrazione conivolge, come gi\`a evidenziato, tutto il semiasse positivo. 
L'insorgere delle divergenze pu\`o essere compreso meglio studiando il canale trasverso, dove esse compaiono nel limite infrarosso $\ell \rightarrow\infty$ e sono dovute alla propagazione di tachioni e di stati di stringa chiusa di massa nulla. I contributi degli stati generici di massa $M_i$ e con degenerazione $c_i$ che si propagano nel canale trasverso sono proporzionali a 
\be
\label{divergenze}
\sum_i c_i\int_0^\infty d\ell e^{-\ell M^2_i} \ = \ \lim_{\ell \to \infty}\left(c_T\frac{e^{|M_T^2|\ell}}{M_T^2}\right) + \sum_i c_i\left.\frac{1}{p_i^2+M_i^2}\right|_{p_i^2=0} \ .
\ee

Nel limite $\ell \to \infty$ il tachione e gli stati a massa nulla danno origine ai contributi dominanti. La divergenza tachionica pu\`o essere regolata formalmente, e in ogni caso \`e assente nelle superstringhe. Al contrario, i contributi divergenti dovuti agli stati di massa nulla sono genericamente inevitabili e hanno implicazioni fisiche, che avremo modo di studiare, e sono dovuti al divergere dei loro propagatori nel limte di impulso nullo, come si \`e messo in evidenza nella (\ref{divergenze}). \`E possibile cancellare queste divergenze notando che nel limite $\ell \rightarrow\infty$ l'ampiezza trasversa fattorizza nella somma di contributi scrivibili come prodotti di due funzioni ad un punto di stati di massa nulla su un bordo o un crosscap e di un propagatore. Si pu\`o quindi eliminare le divergenze dovute ai poli nei propagatori, imponendo che la somma dei residui dei diversi contributi sia nulla, una condizione detta \emph{condizione di tadpole} (vedi figura \ref{figtadpoles}). Alla luce della discussione precedente dovrebbe essere chiaro che questo porta a fissare il segno $\epsilon$ e la dimensione $N$, ovvero a fissare il gruppo di gauge di Chan-Paton.
\begin{figure}
\begin{center}
\epsfbox{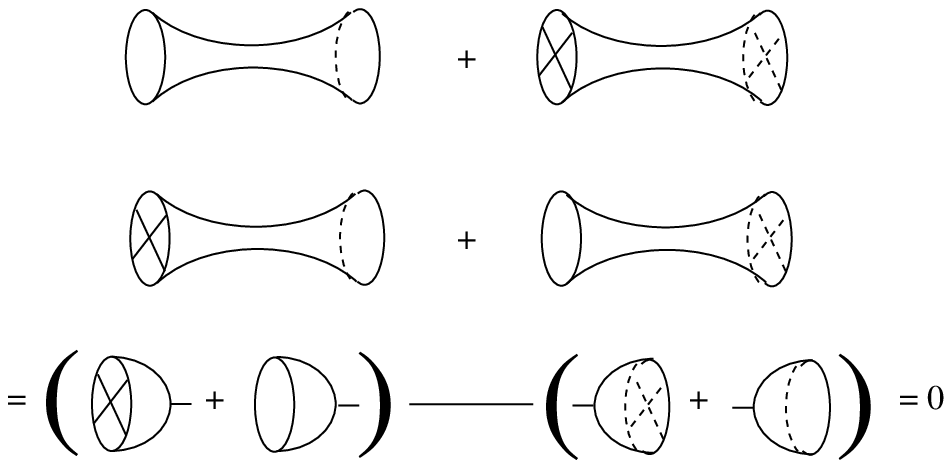}
\end{center}
\caption{condizioni di tadpole}
\label{figtadpoles}
\end{figure}

Nel caso della stringa bosonica, raccogliendo i termini di massa nulla degli sviluppi delle ampiezze (\ref{tildeKI}), (\ref{tildeA}) e (\ref{MtildeI}), si ha
\be
\tilde{\cal K} + \tilde{\cal A} + \tilde{\cal M}\sim 
2^{13} + 2^{-13} \ N^2 - 2 \, \epsilon \, N  = 2^{-13} \
(N - \epsilon \, 2^{13})^2 \, ,
\ee
le condizioni di tadpole fissa $\epsilon=+1$ e $N=2^{13}=8192$, e quindi il gruppo di gauge $SO(8192)$ \cite{douglas1,weinberg,marcus86,bs88}. Posticipiamo per il momento la discussione sul significato fisico della condizione di tadpole, che avremo modo di affrontare in maniera pi\`u generale, limitandoci a dire che essa equivale ad eliminare, nella teoria bosonica effettiva di basse energie, un potenziale per il dilatone $\varphi$,
\be
\label{tad_dilatone}
V \sim (N-\epsilon 2^{13})\int d^{10} x \sqrt{-g}e^{-\varphi} \ .
\ee

\section{Funzioni di partizione di Superstringa}

\subsection{Superstringhe di Tipo II}

Per scrivere la funzione di partizione del toro nel caso della Superstringa in $D=10$, utilizziamo la formula di massa, scritta in termini di $N^\bot=N_B+N_F$ and $\bar N^\bot=\bar N_B+\bar N_F$
\be
M^2 \ = \ \frac{2}{\alpha'} \ \left[N^\bot+\bar N^\bot + a + \bar a \right] \ ,
\ee
dove $a=-1/2$ nel settore NS e $a=0$ nel settore R. Sostituendo nella (\ref{Gamma}) e introducendo, come nel caso bosonico, una funzione $\delta$ che tenga conto della condizione di level-matching, si pu\`o scrivere
\be
\label{torus_susy}
\mathcal{T} \ = \ \int_{\mathcal{F}}\frac{d^2\tau}{\tau_2^2} \
\frac{1}{\tau_2^4} \ \mathrm{Str}\left( q^{ \ N^\bot+ a} \
\bar q^{ \ \bar N^\bot + \bar a}\right) \ .
\ee

Per calcolare la supertraccia occorre tener conto che lo spettro di superstringa \`e somma diretta dei quattro settori distinti NS-NS, R-R, NS-R e R-NS. Procediamo per passi, iniziando con il calcolare le tracce dei settori NS e R
\be
\mathrm{tr}_{NS} \ q^{ \ N_B+N_F-\frac{1}{2}} \ = \
\frac{1}{\sqrt{q}} \
\mathrm{tr}\left(q^{ \ \sum_{n=1}^\infty \alpha_{-n}^i\alpha_n^i}\right)
\ \mathrm{tr}\left(q^{ \ \sum_{r=\frac{1}{2}}^\infty r \ b_{-r}^i b_r^i}\right) \ ,
\ee
\be
\mathrm{tr}_R \ q^{ \ N_B+N_F} \ = \
\mathrm{tr}\left(q^{ \ \sum_{n=1}^\infty \alpha_{-n}^i\alpha_n^i}\right)
\ \mathrm{tr}\left(q^{ \ \sum_{r=1}^\infty r \ b_{-r}^i b_r^i}\right) \ ,
\ee
dove la somma sugli indici $i=1\ldots 8$ \`e sottintesa. La traccia bosonica \`e stata gi\`a calcolata nella (\ref{tr-bose}), e nel calcolare invece la traccia fermionica occorre tener conto che, per il principio di esclusione di Pauli, i possibili numeri di occupazione sono solo $0$ e $1$, per cui nel settore NS si ha
\be
\mathrm{tr}\left(q^{ \ \sum_{r=\frac{1}{2}}^\infty r \ \psi_{-r}^i \psi_r^i}\right) = \
\prod_{i=1}^8\prod_{r=1/2}^\infty(1+q^r) \ .
\ee
Nel settore R si deve tener conto che l'indice $r$ prende valori interi e che, per la presenza dei modi zero anticommutanti
\be
\{ \psi_0^i, \psi_0^j \} = \delta^{ij} \ ,
\ee
il vuoto \`e una rappresentazione dell'algebra di Clifford in $SO(8)$. Di conseguenza la traccia avr\`a un fattore moltiplicativo $2^{(D-2)/2}=16$ che ne conta la degenerazione. 

Mettendo insieme i risultati per la parte bosonica e fermionica e ridefinendo l'indice $r$ nel settore NS si pu\`o scrivere
\be
\mathrm{tr}_{NS}  \left( q^{ \ N_B+N_F-\frac{1}{2}} \right) = \
\frac{\prod_{n=1}^\infty(1+q^{n-1/2})^8}
{\sqrt{q} \ \prod_{n=1}^\infty(1-q^n)^8} \ ,
\ee 
\be
\mathrm{tr}_R  \left( q^{ \ N_B+N_F} \right) = \
16 \ \frac{\prod_{n=1}^\infty(1+q^n)^8}
{\prod_{n=1}^\infty(1-q^n)^8}  \ .
\ee
Lo spettro della superstringa deve essere proiettato usando i proiettori GSO. Nel settore di Ramond si ha che $\mathrm{tr}_{R} \ \left[\Gamma_9(-1)^F \right]= 0$ dal momento che ogni stato \`e accompagnato da un altro di chiralit\`a opposta, e quindi
\be
\label{tracceGSO}
\mathrm{tr}_{NS} \ \left[q^{ \ N_B+N_F-\frac{1}{2}} \ 
\frac{1-(-1)^F}{2}\right]=\frac{\prod_{n=1}^\infty(1+q^{n-1/2})^8 - \prod_{n=1}^\infty(1-q^{n-1/2})^8}
{2\sqrt{q} \ \prod_{n=1}^\infty(1-q^n)^8} \ , 
\ee
\be
\mathrm{tr}_{R} \ \left[q^{ \ N_B+N_F} \ 
\frac{1\pm \Gamma_9(-1)^F}{2}\right] \ = \ 
16 \ \frac{\prod_{n=1}^\infty(1+q^n)^8}
{\prod_{n=1}^\infty(1-q^n)^8}  \ .
\ee
Le espressioni trovate possono essere scritte in termini di funzioni $\theta$ di Jacobi di argomento $z=0$ e caratteristiche $\alpha$ e $\beta$ uguali a 0 o a $\frac{1}{2}$, dove le funzioni $\theta$ di Jacobi \cite{ww} sono definite come
\be
\vartheta \left[{\textstyle {\alpha \atop \beta}} \right] (z|\tau) = \sum_n \
q^{\frac{1}{2} (n+ \alpha)^2} \ e^{2 \pi i (n + \alpha)(z+\beta)} \ ,
\ee
o in forma di prodotto come
\ba
\label{theta}
\vartheta \left[ {\textstyle {\alpha \atop \beta}} \right] (z|\tau) &=&
e^{2 i \pi \alpha (z+\beta)} \ q^{\alpha^2/2} \prod_{n=1}^\infty \
( 1 - q^n)\prod_{n=1}^\infty \ (1 + q^{n + \alpha - 1/2} e^{2 i \pi (z+\beta)} )\nonumber\\
&& \times \prod_{n=1}^\infty (1 + q^{n - \alpha - 1/2} e^{-2 i \pi (z+\beta)} ) \ .
\ea 
Utilizzando le definizioni (\ref{theta}) e (\ref{eta}) si possono costruire 
\ba
& &\frac{\vartheta^4 \left[ {1/2} \atop { 1/2} \right] (0|\tau)}
{\eta^{12}(\tau)} = \frac{\vartheta_1^4(0|\tau)}{\eta^{12}(\tau)} = 0 \, , \\
& &\frac{\vartheta^4 \left[ {1/2} \atop {0} \right] (0|\tau)}
{\eta^{12}(\tau)} = \frac{\vartheta_2^4(0|\tau)}{\eta^{12}(\tau)} =
16 \frac{\prod_{n=1}^{\infty} (1 + q^{n})^8}
{\prod_{n=1}^{\infty} (1 - q^n)^8} \, , \\
& &\frac{\vartheta^4 \left[ {0} \atop {0} \right] (0|\tau)}
{\eta^{12}(\tau)} = \frac{\vartheta_3^4(0|\tau)}{\eta^{12}(\tau)} =
\frac{\prod_{n=1}^{\infty} (1 + q^{n-1/2})^8}{
q^{1/2} \prod_{n=1}^{\infty} (1 - q^n)^8} \, , \\
& &\frac{\vartheta^4 \left[ {0} \atop {1/2} \right] (0|\tau)}
{\eta^{12}(\tau)} = \frac{\vartheta_4^4(0|\tau)}{\eta^{12}(\tau)} =
\frac{\prod_{n=1}^{\infty} (1 - q^{n-1/2})^8}{
q^{1/2} \prod_{n=1}^{\infty} (1 - q^n)^8} \ , 
\label{vacuumd=10}
\ea
che come si \`e visto compaiono nelle (\ref{tracceGSO}), che possono essere riscritte come
\ba
\mathrm{tr}_{NS} \ \left[q^{ \ N_B+N_F-\frac{1}{2}} \
    \frac{1-(-1)^F}{2}\right] \ &=& \ \frac{1}{\eta^8} \
\frac{\vartheta_3^4(0|\tau)
-\vartheta_4^4(0|\tau)}{2\eta^4(\tau) }  \ , \nonumber\\ 
\mathrm{tr}_{R} \ \left[q^{ \ N_B+N_F} \
    \frac{1\pm \Gamma_9(-1)^F}{2}\right] \ &=& \ \frac{1}{\eta^8} \
\frac{\vartheta_2^4(0|\tau)
\pm\vartheta_1^4(0|\tau)}{2\eta^4(\tau) }  \ ,
\ea
dove il segno nel settore di Ramond indica la chiralit\`a del vuoto. A questo punto siamo quindi in grado di scrivere esplicitamente la (\ref{torus_susy}), e tenendo conto del segno negativo per i contributi fermionici
\ba
\label{torus2_susy}
\mathcal{T} \ = \ \int_{\mathcal{F}}\frac{d^2\tau}{\tau_2^6} \
\frac{1}{\eta^8(\tau) \eta^8(\bar \tau)}\left( \frac{\vartheta_3^4(0|\tau)
-\vartheta_4^4(0|\tau)}{2\eta^4(\tau) } - \frac{\vartheta_2^4(0|\tau)
\pm\vartheta_1^4(0|\tau)}{2\eta^4(\tau) } \right) \times \\ \nonumber
\left( \frac{ \vartheta_3^4(0|\bar \tau)
- \vartheta_4^4(0| \bar \tau)}{2 \eta^4(\bar \tau) } - \frac{\vartheta_2^4(0|\bar \tau)
\pm\vartheta_1^4(0|\bar \tau)}{2\eta^4(\bar \tau) } \right) \ .
\ea
I due segni $\mp$ vengono fissati indipendentemente con la scelta della chiralit\`a del vuoto R nei settori sinistro e destro, e si hanno due differenti funzioni di partizione a seconda che i due segni siano concordi o discordi. Le due teorie hanno spettri supersimmetrici: infatti l'identit\`a
\be
\vartheta_3^4-\vartheta_4^4-\vartheta_2^4 \ = \ 0 \ ,
\ee
conosciuta come \textit{aequatio identica satis abstrusa} di Jacobi \cite{ww}, alla luce delle considerazioni fatte, implica che ad ogni livello lo spettro contiene lo stesso numero di gradi di libert\`a bosonici e fermionici.

Per scrivere le funzioni di partizione delle due teorie supersimmetriche introduciamo la notazione dei caratteri, che \`e un utile modo per codificare l'intero contenuto di una rappresentazione. Nelle Teorie Conformi un carattere pu\`o essere scritto come
\be
\label{form}
\chi(q) \ = \ q^{h-c/24}\sum_k d_k q^k ,
\ee
dove $h$ \`e il peso conforme del campo primario e $c$ la carica centrale. I caratteri dell'estensione affine dell'algebra di $so(8)$ di livello $k=1$ sono esprimibili come 
\ba
\label{characters}
O_{8} \ &=& \ \frac{\vartheta_3^4+\vartheta_4^4}{2\eta^4} \ , \qquad\qquad 
V_{8} \ = \ \frac{\vartheta_3^4-\vartheta_4^4}{2\eta^4} \ ,\qquad\qquad 	(NS)\nonumber\\
S_{8} \ &=& \ \frac{\vartheta_2^4+\vartheta_1^4}{2\eta^4}\ , \qquad\qquad 
C_{8} \ = \ \frac{\vartheta_2^4-\vartheta_1^4}{2\eta^4} \ . \qquad\qquad 
(R) 
\ea
Se si tiene conto anche dei gradi di libert\`a bosonici, dividendo i caratteri di $so(8)$ per $\eta^8$ si trova che gli sviluppi in $q$ di $O_8$, $V_8$, $S_8$ e $C_8$ contengono all'ordine pi\`u basso rispettivamente un tachione, un vettore, uno spinore sinistro e uno destro di Majorana-Weyl.

Le funzioni di partizione di superstringa chiusa possono ora essere scritte in forma molto semplice. Scegliendo i vuoti di Ramond destro e sinistro concordi si ottiene la funzione di partizione della teoria IIB
\be
\label{IIB}
\mathcal{T}_{IIB} \ = \ \int \frac{d^2\tau}{\tau_2^2} \
\frac{1}{\tau_2^4 \ (\eta \ \bar\eta)^8} \ |V_8-S_8|^2 \ ,
\ee
altrimenti si ottiene la teoria IIA
\be
\label{IIA}
\mathcal{T}_{IIA} \ = \ \int \frac{d^2\tau}{\tau_2^2} \
\frac{1}{\tau_2^4 \ (\eta \ \bar\eta)^8} \ (\bar V_8-\bar S_8) \ ( V_8-
C_8) \ .
\ee
Lo spettro di basse energie della IIA e della IIB pu\`o essere facilmente ritrovato dalle funzioni di partizione ricordando le propriet\`a dei caratteri. Il termine $V_8\bar V_8$ del settore NS-NS \`e comune alle due teorie e contiene il gravitone  $G_{\mu \nu}$, il dilatone $\phi$ e la due forma  $B_{\mu \nu}$. Per la IIA nel settore R-R $S_8\bar C_8$ porta un vettore abeliano e una 3-forma, mentre i temini NS-R e R-NS, $V_8\bar C_8$ e $S_8\bar V_8$ due gravitini e due dilatini di chiralit\`a opposta. Nella teoria IIB dal settore RR si ha un secondo scalare, un'altra 2-forma e una 4-forma con curvatura autoduale da $S_8\bar S_8$; due gravitini e due dilatini di stessa chiralit\`a dai termini misti $V_8\bar S_8+S_8\bar V_8$.

Per verificare l'invarianza modulare delle funzioni di partizione trovate \`e utile determinare le matrici che implementano le trasformazioni $T$ ed $S$ sui caratteri dell'algebra di $so(8)$. Per farlo occorre partire dalle trasformazioni modulari sulle funzioni $\vartheta$
\be
\vartheta \left[ {\textstyle {\alpha \atop \beta}}
 \right] (z|\tau+1)
= e^{-i \pi \alpha
(\alpha -1)} \vartheta \left[ {\textstyle {\alpha \atop \beta +\alpha - 1/2}}
\right]
(z|\tau) \ ,
\ee
\be
\vartheta \left[ {\textstyle {\alpha \atop \beta}} \right]
\left(\frac{z}{\tau}\right|\left.-\frac{1}{\tau}\right) =
(-i \tau)^{1/2} \ e^{2 i \pi \alpha \beta + i \pi z^2/\tau} \ \vartheta
\left[{\textstyle {\beta \atop -\alpha}}\right] (z|\tau) \ .
\ee
Sostituendo nelle (\ref{characters}) si trova 
\be
\label{Tcharacters}
T = e^{ -i \pi/3} \ {\rm diag} \left( 1,-1,-1,-1 \right) \ ,
\ee
\ba
\label{Scharacters}
S = \frac{1}{2} \ \left( \begin{array}{rrrr}
1 & 1 & 1 & 1 \\
1 & 1 & -1 & -1 \\
1 & -1 & 1 & -1 \\
1 & -1 & -1 & 1
\end{array} \right) \ .
\ea
Ricordando che, come si \`e visto per la stringa bosonica, la misura di integrazione e il fattore $\tau_2^4 \ (\eta\bar\eta)^8$ sono invariati modulari, \`e immediato, usando le trasformazioni (\ref{Tcharacters}) e (\ref{Scharacters}), verificare l'invarianza modulare delle funzioni di partizione trovate.

\subsection{Superstringa di Tipo I}

Come si \`e visto nel caso bosonico, si pu\`o costruire una teoria di stringhe aperte e chiuse non orientate proiettando su stati invarianti sotto $\Omega$. Per avere una teoria consistente occorre che la teoria di partenza sia invariante sotto $\Omega$, e questo avviene solo per la IIB, dove i settori destro e sinistro hanno la stessa proiezione GSO \cite{cargese}. La teoria risultante \`e detta superstringa di tipo I. Al contrario la teoria IIA non pu\`o essere proiettata rispettando l'invarianza di Lorentz in $D=10$. 

Procedendo esattamente come nel caso bosonico e ricordando i calcoli fatti per le tracce fermioniche, si trovano nel canale diretto le ampiezze
\be
\label{Klein_susy}
{\cal K} = \frac{1}{2} \int_0^{\infty} \ \frac{ d \tau_2}{\tau_2^6} \
\frac{ (V_8 - S_8 ) ( 2 i \tau_2)}{\eta^8(2 i \tau_2)} \ ,
\ee
\be
\label{Annulus_susy}
{\cal A} = \frac{N^2}{2} \int_0^{\infty} \ \frac{ d \tau_2}{\tau_2^6} \
\frac{ (V_8 - S_8 ) ( \frac{1}{2} i \tau_2 )}{\eta^8( {\textstyle{1\over 2}} i \tau_2)} \ ,
\ee
\be
\label{Mobius_susy}
{\cal M} = \frac{\epsilon N}{2} \int_0^{\infty} \
\frac{ d \tau_2}{\tau_2^6} \
\frac{ (\hat{V}_8 - \hat{S}_8 ) ( {\textstyle{1\over 2}}
i \tau_2 + {\textstyle{1\over 2}})}{\hat{\eta}^8(
{\textstyle{1\over 2}} i \tau_2 + {\textstyle{1\over 2}})} \ .
\ee
Nella (\ref{Mobius_susy}) per avere un integrando reale si sono introdotti caratteri $\hat \chi$ definiti in generale come
\be
\label{hatcharacter}
\hat{\chi}(i \tau_2 +{\textstyle{1\over 2}}) =
q^{h - c/24} \sum_k (-1)^k d_k q^k \ , \qquad q=e^{-2\pi\tau_2} \ ,
\label{chireal}
\ee
che differiscono da $\chi(i\tau_2+1/2)$ per una fase  $e^{-i\pi(h-c/24)}$. 

La bottiglia di Klein simmetrizza il settore $NS-NS$ e antisimmetrizza il settore $R-R$. Lo spettro di massa nulla nel settore chiuso risultante dalla proiezione  $\mathcal{T}\rightarrow \mathcal{T}/2+\mathcal{K}$ si ottiene dalla IIB eliminando la $2$-forma dal settore $NS-NS$ e lo scalare e la $4$-forma autoduale dal settore $R-R$, e dimezzando i settori misti. Gli stati che sopravvivono alla proiezione formano un multipletto $\mathcal{N}=1$ di supergravit\`a minimale in $D=10$ dimensioni. Nel settore aperto proiettato si hanno $(N^2-\epsilon N)/2$ vettori di massa nulla e i corrispettivi partner supersimmetrici, fermioni di Majorana-Weyl. Si ha quindi un multipletto di $N=1$ super Yang-Mills nell'aggiunta del gruppo $SO(N)$ per $\epsilon=1$, e nell'aggiunta di $USp(N)$ per $\epsilon=-1$.

Il passaggio nel canale trasverso nel caso della bottiglia di Klein e dell'anello non comporta nessuna novit\`a rispetto al caso bosonico, dal momento che come si \`e visto la combinazione $(V_8-S_8)$ \`e invariante sotto S. Le ampiezze trasverse sono quindi:
\be
\label{tildeK_susy}
\tilde{\cal K} = \frac{2^5}{2} \int_0^{\infty} \ d \ell  \
\frac{ (V_8 - S_8 ) (i \ell )}{\eta^8(i \ell)} \, .
\ee
\be
\tilde{\cal A} = \frac{2^{-5} \ N^2}{2} \int_0^{\infty} \
d \ell \
\frac{ (V_8 - S_8 ) ( i \ell )}{\eta^8( i \ell )} \ .
\ee
I coefficienti dei caratteri $V_8$ e $-S_8$ sono da interpretare in termini di quadrati delle funzioni ad un punto su un bordo. Il passaggio al canale trasverso per la M\"obius richiede qualche precisazione. La ridefinzione dei caratteri (\ref{hatcharacter}) porta a ridefinire anche la trasformazione P, che come gi\`a osservato diventa \cite{gpp}
\be
\label{hatP}
P=T^{1/2}ST^2ST^{1/2} \ .
\ee
Sulla base 
\be
\frac{O_8}{\tau_2^4 \eta^8} \, , \quad
\frac{V_8}{\tau_2^4 \eta^8} \, , \quad \frac{S_8}{\tau_2^4
\eta^8} \, , \quad \frac{C_8}{\tau_2^4 \eta^8}
\ee
la matrice T agisce in maniera semplice $T = \ {\rm diag}(-1,1,1,1) \ $, e dal momento che in generale per una teoria conforme vale
\be
S^2=(ST)^3=\mathcal{C} \ ,
\ee 
dove si \`e indicata con $\mathcal{C}$ la matrice di coniugazione, che essendo le rappresentazioni di $so(8)$ autoconiugate, nel nostro caso coincide con l'identit\`a, dalla (\ref{hatP}) si ottiene
\be
P \ = \ T \ = {\rm diag}(-1,1,1,1) \ .
\ee
Quindi l'ampiezza di M\"obius nel canale trasverso risulta essere
\be
\label{Mtilde_susy}
\tilde{\cal M} = 2 \ \frac{\epsilon N}{2} \int_0^{\infty} \ d \ell  \
\frac{ (\hat{V}_8 - \hat{S}_8 ) ( i \ell + {\textstyle{1\over 2}})}
{\hat{\eta}^8(i \ell + {\textstyle{1\over 2}})} \ .
\ee
Le condizioni di tadpole della Tipo I per i settori $NS-NS$ e $R-R$, in conseguenza della supersimmetria, risultano nell'unica condizione
\be
\frac{2^5}{2} + \frac{2^{-5} \ N^2}{2} + 2 \ \frac{\epsilon N}{2} =
\frac{2^{-5}}{2} \ ( N + 32 \epsilon )^2 = 0 \ ,
\ee
che seleziona i valori (N=32, $\epsilon=-1)$, e quindi il gruppo di gauge $SO(32)$ \cite{gsop}.

\`E importante notare che, sebbene per la superstringa di Tipo I i tadpole NS-NS e R-R vengano cacellati simultaneamente, concettualmente le condizioni nei due settori hanno un significato fisico molto differente. Quello che si pu\`o vedere \`e che, dal punto di vista spazio-temporale, i bordi del worldsheet tracciati dalle estremit\`a delle stringhe aperte sono mappati in oggetti aperti, $D9$-brane, che riempiono completamente lo spazio tempo, mentre i crosscap vengono mappati in oggetti non dinamici gli $O$-piani. In generale sia le $Dp$-brane che gli $Op$-piani hanno tensione e portano cariche R-R con potenziali che sono $(p+1)$-forme $C_{p+1}$. Per una $D$-brana tensione e carica sono entrambe positive, mentre come si avr\`a modo di vedere, nel vuoto perturbativo della Tipo I sono presenti due tipi di $O$-piani: gli $O_+$-piani con carica e tensione negative e gli $O_-$-piani con carica e tensione positiva. In pi\`u si hanno $D$-antibrane e $O$-antipiani (indicate con $\bar D$-brane e $\bar O$-piani) con stessa tensione e cariche R-R opposte.

A questo punto dovrebbe essere chiaro che mentre la cancellazione dei tadpole R-R \`e una condizione di di neutralit\`a della totale della carica, necessaria in presenza di compattificazioni dal momento che le linee di Farady del potenziale $C_{p+1}$ si trovano ad essere confinate. Al contrario, la condizione di tadpole NS-NS, come si \`e visto da luogo ha una correzione all'energia di vuoto dipendente dal dilatone, 
\be
V_{\phi} \sim T\int dx \sqrt{-\det G}e^{-\phi} \ ,
\ee
che pu\`o anche non essere cancellata \cite{fs}. Difatti nei modelli con supersimmetria rotta, i tadpole NS-NS non possono essere, in generale, cancellati e sono accompagnati dall'insorgere di divergenze infrarosse. Questo indica che il background Minkowskiano non \`e pi\`u una soluzione della teoria e che \`e necessaria una ridefinizione del vuoto \cite{Dudas:2004nd}.

\section{Proiezione di orientifold}

Si \`e visto sia nel caso di stringa bosonica sia in quello di superstringa come sia possibile costruire, con la proiezione di orientifold, da una teoria di stringhe chiuse orientate, teorie consistenti di stringhe aperte e chiuse non orientate. Diamo ora una formulazione pi\`u generale della costruzione. L'ampiezza di toro \`e in generale scrivibile, lasciando sottintesa l'integrazione, come
(\ref{general torus})
\be
\label{general torus}
{\cal T} \ = \ \sum_{i,j}\bar \chi_i X_{ij} \chi_j \ ,
\ee
con $X_{ij}$ una matrice generica di numeri interi (nei modelli razionali \`e finto dimensionale). Si pu\`o richiedere per semplicit\`a $X_{ij}=0, 1$, dal momento che altrimenti si avrebbe una ambiguit\`a da risolvere nella proiezione. Per avere invariaza modulare si devono imporre i vincoli  
\be
\label{modular_constraints}
S^\dag  \, X  \, S = X \,, \qquad T^\dag  \, X  \, T = X \ .
\ee
La bottiglia di Klein si ottiene identificando modi destri e sinistri, e quindi si propagheranno solo i settori che nel toro siano simmetrici sotto $\Omega$, quindi con $X_{ii}\neq 0$
\be
{\cal K} \ = \ \frac{1}{2}\sum_i{\cal K}^i \chi_i \ , \qquad 
{\cal K}^i = \pm X_{ii} \ .
\ee
Il segno dei ${\cal K}^i$ \`e indeterminato, dal momento nell'ampiezza di toro $X_{ij}$ era associato ai moduli quadri dei caratteri. Nel canale trasverso si pu\`o scrivere
\be
\tilde{\cal K} =  \frac{1}{2} \sum_i  ({\Gamma}^i)^2 \,
\chi_i \ ,
\ee
dove i coefficienti ${\Gamma}^i$ sono funzioni ad un punto (ovvero coefficienti di riflessione) dei caratteri sui crosscap. I segni di ${\cal K}^i$ quindi devono essere scelti in modo da avere una teoria interagente consistente ovvero con coefficienti positivi $({\Gamma}^i)^2=K_jS_{ij}$ nel canale trasverso. 

L'ampiezza di anello nel canale trasverso ha la forma
\be
\tilde{\cal A} = \frac{1}{2} \sum_{i} \, \chi_i
\left( \sum_a B^i_a \, n^a \right)^2 \ ,
\ee
dove $\sum_a B^i_a n^a$ sono le funzioni ad un punto sui bordi e le $n_a$ le molteplicit\`a associate alle cariche di Chan-Paton. Dal momento che un settore che si rifletta su di un bordo subisce una coniugazione di carica, nell'anello trasverso si propagheranno solo i caratteri $\chi_i$ che compaiano nel toro nella forma $\bar \chi_i^C \chi_i $. Nel canale diretto l'ampiezza di anello pu\`o essere scritta come
\be
{\cal A} = \frac{1}{2} \sum_{i,a,b} \, {\cal A}^i_{ab}
\, n^a\, n^b\,
\chi_i \, ,
\ee
con ${\cal A}^i_{ab}=0, 1$ come nel caso del toro. 

Il nastro di M\"obius nel canale trasverso si \`e visto essere un tubo di propagazione fra un bordo e un crosscap, e pertanto i coefficenti dei caratteri possono essere scritti come prodotti delle rispettive funzioni ad un punto su un bordo e su un crosscap che compaiono nell'anello e nella bottiglia di Klein. Un fattore moltiplicativo due tiene conto delle due possibili configurazioni. Riassumendo l'ampiezza trasversa risulta
\be
\tilde{\cal M} = \frac{2}{2} \sum_{i}\,
\hat{\chi}_i \, {\Gamma}^i  \left( \sum_a B^i_a \, n^a \right) \ ,
\ee
mentre nel canale diretto si ha
\be
{\cal M} = {\textstyle \frac{1}{2}}\sum_{i,a} \, {\cal M}^i{}_{a} \, n^a\,
\hat{\chi}_i \ .
\ee
Si pu\`o verificare la consistenza della costruzione controllando che $\cal M$ sia la corretta proiezione di $\cal A$, ovvero che si abbia ${\cal M}^i{}_{a}=\pm {\cal A}_{aa}^i$.


\chapter{Compattificazioni e T-dualit\`a}

I modelli di superstringa sono consistenti in $D=10$, mentre la stringa bosonica ha dimensione critica $D=26$. Dal momento che la nostra esperienza fisica mostra l'esistenza di sole tre dimensioni spaziali estese e del tempo, per costruire modelli realistici occorre introdurre meccanismi che rendano compatte e ``microscopiche'' le dimensioni extra. In altri termini occorre pensare ad uno spazio-tempo Minkowskiano spontaneamente rotto,
\be
{\cal M}^D = {\cal M}^4 \times K^{D-4} \ ,
\ee
dove $K$ \`e una variet\`a compatta.  Il modo pi\`u semplice di introdurre compattificazioni \`e quello di identificare periodicamente $n$ coordinate su di un toro di dimensione $n$. Si tratta di una estensione molto semplice del modello di Kaluza-Klein, che per\`o nel caso di stringa comporta l'insorgere di fenomeni del tutto nuovi rispetto alla teoria dei campi come la presenza degli stati di winding, l'innalzamento delle simmetrie di gauge e la T-dualit\`a \cite{cs}. Si \`e inoltre indotti poi ad introdurre oggetti dinamici estesi, le $D$-brane e gli $O$-piani \cite{pol95}.

Una generalizzazione di questo tipo di compattificazioni sono le compattificazioni su orbifold \cite{dhvw}, che si ottengono identificando i punti della variet\`a interna , sotto l'azione di un gruppo discreto. Un orbifold non \`e una variet\`a liscia dal momento, che in generale, si avranno punti fissi sotto l'azione del gruppo discreto che per\`o possono essere rimossi ottenendo variet\`a lisce di Calabi-Yau.

\section{Compattificazioni toroidali}

\subsection{Compattificazione su $S^1$ per stringhe chiuse}

Partiamo dal caso pi\`u semplice, quello di una teoria di stringa bosonica compattificata su un cerchio unidimensionale $S^1$ di raggio R. Si devono imporre condizioni di periodicit\`a su un singolo campo scalare, 
\be
X(\tau,\sigma) \sim X(\tau,\sigma)+2\pi R  \ ,
\ee
e la periodicit\`a ha due effetti. Anzitutto, dal momento che l'operatore di traslazione deve essere univocamente definito
\be
e^{i p X} \sim e^{ip(X+2\pi R)} \ ,
\ee
l'impulso del centro di massa \`e quantizzato:
\be
p=\frac{n}{R} \ .
\ee
Il secondo effetto \`e peculiare della Teoria delle Stringhe: una stringa chiusa pu\`o avvolgersi intorno alla dimensione compatta,
\be
X(\tau,\sigma + \pi) = X(\tau,\sigma)+2\pi Rw \ , \qquad w \in \mathbb{Z} \ .
\ee

Il  numero intero $w$ \`e il \emph{winding number} ed \`e conservato nelle interazioni di stringa. Le osservazioni fatte possono essere raccolte scrivendo l'espasione del campo $X$ in oscillatori
\be
X \ = \ x \ + \ 2\alpha'\frac{n}{R}\tau+2wR\sigma+
i\frac{\sqrt{2\alpha'}}{2} \ \sum_{k\neq
  0}\left(\frac{\alpha_k}{k} \ e^{-2ik(\tau-\sigma)}+
\frac{\tilde\alpha_k}{k} \ e^{-2ik(\tau+\sigma)}\right) \ .
\ee
Che pu\`o essere scritta come somma dei modi destri e sinistri $X=X_L+X_R$, con
\be
X_{L,R} \ = \ \frac{1}{2}x+\alpha'p_{L,R}(\tau\mp\sigma)+(oscillatori) \ ,
\ee
dove si sono definiti gli impulsi sinistro e destro come
\be
\label{PLPR}
p_L=\frac{n}{R}+\frac{w R}{\alpha'} \ , \qquad \qquad
p_R=\frac{n}{R}-\frac{w R}{\alpha'} \ .
\ee

La massa degli stati di stringa va riscritta tenendo conto che, dal punto di vista di un osservatore che viva in $(D-1)$-dimensioni, $M^2=-p^2$, dove  p \`e il momento degli stati in 25 dimensioni. In altri termini, i momenti interni contribuiscono all'energia a riposo della stringa, e la formula di massa \`e quindi 
\be
\label{mc}
M^2 \ = \ \frac{2}{\alpha'}\left[ \frac{\alpha'}{4}p_L^2+
\frac{\alpha'}{4}p_R^2+ N^\bot+\bar{N}^\bot -2\right] \ , 
\ee
con il vincolo 
\be
\label{lmcmc}
\frac{\alpha'}{4}p_R^2+N^\bot - \left(\frac{\alpha'}{4}p_L^2+\bar N^\bot \right) \ = \ 0 \ .
\ee

\subsubsection{Spettro compattificato ed allargamento della simmetria di gauge}

Usando le (\ref{PLPR}), possiamo riscrivere il vincolo di massa e la condizione di level matching come
\be
\label{compactmass}
M^2 \ = \ \frac{n^2}{R^2}+ \frac{w^2R^2}{\alpha'^2}+ 
\frac{2}{\alpha'}\left[ N^\bot+\bar{N}^\bot-2 \right] \ , \qquad 0=nw +N^\bot-\bar{N}^\bot \ .
\ee
Studiamo i modi di massa nulla, cosiderando una compattificazione lungo la direzione $X^{25}$. A valori generici di R, per avere stati di stringa a massa nulla, deve aversi $m=w=0$ e $N=\bar N=1$. Si hanno $24^2$ stati di massa nulla come per la teoria non compatta, che adesso possiamo scrivere in maniera conveniente, dal punto di vista 25 dimensionale come
\be
\alpha^\mu_{-1}\tilde\alpha^\nu_{-1}|0,0\rangle \ , \qquad 
\alpha^\mu_{-1}\tilde\alpha^{25}_{-1}|0,0\rangle \ , \qquad 
\alpha^{25}_{-1}\tilde\alpha^\mu_{-1}|0,0\rangle \ , \qquad 
\alpha^{25}_{-1}\tilde\alpha^{25}_{-1}|0,0\rangle \ .
\ee
Il primo si decompone in un gravitone $G_{\mu \nu}$, un tensore antisimmetrico $B_{\mu \nu}$ ed un dilatone $\phi$ in 25 dimensioni. Il secondo e il terzo sono vettori di Kaluza-Klein 
\be
A{_\mu(R)}\equiv \frac{1}{2}(G-B)_{\mu,25} \ , \qquad  A{_\mu(L)}\equiv \frac{1}{2}(G+B)_{\mu,25}
\ee
che portano una simmetria di gauge $U(1)_L \times U(1)_R$ . L'ultimo \`e uno scalare che possiamo scrivere come 
\be
\phi'\equiv \frac{1}{2}\log G_{25,25} \ , 
\ee
ed \`e il modulo del raggio effettivo di compattificazione, dal momento che la radice quadrata della componente della metrica $G_{25,25}$ \`e proprio la misura del raggio della direzione compattificata $X^{25}$. Il suo valore di aspettazione non pu\`o essere fissato con un principio di minimo e rimane indeterminato.  

Un nuovo effetto tipicamente di stringa si ha al valore del raggio $R=\sqrt{\alpha'}$, dove il vincolo di massa pu\`o essere scritto come
\be
M^2=\frac{2}{\alpha'}(n+w)^2 + \frac{4}{\alpha'}(N^\bot-1)=
\frac{2}{\alpha'}(n-w)^2 + \frac{4}{\alpha'}(\bar N^\bot-1) \ ,
\ee
con i vincoli
\be
(n+w)^2 + 4N^\bot = 4 \ , \qquad (n-w)^2 + 4\bar N^\bot = 4 \ .
\ee
In aggiunta alle soluzioni di massa nulla gi\`a discusse per $n=w=0$ con $N=\bar N=1$, si hanno anche 
\be
n=w=\pm 1 \ , \ N^\bot=0 \ , \ \bar N^\bot =1 \ ; \qquad n=-w=\pm 1 \ , \ N^\bot= 1 \ , \  \bar N^\bot =0 \ ;
\ee 
\be
n= \pm 2 \ , \ w=0 \ , \ N^\bot=\bar N^\bot =0 \ ; \qquad n=0 \ , \ w=\pm 2 \ , \ N^\bot=\bar N^\bot =0\ .
\ee
Che in termini di stati corrispondono a 8 nuovi scalari
\be
\tilde\alpha_{-1}^{25}|\pm1,\mp1\rangle \ , \qquad
\alpha_{-1}^{25}|\pm1,\pm1\rangle \ , \qquad
|\pm2,0\rangle \ , \qquad
|0,\mp2\rangle \ ,
\ee
e 4 nuovi vettori
\be
\tilde\alpha_{-1}^\mu|\pm1,\mp1\rangle \ , \qquad
\alpha_{-1}^\mu|\pm1,\pm1\rangle \ .
\ee
In totale si hanno quindi 9 particelle scalari, 3 vettori destri e 3 vettori sinistri, il gruppo di simmetria abeliano $U(1)_L \times U(1)_R$ \`e stato promosso a $SU(2)_L\times SU(2)_R$. 

Spostando R dal valore $\sqrt {\alpha'}$ i bosoni di gauge extra acquistano massa
\be
M = \frac{\left| R^2-\alpha' \right|}{R\alpha'}\approx \frac{2}{\alpha'} \left| R-\alpha'^{1/2}\right| \ .
\ee
Come noto, \`e possibile in Teoria dei Campi (e quindi anche nella teoria effettiva di basse energie di stringa) dare massa a bosoni di gauge attraverso una rottura spontanea di simmetria. Infatti per $R=\sqrt {\alpha'}$ ci sono 10 scalari di massa nulla, alcuni dei quali vengono riassorbiti per dare massa a 4 dei sei vettori di gauge.

\subsubsection{T-dualit\`a per stringhe chiuse}

Dalla formula di massa (\ref{compactmass}) si vede che nel limite $R \to \infty$ gli stati di winding diventano infinitamente massivi mentre lo spettro degli impulsi diventa continuo, e pertanto nel limite di decompattificazione si ritrova la fisica conosciuta. Al contrario nel limite $R \to 0$ gli stati di momento compattificato diventano infinitamente massivi, mentre lo spettro di winding diventa continuo. Si trova quindi che anche nel limite di riduzione dimensionale lo spettro di stringa sembra nuovamente diventare quello di una dimensione non compatta, ovvero i limiti $R \to \infty$ e $R \to 0$ sono fisicamente identici. Lo spettro di massa \`e infatti invariante sotto
\be
\label{T-duality close}
T: \qquad R \to R'=\frac{\alpha'}{R}, \qquad n \leftrightarrow w \ .
\ee
La trasformazione definita, detta \emph{T-dualit\`a}, agisce sui momenti come
\be
T: \qquad p_L \to p_L, \  p_R \to - p_R \qquad {\rm ovvero} \qquad 
T: \qquad  \alpha_0 \to \alpha_0 , \ \tilde\alpha_0 \to -\tilde\alpha_0 \ .
\ee
Si pu\`o definire l'azione sugli oscillatori generalizzando quella sugli zero modi, si vede in questo modo che la T-dualit\`a agisce come una trasformazione di parit\`a sul solo settore destro:
\be
\label{TX}
T:X(\tau, \sigma)= X_L(\tau - \sigma) + X_R(\tau+\sigma) 
\to X^T(\tau, \sigma) = X_L(\tau - \sigma) - X_R(\tau+\sigma)
\ee

Si \`e detto che i limti $R \to \infty$ e $R \to 0$ sono fisicamente equivalenti per la stringa bosonica, e quindi lo spazio delle teorie inequivalenti \`e definito dai dominii $R \geq \alpha'^{1/2}$ oppure $0\leq R \geq \alpha'^{1/2}$. La prima scelta \`e pi\`u naturale, dal momento che \`e pi\`u intuitivo ragionare in termini di momenti continui piuttosto che di windings. 

Dalla teoria effettiva di bassa energia si pu\`o vedere che la T-dualit\`a  \`e anche una simmetria della teoria interagente, ma la sua azione \`e non banale sul dilatone (e pi\`u in generale sui campi di background), e quindi modifica la costante di accoppiamento di stringa:
\be
T: \phi \to \phi'=\phi+log{{\sqrt{\alpha'}}{R}} \ .
\ee
Si pu\`o vedere che al raggio autoduale $R=\sqrt{\alpha'}$, in cui si ha l'enhancement del gruppo di gauge, la simmetria $\mathbb Z_2$ di T-dualit\`a diventa parte della simmetria continua $SU(2)\times SU(2)$. Questo \`e particolarmente importante in quanto che indica che la T-dualit\`a non \`e una simmetria solo della teoria perturbativa, ma pi\`u in generale della teoria esatta.

\subsubsection{Funzione di partizione}

Calcoliamo l'ampiezza di vuoto di stringa chiusa su $S^1$ a partire dalla (\ref{imp}), dove l'integrazione va fatta sui momenti dello spazio non compatto \mbox{($D-1$)dimensionale}, per cui si ottiene una differente potenza del modulo $\tau_2$
\be
\frac{1}{\tau_2^{D/2+1}} \ \rightarrow \ \frac{1}{\tau_2^{(D-1)/2+1}} \ .
\ee
Utilizzando la formula di massa (\ref{mc}) e il vincolo (\ref{lmcmc}) si ottiene inoltre la traccia 
\be
\mathrm{tr}\left(q^{N^\bot+\frac{\alpha'}{4}p_R^2 -1} \
\bar q^{\bar N^\bot +\frac{\alpha'}{4}p_L^2 -1}  \right) = 
\frac{\sum_{m,n} q^{\frac{\alpha'}{4}p_R^2} \ \bar
  q^{\frac{\alpha'}{4}p_L^2}}{[\eta(\tau)  \eta(\bar \tau)]^{24}}\ ,
\ee
che \`e stata calcolata sugli stati $|k; n, w\rangle$, e la funzione di partizione del toro \`e quindi
\be
\label{torus_compact}
\mathcal{T} \ = \ \int_{\mathcal{F}}\frac{d^2\tau}{\tau_2^2} \
\frac{\sum_{m,n} q^{\frac{\alpha'}{4}p_R^2} \ \bar
  q^{\frac{\alpha'}{4}p_L^2}}{\tau_2^{12-1/2}[\eta(\tau)  \eta(\bar \tau)]^{24}}  \ .
\ee
L'ampiezza trovata \`e invariante modulare. Sotto T, essa prende una fase $\exp(i\pi (p_L^2-p_R^2))$ che si vede essere unitaria ricordando $p_L^2-p_R^2=2nw$. Per verificare l'invarianza sotto S occorre la formula di rissomazione di Poisson
\be
\label{poissonsum}
\sum_{\{n_i\} \in \mathbb{Z}} \, e^{-\pi \, n^{\rm T}\, A \,n \ + \ 2 \, i
\, \pi \, b^{\rm T} \, n} \ = \
\frac{1}{\sqrt{{\rm det}(A)}} \, \sum_{\{m_i\} \in \mathbb{Z}}
 \, e^{-\pi \, (m - b)^{\rm T} \, A^{-1} \, (m - b)} \, , 
\ee
e dopo una trasformazione $S$ la sommatoria in (\ref{torus_compact}) usando le espressioni (\ref{PLPR}) si scrive
\be
\sum_{n,w}e^{-\pi n^2\left(\frac{\alpha'\tau_2}{|\tau|^2 R^2}\right)} \ 
e^{2\pi in\left(\frac{w\tau_1}{|\tau|^2}\right)} \
e^{\frac{-\pi\tau_2 w^2 R^2}{\alpha'|\tau|^2}} \ .
\ee
Si pu\`o usare la formula di risommazione per $n$ con $A= \alpha'\tau_2 / |\tau|^2 R^2/$ e $b=w\tau_1/|\tau|^2$ ottenendo
\be
\frac{|\tau|R}{\sqrt{\alpha'\tau_2}}\sum_{n',w}e^{-\frac{\pi|\tau|^2
    R^2}{\alpha'\tau_2}\left(n'-\frac{w\tau_1}{|\tau|^2}\right)^2}
e^{\frac{-\pi\tau_2 w^2 R^2}{\alpha'|\tau|^2}} \ ,
\ee
che pu\`o essere riscritta nella forma
\be
\frac{|\tau|R}{\sqrt{\alpha'\tau_2}}\sum_{n',w}e^{-\pi
  w^2\left(\frac{R^2}{\alpha'\tau_2}\right)}
 \ e^{2\pi i n'\frac{iR^2 w \tau_1}{\alpha'\tau_2}} \ e^{-\pi
   n'^2\frac{|\tau|^2R^2}{\alpha'\tau_2}} \ .
\ee
Utilizzando nuovamente la formula di risommazione rispetto a $w$ con $A=R^2/\alpha'\tau_2$ e $b=iR^2 n
\tau_1/\alpha'\tau_2$ si arriva a scrivere 
\be
|\tau|\sum_{n',w'}e^{-\pi\frac{\alpha'\tau_2}{R^2}\left(w'+i\frac{R^2 n'\tau_1}{\alpha'\tau_2}\right)^2}
\ e^{-\pi\frac{|\tau|^2 R^2 n'^2}{\alpha'\tau_2}} \ ,
\ee
che si vede subito essere l'espressione di partenza con winding e momenti scambiati e con un fattore moltiplicativo $|\tau|$ che viene riassorbito dalla trasformazione S su $1/\eta\bar\eta$.

\subsection{Compattificazione su $S^1$ per stringhe aperte}

Le stringhe aperte non possono allacciarsi lungo una direzione compatta, e non hanno quindi numero di winding, e pertanto il loro comportamento nel limite $R \to 0$ deve per questo essere differente da quello delle stringhe chiuse e pi\`u simile  a quello di una particella. Per $R \to 0$ gli stati di momento diventano infinitamente massivi ma, mancando uno spettro di winding che diventi continuo, si ha una riduzione dimensionale. Si \`e visto che una teoria di stringhe aperte deve contenere anche stringhe chiuse, e sembrerebbe quindi che nel limite di ``riduzione dimensionale'' si trovi una teoria in cui le stringhe aperte vivono in  $D-1$ dimensioni e quelle chiuse in $D$. In realt\`a i punti che distinguono una stringa chiusa da una aperta sono gli estremi. Con questa osservazione si pu\`o arrivare a capire che nel limite $R \to 0$ entrambe le stringhe oscillano in spazi $D$-dimensionali, ma gli estremi di stringa aperta si trovano vincolati ad un iperpiano $D-1$-dimensionale.

Scrivendo l'espansione dei modi di una stringa aperta nella forma $X=X_L+X_R$ con,
\ba
X_L &=& \frac{1}{2}(x+q)+\alpha'\frac{n}{R}(\tau+\sigma)+
i\frac{\sqrt{2\alpha'}}{2}\sum_k\frac{\alpha_k}{k}e^{-ik(\tau+\sigma)}\ ,\nonumber\\
X_R &=& \frac{1}{2}(x+q)+\alpha'\frac{n}{R}(\tau-\sigma)+
i\frac{\sqrt{2\alpha'}}{2}\sum_k\frac{\alpha_k}{k}e^{-ik(\tau-\sigma)}
\ ,
\ea
dove $q$ indica un termine arbitrario che si cancella nella somma, e ricordando la definizione della T-dualit\`a  sui campi X nella (\ref{TX}), per la coordinata duale si ottiene 
\be
X^T = X_L - X_R = q + 2\alpha'\frac{n}{R}\sigma+
i(\sqrt{2\alpha'})\sum_k\frac{\alpha_k}{k}e^{-ik\tau}\sin{(k\sigma)}\ .
\ee
Nella coordinata T-duale non c'\`e pi\`u dipendenza da $\tau$ negli zero modi, e per questo il momento del centro di massa \`e nullo. Infatti, dal momento che i termini oscillanti si annullano per $\sigma = 0, \pi$, si vede che la stringa ha estremi fissati nella direzione compattificata:
\be
X^T(\pi)=X^T(0)=\frac{2\pi \alpha' n}{R}=2\pi nR' \ .
\ee
Si hanno pertanto condizioni di Dirichlet agli estremi $\partial_\tau X=0$, in luogo delle usuali condizioni di Neumann $\partial_\sigma X=0$ . \`E immediato rendersi conto di cosa \`e avvenuto osservando che 
\ba
\partial_\sigma X &=& X'_L(\tau+\sigma)-X'_R(\tau-\sigma)=\partial_\tau X^T \ , \nonumber\\
\partial_\tau X &=& X'_L(\tau+\sigma)+X'_R(\tau-\sigma)=\partial_\sigma X^T \ .
\ea
Naturalmente questo vale solo nella coordinata T-dualizzata, mentre gli estremi rimangono liberi di muoversi liberamente nelle altre dimensioni, che definiscono un iperpiano detto $D$-brana \cite{pol95,D-Branes}. Le $D$-brane, come si vedr\`a, sono in realt\`a oggetti dinamici.

\subsubsection{Cariche di Chan-Paton e linee di Wilson}

Con le cariche di Chan-Paton, la T-dualit\`a di stringa aperta si arricchisce di nuovi aspetti. \`E infatti possibile rompere il gruppo di gauge delle cariche introducendo \emph{linee di Wilson} preservandone per\`o il rango \cite{bps}. Consideriamo, per semplicit\`a, il caso della stringa bosonica orientata con gruppo di gauge $U(N)$. Un campo costante di background, della direzione compattificata $X^{25}$, pu\`o essere diagonalizzato nella forma
\be
A_{25}= diag \left\{\theta_1\ , \theta_2\ , \dots \ , \theta_N \right\}/2\pi R \ .
\ee
In questa forma il campo si trova nel sottogruppo diagonale $U(1)^n$ di $U(N)$. Localmente \`e un campo di pura gauge, la cui curvatura \`e nulla e quindi le equazioni del campo sono soddisfatte in maniera banale da
\be
A_{25}=-i\Lambda^{-1}\partial \Lambda \ ,\qquad
 \Lambda=diag \left\{e^{iX^{25}\theta_1/2\pi R} \ , e^{iX^{25}\theta_2/2\pi R} \ , \dots \ ,  e^{iX^{25}\theta_N/2\pi R} \right\} \ .
\ee

Dal momento che il parametro di gauge $\Lambda$ non \`e periodico nella dimensione compatta si hanno interessanti effetti fisici. Infatti, anche ponendo il campo di background a zero con una trasformazione di gauge $\Lambda^{-1}$, si ha che i campi carichi sotto una trasformazione $X^{25} \to X^{25} +2\pi R$ acquistano una fase
\be
diag \left\{e^{i\theta_1} \ , e^{i\theta_2} \ , \dots \ ,  e^{i\theta_N} \right\} \ .
\ee
Si pu\`o definire una quantit\`a gauge invariante detta linea di Wilson, che tenga conto degli effetti fisici introdotti dal campo di background
\be
W=\exp{\left( iq \oint dX^{25}A_{25} \right) = \exp(-iq\theta)} \ .
\ee
L'azione di stringa viene modificata dall'introduzione di $A_{25}$ con la comparsa di termini di bordo
\be
S=-\frac{T}{2}\int d\tau d\sigma (\dot{X}^2-X'^2)+ q_i\int d \tau
\left.A_{25}\dot{X}^{25}\right|_{\sigma=0} + q_j\int d \tau
\left.A_{25}\dot{X}^{25}\right|_{\sigma=0}\ .
\ee
Il momento canonico risulta ``shiftato'', per una stringa $|ij\rangle$ che ha carica $+1$ sotto $U(1)_i$ , $-1$ sotto $U(1)_j$ si ha
\be
p_{25}= \frac{(2\pi n - \theta_j + \theta_i)}{2\pi R} \ ,
\ee
e lo spettro di massa diventa
\be
M^2= \frac{(2\pi n - \theta_j + \theta_i)^2}{4\pi^2 R^2} + \frac{1}{\alpha'}(N-1) \ .
\ee
Per i bosoni di gauge a massa nulla, con n=0 e N=1
\be
M^2= \frac{(\theta_j - \theta_i)^2}{4\pi^2 R^2}  \ ,
\ee
se tutti i $\theta_i$ sono distinti, gli unici vettori di massa nulla sono quelli con $i=j$, e il gruppo di gauge \`e rotto a $U(1)^n$.  Se invece $r$ dei $\theta_i$ sono uguali, la corrispondente matrice $r \times r$ di vettori \`e di massa nulla, e il gruppo di simmetria \`e rotto a $U(r)\times U(1)^{N-r}$.

Per capire cosa stia succedendo passiamo nella picture T-duale, dove, dal momento che gli impulsi diventano winding, si avranno winding frazionari ovvero,
\be
X^{25,T}(\pi)-X^{25,T}(0)= (2\pi n + \theta_i - \theta_j)R' \ ,
\ee
e, a meno di costanti additive, l'estremo della stringa nello stato $i$ \`e nella posizione
\be
X^{25,T}= \theta_iR'=2\pi \alpha' A_{25, ii} \ .
\ee
Gli estremi della stringa giacciono quindi su due iperpiani ($D$-brane) che si trovano in posizioni differenti. Nella picture T-duale i $\theta_i$ sono quindi angoli che definiscono nella dimensione compattificata la posizione degli iperpiani. In generale si hanno N iperpiani in posizioni differenti lungo il cerchio di compattificazione, e in questo caso gli stati di massa nulla sono gli scalari $\alpha_{-1}^{25}|i,i\rangle$ e i vettori $\alpha_{-1}^i|i,i\rangle$ nell'aggiunta di U(1). Quando si fanno coincidere $r$ delle N $D$-brane, si hanno $r^2$ vettori di massa nulla (e lo stesso numero di scalari) dati dagli stati $\alpha_{-1}^i|i,j\rangle$ per $i,j\in 1,\ldots r$, che corrispondono agli stati di stringhe con gli estremi attaccati a brane coincidenti.

\subsection{T-dualit\`a per stringhe non orientate}

Consideriamo una teoria di stringhe chiuse non orientate. Per ottenerla si \`e proiettato lo spazio degli stati della teoria orientata sugli autovettori di $\Omega$ con autovalore $+1$. La T-dualit\`a si \`e vista essere una parit\`a sui soli modi destri della stringa. Quindi nella picture T-duale, l'azione di $\Omega$ \`e
\be
\Omega:X^T=(X_L-X_R) \ \rightarrow \ (X_R-X_L)=-X^T \ .
\ee
Dal momento che, proiettando lo spettro con $\Omega$ si sono ottenute \emph{teorie non orientate}, mentre, come si vedr\`a nei prossimi paragrafi, proiettando con una riflessione spazio temporale ${\mathbb Z}_2$ si ottengono gli \emph{orbifold}, il risultato combinato prende il nome di \emph{orientifold}. 

Nella picture T-duale il cerchio di compattificazione di raggio R' \`e mappato nel segmento $[0, \pi R']$, i cui estremi sono punti fissi dell'involuzione. Questi piani detti $O$-piani sono all'origine della natura non orientata dello spettro. Al contrario, lontano dagli $O$-piani, la fisica locale \`e quella delle stringhe orientate. A differenza della $D$-brane, gli $O$-piani non sono oggetti dinamici. Nel caso di una teoria con $k$ dimensioni compatte, lo spazio tempo duale sar\`a naturalmente il toro $T^k$ con identificazioni ${\mathbb Z}_2$ nelle direzioni compatte, ovvero un ipercubo con $2^k$ O(25-k)-piani sui vertici.

Nel caso di stringhe aperte, cosiderando una sola dimensione compatta, si hanno due $O$-piani nei punti $X=0, \pi R'$. Introducendo cariche di Chan-Paton con simmetria di gauge SO(2N) (il caso del gruppo simplettico \`e identico), una linea di Wilson generica pu\`o essere ridotta alla forma diagonale per 2N autovalori  
\be
W = diag \left\{e^{i\theta_1} \ , e^{-i\theta_1} \ , e^{i\theta_2} \ , e^{-i\theta_2} \ ,\dots \ ,  e^{i\theta_N} \ , e^{-i\theta_N} \ , \right\} \ .
\ee
Si hanno cio\`e nella picture T-duale $N$ $D$-brane sul segmento fondamentale $[0, \pi R']$ ed altre $N$ $D$-brane immagine nei punti identificati dall'involuzione di orientifold. Le stringhe si possono stendere fra le $D$-brane acquistando massa, e il gruppo di gauge si rompe, nel caso pi\`u generale, a $U(1)^n$. Come nel caso orientato, un gruppo di $r$ $D$-brane coincidenti ricompone un fattore $U(r)$. Adesso per\`o c'\`e una nuova possibilit\`a: che $r$ $D$-brane coincidano con un $O$-piano e quindi con le loro brane immagine: in questo modo si ottiene l'ampliamento della simmetria di gauge a $SO(2r)$. La simmetria massimale si recupera naturalmente per tutte le brane coincidenti con uno dei due $O$-piani, e in questo caso il gruppo \`e $SO(2N)$. La rottura della simmetria di gauge, introdotta con questo meccanismo preserva il rango.

\subsection{Compattificazioni di Superstringa su $S^1$}

\subsubsection{Funzioni di partizione di Superstringa}

Per illustrare la compattificazione toroidale nel caso di superstringa studiamo il caso della IIB e la costruzione di orientifold della Tipo I. Nel caso bosonico si \`e visto che, se una dimensione viene compattificata, l'integrazione sui momenti interni \`e sostituita da una sommatoria sul $n$ e $w$. L'effetto della compattificazione sull'ampiezza di vuoto pu\`o essere riassunto con la sostituzione
\be
\frac{1}{\eta\bar \eta \ \sqrt{\tau_2}} \ \rightarrow \
\sum_{n,w}\frac{q^{\frac{\alpha'}{4}p_R^2} \ \bar
  q^{\frac{\alpha'}{4}p_L^2}}{\eta \bar \eta} \ ,
\ee
ed \`e immediato rendersi conto che tale sostituzione \`e valida anche nel caso di superstringa. La funzione di partizione della IIB, dando per intesa l'integrazione su $\tau_2$ (e la misura di integrazione) e i gradi di libert\`a bosonici non compatti, si scrive
\be
{\cal T} = |V_8 - S_8 |^2 \sum_{n,w} \frac{ q^{\alpha'
p_ R^2/4} \; \bar{q}^{\alpha'
p_L^2/4}}{\eta(\tau) \eta(\bar{\tau})} \ .
\ee
Per costruire i discendenti aperti iniziamo con la bottiglia di Klein. Nella bottiglia di Klein possono propagarsi solo gli stati simmetrici sotto scambio dei modi sinistri e destri, che, come si vede facilmente dalle (\ref{PLPR}), sono gli stati con $w=0$
\be
\label{k1s}
{\cal K} = \frac{1}{2} \, (V_8 - S_8) (2i\tau_2) \sum_{n}
\frac{ \left(e^{-2\pi\tau_2}\right)^{\alpha' n^2/2 R^2}}{\eta(2 i \tau_2)} \ .
\ee
C'\`e un'altra possibilit\`a nella definizione di $\Omega$, perch\'e \`e possibile  assegnare differenti autovalori agli stati di momento pari e dispari, compatibilmente con la teoria interagente, ottenendo
\be
{\cal K'} = \frac{1}{2} \, (V_8 - S_8) (2i\tau_2) \sum_{n}(-)^n
\frac{ \left(e^{-2\pi\tau_2}\right)^{\alpha' n^2/2 R^2}}{\eta(2 i \tau_2)} \ ,.
\ee
L'ampiezza nel canale trasverso la si ottiene con la sostituzione $t=2\tau_2$, effettuando una trasformazione modulare $S$ ed infine utilizzando la formula di risommazione di Poisson (\ref{poissonsum})
\be
\tilde{\cal K} = \frac{2^5}{2} \frac{R}{\sqrt{\alpha'}} \
(V_8 - S_8) (i \ell) \sum_{w} \frac{ \left(e^{-2\pi\ell}\right)^{(2w
R)^2/4\alpha'}}{\eta(i \ell)} \ ,
\ee
\be
\tilde{\cal K'} = \frac{2^5}{2} \frac{R}{\sqrt{\alpha'}} \
(V_8 - S_8) (i \ell) \sum_{w} \frac{ \left(e^{-2\pi\ell}\right)^{(2w+1)^2
R^2/4\alpha'}}{\eta(i \ell)} \ ,
\ee
dove le potenze di 2 si ottengono considerando l'ampiezza completa, con il fattore $\tau_2^{4-1/2}$. E'interessante notare come nel canale trasverso i momenti siano diventati winding. Inoltre la formula di Possion ha dato origine ad un fattore $R/\sqrt{\alpha'}$ che tiene conto del volume della dimensione trasversa. 

Nella seconda proiezione si vede come il fattore $(-)^n$ dopo la risommazione porti uno shift del winding. In questo modello, nella bottiglia di Klein non si propagano stati di massa nulla, dal momento che lo shift nei winding fa si che manchi il termine $q^0$, e non ci siano pertanto divergenze. Si tratta di un modello consistente di sole stringhe chiuse, mentre il primo modello richiede l'introduzione dei settori aperti. Si \`e detto che gli stati che si riflettono su di un bordo subiscono una coniugazione, e nell'anello trasverso si propagheranno quindi solo stati di momento nullo, quelli per i quali $p_L=-p_R$ 
\be
\tilde{\cal A} = \frac{2^{-5}}{2}\, N^2 \, \frac{R}{\sqrt{\alpha'}}
\ (V_8 - S_8) \ (i \ell) \sum_{w} \
\frac{ \left(e^{-2\pi\ell}\right)^{w^2 R^2/ 4 \alpha'}}{\eta(i \ell)} \ .
\ee
L'ampiezza di anello diretto si trova, come di consueto, con una trasformazione S e una risommazione di Poisson
\be
{\cal A} = \frac{1}{2} \, N^2 \, (V_8 - S_8)
({\textstyle{1\over 2}} i\tau_2)
\sum_{n} \ \frac{ \left(e^{-2\pi\tau_2}\right)^{\alpha' n^2/2 R^2}}{
\eta({\textstyle{1\over 2}} i \tau_2)} \ ,
\ee
dalle ampiezze trasverse di anello e bottiglia di Klein si pu\`o scrivere l'ampiezza della striscia di M\"obius nel canale trasverso
\be
\tilde{\cal M} = - \frac{2}{2} \, N \, \frac{R}{\sqrt{\alpha'}} \
(\hat{V}_8 - \hat{S}_8) (i \ell+ {\textstyle\frac{1}{2}}) \sum_{w} \
\frac{ \left(e^{-2\pi\ell}\right)^{(2w R)^2/4\alpha'}}{\hat{\eta}(i \ell+ {\textstyle\frac{1}{2}}
)} \ ,
\ee
ed infine operando una trasformazione $P$ ed una risommazione, si ottiene l'ampiezza nel canale diretto
\be
{\cal M} = - \frac{1}{2} \, N \, (\hat{V}_8 - \hat{S}_8)
({\textstyle{1\over 2}} i \tau_2 + {\textstyle{1\over 2}}) \sum_{n} \
\frac{ \left(e^{-2\pi\tau_2}\right)^{\alpha' n^2/2 R^2}}{\hat{\eta}
({\textstyle{1\over 2}} i \tau_2+{\textstyle{1\over 2}})} \ .
\ee
Le condizioni di tadpole NS-NS e R-R sono entrambe soddisfatte fissando $N=32$, $\epsilon = +1$, ottenendo il gruppo di gauge $SO(32)$. In altri termini si hanno 32 $D9$-brane su un $09_+$-piano, e questo garantisce la cancellazione delle divergenze.

Si \`e detto che le stringhe aperte permettono l'introduzione di linee di Wilson che rompono il gruppo di simmetria delle cariche di Chan-Paton preservandone il rango.  Possiamo vedere un esempio di questo fenomeno nella superstringa Tipo I. Una linea di Wilson produce uno shift sui momenti $n\to n+a_i+a_j$, l'ampiezza di anello nel canale diretto si scrive
\be
{\cal A} = \frac{1}{2} \ (V_8 - S_8)
({\textstyle{1\over 2}} i\tau_2)
\sum_{n,i,j} \ \frac{ \left(e^{-2\pi\tau_2}\right)^{\alpha' (n+a_i+a_j)^2/2 R^2}}{
\eta({\textstyle{1\over 2}} i \tau_2)} \ .
\ee
Nel canale trasverso, come si vede dalla formula di risommazione di Possion, lo shift sui momenti si traduce in termini lineari in $w$ nell'esponenziale e, di conseguenza, in una fase che pu\`o essere scritta in maniera conveniente definendo una matrice costante diagonale
\be
W=\rm{diag}\left(e^{2\pi i a_1},  e^{2\pi i a_2}, \  \ldots \ , e^{2\pi i a_{32}}\right) \ ,
\ee
con $a_2=-a_1, \ a_4=-a_3, \ \ldots a_{32}=-a_{31}$, $ \ 0 < |a_{i}| < 1$. Non \`e difficile rendersi conto che i coefficienti cercati corrispondono a 
\be
(\rm{tr}W^w)^2 \ = \ \sum_{i,j}e^{2\pi i w (a_i+a_j)} \ .
\ee
L'ampiezza di anello nel canale trasverso si scrive
\be
\tilde{\cal A} = \frac{2^{-5}}{2} \ \frac{R}{\sqrt{\alpha'}}
\ (V_8 - S_8) \ (i \ell) \sum_{w} \
\frac{ \left(\rm{tr}W^w\right)^2\left(e^{-2\pi\ell}\right)^{w^2 R^2/ 4 \alpha'}}{\eta(i \ell)} \ ,
\ee
in le fasi introdotte dalle linea di Wilson corrispondono a coefficienti di riflessione diferenti nei diversi settori di winding. L'ampiezza M\"{o}bius trasversa si ricava come di solito da quella di anello e di Klein. Nella Klein trasversa si propagavano solo i settori con winding pari, quindi si ottiene
\be
\tilde{\cal M} = - \frac{2}{2} \ \frac{R}{\sqrt{\alpha'}} \
(\hat{V}_8 - \hat{S}_8) (i \ell+ {\textstyle\frac{1}{2}}) \sum_{w} \
\frac{ \rm{tr}W^{2w} \left(e^{-2\pi\ell}\right)^{(2w
    R)^2/4\alpha'}}{\hat{\eta}(i \ell+ {\textstyle\frac{1}{2}})} \ .
\ee
Passando al canale diretto si vede che si propagano solo i settori con $a_i=a_j$, che \`e un risultato atteso tenendo conto della non orientabilit\`a delle superficie
\be
{\cal M} = - \frac{1}{2} \ (\hat{V}_8 - \hat{S}_8)
({\textstyle{1\over 2}} i \tau_2 + {\textstyle{1\over 2}}) \sum_{n,i} \
\frac{ \left(e^{-2\pi\tau_2}\right)^{\alpha' (n+2a_i)^2/2 R^2}}{\hat{\eta}
({\textstyle{1\over 2}} i \tau_2+{\textstyle{1\over 2}})} \ .
\ee
Studiando lo spettro di massa nulla di anello di vede che, in generale, sono presenti 16 vettori che corrispondono a prendere $n=0$, $a_i=-a_j$. Il gruppo di gauge \`e quindi rotto a $U(1)^{16}$. Nella rappresentazione T-duale questa scelta corrisponde a posizionare tutte le brane in punti differenti.

Naturalmente ponendo a zero un certo numero di $a_i$ e segliendo i rimanenti $a_1\ldots a_{2M}$ tutti con valori differenti si ha un parziale recupero della simmetria di gauge: $SO(32-2M)\times U(1)^M$. Questo nel mondo T-dualizzato corrisponde ad avere $M$ D$8$-brane e le rispettive brane-immagine separate e  $(32-2M)$ D$8$-brane coincidenti con l'$O$-piano nell'origine della dimensione compatta.

Specializziamo ore le ampiezze trovate in un caso di particolare interesse, per $a_1=a_3=\ldots a_{2M-1}=A$, $a_2=a_4=\ldots a_{2M}=-A$, e $a_{2M+1}=\ldots a_{32}=0$. Definendo $N=32-2M$, si pu\`o scrivere
\ba
{\cal A} &=&  (V_8 - S_8) ({\textstyle\frac{1}{2}}i \tau_2) \sum_{n} \
\left\{
\left( M \bar{M} + \frac{1}{2} \, N^2 \right)
\frac{q^{\alpha'
n^2/2 R^2}}{\eta({\textstyle\frac{1}{2}} i \tau_2)} \right. \nonumber\\
& & \left.+ M N \
\frac{q^{\alpha' (n+A)^2/2 R^2}}{\eta(
{\textstyle\frac{1}{2}} i \tau_2)} +  \bar{M} N \
\frac{q^{\alpha' (n-A)^2/2 R^2}}{\eta({\textstyle\frac{1}{2}}i \tau_2)}
\right.\nonumber \\
& &\left.+
\frac{1}{2} \, M^2 \ \frac{q^{\alpha' (n+2A)^2/2
R^2}}{\eta({\textstyle\frac{1}{2}} i \tau_2)} +
\frac{1}{2}\, \bar{M}^2
\ \frac{q^{\alpha'
(n-2A)^2/2 R^2}}{\eta({\textstyle\frac{1}{2}} i \tau_2)} \right\} \, ,
\ea
\ba
{\cal M} &=& - (\hat{V}_8 - \hat{S}_8)
({\textstyle\frac{1}{2}} i \tau_2 + {\textstyle\frac{1}{2}})
\sum_{n} \ \left\{
 \frac{1}{2} \ N  \frac{q^{\alpha' n^2/2
R^2}}{\hat{\eta}(
{\textstyle\frac{1}{2}} i \tau_2+ {\textstyle\frac{1}{2}})}
\right.\nonumber \\
& &\left.+
\frac{1}{2} \ M \ \frac{q^{\alpha' (n+2A)^2/2
R^2}}{\hat{\eta}
({\textstyle\frac{1}{2}}i \tau_2+{\textstyle\frac{1}{2}})}
+ \frac{1}{2} \, \bar{M} \
\frac{q^{\alpha' (n-2A)^2/2 R^2}}{\hat{\eta}
({\textstyle\frac{1}{2}} i \tau_2+{\textstyle\frac{1}{2}})}
\right\} \, .
\ea
Naturalmente $M$ e $\bar M$ sono numeri uguali, ma la notazione \`e utile per sottolineare che si riferiscono a cariche coniugate. Si hanno $M\bar M$ e $N^2/2$ stati di massa nulla nell'ampiezza di anello e $-N/2$ dalla M\"{o}bius, che  formano la rappresentazione aggiunta del gruppo di gauge $SO(N)\times U(M)$. Nel canale trasverso
\ba
\tilde{\cal A} &=& \frac{2^{-5}}{2} \
(V_8 - S_8) (i \ell) \ \frac{R}{\sqrt{\alpha'}}
\sum_{w} \
\frac{ q^{w^2 R^2/ 4 \alpha'}}{\eta(i \ell)} \ \left( N + M
\, e^{2 i \pi A w} +
\bar{M} \, e^{- 2 i \pi A w}    \right)^2 \ ,\nonumber\\
\tilde{\cal M} &=& - \frac{2}{2} \
(\hat{V}_8 - \hat{S}_8)  (i \ell + {\textstyle\frac{1}{2}}) \
\frac{R}{\sqrt{\alpha'}}
 \sum_{w} \
\frac{ q^{(2w R)^2/ 4\alpha'}}{\eta(i \ell+ {\textstyle\frac{1}{2}})}
\ \left( N + M\,
e^{4 i \pi A w} +
\bar{M}\, e^{- 4 i \pi A w} \right) \ ,\nonumber\\
\ea
e le condizioni di tadpole fissano $N=32-2M$. Nella picture T-duale si hanno $32-2M$ D$8$-brane sull'$O$-piano posto nell'origine e le altre $M$ brane e le loro immagini in un punto arbitrario della dimensione compatta. Anche in questo caso si pu\`o verificare un fenomeno di estensione della simmetria di gauge. Per la scelta $A=1/2$, $M$ e $\bar M$ hanno uguali coefficienti di riflessione nel canale trasverso, mentre nel canale diretto l'ampiezza dipende solo dalla loro somma. Questa scelta delle $a_i$ corrisponde nella descrizione T-duale ad avere N $D8$-brane coincidenti con l'$O$-piano all'origine ed M $D8$-brane coincidenti con l'$O$-piano posto a $\pi R'$.

\subsubsection{T-dualit\`a delle superstringhe  di Tipo II}

Nel caso delle superstringhe  di Tipo II la T-dualit\`a mostra l'esistenza di una relazione molto profonda che lega le due teorie IIA e IIB. Compattificare una singola coordinata $X^9$ in una delle due teorie considerarne il limite $R \to 0$ equivale a considerare il limite $R \to \infty$ nella coordinata duale, la cui componente destra \`e riflessa
\be
X^{{ }9, T}_R=-X^{{}9}_R \ ,
\ee
come nel caso bosonico. Per l'invarianza superconforme si deve avere anche
\be
\tilde \psi^{{ } 9, T}= - \tilde \psi^{{ }9} \ ,
\ee
questo implica che la chiralit\`a del vuoto del settore destro di Ramond \`e invertita. Se ad esempio si parte dalla teoria IIA, si compattifica una direzione e si prende il limite di $R$ piccolo, si ottiene esattamente la teoria  IIB con $R$ grande, e viceversa.

\subsection{Compattificazioni toroidali in pi\`u dimensioni}

Il caso della compattificazione sul cerchio pu\`o essere generalizzato a compattificazioni su tori d-dimensionali \mbox{$T^d\simeq (S^1)^d$}. Denotando con $X^m$ le dimensioni compatte, con $m=1, \dots, d$, la condizione di periodicit\`a \`e data da
\be
X^m \sim X^m+2\pi R^{(m)}n^m \ ,
\ee
dove $n^m$ sono interi e $R^{(m)}$ \`e il raggio del m-esimo cerchio. La metrica sul toro pu\`o essere scritta, introducendo i vielbein $e^a_m$ nella forma diagonale
\be
g_{mn}=\delta_{ab}e^a_me^b_n \ ,
\ee
ed \`e conveniente definire coordinate nello spazio tangente $X^a=X^me^a_m$. Pi\`u in generale con una matrice $e^a_m$ generica la relazione di equivalenza diventa
\be
X^a\sim X^a+2\pi e^a_m n^m \ .
\ee
Si \`e quindi definito un reticolo $\Lambda=\{e^a_mn^m, n^m \in \mathbb Z\}$. Il toro pu\`o pertanto essere pensato come quoziente dello spazio piatto d-dimensionale sul reticolo
\be
T \equiv \frac{\mathbb R^d}{2\pi \Lambda} \ .
\ee
I momenti coniugati delle coordinate $X^a$, che indichiamo con $p^a$ saranno quantizzati. Infatti spostandoci da un punto del reticolo ad un altro, producendo una variazione di $X$ di $\delta X \in 2\pi \Lambda$, si deve avere 
\be
e^{ip\cdot X}=e^{ip\cdot (X+ \delta X)} \ ,
\ee
che impone $p \cdot \delta X \in 2 \pi \mathbb{Z}$, ovvero,
\be
\label{multip}
p^n=g^{mn}n_m \ ,
\ee
dove $n_m$ sono interi. Moltiplicando la (\ref{multip}) per $e^a_n$, si trova che gli impulsi $p^a$ vivono nel reticolo duale 
\be
\Lambda^\ast\equiv\{e^{\ast am}n_m, n_m \in \mathbb Z\} \ ,
\ee
dove i vielbein inversi sono definiti con la metrica inversa
\be
e^{\ast am}\equiv e^a_mg^{mn} \ , \qquad e^{\ast am}e^b_m=\delta^{ab} \ .
\ee  
Naturalmente si possono avere anche settori di winding. Sul modello di quanto visto nel caso unidimensionale, definiamo un momento destro ed uno sinistro
\be
p_L^a=p^a+\frac{w^aR^{(a)}}{\alpha'}\equiv e^{\ast am}n_m+\frac{1}{\alpha'}e^a_mw^m \ ,
\ee
\be
p_R^a=p^a-\frac{w^aR^{(a)}}{\alpha'}\equiv e^{\ast am}n_m-\frac{1}{\alpha'}e^a_mw^m \ .
\ee

Per studiare le propriet\`a del reticolo definiamo una base ``pi\`u grande'' che contenga i due reticoli distinti $\Lambda, \Lambda^\ast$:
\be
\hat e_m=\frac{1}{\alpha'}\left(e^a_m \atop -e^a_m \right) \ , 
\qquad \hat e^{\ast m}=\left(e^{\ast am} \atop e^{\ast am} \right) \ .
\ee
Nella nuova base si ha
\be
\hat p=\left(p^a_L \atop  p^a_R \right)= \hat e_mw^m + \hat e^{\ast m}n_m \ .
\ee
Si \`e definito un reticolo in $(d+d)$ dimensioni che chiamiamo $\Gamma_{d,d}$. Possiamo definire una metrica Lorentziana di segnatura $(d, d)$
\be
G=\left( {\delta_{ab} \atop 0} {0 \atop -\delta_{ab}} \right) \ ,
\ee
da cui si vede che
\ba
&&\hat e_m \cdot \hat e_n = 0 = \hat e^{\ast m} \cdot \hat e^{\ast n} \ , \nonumber \\
&&\hat e_m \cdot \hat e^{\ast n}=\frac{2}{\alpha'}\delta_n^m \ ,
\ea
che mostrano che il reticolo \`e \emph{autoduale} dal momento che, a meno di un riscalamento, si ha $\Gamma^\ast_{d,d}=\Gamma_{d,d}$. Inoltre, il prodotto interno fra due momenti \`e
\be
(\hat e_mw^m + \hat e^{\ast m}n_m)\cdot(\hat e_nw'^n + \hat e^{\ast n}n'_n)=\frac{2}{\alpha'}(w^mn'_m+n_mw'^m)\ ,
\ee
e quindi il reticolo \`e \emph{pari}. Queste due propriet\`a del reticolo sono necessarie per garantire l'invarianza modulare della teoria. Consideriamo infatti la generalizzazione della funzione di partizione per una compattificazione in $d$ dimensioni, che pu\`o essere scritta nella forma (per i gradi di libert\`a compatti)
\be
{\cal T}=\frac{\sum_{\Gamma_{d,d}} q^{\frac{\alpha'}{4}p_L\cdot p_L} \ \bar
  q^{\frac{\alpha'}{4}p_R \cdot p_R}}{[\eta(\tau)  \eta(\bar \tau)]^{d}}\ ,.
\ee
Sotto T si ha un fattore di fase nella funzione di partizione $\exp{(i\pi \alpha'(p_L\cdot p_L-p_R\cdot p_R)/2)}$, che \`e uguale ad 1 per la parit\`a del reticolo, mentre sotto una trasformazione S, generalizzando i calcoli visti nel caso della funzione di partizione di una compattificazione unidimensionale, si trova
\be
{\cal T}_{\Gamma}\left(-\frac{1}{\tau}\right)=vol(\Gamma^\ast){\cal T}_{\Gamma^\ast}(\tau) \ .
\ee
Nel caso di reticoli autoduali il volume della cella fondamentale \`e unitario, infatti in generale vale $vol(\Gamma)vol(\Gamma^\ast)=1$ e, per i reticoli autoduali, deve aversi anche ovviamente $vol(\Gamma)=vol(\Gamma^\ast)$. La funzione di partizione risulta quindi invariante modulare.

Ogni reticolo con metrica Lorentziana con segnatura $(d, d)$ pu\`o essere trasformato in un altro reticolo Lorenziano mediante una trasformazione di $O(d, d, \mathbb R)$. Il gruppo $O(d, d, R)$ non \`e una simmetria della teoria  perch\'e cambia lo spettro. Lo spettro di massa dipende per\`o dai prodotti scalari $p_L\cdot p_L$ e $p_R\cdot p_R$, ed \`e quindi invariante sotto $O(d,\mathbb R)\times O(d,\mathbb R)$. Le teorie inequivalenti sono date dalle descritte
\be
\label{group}
\frac{O(d, d, \mathbb R)}{O(d,\mathbb R)\times O(d,\mathbb R)} \ ,
\ee
il cui numero di parametri \`e
\be
\frac{2d(2d-1)}{2}-d(d-1)=d^2 \ .
\ee
Dal punto di vista micorscopico i $d^2$ gradi di libert\`a possono essere identificati con la combinazione del tensore simmetrico $g_{mn}$ e di quello antisimmetrico $B_{mn}$. Infatti \`e possibile considerare una naturale generalizzazione dell'azione di stringa in cui si considerano come campi di background oltre ai gravitoni anche i tensori antisimmetrici e i dilatoni tutti a valori costanti. Per gli impulsi, in questo caso, si ha, scalando il vielbein,
\be
p_{L,m}=n_m+\frac{1}{\alpha'}(g_{mn}-B_{mn})w^n \ ,
\ee
\be
p_{R,m}=n_m-\frac{1}{\alpha'}(g_{mn}+B_{mn})w^n \ .
\ee
Il gruppo di trasformazioni (\ref{group}) deve essere diviso anche per il gruppo di trasformazioni di T-dualit\`a, che \`e molto pi\`u ampio rispetto al caso di compattificazione su $S^1$, e corrisponde a $O(d, d, \mathbb Z)$:
\be
\frac{O(d, d, \mathbb R)}{O(d,\mathbb R)\times O(d,\mathbb R)\times O(d, d, \mathbb Z)} \ ,
\ee 
Questo gruppo contiene, rispetto ai campi di background, trasformazioni di dualit\`a sui singoli assi $R \to \alpha'/R$, shift del tensore antisimmetrico $B_{mn}\to B_{mn}+N_{mn}$ con $N_{mn}$ interi e trasformazioni che rispettano la periodicit\`a della forma $x'=L^m_nx^n$, con L matrice a valori interi e determinante unitario.

La generalizzazione delle funzioni di partizione aperte nel caso di compattificazioni su $S^1$ al caso $T^d$ \`e abbastanza immediata, ma esiste un'importante novit\`a. \`E interessante infatti notare come la costruzione di orientifold, imponga la quantizzazione del tensore antisimmetrico $B_{mn}$. Imponendo la simmetria sotto $\Omega$ della teoria si trova infatti la condizione
\be
n_m+\frac{1}{\alpha'}(g_{mn}-B_{mn})w^n=n'_m-\frac{1}{\alpha'}(g_{mn}+B_{mn})w'^n \ ,
\ee
che porta a richiedere $n=n'$, $w=-w'$ e
\be
\frac{2}{\alpha'}B_{mn} \in \mathbb Z \ .
\ee
La presenza del tensore antisimmetrico quantizzato permette di rompere il gruppo di gauge riducendone il rango \cite{bps, wittor, bianchitor}. Essa pu\`o essere interpretata come una manifestazione della presenza simultanea di due tipi di $O$-piani, $O_+$ e $O_-$, con cariche e tensioni opposte.

\section{Compattificazioni su orbifold}

Gli orbifold toroidali \cite{dhvw} sono una classe di compattificazioni di grande interesse in teoria delle stringhe, dal momento che introducono in modo naturale rotture di simmetrie. Prima di vedere alcuni esempi, diamone una formulazione geometrica.

In generale un orbifold \`e lo spazio quoziente di una variet\`a $\cal M$ su un gruppo discreto $G$, la cui azione sia definita sulla variet\`a  $G:\cal M \to M$. Lo spazio quoziente $\Gamma \equiv \cal M/G$ \`e quindi costruito identificando i punti con la relazione di equivalenza $x\sim gx$ per tutti gli elementi del gruppo $g \in G$.  In generale l'azione del gruppo definisce dei punti fissi, ovvero punti che non trasformano sotto l'azione di $G$: dato un punto $x \in \cal M$ si ha $gx=x$ per $g\in G$. I punti fissi sono singolari e un orbifold in generale non \`e una variet\`a. Da un punto di vista fisico nei punti singolari (si comportano genericamente come il vertice di un cono) la dinamica delle particelle \`e mal definita, ma nel caso di stringhe chiuse orientate l'invarianza modulare determina completamete gli spettri. \`E anche possibile rimuovere le singolarit\`a (tecnica di ``blow up'') ottenendo variet\`a lisce di Calabi-Yau e loro generalizzazioni.

Vediamo la costruzione di una teoria di stringa su un orbifold \cite{Cft, Ginsp}. Consideriamo una teoria invariante modulare, il cui spazio di Hilbert ammetta una simmetria discreta G. A partire da questa teoria se ne pu\`o definire un'altra i cui stati siano invarianti sotto l'azione del gruppo e che sia ancora invariante modulare. Per costruire la nuova funzione di partizione si comincia dal definire funzioni di partizione ``twistate'', che chiameremo $Z_{g, h}$, ovvero si impongono sui campi condizioni al bordo definite come
\be
\phi(z+1)=g\phi(z)\ , \qquad \phi(z+\tau)=h\phi(z)\ ,
\ee 
dove $g, h \in G$ e $|G|$ \`e il numero di elementi del gruppo. La funzione di partizione della nuova teoria invariante modulare si pu\`o a questo punto scrivere nella forma
\be
\label{Zorbifoldgen}
Z_{orb}=\frac{1}{|G|}\sum_{g, h\in G}Z_{g, h} \ ,
\ee
dove nel caso di gruppi non abeliani la somma coinvolge in realt\`a solo coppie $(g, h)$ commutanti. La somma su $g$ \`e riferita ai vari settori ``twistati'' dello spazio di Hilbert della teoria, cio\`e a fissate scelte del tempo $\tau$ sul world-sheet, mentre la somma su $b$ produce una proiezione invariante sotto l'azione del gruppo in ogni settore. La (\ref{Zorbifoldgen}) \`e effettivamente invariante modulare, come si verifica facilmente dal momento che
\be
T: Z_{g, h} \to Z_{g, gh} \ , \qquad S: Z_{g, h} \to Z_{h, g} \ ,
\ee
da cui si comprende bene che la richiesta di commutativit\`a degli elementi $(g, h)$ garantisce che l'azione di $T$ sia ben definita. \\
In forma operativa la costruzione di un orbifold avviene con una proiezione dello spazio di Hilbert degli stati su un sottospazio G-invariante, per mezzo del proiettore
\be
P=\frac{1}{|G|}\sum_{h\in G} h \ ,
\ee
da cui si ottiene la funzione di partizione
\be
Z_{proj}=\frac{1}{|G|}\sum_{h\in G}Z_{1, h} \ ,
\ee
che \`e chiaramente non invariante modulare, dal momento che ad esempio una trasformazione S manda $Z_{1, h}$ in $Z_{h, 1}$. L'invarianza modulare viene recuperata sommando su tutti i possibili ``twist'' spaziali $g$ che commutino con $h$, e in questo modo si ottiene la (\ref{Zorbifoldgen}).

\subsection{Orbifold $S^1/ \mathbb Z_2$ di stringa bosonica}

Studiamo, come caso introduttivo, una teoria di stringa bosonica compattificata su un orbifold $S^1/ \mathbb Z_2$. Il cerchio $S^1$, parametrizzato da una coordinata $X$, ha una simmetria $\mathbb Z_2$, \mbox{$g: X \to -X$}. Questa simmetria si estende allo spettro degli stati e agli operatori della teoria di stringa. Alcuni stati sono pari sotto $g$, mentre altri sono dispari. Come nel caso di $\Omega$, si pu\`o definire una nuova teoria proiettando gli stati sul settore pari. Questa operazione \`e il punto di partenza per definire una teoria di stringa sullo spazio di orbifold {$S^1/\mathbb Z_2$}.  Lo spazio in cui vive la stringa ora \`e un segmento i cui estremi sono punti fissi dell'azione di $\mathbb Z_2$, $X=0, \pi R$. Per scrivere la funzione di partizione dobbiamo inserire nella traccia sugli stati della funzione di partizione il proiettore
\be
P=\frac{(1+g)}{2} \ .
\ee
L'azione di $g$ sugli operatori \`e naturalmente 
\be
g: \alpha_k \to -\alpha_k \ , \qquad g: \tilde \alpha_k \to -\tilde \alpha_k \ .
\ee
mentre su uno stato generico di momento $n$ e winding $w$, l'azione di g \`e
\be
 g\prod_{i=1}^{N}\alpha_{-k_i}\prod_{j=1}^{\bar N} \bar \alpha_{-k_j}
\ |n,w\rangle\ = \ (-)^{N+\bar N}\prod_{i=1}^{N}\alpha_{-k_i} \ \prod_{j=1}^{\bar N} \bar \alpha_{-k_j}
\ |-n,-w\rangle \ .
\ee
Nel calcolare la traccia sopraviveranno solo gli stati con $n=w=0$, e denotando con $N_k$ il numero di oscillatori con frequenza $k$, si ha 
\be
\frac{1}{(q\bar q)^{\frac{1}{24}}}{\rm tr} \left(q^{N^\bot+\frac{\alpha'}{4}p_R^2}g\right) =
\frac{1}{(q\bar q)^{\frac{1}{24}}}\prod_{k=1} \sum_{N_k=0} (-q^k)^{N_k}=
\frac{1}{(q\bar q)^{\frac{1}{24}}}\prod_{k=1}\frac{1}{1+q^k} \ .
\ee 
La funzione di partizione nel settore non twistato \`e, dando ancora una volta per intesi l'integrazione e la sua misura e i gradi di libert\`a non compatti,
\ba
{\cal T}_{untwisted} &=& \frac{1}{2}{\cal T}_{S^1}(R)+\frac{1}{2}\frac{1}{(q\bar
  q)^{\frac{1}{24}}}\frac{1}{\prod_{k=1}(1+q^k)(1+\bar q^k)}\nonumber\\
 &=& \frac{1}{2}{\cal T}_{S^1}(R)+\left|\frac{\eta}{\vartheta_2}\right| \ ,
\ea
dove naturalmente si \`e chiamata ${\cal T}_{S^1}$ la funzioni di partizione della stringa sul cerchio (\ref{torus_compact}). Ora occorre completare la funzione di partizione con i settori ``twistati'' per recuperare l'invarianza conforme. Il settore twistato \`e quello in cui (stringa di lunghezza $\pi$)
\be
X(\sigma+\pi)=-X(\sigma) \ ,
\ee
la condizione di antiperiodicit\`a porta ad una espansione in modi semi-interi:
\be
X \ = \ x_0 \ + \
i\sqrt{\frac{\alpha'}{2}} \ \sum_{k\in \mathbb{Z}}
\left(\frac{\alpha_{k+1/2}}{k+1/2} \ e^{-2i(k+1/2)(\tau-\sigma)}+
\frac{\tilde\alpha_{k+1/2}}{k+1/2} \ e^{-2i(k+1/2)(\tau+\sigma)}\right) \ ,
\ee
dove $x_0=0$ o $\pi R$. L'assenza di zero modi indica che il momento \`e nullo e non ci sono winding. L'energia di punto zero \`e corrisondentemente spostata da $-1/24$ per un bosone periodico a $+1/48$ per un bosone antiperiodico, e quindi l'effetto netto \`e $+1/16$. Le condizioni di massa nel settore twisted sono
\be
M^2= \frac{4}{\alpha'}\left(N^\bot-\frac{15}{16}\right) \ , \qquad N^\bot=\bar N^\bot \ .
\ee
Calcoliamo il contributo del settore twistato alla funzione di partizione
\ba
(q\bar q)^{\frac{1}{48}}{\rm tr_{twisted}} \left(q^{N^\bot} \bar q^{\bar N^\bot}\frac{(1+g)}{2}\right) 
&=& (q\bar q)^{\frac{1}{48}}\left[\prod_{k=1}^\infty \frac{1}{\left|1-q^{k-1/2}\right|^2}+  
\prod_{k=1}^\infty \frac{1}{\left|1+q^{k-1/2}\right|^2}\right] \nonumber \\
&=& \left|\frac{\eta}{\vartheta_4}\right| + \left|\frac{\eta}{\vartheta_3}\right|\ .
\ea 
La funzione di partizione dell'orbifold $S^1/ \mathbb{Z}_2$ \`e quindi data da
\be
{\cal T}_{S^1/\mathbb{Z}_2} \ = \ \frac{1}{2}\left(
{\cal T}_{S^1}(R)+2\left|\frac{\eta}{\vartheta_2}\right|
+2\left|\frac{\eta}{\vartheta_4}\right|
+2\left|\frac{\eta}{\vartheta_3}\right|\right) \ ,
\ee
che pu\`o essere riscritta, utilizzando l'identit\`a $\vartheta_2\vartheta_3\vartheta_4=2\eta^3$, come
\be
{\cal T}_{S^1/\mathbb{Z}_2} \ = \ \frac{1}{2}\left(
{\cal T}_{S^1}(R)+\frac{\left|\vartheta_3\vartheta_4\right|}{\eta\bar\eta}
+\frac{\left|\vartheta_2\vartheta_3\right|}{\eta\bar\eta}
+\frac{\left|\vartheta_2\vartheta_4\right|}{\eta\bar\eta}
\right) \ .
\ee
La funzione di partizione \`e ora invariante modulare, come si  pu\`o verificare direttamente, o  ricordando che \`e stata costruita come somma di termini con tutte le possibili condizioni al bordo periodiche e antiperiodiche.

\subsection{Orbifold $T^4/ \mathbb Z_2$ di superstringa}

Studiamo ora due modelli in sei dimensioni, la Tipo IIB e la Tipo I su un orbifold $T^4/ \mathbb Z_2$. \`E bene distinguere i gradi di libert\`a fermionici che vivono nelle dimensioni compatte da quelli definiti sullo spazio-tempo esteso 6-dimensionale. Per far questo \`e opportuno decomporre i caratteri di $SO(8)$ in  rappresentazioni di $SO(4)\times SO(4)$, riferendo il primo $SO(4)$ al cono di luce di ${\cal M}_6$ ed il secondo alla variet\`a interna:
\ba
V_8 &= V_4 O_4 + O_4 V_4 \, , \qquad O_8 &= O_4 O_4 + V_4 V_4 \, ,
\nonumber \\
S_8 &= C_4 C_4 + S_4 S_4 \, , \qquad C_8 &= S_4 C_4 + C_4 S_4 \, .
\ea
Definiamo anche le seguenti combinazioni, che torneranno presto utili:
\ba
\label{so4}
& Q_o = V_4 O_4 - C_4 C_4 \, , \qquad \quad
& Q_v = O_4 V_4 - S_4 S_4 \, ,
\nonumber \\
& Q_s = O_4 C_4 - S_4 O_4 \, , \qquad \quad
& Q_c = V_4 S_4 - C_4 V_4 \, .
\ea
Per avere consistenza fra l'azione di $\mathbb Z_2$ e la supersimmetria di world-sheet, $O_4$ e $C_4$ devono essere pari mentre $V_4$ e $S_4$ dispari sotto $\mathbb Z_2$ \cite{dhvw,4d}. Quindi $Q_o$ e $Q_c$
sono gli autovettori positivi e $Q_v$ e $Q_s$ quelli negativi.

L'ampiezza di toro con una compattificazione su $T^4$ \`e
\be
\label{T4torus}
\mathcal{T}_{++} \ = \ |V_8-S_8|^2 \Sigma_{n,w} \ ,
\ee
dove si sono dati per sottintesi i gradi di libert\`a non compatti e si \`e indicata con $\Sigma_{n,w}$, la sommatoria sul reticolo delle variet\`a interne di metrica $g$
\be
\Sigma_{n,w}= \sum_{n,w} {q^{{\alpha' \over 4} p_{L}^{{\rm T}} g^{-1} p_{ L}}
\bar q^{{\alpha' \over 4} p_{R}^{{\rm T}} g^{-1} p_{R}}
\over \eta^4 \bar \eta ^4 } \ .
\ee
Utilizzando le (\ref{so4}) e separando nella sommatoria i modi zero dagli stati pi\`u alti in momento e winding l'ampiezza di toro (\ref{T4torus}) diventa 
\be
\label{T4torus_2}
\mathcal{T}_{++} \ = \ |Q_o-Q_v|^2\left[\Sigma_{n,w}'+\frac{1}{(\eta\bar \eta)^4}\right] \ ,
\ee
si \`e indicato con $\Sigma'$ la sommatoria senza gli zero modi. Come si \`e visto nel caso $S^1/\mathbb Z_2$, solo gli zero modi vengono proiettati mentre gli altri stati vengono dimezzati.

Per scrivere la funzione di partizione di orbifold si pu\`o proiettare l'ampiezza (\ref{T4torus_2}) con $\mathbb Z_2$, effettuare una trasformazione S e poi proiettare ancora con $\mathbb Z_2$. Infatti la prima proiezione pone condizioni antiperiodiche nella direzione `temporale' sul worldsheet mentre la trasformazione $S$ e la nuova proiezione $\mathbb Z_2$ individuano, rispettivamente, il settore con condizioni antiperiodiche nella direzione `spaziale' ed il settore con condizioni antiperiodiche sia nella direzione `spaziale' che `temporale'. \\
Come si \`e gi\`a trovato nel caso $S^1/\mathbb Z_2$, si ha
\be
\frac{1}{\eta^4} \stackrel{\mathbb{Z}_2}{\longrightarrow}\left(\frac{2\eta}{\vartheta_2}\right)^2
\stackrel{S}{\longrightarrow}4\left(\frac{\eta}{\vartheta_4}\right)^2
\stackrel{\mathbb{Z}_2}{\longrightarrow}4\left(\frac{\eta}{\vartheta_3}\right)^2 \ .
\ee
Utilizzando la trasformazione S definita sui caratteri di $SO(4)$
\ba
S = \frac{1}{2} \ \left( \begin{array}{rrrr}
1 & 1 & 1 & 1 \\
1 & 1 & -1 & -1 \\
1 & -1 & -1 & 1 \\
1 & -1 & 1 & -1
\end{array} \right) \ ,
\ea
si trova 
\be
Q_o+Q_v \stackrel{\mathbb{Z}_2}{\longrightarrow}Q_o-Q_v
\stackrel{S}{\longrightarrow}Q_s+Q_c
\stackrel{\mathbb{Z}_2}{\longrightarrow}Q_s-Q_c \ .
\ee
La funzione di partizione di toro pu\`o essere scritta nella forma
\be
{\cal T} = \frac{1}{2} \Biggl[ |Q_o + Q_v|^2
\Sigma_{n,w}
 + |Q_o - Q_v |^2 \left| {2 \eta \over
\vartheta_2} \right|^4
+ 16 |Q_s + Q_c |^2 \left| {\eta \over \vartheta_4}\right|^4
+ 16 |Q_s - Q_c |^2 \left| {\eta \over \vartheta_3} \right|^4 \Biggr]\, ,
\nonumber\\
\ee
dove il fattore $16=2^4$ tiene conto del numero di punti fissi dell'orbifold.

Lo spettro della teoria chiusa \`e organizzato in multipletti di supersimmetria ${\mathcal N}(2, 0)$ in $D=6$. Si ha un multipletto gravitazionale, che contiene il gravitone, cinque 2-forme autoduali e due gravitini sinistri e 21 multipletti tensoriali, 16 dei quali dal settore twisted, contenenti una 2-forma anti-autoduale, cinque scalari e due spinori destri. Questo \`e l'unico spettro di questo tipo privo di anomalie.

Costruiamo ora i discendenti aperti, inziando come di consueto dalla bottiglia di Klein. L'azione di $\Omega$ scambia $p_L$ e $p_R$, ma grazie alla identificazione $X \sim -X$ si ha anche $p_{L,R} \sim - p_{L,R}$. Questo comporta che nella bottiglia di Klein si propaghino non solo gli stati di momento ma anche quelli di winding. Con le solite regole si scrive
\be
\label{Klein_orb}
{\cal K} = \frac{1}{4} \Biggl[ (Q_o + Q_v) 
\left(\sum_n \; {q^{{\alpha'\over 2} n^{{\rm T}} g^{-1} n} \over \eta(q)^4} + 
\sum_w \; {q^{{1\over 2\alpha'} w^{{\rm T}} g w} \over \eta(q)^4}
\right) + 2 \times 16 \  (Q_s + Q_c) \left(
\frac{\eta}{\vartheta_4} \right)^2 \Biggr] \ ,
\ee
dove si \`e risolta in maniera diagonale l'ambiguit\`a nel settore twistato. Nel canale trasverso si trova
\be
\tilde{\cal K} = \frac{2^5}{4} \left[ (Q_o + Q_v) \left(
v_4 \sum_w \; {q^{{1\over 4\alpha'} w^{{\rm T}} g w} \over \eta(i\ell)^4}
+ \frac{1}{v_4} \sum_n \; {q^{{\alpha'\over 4} n^{{\rm T}} g^{-1} n} \over \eta(i\ell)^4} \right) +
2 (Q_o - Q_v)  \left(
\frac{2 \eta}{\vartheta_2} \right)^2 \right] \, ,
\ee
dove $v_4=\sqrt{{\rm det} g /(\alpha')^4}$  \`e proporzionale al volume delle dimensioni compatte. \\
Lo spettro di massa nulla si ottiene da quello del toro eliminando gli stati antisimmetrici sotto $\Omega$ nel settore NS-NS e gli stati simmetrici nel settore R-R, mentre i settori misti vengono dimezzati. Lo spettro di massa nulla risultante corrisponde ad un multipletto gravitazionale, contenente il gravitone, una 2-forma autoduale e un gravitino sinistro; ad un multipletto tensoriale, contenente una 2-forma anti-autoduale, uno scalare e uno spinore destro, e a 20 ipermultipletti dei quali 16 dal settore twistato, contenenti 4 scalari e uno spinore destro.\\
Al livello di massa nulla la bottiglia di Klein trasversa \`e data da
\be
\tilde{\cal K}_0 = \frac{2^5}{4} \left[ Q_o \left(
\sqrt{v_4} + \frac{1}{\sqrt{v_4}} \right)^2  +
Q_v \left(
\sqrt{v_4} - \frac{1}{\sqrt{v_4}} \right)^2
 \right] \, ,
\ee
da cui si legge il contenuto in termini di $O$-piani, la cui tensione e carica di R-R pu\`o essere letta da $Q_o=V_4O_4-C_4C_4$. Si vede subito che, insieme ai soliti $O9$-piani, sono presenti anche degli $O5$-piani. Quattro T-dualit\`a lungo le direzioni compatte mappano $O9$-piani, in $O5$-piani e $v_4$ in $1/v_4$. 

Sono possibili anche altre proiezioni di Klein; \`e possibile infatti proiettare momenti e windings pari e dispari in maniera diversa, ottenendo 
\ba
{\cal K}' &=& \frac{1}{4} \Biggl[ (Q_o + Q_v) \left(
\sum_n (-1)^n \, {q^{{\alpha'\over 2} n^{{\rm T}} g^{-1} n} \over \eta^4}
+ \sum_w (-1)^w \,
{q^{{1\over 2\alpha'} w^{{\rm T}} g w} \over \eta^4}
\right)
\nonumber \\
& & + 2 \times (8-8) (Q_s + Q_c)
\left(
\frac{\eta}{\vartheta_4} \right)^2 \Biggr] \, ,
\ea
dove $(-1)^n$ e $(-1)^w$, stanno ad indicare simbolicamente diverse possibili scelte disponibili per introdurre, in uno o pi\`u tori, segni alternati. In questo caso si ha una teoria consistente di sole stringhe chiuse non orientate, dal momento che i segni nel canale trasverso si traducono in shift dei momenti e dei winding che eliminano i modi a massa nulla. Una terza possibilit\`a di grande interesse \`e
\be
{\cal K}'' = \frac{1}{4} \left[ ( Q_o + Q_v ) ( \sum_n \; {q^{{\alpha'\over 2} n^{{\rm T}} g^{-1} n} \over \eta(q)^4}  + \sum_w \; {q^{{1\over 2\alpha'} w^{{\rm T}} g w} \over \eta(q)^4} ) - 2
\times 16 ( Q_s + Q_c ){\left(\frac{\eta}{\vartheta_4}\right)}^2 \right]
\ ,
\ee
dal momento che porta un settore aperto non supersimmetrico. Si avr\`a modo di studiare in dettaglio questo modello nel capitolo sulla rottura di supersimmetria.

La scelta pi\`u semplice per l'ampiezza di anello risulta essere
\ba
\label{anellosuperorbifold}
{\cal A} &=& \frac{1}{4} \Biggl[ (Q_o + Q_v) \left(
N^2 \sum_n \; {q^{{\alpha'\over 2} n^{{\rm T}} g^{-1} n} \over \eta(q)^4} 
+ D^2 \sum_w \; {q^{{1\over 2\alpha'} w^{{\rm T}} g w} \over \eta(q)^4}\right) \nonumber\\ 
& &+ \left(R_N^2 + R_D^2 \right) (Q_o - Q_v) \left( {2\eta \over \vartheta_2}\right)^2 \nonumber\\
& & + 2 N D \, (Q_s + Q_c ) \left( {\eta \over \vartheta_4}\right)^2
+ 2 R_N R_D \, (Q_s - Q_c ) \left( {\eta \over \vartheta_3}\right)^2
\Biggr] \, ,
\ea
dove $N$ e $D$ contano la molteplicit\`a degli estremi della stringa aperta, rispettivamente con condizioni di Neumann e di Dirichlet, e $R_D$ e $R_N$ definiscono l'azione dell'orbifold sulle cariche di Chan-Paton \cite{ps}. Nel canale trasverso si ha 
\ba
\label{Annulus_orb}
\tilde {\cal A} &=& \frac{2^{-5}}{4} \Biggl[ (Q_o + Q_v) \left(
N^2 v_4 \sum_w \; {q^{{1\over 4\alpha'} w^{{\rm T}} g w} \over \eta(i\ell)^4}  
\frac{D^2}{v_4}\sum_n \; {q^{{\alpha'\over 4} n^{{\rm T}} g^{-1} n}\over \eta(i\ell)^4} \right) \nonumber \\
& &+ 2 N D \, (Q_o - Q_v ) \left( {2 \eta \over \vartheta_2}\right)^2 \nonumber \\
& & + 16 \left(R_N^2 + R_D^2 \right) (Q_s + Q_c) \left( {\eta \over
\vartheta_4}\right)^2 \nonumber \\
& &- 2 \times 4 R_N R_D \, (Q_s - Q_c )
\left( {\eta \over \vartheta_3}\right)^2 \Biggr] \, , 
\ea
che porta al contributo alla condizione di tadpole
\ba
\tilde {\cal A}_0 &=& \frac{2^{-5}}{4} \Biggl\{ Q_o \left(
N \sqrt{v_4} + \frac{D}{\sqrt{v_4}} \right)^2 +
Q_v \left( N \sqrt{v_4} - \frac{D}{\sqrt{v_4}} \right)^2
\nonumber \\
& + & Q_s \left[ 15 R_N^2 + \left(R_N - 4 R_D \right)^2 \right]+
Q_c \left[ 15 R_N^2 + \left(R_N + 4 R_D \right)^2 \right]
\Biggr\} \, ,
\ea
in cui si distinguono i contributi delle stringhe aperte con le diverse possibili combinazioni di condizioni al bordo di Neumann e di Dirichlet che corrispondono alle configurazioni $D9$-$D9$, $D9$-$D5$ e $D5$-$D9$, $D5$-$D5$. \\
L'ampiezza della striscia di M\"{o}bius nel canale trasverso, calcolata come di consueto, porta a 
\ba
\label{MOebiusSuperOrbifoldtran}
\tilde{\cal M} &=& -{2\over 4} \left[
(\hat Q_o + \hat Q_v ) \left(
N v_4 \sum_w \; {q^{{1\over 4\alpha'} w^{{\rm T}} g w} \over \eta(i\ell)^4}
+ D\frac{1}{v_4}\sum_n \; {q^{{\alpha'\over 4} n^{{\rm T}} g^{-1} n}\over \eta(i\ell)^4} \right) \right.\nonumber\\
& &\left.+ \left( N + D\right)
(\hat Q_o - \hat Q_v ) \left( {2\hat \eta \over \hat \vartheta_2}\right)^2
\right] \, ,
\ea
dove si sono fissati i segni in maniera opportuna per cancellare il tadpole R-R. Il contributo di massa nulla \`e
\ba
\label{M_0}
\tilde{\cal M}_0 &=& -\frac{2}{4}\left[\hat{Q}_o \left(
\sqrt{v_4} + \frac{1}{\sqrt{v_4}} \right) \left(
N\sqrt{v_4} + \frac{D}{\sqrt{v_4}} \right)\right.\nonumber\\
& &\left.+\hat{Q}_v \left(
\sqrt{v_4} - \frac{1}{\sqrt{v_4}} \right)\left(
N\sqrt{v_4} - \frac{D}{\sqrt{v_4}} \right)
\right] \ .
\ea
Per passare nel canale diretto occorre definire una trasformazione $P$ sui caratteri di $SO(4)$, che si trova essere $P={\rm diag}(\sigma_1, \sigma_1)$. L'azione di $P$ quindi scambia $\hat{Q}_v$ con $\hat{Q}_o$, si ottiene
\ba
\label{MOebiusSuperOrbifold}
{\cal M} &=& -{1\over 4} \left[ (\hat Q _o + \hat Q _v ) \left(
N \sum_n \; {q^{{\alpha'\over 2} n^{{\rm T}} g^{-1} n} \over \eta(q)^4} 
+ D \sum_w \; {q^{{1\over 2\alpha'} w^{{\rm T}} g w} \over \eta(q)^4} \right)\right.\nonumber\\
& &\left.- \left( N + D\right)
(\hat Q _o - \hat Q _v )
\left( {2\hat\eta\over \hat\vartheta_2} \right)^2 \right] \, .
\ea
La condizione di tadpole impone la cancellazione della somma dei termini $\tilde{\cal A}_0$, $\tilde{\cal K}_0$, $\tilde{\cal M}_0$. Dal momento che i termini $R_D$ e $R_N$ compaiono solo in $\tilde{\cal A}_0$, occorre che siano entrambe nulle:
\be
\label{cptw}
R_N=R_D=0 \ ,
\ee
e si deve inoltre avere
\be
\left(N\sqrt{v_4}\pm D\frac{1}{\sqrt{v_4}}\right)=32\left(\sqrt{v_4}\pm\frac{1}{\sqrt{v_4}}\right) \ ,
\ee
che fissa 
\be
N=32 \ , \qquad \qquad D=32 \ .
\ee

La corretta parametrizzazione per le molteplicit\`a di Chan-Paton \`e
\ba
N &=& n+\bar n \ , \qquad \qquad n=\bar n = 16 \ , \nonumber\\
D &=& d+\bar d \ , \qquad \qquad d=\bar d  =16 \ ,
\ea
che in accordo con (\ref{cptw}) fissa
\be
R_N=i(n-\bar n) \ , \qquad\qquad R_D=i(d-\bar d) \ .
\ee
Con la parametrizzazione data, le ampiezze del settore aperto, per il livello di massa nulla si scrivono
\ba
{\cal A}_0 &=& ( n \bar{n} + d \bar{d} ) Q_0 + {\textstyle{1\over 2}} (n^2 + \bar{n}^2
+ d^2 + \bar{d}^2 ) Q_v + (n \bar{d} + \bar{n} d) Q_s \nonumber \\
{\cal M}_0 &=& -  {\textstyle{1\over 2}} (n + \bar{n} + d + \bar{d}) \hat{Q}_v \, ,
\ea
da cui si legge lo spettro di bassa energia, che risulta privo di anomalie \cite{bs2,gp}. $Q_o$  porta un multipletto di gauge di $\mathcal{N}=(1,0)$, un vettore e uno spinore sinistro, nella rappresentazione aggiunta di $U(16)_{D9}\times U(16)_{D5}$; $Q_v$ contribuisce con ipermultipletti nella $(16\times 15/2,1)$ e $(1,16\times 15/2)$ e nelle loro complesse coniugate; $Q_s$ contribuisce con met\`a di un ipermultipletto nella $(16,\overline {16})$ e con il suo coniugato nella $(\overline{ 16},16)$, e pertanto ancora con un ipermultipletto completo.


\chapter{Rottura di Supersimmetria}

La costruzione di modelli di stringa che riproducano a basse energie la fisica conosciuta richiede l'introduzione di opportuni meccanismi di rottura della supersimmetria, dal momento che la degenerazione fra bosoni e fermioni tipica della supersimmetria esatta non \`e osservata nei processi ordinari. 

In teoria delle stringhe esistono differenti possibilit\`a di implementare la rottura di supersimmetria. In questo capitolo si vedranno prima le teorie in dieci dimensioni di Tipo 0, che forniscono esempi di rottura ``esplicita'' della supersimmetria, e in seguito modelli con rottura ``spontanea'', nei quali la simmetria pu\`o essere recuperata scegliendo opportunamente un parametro continuo: \emph{Scherk-Schwarz supersymmetry breaking} \cite{ss, ads1}, \emph{Brane supersymmetry breaking} e modelli in cui la rottura \`e indotta da deformazioni magnetiche.

\section{Tipo 0}

\`E possibile costruire, a partire dalla forma generale della funzione di partizione (\ref{general torus}) due modelli in dieci dimensioni invarianti modulari, e non supersimmetrici, detti di Tipo 0 \cite{dhsw}. Lasciando intesi i gradi di libert\`a bosonici e l'integrazione, le funzioni di partizione corrispondenti sono
\ba
\label{torus0}
{\cal T}_{0A} &=& |O_8|^2 + |V_8|^2 + \bar{S}_8 C_8 + \bar{C}_8 S_8 \, , \nonumber \\
{\cal T}_{0B} &=& |O_8|^2 + |V_8|^2 + |S_8|^2 + |C_8|^2 \ .
\label{0AB}
\ea

Ricordando il contenuto dei caratteri coinvolti in termini di particelle non \`e difficile studiare lo spettro di bassa energia delle due teorie. I due modelli non contengono fermioni, dal momento che essi non hanno settori misti NS-R e R-NS. Nel settore NS-NS il termine $|O_8|^2$ rende le due teorie tachioniche, mentre $|V_8|^2$ porta in entrambi i casi un gravitone, un tensore antisimmetrico e un dilatone ($G_{\mu \nu}, B_{\mu \nu}, \phi$). Il settore di R-R distingue le due teorie. Nella 0A si trova $\bar{S}_8 C_8$  e $\bar{C}_8 S_8$, che portano due vettori $A_{\mu}$ e due 3-forme $C_{\mu \nu \rho}$, mentre nella OB $|S_8|^2$ e $|C_8|^2$ danno due scalari, altri due tensori antisimmetrici ed una 4-forma completa $D_{\mu \nu \rho \sigma}$. 

\subsection{Discendenti aperti}

\subsubsection{Orientifold della OA}

Vediamo la costruzione di orientifold della OA \cite{bs}. L'ampiezza di bottiglia di Klein nel canale diretto si ottiene, in base alle regole note, dai soli termini simmetrici sotto lo scambio dei modi sinistri e destri. L'ampiezza nel canale traverso \`e invece ottenuta tramite una trasformazione S da quella diretta, dopo essere passati al modulo del toro doppiamente ricoprente. Si pu\`o quindi scrivere
\be
{\cal K} \ = \ \frac{1}{2}(O_8+V_8) \ ,
\ee
\be
\tilde{\cal{K}} \ = \ \frac{2^5}{2}(O_8+V_8) \ .
\ee
La bottiglia di Klein proietta via il tensore antisimmetrico dal settore NS-NS e dimezza il settore R-R. 

Ricordando che in $so(8)$ le rappresentazioni sono autoconiugate, nell'anello trasverso si propagheranno solo i termini che compaiono nella (\ref{torus0}) in forma diagonale, ovvero  $O_8$ e $V_8$. Si possono associare ai due caratteri differenti coefficienti di riflessione, e quindi
\be
\tilde{\cal{A}} \ = \ \frac{2^{-5}}{2}\left[(n_b+n_f)^2V_8+(n_b-n_f)^2O_8\right] \ .
\ee
Dopo aver ridefinito il modulo, si ottiene l'ampiezza nel canale diretto con una trasformazione S 
\be
{\cal A} \ = \ \frac{1}{2}\left[(n_b^2+n_f^2)(O_8+V_8)-2n_bn_f(S_8+C_8)\right] \ .
\ee

L'ampiezza di M\"{o}bius trasversa si trova ricordando che i suoi coefficienti dei caratteri sono medie geometriche dei rispettivi coefficienti in $\tilde{\cal{K}}$ e $\tilde{\cal{A}}$ moltiplicati per un fattore combinatorio 2
\be
\tilde{\cal M} \ = \ \epsilon \frac{2}{2}\left[(n_b+n_f)\hat{V}_8+(n_b-n_f)\hat{O}_8\right] \ .
\ee
L'ampiezza di M\"{o}bius nel canale diretto si trova con un riscalamento del modulo e una trasformazione P
\be
{\cal M} \ = \ \epsilon \frac{1}{2}\left[(n_b+n_f)\hat{V}_8-(n_b-n_f)\hat{O}_8\right] \ . 
\ee

Nelle ampiezze del canale trasverso non compaiono stati di R-R, e l'unico tadpole che compare \`e quello di NS-NS. Se si decide di imporre la tadopole condition, si trova $\epsilon=-1$ e $n_b+n_f=32$, selezionando il gruppo di gauge $SO(n_b)\times SO(n_f)$. Lo spettro aperto contiene un vettore nell'aggiunta del gruppo, un fermione di Majorana ($S_8+C_8$) nella bifondamentale $(n_b,n_f)$ e tachioni nella $(\frac{n_b^2+n_b}{2},1)$ e nella $(1,\frac{n_f^2-n_f}{2})$. Se invece non si impone la condizione di tadpole NS-NS, si pu\`o anche segliere il segno $\epsilon=+1$, selezionando il gruppo di gauge $USp(n_b)\times USp(n_f)$.

\subsubsection{Orientifold della OB}

Nel caso della OB \cite{bs,susy95} sono possibili tre differenti scelte per l'ampiezza di bottiglia di Klein \cite{fps,pss,pss2}. La prima scelta simmetrizza tutti i caratteri (considerando una base che tenga conto del segno dei fermioni $O_8, V_8, -S_8, -C_8$)
\be
\label{klein0B1}
{\cal K}_1 = \frac{1}{2}(O_8+V_8-S_8-C_8) \ ,
\ee
e lo spettro chiuso risultante contiene un tachione, il gravitone, e il dilatone nel settore NS-NS ed una coppia di 2-forme nel settore di R-R. Passando al modulo del toro doppiamente ricoprente e facendo una trasformazione modulare si ottiene l'ampiezza nel canale trasverso
\be
\tilde{\cal{K}}_1 = \frac{2^6}{2}V_8 \ .
\ee

Nell'anello trasverso si propagheranno tutti i caratteri dal momento che sono autoconiugati ($\mathcal{C}=1$), e compaiono nel toro con il proprio carattere coniugato. Si pu\`o quindi scrivere
\ba
\tilde{\cal A}_1 &=& \frac{2^{-6}}{2} \left[
(n_o + n_v + n_s + n_c)^2 V_8 + (n_o + n_v - n_s - n_c)^2 O_8 \right.\nonumber \\
& &\left. -  (-n_o + n_v + n_s - n_c)^2 S_8 - (-n_o + n_v - n_s + n_c)^2 C_8
\right] \ ,
\ea
e nel canale diretto si trova
\ba
\label{A1}
{\cal A}_1 &=& {\textstyle\frac{1}{2}} \left[( n_o^2 + n_v^2 + n_s^2 + n_c^2) V_8 + 2(n_o n_v + n_s n_c)O_8\right.\nonumber \\
& & \left. - 2(n_v n_s + n_o n_c) S_8 - 2(n_v n_c + n_o n_s) C_8\right] \ .
\ea

L'ampiezza di anello trovata corrisponde all'\emph{ansatz di Cardy} \cite{cardy2}. Nei casi in cui, nella ampiezza di toro (\ref{general torus}), $X_{ij}= \mathcal C$, tutti i caratteri si riflettono sui bordi. Si hanno quindi tante condizioni al bordo quanti sono i settori nello spettro, con una corrispondenza uno ad uno tra indici di bordo e indici dei settori nel bulk. Le regole di fusione, ovvero le regole che determinano il prodotto fra caratteri, si scrivono nella forma
\be
[\chi_i]\times[\chi_j]=\sum_{k} {\mathcal{N}_{ij}^k[\chi_k]} \ ,
\ee
dove si \`e indicato con $\chi_i$ il generico carattere. L'ansatz di Cardy consiste nell'utilizzare la matrice delle regole di fusione $\mathcal{N}_{ij}^k$, espressa in termini della matrice $S$ dalle formula di Verlinde \cite{verlinde}
\be
{\cal N}_{ij}{}^k = \sum_l \, \frac{S_i^l \, S_j^l \, {S^\dag}{}_l^k}{S_1^l} \ ,
\label{Verlinde}
\ee
per definire il contenuto del $k$-esimo stato con condizioni al bordo $i$ e $j$. In questa forma l'espressione dell'ampiezza di anello \`e
\be
\label{ansatz}
{\cal A} \ = \ \frac{1}{2}\sum_{i,j,k}\mathcal{N}_{ij}^k n^i n^j \chi_k \ .
\ee
Per i caratteri di $so(8)$ ($O_8, V_8, -C_8, -S_8$) l'algebra di fusione dice che un generico carattere fonde con $V_8$ dando se stesso, $V_8$ \`e quindi l'identit\`a dell'algebra, mentre la fusione di $-C_8$ e $-S_8$ da $O_8$. Si ritrova  cos\`i dalla (\ref{ansatz}) l'ampiezza di anello (\ref{A1}).
Dalle ampiezze trasverse $\tilde{\cal K}_1$ e $\tilde{\cal A}_1$ si ottiene l'ampiezza di M\"{o}bius trasversa
\be
\tilde{\cal M}_1 \ = \ - \frac{2}{2}(n_o+n_v+n_s+n_c)\hat{V}_8 \ ,
\ee
e da quest'ultima con una trasformazione $P$, l'ampiezza nel canale diretto
\be
\label{M1}
{\cal M}_1 \ = \ - \frac{1}{2}(n_o+n_v+n_s+n_c)\hat{V}_8 \ ,
\ee
che \`e una corretta simmetrizzazione dell'anello. Il segno della M\"{o}bius \`e stato fissato per imporre le condizioni di tadpole che, per i tre settori contenenti modi di massa nulla, $V_8, S_8$ e $C_8$ risultano essere
\ba
n_o + n_v + n_s + n_c &=& 64 \, ,\nonumber \\
n_o - n_v - n_s + n_c &=& 0 \, ,\nonumber \\
n_o - n_v + n_s - n_c &=& 0 \, , \
\ea
e fissano $n_0=n_v$ e $n_s=n_c$, determinando il gruppo di gauge $SO(n_o)\times SO(n_v)\times SO(n_s)\times SO(n_c) \ $. Lo spettro di basse energie ha vettori nell'aggiunta, tachioni nelle diverse rappresentazioni bifondamentali $(n_o,n_v,1,1) \ $, $(1,1,n_s,n_c)$, fermioni sinistri nella $(1,n_v,n_s,1) \ $ e nella $(n_o,1,1,n_c) \ $, fermioni destri nella $(1,n_v,1,n_c)$ e nella $(n_o,1,n_s,1) \ $. Lo spettro \`e chirale in ragione delle diverse rappresentazioni dei fermioni destri e sinistri, ma \`e possibile verificare che le condizioni di tadpole RR eliminano tutte le anomalie di gauge.

Vediamo ora le altre due possibili scelte per l'ampiezza di bottiglia di Klein
\ba
{\cal K}_2 &=& \frac{1}{2}(O_8+V_8+S_8+C_8) \ ,\nonumber\\
{\cal K}_3 &=& \frac{1}{2}(-O_8+V_8+S_8-C_8) \ ,
\ea
che nel canale trasverso portano
\ba
\tilde{\cal{K}}_2 &=& \frac{2^6}{2}O_8 \ , \nonumber\\
\tilde{\cal{K}}_3 &=& \frac{2^6}{2}(-C_8) \ .
\ea
La seconda proiezione simmetrizza il settore NS-NS dando lo stesso spettro del caso precedente e antisimmetrizza il settore di R-R, dando uno scalare complesso e una 4-forma completa. La terza ampiezza di Klein \cite{susy95,c0b,bfl1} elimina dallo spettro il tachione ed il tensore antisimmetrico dal settore NS-NS, mentre nel settore R-R proietta via uno scalare e una 4-forma autoduale dagli stati di $|-C_8|^2$ ed il tensore antisimmetrico da quelli di $|-S_8|^2$. Lo spettro \`e naturalmente chirale a causa delle diverse proiezioni su $-C_8$ e $-S_8$. 

Le ampiezze di anello e M\"{o}bius compatibili con la seconda proiezione possono essere ottenute da quella del primo modello fondendo i vari termini con $O_8$, e infatti non \`e difficile rendersi conto che questa operazione, nel caso della bottiglia di Klein, porta a derivare il secondo modello dal primo. Si ottengono cos\`i le ampiezze nel canale diretto
\ba
{\cal A}_2 &=& {\textstyle\frac{1}{2}} \left[( n_o^2 + n_v^2 +
n_s^2 + n_c^2) O_8 + 2(n_o n_v + n_s n_c)
V_8\right. \nonumber \\
& &\left.- 2(n_v n_s + n_o n_c) C_8 - 2(n_v n_c + n_o n_s) S_8\right] \\
{\cal M}_2 &=& \mp \frac{1}{2}
(n_o + n_v - n_s - n_c) \hat{O}_8 \ ,
\ea
che nel canale trasverso diventano
\ba
\tilde{\cal A}_2 &=& \frac{2^{-6}}{2} \left[
(n_o + n_v + n_s + n_c)^2 V_8 + (n_o + n_v - n_s - n_c)^2 O_8 \right.\nonumber \\
& & \left.+ (n_o - n_v + n_s - n_c)^2 C_8 + (n_o - n_v - n_s + n_c)^2 S_8 \right] \\
\tilde{\cal M}_2 &=& \pm \frac{2}{2}
(n_o + n_v - n_s - n_c) \hat{O}_8 \ .
\ea
Il segno della M\"{o}bius rimane indeterminato, dal momento che non viene fissato da nessuna condizione di tadpole. Visto che in $\cal M$ non compare $V_8$, possiamo interpretare le cariche in termini di gruppi unitari, richiedendo che $n_b=n_o\ $, $\bar n_b = n_v\ $, $n_f = n_s\ $ e $\bar n_f = n_c$. Il tadpole di R-R, che si trova da $\tilde{\cal A}_2$, porta le condizioni $n_b=\bar n_b$ e $n_f=\bar n_f$. Il gruppo di gauge \`e quindi $U(n_b)\times U(n_f)$, e lo spettro di bassa energia contiene vettori nell'aggiuta del gruppo, fermioni sinistri di Majorana-Weyl nella $(1,\bar n_b,1,\bar n_f)$ e nella $(n_b,1,n_f,1)$, fermioni destri di Majorana-Weyl nella $(1,\bar n_b,n_f,1)$ e nella $(n_b,1,1,\bar n_f)$ e tachioni in diverse rappresentazioni destre e sinistre. Lo spettro risulta chirale ma privo di anomalie.

Le rimanenti ampiezze del terzo modello si trovano fondendo i vari termini del primo modello con il carattere $-C_8$,  
\ba
\label{A3}
{\cal A}_3 &=& - \frac{1}{2} \left[
(n_o^2 + n_v^2 + n_s^2 + n_c^2 ) C_8 -
2(n_o n_v + n_s n_c) S_8\right.
\nonumber \\
& &\left. +2(n_v n_s + n_o n_c) V_8 + 2(n_v n_c + n_o n_s) O_8\right] \ ,  \\
\label{M3}
{\cal M}_3 &=& \frac{1}{2} (n_o - n_v - n_s + n_c) \hat{C}_8 \ ,
\ea
che nel canale trasverso divengono
\ba
\label{tildeA3}
\tilde{\cal A}_3 &=& \frac{2^{-6}}{2} \left[
(n_o + n_v + n_s + n_c)^2 V_8 - (n_o + n_v - n_s - n_c)^2 O_8\right. \nonumber \\
& &\left.- (n_o - n_v - n_s + n_c)^2 C_8 + (n_o - n_v + n_s - n_c)^2 S_8\right] \ , \\
\tilde{\cal M}_3 &=& \frac{2}{2}
(n_o - n_v - n_s + n_c) \hat{C}_8 \ .
\label{tildeM3}
\ea
Anche in questo caso il gruppo di gauge \`e unitario, dal momento che in $\cal M$ non compare $V_8$. Ponendo $n_v=n \ $, $n_s=\bar n \ $, $n_o=m \ $ e $n_c=\bar m$, la condizione di tadpole R-R su $S_8$ fissa $m=\bar m$ e $n=\bar n$, mentre quella su $C_8$ fissa $m-n=32$, determinando il gruppo di gauge $U(m)\times U(n) \ $. La particolare scelta $n=0$ elimina i tachioni dal settore aperto, e il modello ottenuto \`e noto come $0'B$. Il gruppo di gauge diventa in realt\`a $SU(32)$, e dal momento che il fattore $U(1)$ \`e anomalo e il corrispondente vettore diventa massivo  \cite{wito32,dsw}. Lo spettro di massa nulla contiene un vettore nell'aggiunta e un fermione destro nelle rappresentazioni antisimmetriche $\frac{m(m-1)}{2}$ e nella $\frac{\bar m(\bar m -1)}{2} \ $.
\begin{table}
\label{tab1}
\begin{center}
\begin{tabular}{c c c c | c}
$O_8$ & $V_8$ & $-S_8$ & $-C_8$ \\
\hline
+ & + & + & + & D9$^{(1)}$ \\
+ & + & $-$ & $-$ & $\overline{\rm{D9}}^{(1)}$\\
$-$ & + & + & $-$ & D9$^{(2)}$ \\
$-$ & + & $-$ & + & $\overline{\rm{D9}}^{(2)}$
\end{tabular}
\hspace {2 cm}
\begin{tabular}{c c c c | c}
$O_8$ & $V_8$ & $-S_8$ & $-C_8$ \\
\hline
$\mp$ & $\mp$ & $\mp$ & $\mp$ & O9$_{\pm}^{(1)}$ \\
$\mp$ & $\mp$ & $\pm$ & $\pm$ & $\overline{\rm{O9}}_{\pm}^{(1)}$\\
$\pm$ & $\mp$ & $\mp$ & $\pm$ & O9$_{\pm}^{(2)}$ \\
$\pm$ & $\mp$ & $\pm$ & $\mp$ & $\overline{\rm{O9}}_{\pm}^{(2)}$
\end{tabular}
\end{center}
\caption{convenzioni per le $D$-brane e gli $O$-piani.}
\end{table}

\`E possibile leggere dalle ampiezze dei tre modelli discussi il contenuto in termini di $O$-piani e $D$-brane. Dal momento che si sono introdotte due differenti cariche di R-R, si hanno due tipi differenti di $O$-piani e $D$-brane che hanno la stessa tensione ma cariche di R-R opposte (vedi Tabella \ref{tab1}). Si vede che le ampiezze trasverse di Klein contengono le seguenti combinazioni di $O$-piani
\be
\tilde{\cal K}_1 \to  \rm{O9}^{(1)}_\pm \oplus \rm{O9}^{(2)}_\pm \oplus
\overline{\rm{O9}}^{(1)}_\pm \oplus \overline{\rm{O9}}^{(2)}_\pm \ , 
\ee
\be
\tilde{\cal K}_2 \to  \rm{O9}^{(1)}_\mp \oplus \rm{O9}^{(2)}_\pm \oplus
\overline{\rm{O9}}^{(1)}_\mp \oplus \overline{\rm{O9}}^{(2)}_\pm \ , 
\ee
\be
\tilde{\cal K}_3 \to  \rm{O9}^{(1)}_\mp \oplus \rm{O9}^{(2)}_\pm \oplus
\overline{\rm{O9}}^{(1)}_\pm \oplus \overline{\rm{O9}}^{(2)}_\mp \ , 
\ee
dove la diversa scelta del segno \`e possibile modificando il segno della M\"{o}bius e non cambiando le condizioni di tadpole nel settore R-R. Da $\tilde{\cal A}_1$ si pu\`o identificare direttamente il tipo di $D$-brane, ottenendo
\be
n_0 \to \overline{\rm{D9}}^{(1)} \ , \qquad n_v \to D9^{(1)} \ , \qquad n_s \to D9^{(2)} \ , 
\qquad n_c \to \overline{\rm{D9}}^{(2)} \ .
\ee

Il secondo ed il terzo caso sono meno immediati, dal momento che le loro brane sono sovrapposizioni di quelle del primo modello. Nel secondo modello si vede che, per avere coefficienti positivi in $\tilde{\cal{A}}_2$ per i caratteri $-S_8$ e $-C_8$, occorre assorbire il segno negativo nel quadrato delle cariche. In questo modo $n_b$ ed $n_f$ devono riferirisi ad oggetti con cariche rispettivamente $(1,1,e^{-i\pi/2},e^{-i\pi/2}) \ $ e  $(-1,1,e^{i\pi/2},e^{-i\pi/2}) $. Questo nel caso di una scelta di segno $+$ per $\tilde{\cal M}_2$. Si ottengono queste propriet\`a combinando con coefficienti complessi $n_o \ $ $ \ \overline{\rm{D9}}^{(1)}$ e $n_v \ $ D9$^{(1)}$ $ \ (n_o=n_v) $, per dare
\be
n_b \ = \ \frac{n_o e^{i\pi/4}+n_ve^{-i\pi/4}}{\sqrt{2}} \qquad {\rm and}\qquad
\bar{n_b} \ = \ \frac{n_o e^{-i\pi/4}+n_ve^{+i\pi/4}}{\sqrt{2}} \ ,
\ee
e $n_s \ $ D9$^{(2)}$ con $n_c \ $ $ \ \overline{\rm{D9}}^{(2)} \ $
$(n_s=n_c)$, ottenendo
\be
n_f \ = \ \frac{n_s e^{i\pi/4}+n_c e^{-i\pi/4}}{\sqrt{2}} \qquad {\rm and}\qquad
\bar{n_f} \ = \ \frac{n_s e^{-i\pi/4}+n_c e^{+i\pi/4}}{\sqrt{2}} \ .
\ee
Nel terzo modello si pu\`o vedere che le giuste combinazioni (e le loro coniugate) sono invece
\be
n \ = \ \frac{n_v e^{i\pi/4}+n_c e^{-i\pi/4}}{\sqrt{2}} \qquad \rm{and} \qquad
m \ = \ \frac{n_o e^{i\pi/4}+n_s e^{-i\pi/4}}{\sqrt{2}} \ .
\ee

\section {Deformazioni di Scherk-Schwarz} 

In compattificazioni di teorie di campo supersimmetriche \`e possibile introdurre shift dei momenti di Kaluza-Klein dei vari campi proporzionali alle loro cariche, producendo differenze di massa fra fermioni e bosoni e rompendo quindi la supersimmetria. Questo meccanismo, detto di Scherk-Schwarz \cite{ss}, in teorie delle stringhe si arricchisce della possibilit\`a di introdurre shift non solo nei momenti ma anche nei winding \cite{ads1}. Per teorie di stringhe chiuse orientate le due possibilit\`a sono legate da una T-dualit\`a e descrivono essenzialmente lo stesso fenomeno. 

Costruendo discendenti aperti si ottengono invece risultati molto diversi. Ci si riferisce a questi due fenomeni come \emph{Scherk-Schwarz supersymmetry breaking} ed \emph{M-theory breaking}, dal momento che il secondo pu\`o essere collegato al primo via T-dualit\`a lungo l'undicesima coordinata \cite{hwi}. Il modello di M-theory breaking presenta un interessante aspetto detto \emph{brane supersymmetry}, ovvero le eccitazioni di bassa energia di brane immerse in un bulk non supersimmetrico possono essere supersimmetriche. In questo caso la rottura di supersimmetria viene considerata a meno di correzioni radiative. Mentre la scala di rottura della supersimmetria via deformazioni di Scherk-Schwarz \`e data dall'inverso del raggio di compattificazione, nel caso di \emph{brane supersymmetry} ci si aspetta contributi radiativi dell'ordine $\frac{1}{R^2}\frac{1}{M_Pl}$. L'analisi di questo fenomeno \`e la motivazione principale degli ultimi capitoli di questa Tesi. 

\subsection{Scherk-Schwarz supersymmetry breaking}

Partiamo dalla teoria IIB e consideriamone un orbifold, proiettando lo spettro con i generatori di $\mathbb Z_2$ dati da $(-)^F\delta$, dove $F=F_L+F_R$ conta i fermioni spazio-temporali e l'azione di $\delta$ \`e lo shift $\delta:X\rightarrow X+\pi R$ lungo la direzione spaziale compattificata su $S^1$ con raggio R. Scrivendo la somma su impulsi e su windings sul cerchio come
\be
\Lambda_{n+a,w+b} =
\sum_{n,w}{q^{{\alpha ' \over 4} \left( {(n+a)\over R} +
{(w+b)R \over \alpha '}\right)^2}\; \bar q ^{{\alpha ' \over 4} \left(
{(n+a)\over R} - {(w+b)R \over \alpha '} \right)^2} \over \eta (q) \,
\eta (\bar q)} \, ,
\ee
la funzione di partizione della IIB \`e
\be
{\mathcal T}_{IIB} = |V_8-S_8|^2\Lambda_{n,w} \ .
\ee

Studiamo per passi l'azione dell'orbifold sulla funzione di partizione. Nel settore untwisted, l'azione $(1 + \delta)/2$ manda $\Lambda_{n,w} \to \left(\Lambda_{n,w}+(-)^m\Lambda_{n,w}\right)/2$. Lo shift $\delta$ identifica le due met\`a del cerchio di compattificazione, e dal punto di vista dell'impulso equivale ad avere un raggio R/2, ovvero a sommare sui soli momenti pari mantenendo invariato il raggio (non cos\`i per il winding). Per recuperare l'invarianza modulare occorre introdurre anche settori twisted, procedendo come gi\`a fatto nel caso dell'orbifold $S^1/\mathbb Z_2$, e si trova che
\be
\Lambda_{n,w} \stackrel{\delta}{\longrightarrow}\Lambda_{n+\frac{1}{2},w}=(-)^m\Lambda_{n,w}
\stackrel{S}{\longrightarrow}\Lambda_{n,w+\frac{1}{2}}
\stackrel{\delta}{\longrightarrow}\Lambda_{n+\frac{1}{2},w+\frac{1}{2}}=(-)^m\Lambda_{n,w+\frac{1}{2}} \ .
\ee

L'operatore $(-)^F\delta$ agisce sui caratteri invertendo il segno dei settori fermionici, e anche in questo caso occorre considerare l'azione della trasformazione modulare S (\ref{Scharacters}) per ricostruire una funzione di partizione invariante modulare
\be
|V_8-S_8|^2 \stackrel{\delta}{\longrightarrow}|V_8+S_8|^2
\stackrel{S}{\longrightarrow}|O_8-C_8|^2
\stackrel{\delta}{\longrightarrow}|O_8+C_8|^2 \ .
\ee
La funzione di partizione invariante modulare corrispondente alla proiezione $(-)^F\delta$ \`e quindi
\ba
{\cal T}_{{\rm KK}} &=&
\frac{1}{2} \biggl[ |V_8 - S_8 |^2 \; \Lambda_{n,w}
+ |V_8 + S_8 |^2 \; (-1)^n \Lambda_{n,w}\nonumber\\
& &  + |O_8 - C_8 |^2 \; \Lambda_{n,w+{1\over 2}} +
|O_8 + C_8 |^2 \; (-1)^n \Lambda_{n,w+{1\over 2}} \biggr] \, ,
\ea
che pu\`o essere riscritta nella forma 
\ba
{\cal T}_{{\rm KK}} &=& (V_8 \bar V_8 + S_8 \bar S_8 ) \Lambda_{2n,w}
+ (O_8 \bar O_8 + C_8 \bar C_8) \Lambda_{2n,w+{1\over 2}} \nonumber \\
& & - (V_8 \bar S_8 + S_8 \bar V_8 ) \Lambda_{2n+1,w}
- (O_8 \bar C_8 + C_8 \bar O_8 ) \Lambda_{2n+1,w+{1\over 2}} \, .
\label{torus_KK}
\ea
Nel limite di decompattificazione si ritrova la funzione di partizione della $IIB$., mentre per raggi piccoli ($R \leq {\mathcal O}(\sqrt{\alpha'})$) si sviluppano instabilit\`a tachioniche. Per $n=w=0$ il carattere $O_8$, che parte con $h=-1/2$, \`e moltiplicato per un fattore $q^{\frac{\alpha'}{4}\left(\frac{R}{2\alpha'}\right)^2}$. Si ha quindi un tachione nello spettro per $-\frac{1}{2}+\frac{R^2}{16\alpha'}< 0 $ ovvero $R<2\sqrt{2\alpha'}$. Nella discussione che segue ci limiteremo per\`o al caso di spettro non tachionico.

Studiamo i discendenti aperti. Nella bottiglia di Klein si propagano solo i settori di impulso nullo, e indicando le somme sui momenti e sugli impulsi come
\be
P_n(q) = \sum_n \; {q^{{\alpha'\over 2} n^{{\rm T}} g^{-1} n} \over \eta(q)^4} \, , \qquad
W_w(q) = \sum_w \; {q^{{1\over 2\alpha'} w^{{\rm T}} g w} \over \eta(q)^4} \, ,
\ee
l'ampiezza si scrive
\be
{\cal K}_{{\rm KK}} = \frac{1}{2} (V_8 - S_8) \; P_{2n} \, ,
\ee
che nel canale trasverso diventa
\be
\tilde{\cal K}_{{\rm KK}}
= {2^5 \over 4} \, v  \, (V_8 - S_8)\;
W_{w} \, ,
\ee
dove  $v=R/\sqrt{\alpha '}$. Nel canale trasverso dell'anello si propagano solo windings, e quindi
\ba
\tilde{\cal A}_{{\rm KK}}
&=& {2^{-5}\over 4} \, v \,
 \Bigl\{
\bigl[ (n_1 + n_2 + n_3 + n_4 )^2 V_8 \ - \ (n_1 + n_2 - n_3 - n_4 )^2 S_8 \bigr] \, W_w
\nonumber\\
& & + \bigl[ (n_1 - n_2 + n_3 - n_4 )^2 O_8 \ - \ (n_1 - n_2 - n_3 + n_4 )^2 C_8 \bigr] \, W_{w+{1\over 2}}
\Bigr\} \, , \nonumber\\
\ea
dove si sono parametrizzati come di solito i coefficienti di riflessione dei caratteri. Dai segni relativi degli $n_i$ si vede che $n_1$ e $n_2$ contano il numero di D9-brane mentre $n_3$ e $n_4$ contano le $\overline{\rm{D}9}$-brane. Una trasformazione modulare S e una risommazione di Poisson danno l'ampiezza nel canale diretto
\ba
{\cal A}_{{\rm KK}} &=&
\frac{1}{2} (n_1^2 + n_2^2 + n_3^2 + n_4^2 )
\left[ V_8\, P_{2n} - S_8 \, P_{2n+1} \right]+\nonumber\\
& & + (n_1 n_2 + n_3 n_4 ) \left[ V_8 \, P_{2n+1} - S_8 \, P_{2n}
  \right]\nonumber\\
& & + (n_1 n_3 + n_2 n_4 ) \left[ O_8 \, P_{2n} - C_8 \, P_{2n+1} \right]\nonumber\\
& & + (n_1 n_4 + n_2 n_3 ) \left[ O_8 \, P_{2n+1} - C_8 P_{2n}
  \right] \ .
\ea

Infine da $\tilde{\cal A}_{{\rm KK}}$ e $\tilde{\cal K}_{{\rm KK}}$ si trova
\be
\tilde{\cal M}_{{\rm KK}} = - \, \frac{v}{2} \,
\Bigl[ (n_1 + n_2 + n_3 + n_4 ) \, \hat V _8 \; W_{w}
 - (n_1 + n_2 - n_3 - n_4 ) \, \hat S _8 \; (-1)^w W_{w}
\Bigr] \, ,
\ee
dove i segni $(-)^w$ per la sommatoria sui winding davanti a $\hat S _8 $ sono stati scelti per avere nel canale diretto un termine del tipo $P_{2n+1}\hat S _8 $. Si trova quindi
\be
{\cal M}_{{\rm KK}}
= - \frac{1}{2} (n_1 + n_2 + n_3 + n_4 )
\, \hat V _8 \, P_{2n} + \frac{1}{2}(n_1 + n_2 - n_3 - n_4 )\, \hat S _8 \,
P_{2n+1} \ ,
\ee
un risultato consistente con l'anello, dove i termini misti nelle cariche corrispondono a contributi orientati che, pertanto, non vengono proiettati. Le condizioni di tadpole sono
\ba
\hbox{{\rm NS-NS:}} & \quad n_1 + n_2 + n_3 + n_4 =& 32 \  , \\
\hbox{{\rm R-R:}} & \quad n_1 + n_2 - n_3 - n_4 =& 32\, .
\label{KKtad}
\ea

La condizione di tadpole R-R fissa il numero totale di brane. Imponendo anche la condizione di tadpole NS-NS vengono rimosse le anti-brane $(n_3 = n_4=0)$, e si ottiene in questo modo la famiglia di gruppi di gauge $SO(n_1)\times SO(32-n_1)$, con vettori nell'aggiunta e spinori chirali nella bi-fondamentale. Nel settore chiuso la supersimmetria \`e rotta alla scala di compattificazione e lo stesso avviene nel settore aperto.

\subsection{M-theory breaking e ``brane supersymmetry''}

Studiamo ora il caso in cui ci sia uno shift sui winding. Ci si aspetta che, rispetto alle eccitazioni di bassa energia nella direzione ortogonale alle D8-brane (T-duali delle D9-brane nel primo modello), lo shift non produca effetti e che quindi la supersimmetria sia preservata nel settore aperto. In realt\`a quello che succede \`e che la supersimmetria viene rotta nelle eccitazioni massive e, attraverso correzioni radiative, anche nel settore aperto di massa nulla. 

L'ampiezza di toro della IIB con shift nei winding pu\`o essere ricavata semplicemente scambiando momenti e winding nella (\ref{torus_KK}), e quindi si ottiene
\ba
{\cal T}_{{\rm W}} &=& ( V_8 \bar V_8 + S_8 \bar S_8 )
\Lambda_{n,2w} + ( O_8 \bar O_8 + C_8 \bar C_8 )
\Lambda_{n+{1\over 2},2w} \nonumber \\
& & - (V_8 \bar S_8 + S_8 \bar V_8)
\Lambda_{n,2w+1} - (O_8 \bar C_8 + C_8 \bar O_8 )
\Lambda_{n+{1\over 2},2w+1} \, .
\ea
Anche in questo caso per $R\sim \sqrt {\alpha'}$ si sviluppano instabilit\`a tachioniche, mentre nel limite di riduzione dimensionale $R \to 0$ si ritrova lo spettro supersimmetrico. Anche in questo caso ci limiteremo a trattare il caso non tachionico.
Nell'ampiezza di bottiglia di Klein nel canale diretto si propagano solo i modi con $w=0$
\be
{\cal K}_{{\rm W}} = \frac{1}{2} (V_8 - S_8 ) \; P_{n}
+ \frac{1}{2} (O_8 - C_8 ) \; P_{n+\frac{1}{2}} \, ,
\ee
e da questa si trova l'ampiezza nel canale trasverso 
\be
\tilde{\cal K}_{{\rm W}} = \frac{2^5}{2}\, 2 \,  v \,
\left( V_8\; W_{4w} - S_8 \; W_{4w+2} \right) \, .
\ee
Si vede che nel canale trasverso gli unici modi di massa nulla vengono dal  settore NS-NS, e ci si aspetta per questo che la carica di R-R totale sia nulla  e che il modello contenga $O-9$-piani e $\overline{\rm O9}$-piani. Al contrario nell'anello trasverso si propagheranno solo stati di momento nullo, che sono associati ai caratteri $V_8$ e $-S_8$. Parametrizzando i differenti coefficienti di riflessione si ottiene
\ba
\tilde{\cal A}_{{\rm W}} &=& \frac{2^{-5}}{2} \, 2 \, v \,
\Bigl\{ \Bigl[ (n_1+n_2+n_3+n_4)^2 \; V_8  - \; (n_1+n_2-n_3-n_4)^2 \; S_8 \Bigl]
W_{4w} \nonumber\\
& & + \Bigl[ (n_1-n_2+n_3-n_4)^2 \; V_8 - \; (n_1-n_2-n_3+n_4)^2 \; S_8
\Bigr]  W_{4w+2} \Bigr\} \, ,\nonumber\\
\ea
dove si \`e fattorizzata la somma sui winding nella forma $W_{2w}=W_{4w}+W_{4w+2}$ per poter poi agevolmente comparare $\tilde{\cal A}_{{\rm W}}$ con $\tilde{\cal K}_{{\rm W}}$. Anche in questo caso $n_1$ e $n_2$ contano il numero di D9-brane, mentre $n_3$ e $n_4$ contano il numero di $\overline{{\rm D}9}$-brane. Come di consueto l'ampiezza nel canale diretto si ottiene tramite una trasformazione modulare S
\be
{\cal A}_{{\rm W}} = \frac{1}{2} (n_1^2 + n_4^2)\,
(V_8 - S_8 )\;
(P_{n} +  P_{n+\frac{1}{2}}) + n_1 n_4 \, (O_8 - C_8) \;
(P_{n+ \frac{1}{4}} + P_{n+ \frac{3}{4}}) \ ,
\ee
e a questo punto non \`e difficile ricavare l'ampiezza trasversa del nastro di M\"{o}bius da $\tilde{\cal K}_{{\rm W}}$ e $\tilde{\cal A}_{{\rm W}}$, che risulta essere
\be
\tilde{\cal M}_{{\rm W}} = - 2 \, v \,  \Bigl[
(n_1+n_2+n_3+n_4) \, \hat V_8\, W_{4w}
 - (n_1-n_2-n_3+n_4) \, \hat S_8\,
W_{4w+2} \Bigr] \ ,
\ee
e da questa, con una trasformazione modulare P,
\be
{\cal M}_{{\rm W}} = - \frac{1}{2}
 (n_1+n_4) \left[ (\hat V _8  - \hat S_8)\,  P_{n} +
(\hat V _8  + \hat S_8)\, P_{n+ \frac{1}{2}} \right]  \, .
\ee

Le condizioni di tadpole risultano essere
\be
\hbox{{\rm NS-NS:}} \quad n_1+n_2+n_3+n_4 = 32 \, , \qquad \qquad
\hbox{{\rm R-R:}} \quad n_1+n_2=n_3+n_4 \ .
\ee
Nel limite di riduzione dimensionale $R \to 0$, si sviluppano ulteriori tadpole dovuti al fatto che in questo limite gli stati di winding in $W_{4n+2}$ diventano a massa nulla. Le nuove condizioni sono
\be
\hbox{{\rm NS-NS:}} \quad n_1-n_2+n_3-n_4 = 0 \, , \qquad \qquad
\hbox{{\rm R-R:}} \quad n_1-n_2-n_3+n_4=32 \ .
\ee
Tutte le condizioni sono risolte per $n_1=n_4=16$ e $n_2=n_3=0$, fissando cos\`i il gruppo di gauge a $SO(16)\times SO(16)$.\\
Dalla somma di ${\cal A}_{{\rm W}}$ e ${\cal M}_{{\rm W}}$ si vede che nel settore aperto lo spettro aperto di bassa energia \`e supersimmetrico, dal momento che contiene uno spinore e un vettore nella rappresentazione aggiunta del gruppo. La supersimmetria invece \`e rotta nel settore aperto massivo e nel bulk, dove lo shift nei winding alza in massa il gravitino.

\section {Brane supersymmetry breaking} 

In modelli con compattificazioni e proiezioni {$\mathbb Z_2$} si hanno due possibilit\`a per rompere la supersimmetria. La prima consiste nell'invertire tensione e carica dell'O9-piano e dell'O5-piano, ottenendo l'ampiezza standard di bottiglia di Klein, e rompendo la supersimmetria nel settore aperto ma non in quello chiuso. La seconda possibilit\`a consiste nell'invertire carica e tensione solo di un tipo di $O$-piano, ad esempio l'O5-piano. Si ottiene in tal modo una diversa proiezione di Klein e si devono introdurre le corrisponenti D5-antibrane. La supersimmetria, preservata dalle D9-brane, viene cos\`i rotta alla scala di stringa.

\`E possibile studiare questo fenomeno in un modello relativamente semplice, l'orbifold $T^4/{\mathbb Z}_2$. Si \`e gi\`a accennato alla possibile proiezione di Klein 
\be
{\cal K} = \frac{1}{4} \left[ ( Q_o + Q_v ) ( P_m + W_n ) - 2
\times 16 ( Q_s + Q_c ){\left(\frac{\eta}{\vartheta_4}\right)}^2 \right] \ .
\ee
Al livello di massa nulla, lo spettro di stringa chiusa coincide nel settore untwisted con quello del modello supersimmetrico studiato, mentre nel settore twisted il settore NS-NS \`e antisimmetrizzato e quello R-R \`e simmetrizzato. Si hanno cio\`e 16 scalari NS-NS, 16 2-forme anti-autoduali e 16 spinori destri. Lo spettro ha ancora supersimmetria $\mathcal{N}=(1,0)$ e contiene il solito multipletto del gravitone, 17 multipletti tensoriali, 16 dei quali dal settore twisted e 4 iper-multipletti.

Nel canale trasverso si trova
\be
\tilde{\cal K} = \frac{2^5}{4} \left[ (Q_o + Q_v) \left(
v_4 W_n^{(e)} + \frac{1}{v_4} P_m^{(e)} \right) -
2 (Q_o - Q_v)  \left(
\frac{2 \eta}{\vartheta_2} \right)^2 \right] \, ,
\ee
il cui limite di bassa energia \`e
\be
\tilde{\cal K}_0 = \frac{2^5}{4} \left[ Q_o \left( \sqrt{v_4}  -
\frac{1}{\sqrt{v_4}}\right)^2 + Q_v \left( \sqrt{v_4}  +
\frac{1}{\sqrt{v_4}}\right)^2 \right] \ .
\ee
Rispetto al modello supersimmetrico precedentemente studiato si vede che c'\`e un segno relativo fra $\sqrt{v_4}$ e $1/\sqrt{v_4}$ nel coefficiente associato a $Q_o$. Sviluppando il quadrato, questo introduce un segno negativo davanti al termine misto e indica la presenza della configurazione annuciata con $O9_+$ e $O5_-$ piani (o in alternativa la configurazione T-duale con  $O9_-$ e $O5_+$ piani). Nel primo caso ci si attende che nel modello compaiano anche $\overline{\rm D 5}$-brane che neutralizzino la carica di R-R, e quindi che nell'ampiezza di anello trasverso compaia un segno relativo fra le cariche di R-R delle $D9$-brane e delle $\overline{\rm D5}$ brane. 

L'ampiezza d'anello trasverso corrispondente \`e
\ba
\tilde {\cal A} &=& \frac{2^{-5}}{4} \Biggl[ (Q_o + Q_v) \left( N^2 v_4 W_n + \frac{D^2}{v_4}P_m \right)  +
16 \left(R_N^2 + R_D^2 \right) (Q_s + Q_c) \left( {\eta \over \vartheta_4}\right)^2 \nonumber\\
&& + 2 N D \, (V_4O_4+C_4C_4 - O_4V_4-S_4S_4 ) \left( {2 \eta \over \vartheta_2}\right)^2 \nonumber \\
& &  - 2 \times 4 R_N R_D \, (O_4C_4+S_4O_4 - V_4S_4-C_4V_4 ) \left( {\eta \over \vartheta_3}\right)^2 \Biggr] \, ,
\ea
che nel limite di basse energie diventa
\be
\tilde{\cal A}_0 = \frac{2^{-5}}{4} \left[ (V_4 O_4 - S_4 S_4) \left( N
\sqrt{v_4}  +
\frac{D}{\sqrt{v_4}}\right)^2  +  (O_4 V_4 - C_4 C_4) \left( N\sqrt{v_4}  -
\frac{D}{\sqrt{v_4}}\right)^2\right]  \, .
\ee

Si vede che il prodotto delle tensioni,  che \`e legato al coefficiente di $O_4 V_4$, \`e sempre positivo mentre le cariche di R-R, che si leggono dal coefficiente di $C_4 C_4$, sono opposte. Nel canale diretto l'ampiezza di anello diventa
\ba
\label{bsb_direct_A} 
{\cal A} &=& \frac{1}{4} \left[(Q_o + Q_v) ( N^2 P_m  + D^2 W_n ) + (R_N^2 + R_D^2) (Q_o - Q_v) {\left(\frac{2
\eta}{\vartheta_2}\right)}^2 \right.\nonumber \\
&& \left. + 2 N D ( O_4 S_4 - C_4 O_4 + V_4 C_4 - S_4 V_4) {\left(\frac{\eta}{\vartheta_4}\right)}^2 \right. \nonumber\\
&& \left.+ 2 R_N R_D ( - O_4 S_4 - C_4 O_4 + V_4 C_4 + S_4 V_4 ){\left(\frac{ \eta}{\vartheta_3}\right)}^2 \right] \ .
\ea
Infine l'ampiezza di M\"{o}bius trasversa si trova essere
\ba
\tilde{\cal M} &=& -\frac{1}{2}\Bigl[v_4 N
 W_n^{(e)}(\hat{V}_4\hat{O}_4+\hat{O}_4\hat{V}_4
-\hat{S}_4\hat{S}_4-\hat{C}_4\hat{C}_4)\nonumber\\
&&
+\frac{1}{v_4} D P_m^{(e)}(-\hat{V}_4\hat{O}_4-\hat{O}_4\hat{V}_4
-\hat{S}_4\hat{S}_4-\hat{C}_4\hat{C}_4) \nonumber\\
&&+N(-\hat{V}_4\hat{O}_4+\hat{O}_4\hat{V}_4
-\hat{S}_4\hat{S}_4+\hat{C}_4\hat{C}_4)\left( {2\hat \eta \over \hat \vartheta_2}\right)^2\nonumber\\
&&
+D(\hat{V}_4\hat{O}_4-\hat{O}_4\hat{V}_4
-\hat{S}_4\hat{S}_4+\hat{C}_4\hat{C}_4)\left( {2\hat \eta \over \hat \vartheta_2}\right)^2\Bigr] \ ,
\ea
che nel limite di basse energie diventa
\ba
\tilde{\cal M}_0 &=& - \frac{1}{2} \left[ \hat{V}_4 \hat{O}_4 \left( \sqrt{v_4}  -
\frac{1}{\sqrt{v_4}}\right) \left( N \sqrt{v_4}  + \frac{D}{\sqrt{v_4}}\right) \right.\nonumber \\
&& \left.+ \hat{O}_4 \hat{V}_4 \left( \sqrt{v_4}  + \frac{1}{\sqrt{v_4}}\right) \left( N \sqrt{v_4}  -
\frac{D}{\sqrt{v_4}}\right) \right.\nonumber \\
&& \left.-\hat{C}_4 \hat{C}_4 \left( \sqrt{v_4}  - \frac{1}{\sqrt{v_4}}\right) \left( N \sqrt{v_4}  -
\frac{D}{\sqrt{v_4}}\right) \right.\nonumber \\
&& \left. - \hat{S}_4 \hat{S}_4 \left( \sqrt{v_4}  + \frac{1}{\sqrt{v_4}}\right) \left( N \sqrt{v_4}  +
\frac{D}{\sqrt{v_4}}\right) \right] \, ,
\ea
ed infine nel canale diretto, tramite una trasformazione P, si trova
\ba
\label{bsb_directM}
{\cal M} &=& - \frac{1}{4} \Biggl[ N P_m ( \hat{O}_4
\hat{V}_4  + \hat{V}_4 \hat{O}_4  - \hat{S}_4 \hat{S}_4 - \hat{C}_4 \hat{C}_4 ) \nonumber \\
&& -  D W_n ( \hat{O}_4 \hat{V}_4  + \hat{V}_4 \hat{O}_4  + \hat{S}_4 \hat{S}_4 + \hat{C}_4
\hat{C}_4 )  \nonumber \\
&& - N( \hat{O}_4 \hat{V}_4 - \hat{V}_4 \hat{O}_4 - \hat{S}_4 \hat{S}_4
+ \hat{C}_4 \hat{C}_4 )\left( {2{\hat{\eta}}\over{\hat{\vartheta}}_2}\right)^2  \nonumber \\
&& + D( \hat{O}_4 \hat{V}_4 - \hat{V}_4 \hat{O}_4 + \hat{S}_4 \hat{S}_4 - \hat{C}_4 \hat{C}_4)\left(
{2{\hat{\eta}}\over{\hat{\vartheta}}_2}\right)^2  \Biggr] \, .
\ea
Dal momento che nell'ampiezza di M\"{o}bius compare, nello spettro di basse energie, un vettore di massa nulla, si hanno cariche reali di Chan-Paton che vengono parametrizzate nella forma
\ba
N=n_1+ n_2 \, , \qquad D=d_1+ d_2 \, , \nonumber \\
R_N=n_1- n_2 \, , \qquad R_D=d_1- d_2 \, ,
\ea
e in questo modo lo spettro di massa nulla nel settore aperto risulta
\ba
{\cal A}_0 + {\cal M}_0 &=& \frac{n_1(n_1-1) + n_2(n_2-1) +
d_1(d_1+1) +  d_2(d_2+1) }{2} \ V_4 O_4 \nonumber \\
 &&- \frac{n_1(n_1-1) + n_2(n_2-1) + d_1(d_1-1) + d_2(d_2-1) }{2} \ C_4
C_4 \nonumber \\ &&+ (n_1 n_2 + d_1 d_2 ) ( O_4 V_4 - S_4 S_4 ) + (
n_1 d_2 + n_2 d_1 ) \ O_4 S_4 \nonumber \\
&& - (n_1 d_1 + n_2 d_2 ) \ C_4 O_4 \, .
\ea
Le condizioni di tadpole R-R, che si trovano sommando i contributi di massa nulla di $\tilde{\cal A}_0$, $\tilde{\cal K}_0$ e $\tilde{\cal M}_0$,  fissano 
\be
n_1=n_2=d_1=d_2=16 \ .
\ee
e quindi
\be
N=D=32 \ ,\qquad \qquad R_N=R_D = 0 \ ,
\ee
e quindi il gruppo di gauge \`e $[SO(16) \times SO(16) ]_9 \times  [ USp(16) \times USp(16) ]_5$, dove i primi due fattori si riferiscono alle $D9$-brane e i secondi alle $\overline{\rm D5}$ brane.

Il settore con condizioni al bordo NN ha uno spettro di massa nulla ancora supersimmetrico, con un multipletto di gauge nella rappresentazione aggiunta di $SO(16)\times SO(16)$ ed un ipermultipletto nella $(16,16,1,1)$.  Al contrario, il settore DD \`e chiaramente non supersimmetrico, dal momento che contiene vettori nell'aggiunta del gruppo $USp(16)\times USp(16)$, 4 scalari nella $(1,1,16,16)$, fermioni di Weyl destri nella $(1,1,120,1)$ e nella $(1,1,1,120)$, e fermioni di Weyl sinistri nella $(1,1,16,16)$. Anche i settori con condizioni al bordo miste ND sono non supersimmetrici e comprendono un doppietto di scalari nella $(16,1,1,16)$ e nella $(1,16,16,1)$, e fermioni di Majorana-Weyl nella $(16,1,16,1)$ e nella  $(1,16,1,16)$. Questi fermioni sono peculiari dello spazio-tempo in sei dimensioni dove i fermioni di Weyl possono essere soggetti ad una condizione addizionale di Majorana, implementata con una matrice di coniugazione di carica pseudoreale.

La rottura di supersimmetria lascia non cancellati i tadpole nel settore NS-NS, che risultano essere 
\be
\left[ (N-32){\sqrt v_4}+{D+32 \over{\sqrt
v_4}}\right]^2 \ V_4O_4 +\left[ (N-32){\sqrt v_4}-{D+32 \over{\sqrt
v_4}}\right]^2 \ O_4V_4 \ .
\ee
Per questa ragione si genera un potenziale del dilatone localizzato sulle $\overline{\rm D5}$ brane. I coefficienti di $V_4O_4$ e di $O_4V_4$ sono proporzionali ai quadrati delle funzioni ad un punto sui bordi e sui crosscap. Dal punto di vista dell'azione di basse energie i due contributi NS-NS hanno origine dai termini
\be
\Delta S \sim  (N-32)\sqrt{v_4} \, \int d^6 x \, \sqrt{-g} \, e^{-\varphi_6} +
\frac{D+32}{\sqrt{v_4}} \, \int d^6 x \, \sqrt{-g} \, e^{-\varphi_6} \ ,
\ee
e in particolare dalle variazioni del dilatone in sei dimensioni $\varphi_6$ e del volume interno $v_4$ rispetto ai loro valori di background. 
\be
\varphi_6 \to \varphi_6 \delta \varphi_6 \ , \qquad {\sqrt v_4}\to (1+\delta h){\sqrt v_4} \ .
\ee
Nell'azione scritta il primo termine si riferisce al contributo del sistema di D9-brane e dell'O9-piano, e, come atteso, si cancella in conseguenza della condizione di tadpole nel settoe R-R, che fissa $N=32$ preservando la supersimmetria. Al contrario, il secondo termine \`e dovuto ai contributi delle $\overline{\rm D5}$ brane e dell'O5$_-$-piano che non viene cancellato. 

\section{Deformazioni magnetiche}

L'introduzione di campi magnetici interni (nello spazio tempo compatto) da luogo genericamente alla rottura della supersimmetria \cite{magnetic, bachasmag, Larosa}. La ragione risiede nel fatto che gli estremi carichi delle stringhe aperte si accoppiano al campo, producendo shift nelle masse degli stati di stringa differenti a seconda dei loro momenti di dipolo. 

\subsection{Campi abeliani e stringhe bosoniche aperte}

Un campo abeliano che viva in un sottospazio (una $D$-brana in uno spazio esteso o compattificato), che indichiamo con gli indici $m, n =0,...d$ dove $d$ \`e la dimensione del sottospazio, si accoppia alle cariche di Chan-Paton di una stringa aperta. L'azione di stringa (consideriamo il caso bosonico) sar\`a  modificata dall'aggiunta di termini al bordo di accoppiamento minimale 
\be
S =   \int d\tau d\sigma
{\mathcal L}(\dot X, X')  + \left. q_{\rm R} \int d\tau A_m \partial_\tau X^{m} \right|_{\sigma =\pi} 
+ \left. q_{\rm L} \int d\tau A_m \partial_\tau X^{m} \right|_{\sigma =0} \, .
\ee
Limitiamoci a considerare un campo costante che possiamo scrivere nella forma
\be
A_m=\textstyle{\frac{1}{2}}F_{mn}X^n \ ,
\ee
per cui l'azione diventa
\ba
S =   \int d\tau d\sigma
{\mathcal L}(\dot X, X')  &+& \frac{1}{2}\left. q_{\rm R} \int d\tau F_{mn} X^n \partial_\tau X^{m} \right|_{\sigma =\pi} \nonumber \\ 
&+& \frac{1}{2} \left. q_{\rm L} \int d\tau F_{mn} X^n \partial_\tau X^{m} \right|_{\sigma =0} \, .
\ea
Le equazioni del moto che si ottengono dalla lagrangiana risultano invariate dal momento che non vengono influenzate dalla presenza dei termini di bordo. Cambiano, invece, le condizioni al bordo, che diventano
\be
\partial_\sigma X_m \pm 2\pi \alpha' q_{\rm L, R} F_{mn} \,
\partial_\tau X^n = 0 \, , \qquad \sigma=0, \pi\ \ ,
\ee
e sono quindi intermedie tra condizioni di Neumann e di Dirichlet. Consideriamo ora in caso di un campo magnetico con tutte le componenti nulle ad eccezione $F_{23}=-F_{32}\equiv H$.  Le coordinate generiche $\mu$ avranno condizioni al bordo di Neumann, mentre $X^2$ e $X^3$ devono soddisfare
\ba
\partial_\sigma X_2 \pm 2\pi \alpha' q_{\rm L, R} H \,
\partial_\tau X^3 &=& 0 \, , \qquad \sigma=0, \pi\nonumber\\
\partial_\sigma X_3 \mp 2\pi\alpha' q_{\rm L, R} H \,
\partial_\tau X^2 &=& 0 \, , \qquad \sigma=0, \pi \ .
\ea
Se la coordinata $X^3$ \`e compattificata, \`e possibile effettuare una trasformazione di T-dualit\`a modificando le condizioni al bordo con il risultato che $\partial_\sigma \leftrightarrow \partial_\tau$. In questo modo si ottiene 
\ba
\label{partial}
\partial_\sigma (X^2 \pm 2\pi \alpha' q_{\rm L, R} H \, X^{3,T}) &=& 0 \, , \qquad \sigma=0, \pi \ ,\nonumber\\
\partial_\tau (X^{3,T} \mp 2\pi\alpha' q_{\rm L, R} H \, X^2) &=& 0 \, , \qquad \sigma=0, \pi \ .
\ea
Introducendo gli angoli definiti da $\tan(\theta_{L, R}) =  - 2 \pi \alpha' q_{L,R} H$, la (\ref{partial}) si scrive
\ba
\partial_\sigma (\cos(\theta_{L, R})X^2 \mp  \sin(\theta_{L, R})X^{3,T}) &=& 0 \, , \qquad \sigma=0, \pi \ ,\nonumber\\
\partial_\tau ( \sin(\theta_{L, R}) \, X^2 \pm \cos(\theta_{L, R})X^{3,T}) &=& 0 \, , \qquad \sigma=0, \pi \ ,
\ea
che suggerisce un'interpretazione geometrica: l'introduzione di campi magnetici ruota le $D$-brane nella picture T-duale. Nelle nuove coordinate 
\ba
Y^1=\cos(\theta_{L, R})X^2 \mp  \sin(\theta_{L, R})X^{3,T} \ ,\\
Y^2=\sin(\theta_{L, R}) \, X^2 \pm \cos(\theta_{L, R})X^{3,T} \ ,
\ea
si hanno condizioni di Neumann nelle direzioni tangenziali e di Dirchlet in quelle ortogonali alle brane. Nel caso di stringhe con cariche uguali ed opposte agli estremi la rotazione delle brane non ha effetti sulla massa, ma se le cariche differenti, le brane su cui sono posti gli estremi vengono ruotate di angoli differenti e i modi delle stringhe acquistano massa.

Si pu\`o a questo punto quantizzare la stringa. \`E utile introdurre le coordinate complesse
\be
X_{\pm} = \frac{1}{\sqrt{2}} (X^1 \pm i X^2)\,,
\ee
i cui momenti coniugati sono
\be
P_\mp(\tau,\sigma) = \frac{1}{2 \pi \alpha'}
\left\{ \partial_\tau  X_\mp (\tau,\sigma) +
i X_\mp (\tau,\sigma) \ 2\pi\alpha' H \ \left[ q_L \delta(\sigma) +
q_R \delta(\pi - \sigma) \right]
\right\} \, .
\ee
Risolvendo le equazioni del moto si ottengono espansioni nei modi che dipenderanno dalla carica totale $q_L+q_R$. Se $q_L+q_R\neq 0$, $X_+(\tau,\sigma)$, $ \ X_-=X_+^\dag$, si trova
\be
X_+(\tau,\sigma) = x_+ + i \sqrt{2 \alpha'}
\left[ \sum_{n=1}^\infty a_n \psi_n(\tau,\sigma) -
\sum_{m=0}^\infty b^\dag_m \psi_{-m}(\tau,\sigma) \right] \, ,
\ee
con
\be
\psi_n(\tau,\sigma) = \frac{1}{\sqrt{|n-z|}} \, \cos\left[
(n - z)\sigma + \gamma \right]\, e^{-i(n-z)\tau} \, ,
\ee
dove $z$, $\gamma$ e $\gamma'$ definite come
\be
z = {1\over \pi} (\gamma + \gamma' ) \,,
\qquad
\gamma= \tan ^{-1} (2\pi\alpha'q_L H ) \, , \quad \gamma' =\tan^{-1}
(2\pi\alpha' q_R H ) \, .
\ee
Per $a_n, a^\dag_m$ e  $b_n, b^\dag_m$ si hanno le solite relazioni di commutazione mentre, gli zero modi non commutano
\be
[x_+,x_-]=\frac{1}{H(q_L+q_R)} \, .
\ee
Essi sono infatti l'analogo degli usuali operatori di creazione e di distruzione per i livelli di Landau che, nel limite di campi deboli, danno correzioni alla massa della forma
\be
\Delta M^2 = (2n+1)(q_L+q_R)H \ .
\ee
Nel caso di carica totale nulla si ha invece $z=0$, e gli oscillatori non risentono pi\`u del campo magnetico, il cui effetto si manifesta negli zero modi 
\be
X_+(\tau,\sigma) = \frac{x_+ + p_-\left[ \tau - i2\pi\alpha'qH(\sigma - \frac{1}{2}
\pi) \right]}{\sqrt{1 + (2\pi\alpha'qH)^2}} + i \sqrt{2 \alpha'} \sum_{n=1}^\infty
\left[ a_n \psi_n(\tau,\sigma) -
b^\dag_n \psi_{-n}(\tau,\sigma) \right] .
\label{dip}
\ee

Consideriamo ora un campo magnetico costante che viva in uno spazio compatto. Ad esempio, compattifichiamo due dimensioni su un toro $T^2$. Se $2\pi R_1$ e $2\pi R_2$ sono i due lati della cella fondamentale, la degenerazione di Landau $k$ \`e data da
\be
k= 2\pi R_1 R_2 q H = 2\pi \alpha' v q H
\label{Land}
\ee
dove si \`e definito $v=R_1R_2/\alpha'$. Si pu\`o vedere che $k$ cos\`i definito \`e anche il numero intero che compare nella condizione di quantizzazione di Dirac, dovuta alla natura di monopolo del campo magnetico uniforme sul toro. Definiamo sulla cella fondamentale un potenziale vettore 
\be
A_2=a_1 \ , \qquad \qquad A_3 = a_2 + H X_2 \ ,
\ee
tale che $F_{23}=H$. Come si \`e visto, le costanti $a_{1,2}$ sono legate all'introduzione di linee di Wilson, e in questo caso le possiamo porre a zero, concentrandoci sul termine che introduce una curvatura. Le trasformazioni di gauge
\be
A_i = A_i -i e^{-i\varphi}\dd_i e^{i\varphi} \ , \qquad \qquad i=1,2
\ee
per
\be
\varphi=2\pi R_1 H X_2 \ ,
\ee
permettono di spostarsi con continuit\`a da $X_1=0$ a $X_1=2\pi R_1$. Imponendo la monodromia della funzione $q\varphi$, si trova la condizione di quantizzazione di Dirac $2\pi\alpha'vqH=k$, da cui si vede chiaramente che l'intero $k$ \`e lo stesso che definisce la degenerazione dei livelli di Landau. Nella picture T-duale $k$ viene interpretato come il numero di avvolgimenti della $D$-brana ruotata sul toro e, come si vede dalla definizione dell'angolo di rotazione, usando $R_2^T= \alpha'/R_2$,
\be
\tan{\theta} = k\frac{{R^T_2}}{R_1} \ .
\ee

\subsection{Stringhe aperte su orbifold magnetizzati}

Consideriamo l'azione effettiva di bassa energia di una D9-brana immersa in un campo abeliano di background,
\ba
S_9 & = & -T_9\sum_{a=1}^{32}\int_{{\cal M}_{10}}d^{10}Xe^{-\phi} \sqrt{-{\rm
    det}(g_{10}+2\pi\alpha'q_a F)} \nonumber\\
&& -\mu_9\sum_{p,a}\int_{{\cal M}_{10}}e^{2\pi\alpha' q_a F}\wedge C_{p+1}
\ ,
\label{BI-WZ}
\ea
dove $a$ identifica le cariche di Chan-Paton che si accoppiano ai campi magnetici. Il primo termine \`e l'azione di Born Infeld ed il secondo \`e il termine di Wess-Zumino di accoppiamento con i campi di R-R $C_{p+1}$. $T_9$ e $\mu_9$ sono rispettivamente la tensione e la carica di R-R della brana, che per una generica brana BPS sono legati dalla relazione
\be
T_p=|\mu_p|=\sqrt{\frac{\pi}{2k^2}}(2\pi\sqrt{\alpha'})^{3-p} ,
\label{tens}
\ee
dove $k^2=8\pi G_N^{(10)}$ definisce la costante di Newton in $10$ dimensioni. Introducendo una compattificazione dello spazio su due doppi tori in cui vivano due campi magnetici abeliani costanti
\be
{\cal M}_{10} = {\cal M}_6 \times T^2(H_1)\times T^2(H_2)  \ ,
\ee
e utilizzando (\ref{tens}) si ottiene per l'azione
\ba
S_9 &=& -T_9 \int_{{\cal M}_{10}}d^{10}Xe^{-\phi}
\sum_{a=1}^{32}\sqrt{-g_6}\sqrt{(1+2\pi\alpha'q_a H_1^2)
(1+2\pi\alpha'q_a H_2^2)}\nonumber\\
&&-32 \ \mu_9\int_{{\cal M}_{10}} C_{10} \ - \
\mu_5 \ v_1 v_2 H_1 H_2 \sum_{a=1}^{32}(2\pi q_a)^2 \int_{{\cal M}_{6}} C_{6}
 \ ,
\ea
dove $v_i=R_i^1R_i^2/\alpha'$ sono i volumi dei due tori di di raggi $R_i^1$ e $R_i^2$. Il termine lineare nell'espansione dell'azione di Wess-Zumino non compare, perch\'e il generatore del gruppo abeliano $U(1)$ \`e a traccia nulla, dal momento che nel nostro caso \`e un sottogruppo del gruppo di gauge totale $SO(32)$ definito sulle D9-brane. Fissando $H_1=\pm H_2$ e usando le condizioni di quantizzazione di Dirac $k_i=2\pi\alpha' v_i q H_i$ per entrambi i campi magnetici, l'azione si semplifica notevolmente e diventa
\ba
S_9 &=& - 32 \int_{{\cal M}_{10}} \left( d^{10}X
\sqrt{-g_6} \ e^{-\phi} \ T_9 \ + \ \mu_9 \ C_{10} \right)\nonumber\\
&&
-\sum_{a=1}^{32}\left(\frac{q_a}{q}\right)^2 \int_{{\cal M}_{6}}\left( d^6 X
\sqrt{-g_6} \ |k_1k_2| \ T_5 \ e^{-\phi} \ + \ k_1 k_2 \ \mu_5 \ C_{6} \right)
 \ .
\ea
L'azione trovata indica che una D-9 brana magnetizzata mima $|k_1k_2|$ D5 brane se $k_1k_2 > 0 \ $ ($H_1=+H_2$), ovvero $\overline{\rm D5}$ brane se $k_1k_2<0 \ $ ($H_1=-H_2$). Questo fenomeno si ritrova nei modelli di orbifold di stringa, in cui la presenza di O5-piani che riassorbano la carica delle D5 brane dovuta alle D-9 brane magnetizzate, permette la costruzione di modelli supersimmetrici \cite{magnetic}.

Un esempio di questo fenomeno \`e dato illustrato modello  $\mathcal{M}_6 \times [T^2(H_1)\times T^2(H_2)]/\mathbb{Z}_2$, in cui si hanno due campi magnetici abeliani nei due tori $T^2$. Per scrivere le ampiezze di vuoto occorre partire dal modello supersimmetrico $T^4/ \mathbb Z_2$. Decomponiamo i caratteri interni in rappresentazioni di $SO(2)\times SO(2)$ 
\begin{eqnarray}
Q_o (z_1 ; z_2) &=& V_4 (0) \left[ O_2 (z_1 ) O_2 (z_2  ) +
V_2 (z_1 ) V_2 (z_2  ) \right] \nonumber\\
&&- C_4 (0) \left[ S_2 (z_1 ) C_2 (z_2  ) +
C_2 (z_1 ) S_2 (z_2  ) \right] \, ,
\nonumber
\\
Q_v (z_1 ; z_2 ) &=&  O_4 (0) \left[ V_2 (z_1 ) O_2 (z_2  ) +
O_2 (z_1 ) V_2 (z_2  ) \right] \nonumber \\
&&- S_4 (0) \left[ S_2 (z_1 )
S_2 (z_2  ) + C_2 (z_1 ) C_2 (z_2  ) \right] \, ,
\nonumber
\\
Q_s (z_1 ;z_2  ) &=& O_4 (0) \left[ S_2 (z_1 ) C_2 (z_2  ) +
C_2 (z_1 ) S_2 (z_2  ) \right] \nonumber \\
&&- S_4 (0) \left[  O_2 (z_1 ) O_2 (z_2  ) +
V_2 (z_1 ) V_2 (z_2  ) \right] \, ,
\nonumber
\\
Q_c (z_1 ; z_2 )
&=&  V_4 (0) \left[ S_2 (z_1 ) S_2 (z_2  ) +
C_2 (z_1 ) C_2 (z_2  ) \right] \nonumber \\
&&- C_4 (0)
\left[  V_2 (z_1 ) O_2 (z_2  ) +
O_2 (z_1 ) V_2 (z_2 ) \right] \, ,
\end{eqnarray}
dove gli argomenti $z_i$ sono gli shift nei modi dovuti all'introduzione dei campi magnetici, che si sono trovati essere 
\be
z_i = {1\over \pi} (\tan ^{-1} (2\pi\alpha'q_L H_i ) + \tan^{-1} (2\pi\alpha' q_R H_i ) ) \, ,
\ee
e i caratteri di livello 1 dell'estensione affine di $O(2n)$ sono legati alle quattro funzioni theta di Jacobi dalle relazioni
\ba
\label{so2n}
O_{2n}(z) &=& \frac{1}{2 \eta^n (\tau)} \left[
\vartheta_3^n(z|\tau)
+ \vartheta_4^n(z|\tau)\right] \, ,
\nonumber \\
V_{2n}(z) &=&\frac{1}{2 \eta^n (\tau)} \left[
\vartheta_3^n(z|\tau)
- \vartheta_4^n(z|\tau) \right] \, , \nonumber \\
S_{2n}(z) &=& \frac{1}{2 \eta^n (\tau)} \left[
\vartheta_2^n(z|\tau) + i^{-n} \vartheta_1^n(z|\tau)\right] \,
, \nonumber \\
C_{2n}(z) &=& \frac{1}{2 \eta^n (\tau)} \left[
\vartheta_2^n(z|\tau)
- i^{-n} \vartheta_1^n(z|\tau) \right] \, .
\ea

Il settore chiuso non viene alterato dall'introduzione dei campi magnetici sulle $D$-brane che interagiscono solo con le cariche di Chan-Paton del settore aperto. L'ampiezza di toro continua quindi ad assumere la forma usuale
\be
{\cal T} = \frac{1}{2} \Biggl[ |Q_o + Q_v|^2
\Sigma_{n,w}
 + |Q_o - Q_v |^2 \left| {2 \eta \over
\vartheta_2} \right|^4
+ 16 |Q_s + Q_c |^2 \left| {\eta \over \vartheta_4}\right|^4
+ 16 |Q_s - Q_c |^2 \left| {\eta \over \vartheta_3} \right|^4 \Biggr]\, .
\nonumber\\
\ee
e indicando le somme sui momenti e sui winding per i due tori come $P_i$ e $W_i$, l'ampiezza di Klein \`e
\be
{\cal K} = \frac{1}{4} \Biggl\{ (Q_o + Q_v) (0;0) \left[ P_1 P_2 +
W_1 W_2 \right] + 16\times 2 (Q_s + Q_c ) (0;0) \left( {\eta \over
\vartheta_4 (0)} \right)^2 \Biggr\} \, ,
\ee

Nel settore aperto, per gruppi di gauge unitari come nel caso del modello originario supersimmetrico $T^4/\mathbb{Z}_2$, indichiamo il numero di D9 brane neutre con $N_0=n+\bar n$, mentre  $m$ e $\bar m$ contano il numero di D9 brane magnetizzate con cariche $U(1)$ uguali a $+1$ o $-1$. Si hanno poi le D5 brane, con la loro molteplicit\`a $d+\bar d$. L'ampiezza di anello coinvolge quindi diversi tipi di stringhe aperte: le stringhe ``dipolari'', con molteplicit\`a di Chan-Paton $m\bar m$; quelle scariche con molteplicit\`a indipendenti da $m$ e $\bar m$; quelle con una singola carica, con moltpelicit\`a linerari in $m$ e $\bar m$; infine quelle doppiamente cariche con molteplicit\`a $m^2$, $\bar m^2$. Per le stringhe ``dipolari'' si \`e visto che l'introduzione di campi magnetici non introduce uno shift, ma modifica gli zero modi (\ref{dip}). I momenti devono quindi essere quantizzati in unit\`a di $1/R\sqrt{1+(2\pi\alpha' \ H_i)^2}$. Per queste stringhe quindi si deve sostituire la somma $P_1P_2$ con $\tilde{P}_1
\tilde{P}_2$, definita in termini dei momenti $m_i/R\sqrt{1+(2\pi\alpha' \ H_i)^2}$. Alla luce delle osservazioni fatte, l'ampiezza di anello si trova essere
\begin{eqnarray}
{\cal A} &=& \frac{1}{4} \Biggl\{ (Q_o + Q_v)(0;0) \left[
(n+\bar n)^2 P_1 P_2 + (d+\bar d)^2 W_1 W_2
+ 2 m \bar{m} \tilde P_1 \tilde P_2 \right]
\nonumber
\\
&-& 2 (m+\bar m) (n + \bar{n}) (Q_o + Q_v )(z_1 \tau ; z_2 \tau
) {k_1 \eta \over
\vartheta_1 (z_1 \tau)} {k_2 \eta \over \vartheta_1 (z_2 \tau)}
\nonumber
\\
&-& ( m^2 + \bar{m}^2 ) (Q_o + Q_v ) (2 z_1 \tau ; 2 z_2 \tau )
{2 k_1 \eta \over
\vartheta_1 (2 z_1 \tau)} {2 k_2 \eta \over \vartheta_1 (2 z_2 \tau)}
\nonumber
\\
&-& \left[ (n-\bar n)^2 -2 m\bar m + (d-\bar d)^2 \right] (Q_o - Q_v ) (0;0)
\left( {2\eta \over \vartheta_2 (0)}\right)^2
\nonumber
\\
&-& 2 (m-\bar m) (n - \bar{n}) (Q_o - Q_v ) (z_1 \tau ; z_2 \tau)
{2\eta \over \vartheta_2
(z_1 \tau)} {2\eta \over \vartheta_2 (z_2 \tau)}
\nonumber
\\
&-& (m^2 + \bar{m}^2) (Q_o - Q_v ) (2z_1 \tau ; 2z_2 \tau)
{2\eta \over \vartheta_2
(2z_1 \tau)} {2\eta \over \vartheta_2 (2z_2 \tau)}
\nonumber
\\
&+& 2 (n+\bar n ) (d+\bar d) (Q_s + Q_c) (0;0) \left({\eta \over
\vartheta_4 (0)}\right)^2
\nonumber
\\
&+& 2 (m + \bar{m})(d+\bar d)(Q_s + Q_c) (z_1 \tau ; z_2 \tau)
{\eta \over \vartheta_4
(z_1 \tau )} {\eta \over \vartheta_4 (z_2 \tau )}
\nonumber
\\
&-& 2 (n-\bar n) (d - \bar d) (Q_s - Q_c )
(0;0) \left( {\eta \over \vartheta_3 (0)}\right)^2 \label{annsusy}
\\
&-& 2  (m - \bar{m})(d-\bar d) (Q_s - Q_c) (z_1 \tau ; z_2 \tau)
 {\eta \over \vartheta_3
(z_1 \tau )} {\eta \over \vartheta_3 (z_2 \tau )} \Biggr\} \, ,
\nonumber
\end{eqnarray}
e l'ampiezza di M\"{o}bius risulta
\begin{eqnarray}
{\cal M} &=& -\frac{1}{4} \Biggl[
(\hat Q_o + \hat Q_v )(0;0) \left[ (n+\bar n) P_1 P_2 + (d+\bar d) W_1
W_2 \right]
\nonumber
\\
&-& ( m + \bar{m}) (\hat Q_o + \hat Q_v ) (2z_1 \tau ; 2z_2
\tau) {2 k_1
\hat\eta \over \hat \vartheta_1 (2z_1\tau)} {2 k_2
\hat\eta \over \hat \vartheta_1 (2z_2\tau)}
\nonumber
\\
&-& \left( n+ \bar n + d + \bar d \right) (\hat Q_o - \hat Q_v )(0;0) \left(
{2\hat\eta \over \hat \vartheta_2 (0)}\right)^2 \label{mobsusy}
\\
&-& (m + \bar{m}) (\hat Q_o - \hat Q_v ) (2 z_1 \tau ; 2 z_2 \tau )
{2\hat\eta \over \hat\vartheta_2 (2z_1\tau)}
{2\hat\eta \over \hat\vartheta_2 (2z_2\tau)} \Biggr] \, ,
\nonumber
\end{eqnarray}
dove si sono raggruppati i termini con cariche $U(1)$ opposte, e con argomenti $z_i$ opposti, utilizzando la simmetria delle funzioni theta di Jacobi e, si sono indicati il modulo di ${\cal A}$ e di ${\cal M}$ con $\tau$. Si pu\`o anche notare come le stringhe con uno o due estremi carichi sono associate rispettivamente a funzioni di argomenti $z_i$ e $2z_i$.

Lo spettro non \`e pi\`u supersimmetrico, e in generale pu\`o sviluppare modi tachionici (instabilit\`a di Nielsen-Olesen) \cite{nole}. Nel settore untwisted, nel limite di campi deboli la formula di massa riceve correzioni della forma
\be
\Delta M^2 = \frac{1}{2 \pi \alpha'} \; \sum_{i=1,2}
\Bigl[  (2 n_i + 1) |2 \pi \alpha ' (q_{\rm L} + q_{\rm R}) H_i| \nonumber\\
+ 4 \pi\alpha' (q_{\rm L} + q_{\rm R}) \Sigma_i H_i \Bigr] \, ,
\ee
dove il primo termine \`e il contributo dei livelli di Landau e il secondo \`e l'accoppiamento dei momenti magnetici di spin $\Sigma_i$ ai campi magnetici. Si vede chiaramente nella formula di $\Delta M^2$ che, per valori generici dei campi, l'accoppiamento dei vettori interni pu\`o abbassare l'energia di Landau del livello pi\`u basso generando tachioni. I modi fermionici di spin semintero possono al limite compensare il contributo dei vettori. Nel settore twistato non ci sono livelli di Landau ma, mentre la parte fermionica di $Q_s$, $S_4O_4$ non sviluppa tachioni dal momento che i caratteri interni sono scalari e gli scalari non hanno accoppiamento magnetico, la parte bosonica $O_4C_4$ ha accoppiamento magnetico e sviluppa tachioni.

Si vede per\`o che un'opprtuna scelta dei campi magnetici rende lo spettro privo di tachioni. Infatti ponendo $H_1=H_2$ si eliminano tutte le instabilit\`a tachioniche. Inoltre di pu\`o vedere che questa scelta porta ad avere ampiezze di anello e di M\"{o}bius identicamente nulle, un segnale che una supersimmetria residua \`e presente nello spettro completo di stringa.

Completiamo questa breve rassegna del modello studiandone le condizioni di tadpole. Nel settore R-R untwisted, per il termine $C_4S_2C_2$, si ha
\be
\left[ n+\bar n + m + \bar{m} - 32 +  (2 \pi \alpha 'q)^2  H_1 H_2 (m +
\bar m ) \right] \sqrt{v_1 v_2}
+ {1\over \sqrt{v_1 v_2}} \left[ d+\bar d
- 32\right]  = 0 \, .
\label{RRtad}
\ee

Si pu\`o vedere che le altre condizioni di tapole di R-R nel settore untwisted sono compatibili con questa o si cancellano dopo l'identificazion $n=\bar n$, $m=\bar m$, $d=\bar d$. La condizione (\ref{RRtad}) \`e legata al termine di Wess-Zumino nell'azione di basse energie. Imponendo la condizione di quantizzazione di Dirac su entrambi i due tori
\be
2 \pi \alpha' q H_i v_i= k_i \qquad (i=1, 2)
\ee
si ottiene
\begin{eqnarray}
& & m+\bar m + n + \bar n = 32 \, ,
\nonumber
\\
& & k_1 k_2 (m + \bar m )  + d + \bar d = 32 \, ,
\label{tadp}
\end{eqnarray}
da cui si vede che le D9 brane magnetizzate acquisiscono la carica di R-R di $|k_1k_2|$ D5 brane se $k_1k_2>0$ o altrettante $\overline{\rm D5}$ antibrane se  $k_1k_2<0$.

Il settore NS-NS untwisted contiene in generale si hanno condizioni di tadpole non cancellate. Per il tadpole del dilatone, da $V_4O_2O_2$ si ottiene
\ba
& & \left[ n+\bar n + (m + \bar m) \sqrt{\left( 1 +  (2 \pi
\alpha' q)^2 H_1^2 \right)
\left( 1 + (2 \pi \alpha 'q) ^2 H_2^2 \right) }  -32 \right] \sqrt{v_1 v_2}
\nonumber \\
& & + {1\over \sqrt{v_1 v_2}} \left[ d +
\bar d - 32 \right]  \ ,
\ea
che  pu\`o essere legato alle derivate del termine di Born-Infeld nell'azione di bassa energia rispetto al campo del dilatone. Scegliendo  $H_1=H_2$ e utilizzando la relazione di quantizzazione di Dirac si ritrova la forma della (\ref{tadp}), e in questo caso quindi il tadpole di NS-NS si annulla identicamente imponendo le condizioni di tadpole nel settore R-R.

Per il termine $O_4V_2O_2$ la condizione di tapole risulta essere
\ba
& & \left[ n+\bar n + (m + \bar m) \, {1 - (2\pi\alpha' q H_1)^2 \over
\sqrt{ 1 + (2\pi\alpha' q H_1)^2 }}\,
\sqrt{ 1 + (2 \pi\alpha' q H_2)^2 } -32 \right] \sqrt{v_1
v_2} \nonumber \\
& & - {1\over \sqrt{v_1 v_2}} \left[ d+\bar
d - 32 \right]  \, ,
\ea
che corrisponde anche alla condizione di tadpole per $O_4O_2V_2$ scambiando $H_1$ e $H_2$. Questi termini sono legati alle derivate del termine di Born-Infeld rispetto al volume dei due tori interni. Nelle condizioni di tadpole trovate non compaiono quadrati perfetti, a causa del comportamento del campo magnetico sotto inversione temporale. Le ampiezze del canale trasverso, della forma $\langle {\cal T}(B)|q^{L_{0}}|B \rangle$, coinvolgono un'operazione di inversione temporale ${\cal T}$ sotto la quale il campo magnetico \`e dispari, questo introduce segni nelle ampiezze che impediscono la formazione di forme sesquilineari. Si pu\`o recuperare la corretta struttura dell'ampiezza di anello aggiungendo all'ampiezza di M\"{o}bius anche i contributi $\langle {\cal T}(B)|q^{L_{0}}|C \rangle$ e $\langle {\cal T}(C)|q^{L_{0}}|B \rangle$.

Anche in questo caso la scelta $H_1=H_2$, insieme alle condizioni di quantizzazione di Dirac, porta alla cancellazione dei tadpole.

La condizione di tadpole R-R nel settore twisted per $S_4O_2O_2$ \`e
\be
15 \left[ {\textstyle{1\over 4}} (m-\bar m + n -\bar n )
\right]^2
+ \left[ {\textstyle{1\over 4}} (m-\bar m + n - \bar n ) - (d-\bar d)
\right]^2 \ ,
\ee
che riflette il fatto che le D5 brane sono tutte coincidenti con il medesimo punto fisso. Si annulla identificando le molteplicit\`a coniugate. Il corrispondente tadpole NS-NS
\be
{2 \pi \alpha ' q \, (H_1 - H_2 )  \over \sqrt{ (1 + (2 \pi \alpha' q H_1)^2 )
(1 + (2\pi \alpha' q H_2)^2 ) }} \,,
\ee
e come nei casi precedenti si annulla solo per la scelta $H_1=H_2$.


\chapter{Superstringhe oltre un loop}

\section{Oltre un loop}

Al livello ad albero e ad un loop le ampiezze per i diversi modelli di superstringa sono state calcolate da tempo, ma una formulazione operativa delle ampiezze a genere pi\`u alto anche, per i casi pi\`u semplici, \`e mancata a lungo. Recentemente E. D'Hoker e D.H. Phong hanno ottenuto una formulazione gauge invariante molto esplicita per le ampiezze di genere due per le superstringhe di Tipo II ed Eterotiche \cite{D'Hoker:2002gw, D'Hoker:2001nj, D'Hoker:2001zp, D'Hoker:2001it, D'Hoker:2001qp}.

La complicazione fondamentale a genere pi\`u alto \`e l'emergere nella procedura di gauge fixing di supermoduli dispari grassmaniani che invece sono del tutto assenti al livello ad albero e ad un loop per strutture di spin pari. Per strutture di spin dispari ad un loop compare invece un modulo dispari ma che non introduce eccessive complicazioni.

Nel corso degli ultimi dieci anni molti sforzi sono stati spesi nel tentativo di dare una formulzione consistente per le ampiezze a genere superiore al primo. In particolare Friedan, Martinec e Shenker \cite{fms} hanno proposto un primo approccio al problema basato sulla Teoria di Campo Conforme sul world-sheet, l'invarianza BRST e l'operatore di \emph{picture changing}, in cui gli effetti dei supermoduli dispari vengono riassunti in termini di inserzioni dell'operatore di picture changing su un world-sheet puramente bosonico definito da soli moduli bosonici \cite{fms}. Calcoli diretti hanno per\`o dimostrato che questo produce ampiezze a due loop dipendenti dalla scelta dell'orbita di gauge.

Per superare questo ostacolo si sono cercate a lungo formulazioni differenti a partire da una grande variet\`a di principi quali l'invarianza modulare, il gauge del cono di luce, la geometria globale dello spazio di Teichmuller, la gauge unitaria, il formalismo operatoriale, metodi gruppali, fattorizzazioni e geometria algebrica. In nessuno di questi modi si \`e per\`o riusciti a derivare ampiezze gauge indipendenti. La difficolt\`a nel definire ampiezze di superstringa a genere pi\`u alto del primo ha portato addirittura a  valutare l'ipotesi che queste ampiezze fossero intrinsecamente ambigue.

Una formulazione delle ampiezze di superstringa di Tipo II e Eterotica, come integrali sullo spazio dei supermoduli, \`e stata proposta infine da D'Hoker e Phong a partire dallo formalismo di superspazio del world-sheet. In particolare \`e stata trovata una procedura  consistente di gauge fixing e di separazione chirale sullo spazio dei supermoduli che permette la costruzione delle ampiezze. I lavori di D'Hoker e Phong consentono la costruzione operativa delle ampiezze in termini di integrali sui soli moduli bosonici.

Inoltre D'Hoker e Phong hanno mostrato che l'inconsistenza della fomulazione mediante inserzioni di operatori di picture changing evidenziata da Verlinde e Verlinde \cite{vv1} nasce da un'eliminazione dei supermoduli anticommutanti inconsistente con la supersimmetria locale sul world-sheet. Per superare questo problema, i due autori hanno proposto una nuova procedura di gauge-fixing basata sulla proiezione delle supergeometrie sulla loro matrice super-periodica, in luogo di quella effettuata sulle loro soggiacenti geometrie bosoniche. A differenza di quest'ultima, la proiezione sulla matrice super-periodica si \`e dimostrata essere invariante sotto supersimmetria locale del world-sheet.

Le prime sezioni di questo capitolo contengono una breve rassegna sulle propriet\`a degli spinori sulle superfici di Riemann e sulla formulazione delle ampiezze di superstringa. Un'introduzione pi\`u dettagliata alla supergeometria e alla formulazione dell'azione di superstringa in termini di supercampi \`e data nell'Appendice B, insieme ad una breve raccolta di ampiezze di superstringa al livello ad albero e ad un loop. La parte centrale del capitolo \`e dedicata alla discussione dei problemi legati alla definizione di ampiezze di superstringa di genere pi\`u alto del primo, in particolare alle difficolt\`a della costruzione mediante picture changing operator, e ai risultati di D'Hoker e Phong per le superstringhe di Tipo II e Eterotiche. Le ultime sezioni sono infine dedicate al lavoro originale di questa Tesi che consiste nella generalizzazione dei risultati ottenuti da D'Hoker e Phong ad altri casi con supersimmetria rotta (modelli di tipo 0 e modelli con ``brane supersymmetry breaking'' in dieci dimensioni).

\section{Spinori su superfici di Riemann}

In superfici di topologia non banale gli spinori devono essere definiti in maniera opportuna. A differenza di un vettore, uno spinore acquista in generale delle fasi a seguito del trasporto parallelo lungo curve chiuse e si hanno pertanto delle ambiguit\`a. Come si vedr\`a, per superfici chiuse orientate di genere $g$, e quindi per teorie di stringhe chiuse orientate, esistono $2^{2g}$ scelte consistenti di fasi per spinori reali. Ognuna di queste scelte \`e detta \emph{struttura di spin}.

Un aspetto importante nella definizione di una teoria perturbativa per le superstringhe \`e quindi l'assegnazione delle strutture di spin. Nel formalismo funzionale la proiezione GSO per le stringhe di tipo II si ottiene separando gli spinori di chiralit\`a destra e sinistra, assegnando a ciascun gruppo strutture di spin indipendenti $\nu$ e $\bar \nu$, e sommando su tutte le strutture di spin. Questa \`e anche la prescrizione pi\`u naturale per evitare la comparsa di anomalie, dal momento che nessuna delle possibili strutture di spin viene privilegiata. La scelta di assegnare a tutti gli spinori di uno stesso gruppo la stessa struttura di spin \`e una richiesta necessaria per preservare l'invarianza spazio-temporale di Lorentz.

In realt\`a il principio di separazione dei gradi di libert\`a destri e sinistri richiede di far fronte e diversi inconvenienti. Come si \`e visto, la formulazione funzionale richiede che le azioni formulate con segnatura minkowskiana siano analiticamente continuate a segnature euclidee. Nello spazio di Minkowski $\psi^\mu$ e $\chi_m$ sono spinori di Majorana-Weyl, ma nella segnatura Euclidea non esistono spinori di Majorana-Weyl: le due componenti chirali di uno spinore di Majorana sono l'una la complessa coniugata dell'altra, e devono quindi avere la medesima struttura di spin. Per aggirare questa difficolt\`a si parte da uno spinore reale (somma di due spinore di Weyl con chiralit\`a opposta) separando i contributi solo nella quantizzazione. In questo modo, ogni fattore pu\`o essere pensato come il contributo di un fermione di Majorana-Weyl. Una difficolt\`a maggiore viene dal campo $X^\mu$ e dal termine $\chi \bar \chi \psi_+ \psi_-$ dell'azione di superstringa (\ref{susylocale}), che si dovr\`a opportunamente separare. Nei prossimi paragrafi si vedr\`a che la separazione chirale dei contributi associati ai momenti interi $p_I^\mu$ pu\`o essere implementata in virt\`u di un teorema \cite{D'Hoker:1988ta} (vedi Appendice B).

\begin{figure}
\begin{center}
\includegraphics[width=10cm,height=4cm]{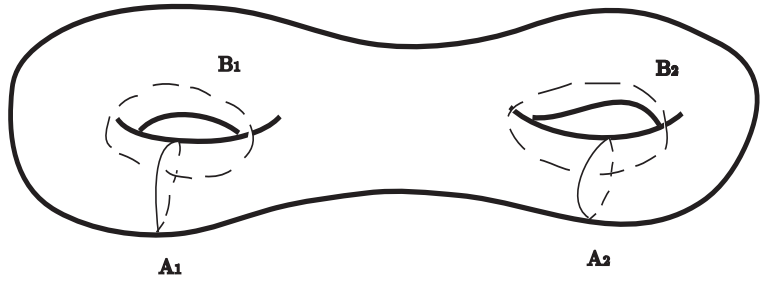}
\end{center}
\caption{Superficie di Riemann di genere $2$ con base canonica in omologia generata dalle curve chiuse $A_I$ e $B_I$}
\label{baseomologia}
\end{figure}

Per definire in maniera appropiata le strutture di spin \`e utile richiamare alcune nozioni della teoria delle superfici di Riemann. Il \emph{primo gruppo di omologia} di una superficie compatta $M$ senza bordi e con $h$ manici \`e 
\be
H^1(M)={\mathbb Z}^{2h} \ .
\ee
Una base canonica per questo gruppo \`e costituite da curve chiuse $A_I$, $B_I$, $I=1, 2, \dots, h$, con forma di intersezione definita da  
\be
\#(A_I, A_J)=0 \ , \qquad \#(A_I, B_J)=\delta_{IJ} \ , \qquad \#(B_I, B_J)=0 \ ,
\ee
dove $\#(A_I, B_J)=-\#(B_I, A_J)$ (si veda la figura \ref{baseomologia}). La scelta di una base non \`e unica: data una base canonica $(A_I, B_I)$, si pu\`o definire una nuova base canonica $(A'_I, B'_I)$ come
\be
B'_I=B_{IJ}A_J+A_{IJ}B_J \ , \qquad A'_I=D_{IJ}A_J+C_{IJ}B_J \ ,
\ee
dove la matrice $(2h\times 2h)$
\be
M=\left(
\begin{array}{cc}
A & B  \\
C & D  \\
\end{array}
\right) \ ,
\ee
appartiene al gruppo simplettico con coefficienti interi $Sp(2h, {\mathbb Z})$, che viene anche detto \emph{gruppo modulare di Siegel}, che preserva la forma d'intersezione. Usando una decomposizione omologica canonica in cicli $A$ e $B$ si pu\`o aprire la superficie, come si \`e visto nel caso del toro, ottenendo una sua rappresentazione come regione semplicemente connessa del piano con metrica opportuna (piatta o di Poincar\'e a seconda che il genere sia uno o maggiore) i cui lati siano identificati a coppie. 

Per definire gli spinori sulla superficie di Riemann \`e conveniente fissare una struttura di spin di riferimento $\nu$. Per fermioni reali il fattore di fase dovuto al trasporto lungo ciascun ciclo della base canonica $A_I$ e $B_I$, di ogni altra struttura di spin, differir\`a da quello di $\nu$ per $0$ o $\pi$, e si avranno complessivamente $2^{2g}$ differenti strutture di spin. 

Ogni struttura di spin definisce una distinta classe di spinori ed un corrispondente operatore di Dirac. Esiste una classsificazione naturale delle strutture di spin in pari e dispari, rispetto alla parit\`a del numero di zero modi dell'operatore di Dirac. In generale, il numero di zero modi dell'operatore di Dirac \`e o $0$ o $1$. Diffeomorfismi sul world-sheet $M$ possono collegare strutture di spin distinte $\nu$ e $\nu'$, ma preserveranno la parit\`a della struttura di spin, dal momento che la parit\`a degli zero modi di Dirac \`e invariante.

\section{Ampiezze di Superstringa}

Come si \`e visto nel primo capitolo, nella formulazione di Ramond--Neveu--Schwarz delle superstringhe i gradi di libert\`a sono la posizione bosonica $X^\mu$ e la sua controparte fermionica $\psi^\mu$, campi sul world-sheet $\Sigma$ che trasformano come vettori sotto le trasformazioni di Lorentz dello spazio piatto Minkowskiano. In questa fomulazione compaiono anche la metrica sulla superficie di universo $g_{mn}$ e il campo del gravitino $\chi_m$, che risultano non dinamici. L'azione costruita in (\ref{susylocale}) \`e invariante sotto diffeomorfismi, trasformazioni di supersimmetria $N=1$, trasformazioni di Weyl e di super Weyl del world-sheet.  

Alla luce del ruolo chiave giocato dalla supersimmetria locale, \`e conveniente riformulare l'azione in termini di un supercampo di materia ${\mathbf X}^\mu$ e di una supergeometria specificata dal riferimento locale $E_M{}^A$
e dal supercampo di connessione $U(1)$, $\Omega _M$. La relazione fra i supercampi e i campi ordinari \`e
\ba
{\mathbf X}^\mu & \equiv & X^\mu + \theta \psi ^\mu _+ + \bar \theta \psi _- ^\mu +
i \theta \bar \theta F^\mu \ ,
\nonumber \\
E_m {}^a & \equiv & e_m {}^a + \theta \gamma ^a \chi _m - {i \over 2}
\theta \bar \theta A e_m {}^a \ ,
\ea
dove $A$ e $F$ sono campi ausiliari. L'azione (\ref{susylocale}) pu\`o essere riscritta in termini di supercampi in forma estremamente compatta
\ba
S  ={1\over 4\pi} \int _\Sigma d^{2|2}{\bf z} \, E\, {\cal D}_+{\bf X}^{\mu}{\cal
D}_-{\bf X}^{\mu} \ , \qquad \quad E \equiv {\rm sdet} E_M{}^A \ ,
\ea
dove ${\cal D}_\pm$ sono derivate supercovarianti la cui forma esplicita pu\`o essere trovata nell'Appendice B ma non \`e necessaria alla presente esposizione.

Nella formulazione perturbativa di stringa alla Polyakov le ampiezze sono della forma
\be
{\bf A}_{\cal O} = \sum _{g=0}^\infty  \int {D(E\Omega) \ \delta (T)
\over {\rm Vol \ (Symm)}} \int DX^\mu {\cal O} \ e^{-S} \ ,
\ee 
dove l'operatore ${\cal O}$ rappresenta sinteticamente l'inserzione di un numero arbitrario di operatori di vertice (figura \ref{seriepert}). Per stringhe critiche in $D=10$, le simmetrie classiche sono tutte preservate anche a livello quantistico:
\be
\label{symmetries}
{\rm Symm} = {\rm sDiff}(\Sigma) \times {\rm sWeyl}(\Sigma) \times {\rm
sU}(1) (\Sigma) \ .
\ee
Nella formulazione di Polyakov delle ampiezze, la misura di integrazione viene opportunamente divisa per il volume di questo gruppo di simmetria. Infine $\delta (T)$ implementa il vincolo sulla torsione della supergeometria $N=1$.

\begin{figure}
\begin{center}
\includegraphics[width=13cm,height=5cm]{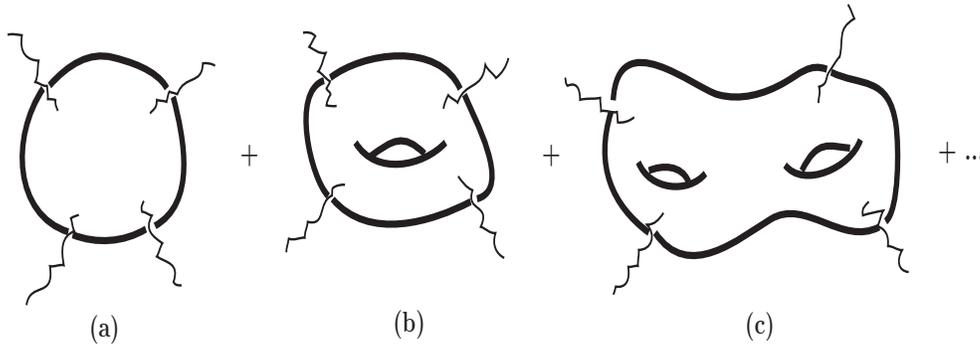}
\end{center}
\caption{Ampiezza di stringa chiusa con quattro vertici di interzione: (a) livello ad albero, (b) un loop e (c) due loop}
\label{seriepert}
\end{figure}

\section{Gauge fixing sul superspazio e separazione chirale}

Una procedura di gauge fixing consistente pu\`o essere derivata a partire da principi primi riducendo l'integrale su tutte le supergeometrie ad un integrale finito dimensionale sullo \emph{spazio dei supermoduli} che \`e definito come il quoziente delle supergeometrie per tutte le simmetrie locali (\ref{symmetries}). Le dimensioni dello spazio dei supermoduli sono
\ba
s{\mathcal M}_h  & \equiv &  \{E_M{}^A, \Omega _M + {\rm torsion \ constraints} \}
\big  / {\rm sDiff } \times {\rm sWeyl} \times {sU(1)} 
\nonumber \\ && \nonumber \\
{\rm dim} (s{\mathcal M}_h) & = & 
\cases{
(0|0) & $h=0$ \ ,\cr 
(1|0)_e \ {\rm or} \ (1|1)_o & $h=1$ \ ,\cr
(3h-3|2h-2) & $h\geq 2$ \ ,\cr} 
\ea
dove $e$ e $o$ indicano rispettivamente i casi di strutture di spin pari e dispari. 

Lo spazio dei supermoduli \`e uno spazio quoziente, e quindi non ammette una parametrizzazione canonica. Occorre scegliere un'orbita di gauge ${\cal S}$ della stessa dimensione di $s{\cal M}_h$, che intersechi tutte le orbite del gruppo di simmetria (\ref{symmetries}). Per superfici di genere $h\geq 2$, si parametrizza ${\cal S}$ con $m^A = (m^a|\zeta ^\alpha)$, dove $a=1, \cdots ,3h-3$ identifica i supermoduli pari e $\alpha=1,\cdots ,2h-2$ identifica i supermoduli dispari. La procedura di gauge fixing \`e stata proposta da E. Verlinde e H. Verlinde \cite{vv1} e derivata da principi primi da D'Hoker e Phong \cite{D'Hoker:1988ta}, e coinvolge i supercampi di ghost $B$ e $C$ la cui espansione in campi ordinari \`e
\ba
B & \equiv & \beta + \theta b + {\rm campi \ ausiliari} \ ,
\nonumber \\
C & \equiv & c + \theta \gamma + {\rm campi \ ausiliari} \ ,
\ea
insieme ai loro complessi coniugati. L'espressione per le ampiezze in seguito alla procedura di gauge fixing \`e
\be
\label{nonchiral}
{\bf A} _{\cal O} = \int_{s{\cal M}} \! |dm^A|^2 \int \! \! D({\bf X} B C) \ 
\big | \prod _A \delta (\langle H_A |B\rangle ) \big |^2 \ {\cal O} \ e^{-S}  \ ,
\ee
dove l'azione di campi di materia e di ghost \`e
\be
I  \equiv  {1\over 2 \pi} \int _\Sigma \! d^{2|2}{\bf z} E \ 
\biggl ( {1\over 2} {\cal D}_+ X^\mu {\cal D}_- X_\mu +
 B {\cal D}_- C + \bar B {\cal D}_+ \bar C \biggr ) \ .
\ee
I differenziali di super-Beltrami sono i vettori tangenti all'orbita di gauge ${\cal S}$ e sono definiti come
\be
\label{superbeltrami}
( H_A ) _- {}^z \equiv (-)^{A(M+1)} E_- {}^M {\partial E_M{}^z \over \partial m^A}
= \bar \theta (\mu_A - \theta \chi _A) \bigg | _{\rm WZ} \ .
\ee

La formula delle ampiezze (\ref{nonchiral}) \`e stata ricavata a partire da una formulazione euclidea sul world-sheet dell'azione, come naturale per la formulazione alla Polyakov della teoria perturbativa. In questa fomulazione i gradi di libert\`a destri e sinistri sono collegati da un operazione di coniugazione. Al contrario, nella teoria originaria con segnatura Minkowskiana sul word-sheet i fermioni destri e sinistri erano indipendenti. Questa indipendenza \`e cruciale nella definizione delle teorie di stringa chiusa, dal momento che le strutture di spin destre e sinistre sono indipendenti e la proiezione GSO deve avvenire in maniera indipendente sui gradi di libert\`a con chiralit\`a destra e sinistra.

Per recuperare l'indipendenza dei gradi di libert\`a occorre applicare una procedura di separazione chirale. Sebbene a prima vista l'azione (\ref{susylocale}) sembrebbe non consentire una procedura di separazione, dal momento che il temine quartico fermionico accoppia chiralit\`a differenti e gli zero modi del campo scalare $X^\mu$ non possano essere separati, questa risulta possibile all'interno di ogni blocco conforme caratterizzato da un momento interno di loop $p_I^{\mu}$, $ I=1,\cdots,h$. \`E conveniente a tal fine scegliere una base canonica per la prima omologia della superficie in termini dei cicli $A_I$ e $B_I$, per $I=1,\cdots ,h$ (vedi fig. \ref{baseomologia}). I momenti di loop possono essere cos\`i pensati come i momenti che attraversano i cicli $A_I$.

La precedura di separazione chirale pu\`o essere riassunta in termini di regole operative. Le funzioni di correlazioni per il supercampo scalare puossono essere scritte come
\be
\label{chispl}
\langle \prod_{i=1}^NV_i(k_i, \epsilon _i )\rangle_{X^{\mu}}
=
\int dp_I^{\mu}\ \bigg | \bigg \langle \prod_{i=1}^N
V_{i}^{chi}(k_i, \epsilon _i;p_I^{\mu})
\bigg \rangle_+ \bigg | ^2 \ ,
\ee
dove $\langle \cdots \rangle_+$ indica che si sono usate le regole effettive per le contrazioni degli operatori di vertice $V_{i}^{chi}(k_i, \epsilon _i ;p_I^{\mu})$ date nella tabella \ref{table:1}.

\begin{table}[htb]
\begin{center}
\begin{tabular}{|c||c|c|} \hline 
& {\rm Originario} & {\rm Effettivo Chirale} 
                \\ \hline \hline
              {\rm Bosoni}  
            & $X^{\mu}(z)$ 
            & $X_+^{\mu}(z)$        
             
 \\ \hline
              Fermioni  
            & $\psi_+ ^\mu (z)$ 
            & $\psi_+ ^\mu (z)$
\\ \hline
 Momenti interni di loop
& None
& ${\rm exp}(p_I^{\mu}\oint_{B_I}dz\partial_zx^{\mu} _+)$       
            
 \\ \hline
              Propagatore $X$  
            & $\langle X^{\mu}(z)X^{\nu}(w)\rangle$ 
            & $-\delta^{\mu\nu}{\rm ln}\,E(z,w)$        
 \\ \hline
              Propagatore $\psi_+$ 
            & $\langle\psi_+ ^\mu (z)\psi_+^{\nu}(w)\rangle$ 
            & $- \delta^{\mu\nu}S_{\delta}(z,w)$        
 \\ \hline
              Derivate Covarianti 
            & ${\cal D}_+$ 
            & $\partial_{\theta}+\theta\partial_z$        
 
 \\ \hline
\end{tabular}
\end{center}
\caption{Regole per la separazione chirale.}
\label{table:1}
\end{table}

Nella tabelle $E(z,w)$ \`e la forma prima e $S_\delta (z,w)$ \`e il kernel di Szeg\"o. Il punto delle regole effettive \`e che queste coinvolgono solo oggetti meromorfi, a differenza dell'usuale propagatore $\langle x^{\mu}(z)x^{\nu}(w)\rangle$ che \`e dato dalla funzione di Green scalare $\delta ^{\mu\nu}G(z,w)$. Per le ampiezze si ottiene 
\be
\label{smeaschi}
{\bf A}_{\cal O} [\delta] 
=
\int |\prod_A dm^A|^2
\int dp_I^{\mu}\
\bigg | e^{i\pi p_I^{\mu}\hat\Omega_{IJ}p_J^{\mu}}\,
{\cal A} _{\cal O} [\delta]\, \bigg |^2 \ , 
\ee
dove ${\cal A} _{\cal O}[\delta]$ \`e il correlatore chirale effettivo
\be
\label{smea}
{\cal A} _{\cal O} [\delta] 
=
\bigg \langle 
\prod_A\delta(\langle H_A|B\rangle) \ {\cal O}_+ \ 
\exp \biggl \{ \int _\Sigma \! {d^2\! z \over 2 \pi} \chi_{\bar z}{}^+
S(z) \biggr  \}\bigg \rangle _+ \ ,
\ee
e $S(z)$ \`e la supercorrente totale
\be
S(z)
=-{1\over 2}\psi_+^{\mu}\partial_zx_+^{\mu}
+{1\over 2}b\gamma-{3\over 2}\beta\partial_zc
-(\partial_z\beta)c\ ,
\ee

La matrice $\hat \Omega_{IJ}$ \`e detta \emph{matrice dei super periodi} e pu\`o essere definita ad ogni genere $h > 0$. A genere 2 la sua espressione \`e piuttosto semplice,
\be
\label{sper}
\hat\Omega_{IJ}
=
\Omega_{IJ}-{i\over 8\pi}\int _\Sigma \! d^2 \! z \int _\Sigma \! d^2 \! w
\ \omega_I(z) \chi_{\bar z}  S_{\delta}(z,w) \chi_{\bar w} \omega_J(w) \ ,
\ee
dove $\Omega_{IJ}$ \`e la matrice dei periodi corrispondente alla struttura complessa della metrica $g_{mn}$.  Le funzioni $\omega_I(z)$ sono una base di differenziali olomorfi abeliani duali agli $A_I$-cicli, tali che
\be
\oint _{A_I} \omega _J = \delta _{IJ} \ ,
\qquad \qquad 
\oint _{B_I} \omega _J = \Omega _{IJ}\ .
\ee
In analogia con i differenziali Abeliani ordinari si possono introdurre forme $1/2$ superolomorfe  $\hat \omega _I$, che possono essere normalizzate canonicamente sui cicli $A_I$, e che integrate sui cicli $B_I$ danno la matrice dei super periodi, 
\be
{\cal D} _- \hat \omega _I =0 \ ,\hskip 1in
\oint _{A_I} \hat \omega _J =\delta _{IJ} \ ,
\qquad 
\oint _{B_I} \hat \omega _J = \hat \Omega _{IJ} \ .
\ee

L'ampiezza per la superstringa di Tipo II si ottiene tenendo conto dei contributi dei modi destri e sinistri, utilizzando la stessa matrice dei periodi e i momenti interni, ma strutture di spin indipendenti. L'ampiezza per la Tipo II risulta essere
\be
\label{ampiezzeg2}
{\bf A} _{II {\cal O}}  =  \int dp^\mu _I \sum _{\delta , \bar \delta}
\eta_{\delta, \bar \delta}
\int _{s{\cal M} _h} |dm^A|^2 |\exp \{ i \pi p^\mu _I \hat \Omega _{IJ} p^\mu
_J\}| {\cal A}_{\cal O} [\delta ](\hat \Omega) \bar {\cal A}_{\cal O} [\bar \delta
](\hat \Omega ^*) \ ,
\ee
dove ${\cal M}_h$ indica lo spazio dei moduli bosonici delle superfici di Riemann di genere $h$. Le fasi $\eta _{\delta , \bar \delta}$ devono essere scelte in modo da essere consistenti con l'invarianza modulare, e sono necessarie per introdurre la proiezione GSO indipendentemente sui modi sinistri e destri. Le ampiezze per la superstringa eterotica si possono scrivere in maniera analoga. 

\section{Calcolo della misura chirale}

\subsection{Costruzione della misura chirale}

La misura chirale nella formulazione di D'Hoker e Phong fa uso dei supermoduli supersimmetrici $m^A = (\hat \Omega _{IJ} , \zeta ^\alpha)$. Tutte le quantit\`a calcolate originariamente per la metrica $g_{mn}$ con struttura complessa $\Omega _{IJ}$ devono essere riespresse in termini della matrice dei super periodi $\hat \Omega _{IJ}$. Nelle funzioni di correlazione, questo cambiamento si ottiene inserendo il tensore energia-impulso,
\be
\Omega _{IJ} \to \hat \Omega _{IJ} 
\quad 
\left \{ \matrix{
g & \to & \hat g = g + \hat \mu \cr
\partial _{\bar z} & \to & \hat \partial _{\bar z} = \partial _{\bar z} + \hat \mu \partial
_z\cr
\langle \cdots \rangle (g) & = & \langle \cdots \rangle (\hat g) + \int \hat \mu
\langle T \cdots \rangle (\hat g) \cr} \ ,
\right .
\ee
dove il differenziale di Beltrami \`e associato alle deformazioni della struttura complessa $\Omega _{IJ}$ in $\hat \Omega _{IJ}$. Alla luce della relazione che lega le due matrici periodiche, $\hat \mu$ \`e dato da
\be
\int _\Sigma \hat \mu \omega _I \omega _J = 
{1\over 8\pi}\int _\Sigma \! d^2 \! z \int _\Sigma
\! d^2 \! w \ \omega_I(z) \chi_{\bar z}  S_{\delta}(z,w) \chi_{\bar w} \omega_J(w) \ .
\ee

La matrice dei super periodi \`e invariante sotto variazioni dei supermoduli dispari $\zeta $, e questo implica che nessuna delle componenti del super differziale di Beltrami \`e nulla,
\be
\label{complication}
\delta _\zeta \hat \Omega _{IJ} =0 \quad  
\Rightarrow \quad
\left \{ \matrix{H_A = \bar \theta (\mu _A - \theta \chi _A) \cr
\mu _A \not=0 \quad \& \quad \chi _A \not=0\cr} \right . \ .
\ee

Gli oggetti duali dei  super differenziali di Beltrami sono 3/2 forme superolomorfe di tipo dispari $\Phi _{IJ}$ e pari $\Phi _\alpha$. La loro formula esplicita per $\Phi _{IJ}$ \`e 
\be
\label{Phi}
\Phi _{IJ} = -{i\over 2} \bigg ( \hat \omega _I {\cal D} _+ \hat \omega _J + 
\hat \omega _J {\cal D} _+ \hat \omega _I \bigg ) \ ,
\ee
che \`e normalizzata in modo da soddisfare le relazioni
\be
\langle H_a | \Phi _{IJ}\rangle =
\delta _{a,IJ} \qquad {\rm e} \qquad  \langle H_\alpha | \Phi _{IJ}\rangle = 0 \ .
\ee

\subsection{Cambio di base per i super differenziali di Beltrami}

Prima di procedere al calcolo diretto occorre introdurre un cambio di base per i super differenziali di Beltrami. Questo \`e necessario dal momento che l'utilizzo di moduli bosonici supersimmetrici forza tuttte le componenti di $H_A$ ad essere non nulle come indicato nella (\ref{complication}). Senza operare il cambio di base il prodotto dei fattori $\delta (\langle H_A |B\rangle)$ produce una forma per le funzioni di correlazione eccessivamente complicata e impossibile da gestire.

Per operatori di vertice che siano indipendenti dai super campi di ghost $B$ (come nel caso di vertici NS), $H_A$ risulta essere effettivamente accoppiato con $B$. Si pu\`o quindi effettuare un cambio di base da $H_A$ ad un nuovo super differenziale di Beltrami $H_A ^*$, scelto per semplicit\`a della forma
\ba
H^* _a & = & \bar \theta \delta (z,p_a) \hskip 1in a=1,2,3 \ ,
\nonumber \\
H^* _\alpha & = & \bar \theta \theta \delta (z,q_\alpha)
\hskip .9in \alpha =1,2 \ .
\ea

Considerando un insieme completo arbitrario di 3/2 forme superolomorfe pari e dispari $\Phi _C$, si ha
\be
\prod _A \delta (\langle H_A | B \rangle ) 
= {{\rm sdet} \langle H_A | \Phi _C \rangle \over {\rm sdet} \langle H_A ^* | \Phi _C \rangle }
\prod _ a b(p_a) \prod _\alpha \delta (\beta (q_\alpha))
\ee
Questa formula \`e chiaramente indipendente dalla scelta delle  $\Phi _C$. \`E di grande utilit\`a che tutte le funzioni di correlazione siano ora espresse in termini di inserzioni di campi ordinari e che tutte le complicazioni presenti in $H_A$ sono confinate al fattore moltiplicativo.

Esistono due scelte naturali della base per $\Phi_C$. La prima, che indichiamo con $\Phi _C$, \`e duale di $H_A$, mentre la seconda, che indichiamo con $\Phi^* _C$, \`e duale di $H^* _A$,
\be
\langle H_A | \Phi _C \rangle = \langle H_A ^* | \Phi ^* _C \rangle = \delta _{AC}
\ee
Mentre la forma esplicita di $\Phi ^* _C$ \`e nota, per  $\Phi _C$ la forma esplicita \`e nota solamente per le componenti dispari (\ref{Phi}). Per le componenti pari non si ha un espressione canonica. In generale esse possono essere espresse come una combinazione lineare
\be
\Phi _\gamma ({\bf z}) = \Phi _\epsilon ^* ({\bf z}) C^\epsilon {} _\gamma + 
\Phi _{IJ} ({\bf z}) D^{IJ} {}_\gamma \ ,
\ee
dove $C$ e $D$ sono matrici indipendenti da $\bf z$ ma dipendenti dai moduli. Accoppiando con $H_\alpha$ e usando il fatto che $\langle H_\alpha |\Phi_{IJ}\rangle=0$, si ha $\det C \times \det \langle H_\alpha |\Phi ^*
_\gamma \rangle=1$, e tenendo conto di tutti i fattori
\be
{{\rm sdet} \langle H_A | \Phi _C \rangle \over {\rm sdet} \langle H_A ^* | \Phi _C \rangle }
=
{1 \over {\det} \Phi _{IJ} (p_a) \times {\det} \langle H_\alpha |\Phi ^* _\gamma\rangle} \ .
\ee
Le componenti $\Phi _{IJ} (p_a)$ sono note in forma esplicita, e le componenti $\mu _\alpha$ e $\chi _\alpha$ di $H_\alpha = \bar \theta (\mu _\alpha - \theta \chi_\alpha)$ sono anch'esse note. L'oggetto $\chi_\alpha$ rappresenta la scelta dell'orbita dei gravitini sul worldsheet, ed \`e pertanto fissato nella procedura di gauge fixing (le ampiezze dovranno risultare indipendenti da questa scelta). L'oggeto $\mu _\alpha$ si pu\`o dimostrare essere dato da $\mu _\alpha = \partial \hat \mu / \partial \zeta ^\alpha$. Tutti i fattori della formula in cui si \`e fissato un gauge sono quindi noti in forma esplicita, e si pu\`o scrivere la completa misura chirale nella forma
\be
{\cal A} [\delta] 
=
{ \langle \prod _a b(p_a) \prod _\alpha \delta (\beta (q_\alpha)) \rangle
\over 
\det \Phi _{IJ+} (p_a) \det \langle H_\alpha | \Phi _\beta ^* \rangle }
\biggl \{ 1 + {1 \over 2 \pi} \int \hat \mu \langle T \rangle
- {1 \over 8 \pi ^2} \int _\Sigma \! \! \int _\Sigma \chi \chi \langle S S \rangle 
\biggr \}  
\ee
Il calcolo diretto dimostra che questa espressione \`e invariante sotto trasformazioni locali di supersimmetria sul world-sheet, come atteso \cite{D'Hoker:2001zp}.

\subsection{Il calcolo in componenti}

Per ottenere una formula operativa per il calcolo delle ampiezze si pu\`o considerare per il gravitino un'orbita con supporto su due punti arbitrari $x_1$ e $x_2$,
\be
\chi _\alpha (z) = \delta (z,x_\alpha) \ .
\ee
La misura chirale pu\`o essere cos\`i espressa interamente in termini di quantit\`a che sono meromorfe sul worldsheet, quali la forma prima $E(z,w)$, il kernel di Szeg\"o $S_\delta (z,w)$, la funzione di Green per i ghost $b-c$ $G_2(z,w)$ (che \`e definita in modo da annullarsi quando $z=p_1,p_2,p_3$ alla luce delle inserzioni di $b$ a $p_a$) la funzione di Green di superghost $G_{3/2}(z,w)$ (che si annulla quando $z=q_1,q_2$ in ragione dell'inserzione $\delta (\beta )$ a $q_\alpha$). Esistono inoltre alcuni differenziali olomorfi, $\psi _\alpha ^* (z)$ e $\bar \psi _\alpha (z)$,  3/2 forme olomorfe normalizzate in modo da avere $\psi_\alpha ^* (q_\beta ) =\bar \psi _\alpha (x_\beta )= \delta _{\alpha \beta}$, e la quantit\`a $\varpi _a (z,w)$ che induce una mappa uno ad uno fra le due forme olomorfe in una variabile e le forme olomorfe di due variabili ciascuna di peso 1. Quest'ultima \`e normalizzata in modo da avere $ \varpi _a (p_b, p_b) =\delta _{ab}$. 

La misura chirale si scrive nella forma
\be
{\cal A} [\delta] 
=
{ \langle\prod _a b(p_a) \prod _\alpha \delta (\beta (q_\alpha))  \rangle  \over  \det \omega _I \omega _J (p_a)  \cdot \det  \psi ^* _\beta  (x_\alpha)} \biggl \{ 1  + {\zeta ^1 \zeta ^2 \over 16 \pi ^2} \sum _{i=1}^6  {\cal X}_i \biggr \} \ .
\ee
Le quantit\`a ${\cal X}_i$ sono definite come 
\ba
\label{bigg}
{\cal X}_1  & = & 
 -10 S_\delta (x_1,x_2) \partial _{x_1} \partial _{x_2} \ln E(x_1,x_2) \ ,
\nonumber \\
&&
-3 \partial _{x_2} G_2 (x_1, x_2) G_{3/2}(x_2,x_1) 
- 2 G_2 (x_1,x_2) \partial _{x_2} G_{3/2}(x_2,x_1)  -(1 \leftrightarrow 2) \ ,
\nonumber \\ && \nonumber \\
{\cal X} _2  &=& 
S_\delta (x_1,x_2) \omega _I(x_1) \omega _J(x_2)  
\partial _I \partial _J \ln \biggl ({\vartheta [\delta ](0)^5 \vartheta (p_1+p_2+p_3-3\Delta )
\over
\vartheta [\delta ](q_1+q_2-2\Delta )} \biggr ) \ ,
\nonumber \\ && \nonumber \\
{\cal X} _3  &=& 
2 S_\delta (x_1,x_2) \sum _a  \varpi _a  (x_1, x_2) \bigl [B_2(p_a)
+ B_{3/2}(p_a) \bigr ] \ ,
\nonumber \\ && \nonumber \\
{\cal X}_4  &=&  2
S_\delta (x_1,x_2) \sum _a 
\partial _{p_a} \partial _{x_1} \ln E(p_a,x_1) \varpi  _a(p_a, x_2)
- (1 \leftrightarrow 2) \ ,
 \\ && \nonumber \\
{\cal X} _5  &=& 
\sum _a  
S_\delta (p_a, x_1) \partial _{p_a} S_\delta (p_a,x_2)  
\varpi _a  (x_1,x_2) -(1 \leftrightarrow 2) \ ,
\nonumber \\ && \nonumber \\
{\cal X}_6  &=& 
3  \partial _{x_2} G_2(x_1,x_2) G_{3/2} (x_2,x_1)
+ 2 f_{3/2} (x_1) G_2(x_1,x_2) \partial \bar \psi _1 (x_2)
    -(1 \leftrightarrow 2) \ 
\nonumber \\
&&  
+ 2 G_{3/2} (x_2,x_1)  G_2(x_1,x_2) \partial \bar \psi _2(x_2) 
+ \partial _{x_2} G_2(x_2,x_1) \partial \bar \psi _2 (x_1)
    -(1 \leftrightarrow 2) \ ,
\nonumber
\ea
dove si \`e utilizzata la seguente notazione,
\ba
f_n(w) & = & \omega_I(w)\partial_I{\rm ln}\,\vartheta [\delta](D_n)
+\partial_w{\rm ln} (\prod_i \sigma (w) E(w,z_i) ) \ ,
\nonumber \\
B_2(w) & = & -27 T_1(w) +
 {1 \over 2} f_2 (w)^2 - {3 \over 2} \partial _w f_2 (w)
-2 \sum _a \partial_{p_a} \partial _w \ln E(p_a,w) \varpi _a (p_a, w) \ ,
\nonumber \\
B_{3/2} (w) & = & 12 T_1(w) - {1 \over 2} f_{3/2}(w)^2 + \partial _w f_{3/2}(w) 
+ {3 \over 2} \partial _{x_1} G_2 (w,x_1) + {3 \over 2} \partial _{x_2} G_2 (w,x_2) 
\nonumber \\
&& 
- {3 \over 2} \partial _w G_{3/2}(x_1,w) \bar \psi _1 (w)
- {3 \over 2} \partial _w G_{3/2}(x_2,w) \bar \psi _2 (w)
- {1 \over 2}  G_{3/2}(x_1,w) \partial \bar \psi _1 (w)
\nonumber \\
&&
- {1 \over 2}  G_{3/2}(x_2,w) \partial \bar \psi _2 (w)
+ G_2 (w,x_1) \partial \bar \psi _1 (x_1) + G_2 (w,x_2) \partial \bar \psi _2 (x_2) \ .
\ea

L'espressione ottenuta per la misura chirale \`e una somma di termini che sono tutti manifestamente funzioni scalari, meromorfe, di $x_\alpha$, $q_\alpha$ e $p_a$. Si pu\`o dimostrare tramite calcolo diretto che il risultato finale \`e indipendente da tutti questi punti, e questo dimostra la consistenza dell'espressione trovata \cite{D'Hoker:2001it}.

\section{Problemi con la definizione precedente dell'ampiezza}

Possiamo a questo punto accennare ai problemi della formulazione via picture changing e quantizzazione BRST. Un'assunzione centrale in questo approccio \`e la scelta di fissare una metrica indipendente dai supermoduli dispari, mentre la forma del gravitino (preso a supporto puntiforme nei punti $z_\alpha$) caratterizza l'orbita di gauge per i supermoduli dispari
\be
\label{ass}
g_{mn} (m^a) \ ,
\qquad \qquad
\chi = \sum _{\alpha =1,2} \zeta ^\alpha \chi _\alpha (m^a) \ .
\ee
Questa scelta porta ad avere, nell'integrale dell'ampiezza, un termine di integrazione sulle variabili grassmaniane del tipo
\be 
\bigg \langle {\cal O} \prod _{a=1}^{3h-3} (\mu _a |b) \prod _{\alpha =1} ^{2h-2}
Y (z_\alpha) \bigg \rangle \prod _{a=1} ^{3h-3} d m^a \ ,
\ee
dove $Y(z_\alpha)$ \`e l'operatore di picture changing. Come si \`e detto un calcolo esplicito dimostra che in quest'ampiezza c'\`e una dipendenza residua dai punti di inserzione $z_\alpha$.

Le ragioni del fallimento di questo approccio risiedono nella procedura di gauge fixing sullo spazio dei super moduli. I super moduli anticommutanti possono essere pensati come fibre sui supermoduli pari. L'operazione di integrazione dei supermoduli, che permette di avere delle ampiezze di superstringa espresse come integrali sui soli moduli pari, equivale ad una proiezione lungo le fibre dello spazio dei supermoduli sulle loro basi pari.

L'originale scelta problematica del gauge equivale alla proiezione
\ba
\label{badproj}
(g_{mn}, \chi _m ) \hskip .25in & \sim & \hskip .3in (g_{mn} ', \chi _m ')
\hskip .5in {\rm sotto \ \ SUSY}
\nonumber \\
\downarrow \hskip .5in  & & \hskip .55in \downarrow
\nonumber \\
g_{mn} \hskip .4in & \sim \! \! \! \! \! / & \hskip .45in g_{mn} '
\hskip .75in {\rm sotto \ \ Diff} \times {\rm Weyl} 
\ea
La scelta di fissare i moduli dispari solo in funzione del gravitino produce un'inconsistenza nelle ampiezze, come si pu\`o capire osservando che, una trasformazione di supersimmetria sulla metrica, 
\be
\delta g_{mn}  = 2 \xi ^+ \chi _{\{m} {}^+ e_{n\}} {}^{\bar z}
\ee
modifica anche i moduli $m^a$ definiti in precedenza. La scelta operata per i supermoduli anticommutanti non \`e quindi invariante sotto supersimmetria. Questo spiega le ambiguit\`a trovate nella definizione delle ampiezze.

Una corretta proiezione deve quindi essere tale da ottenere dei supermoduli $m^a$ definiti in maniera invariante sotto l'azione della supersimmetria locale,
\ba
\label{consproj}
(g_{mn}, \chi _m ) \hskip .3in & \sim & \hskip .3in (g_{mn} ', \chi _m ')
\hskip .5in {\rm sotto \ \ SUSY}
\nonumber \\
\downarrow \hskip .6in  & & \hskip .6in \downarrow
\nonumber \\
\hat g_{mn} (m^a) \hskip .25in & \sim & \hskip .35in 
\hat g_{mn} ' (m^a)
\hskip .5in {\rm sotto \ \ Diff} \times {\rm Weyl} \ .
\ea 
Come si \`e visto nell'approccio alla D'Hoker e Phong questo avviene grazie all'utilizzo di una matrice dei super periodi $\hat \Omega_{IJ}$ invariante sotto supersimmetria locale.

\section{Formule esplicite in termini di funzioni $\vartheta$}

L'indipendenza della misura da tutti i punti consente di fissare $x_\alpha = q_\alpha$. Dal momento che i termini ${\cal X}_2$, ${\cal X}_3$, ${\cal X}_4$ sono proporzionali a $S_\delta (x_1,x_2)$ essi si annullano scegliendo la \emph{split gauge} $S_\delta (q_1,q_2)=0$. Questa scelta risulta particolarmente naturale dal momento che implica $\hat \Omega _{IJ} = \Omega _{IJ}$.  

\`E infine vantaggioso scegliere come punti $p_a$ i tre zeri di una 3/2 forma olomorfa $\psi _A (z)$. Questa scelta porta ad una forma particolarmente utile per la funzione di Green $G_2$ dei campi $b-c$ in termini di $\psi _A$ e del kernel di Szeg\"o, 
\be
G_2 (z,w) = S_\delta (z,w) \psi_A (z)/\psi_A (w)  \ ,
\ee
tale che in \emph{split gauge} $G_2 (q_1, q_2)=0$. Combinando tutti i contributi, si ottiene ${\cal X}_1+{\cal X}_6={\cal X}_2= {\cal X}_3 = {\cal X}_4=0$ mentre il solo ${\cal X}_5$ \`e diverso da zero. Il calcolo esplicito di ${\cal X}_5$ \`e piuttosto complesso \`e pu\`o essere trovato in \cite{D'Hoker:2001qp}. 

La formula finale della misura chirale di superstringa 
\be
d \mu [\delta] (\Omega ) =   {\Xi _6 [\delta ](\Omega ) \ \vartheta 
[\delta ]^4  (0,\Omega) \over
16 \pi ^6 \ \Psi _{10} (\Omega) }  \  d^3 \Omega _{IJ} \ ,
\ee
\`e invece estremamente semplice ed \`e espressa in termini una forma modulare, di funzioni $\vartheta$ a genere due, e di una funzione $\Xi _6 [\delta](\Omega)$ su cui torneremo a breve.

A genere due esistono 16 strutture di spin, che possono essere descritte mediante caratteristiche semi-intere
\be
\kappa = (\kappa ' | \kappa '')
\qquad \qquad 
\kappa ', \kappa '' \in (0,{1 \over 2})^2 \ .
\ee
in questa notazione le due componenti di $\kappa '$ si riferiscono alle strutture di spin sui cicli $A_I$, mentre le componenti di $\kappa ''$ si riferiscono a quelli sui cicli $B_I$. Le strutture di spin pari e dispari possono essere distinte in base al valore pari o dispari del prodotto $4\kappa ' \cdot \kappa ''$. Si trova cos\`i che le 16 strutture di spin sono divise in 10 strutture pari, indicate genericamente con $\delta$, e 6 dispari, indicate con $\nu$. Ogni struttura pari, a genere 2, pu\`o essere scritta in due diversi modi come somma di tre distinte strutture di spin dispari,
\be
\delta   = \nu _{i_1}+\nu _{i_2}+\nu _{i_3}  = \nu _{i_4}+\nu _{i_5}+\nu _{i_6} \ .
\ee

Date due strutture di spin $\kappa$ e $\rho$ la loro segnatura \`e definita come
\be
\langle \kappa | \rho \rangle \equiv \exp \{ 4 \pi i (\kappa ' \cdot \rho '' - \rho ' \cdot \kappa '')\} \ ,
\ee
che assume valori $\pm 1$.

Le funzioni $\vartheta$ di genere due con caratteristica $\kappa$ sono definite come
\be
\vartheta [\kappa](v, \Omega) \equiv \sum _{n - \kappa ' \in {\bf Z}^2}
\exp \bigl \{ i \pi n^t \Omega n + 2 \pi i n^t (v+ \kappa '')\bigr \} \ ,
\ee
e sono funzioni di $v \in {\bf C}^2$ pari o dispari a seconda che $\kappa$ sia una struttura di spin pari o dispari. Di grande importanza sono le $\vartheta$-costanti, definite come
\be
\vartheta [\delta] \equiv \vartheta [\delta](0,\Omega) \ ,
\ee
dove $\delta$ \`e una struttura di spin pari. Per strutture di spin dispari, in analogia con il caso di genere 1 si ha,
\be
\vartheta [\nu](0,\Omega) \equiv 0
\ee
L'oggetto indicato con $\Psi _{10}$ \`e una forma modulare di genere due definita come
\be
\Psi _{10} (\Omega ) \equiv \prod _{\delta \ {\rm even}} 
\vartheta [\delta ]^2 (0, \Omega) \ ,
\ee
mentre $\Xi_6 [\delta ](\Omega)$ \`e definita come
\be
\label{xiodd}
\Xi _6 [\delta](\Omega) \equiv \sum _{1 \leq i< j \leq 3} \!
\langle\nu_i | \nu_j\rangle \! \prod _{k=4,5,6} \! \vartheta [\nu_i + \nu _j 
+\nu_k]^4 (0,\Omega) \ .
\ee
\`E importante osservare che $\Xi _6 [\delta](\Omega)$ non \`e una forma modulare. Nella definizione di $\Xi_6 [\delta ](\Omega)$, la struttura di spin $\delta$ \`e scritta come somma di tre distinte strutture di spin dispari $\delta= \nu_1+\nu_2+\nu_3$ mentre $\nu_4$, $\nu_5$ e $\nu_6$ indicano le rimanenti strutture di spin dispari.

\section{Propriet\`a modulari}

Le trasformazioni modulari a cui si \`e precedentemente accennato sono definite come le trasformazioni che lasciano la matrice canonica di intersezione invariante 
\be
\left ( \matrix{A & B \cr C & D \cr} \right ) 
\left ( \matrix{0 & I \cr -I & 0 \cr} \right )
\left ( \matrix{A & B \cr C & D \cr} \right ) ^t
=
\left ( \matrix{0 & I \cr -I & 0 \cr} \right )
\qquad \qquad
\left ( \matrix{A & B \cr C & D \cr} \right ) \in Sp(4,{\bf Z})
\ee
e formano a genere due il gruppo $Sp(4,{\bf Z})$. I generatori di questo gruppo sono,
\ba
\label{generatori}
M_i &=& \left ( \matrix{I & B_i \cr 0 & I \cr} \right ) \qquad 
B_1 = \left ( \matrix{1 & 0 \cr 0 & 0 \cr} \right ) \qquad 
B_2 = \left ( \matrix{0 & 0 \cr 0 & 1 \cr} \right ) \qquad 
B_3 = \left ( \matrix{0 & 1 \cr 1 & 0 \cr} \right ) \qquad \nonumber \\
S &=& \left ( \matrix{0 & I \cr -I & 0 \cr} \right ) \qquad \nonumber \\
\Sigma &=& \left ( \matrix{\sigma & 0 \cr 0 & \sigma \cr} \right ) \qquad 
\sigma = \left ( \matrix{0 & 1 \cr -1 & 0 \cr} \right ) \qquad \nonumber \\
T &=& \left ( \matrix{\tau_+ & 0 \cr 0 & \tau_- \cr} \right ) \qquad 
\tau_+ = \left ( \matrix{1 & 1 \cr 0 & 1 \cr} \right ) \qquad
\tau_- = \left ( \matrix{1 & 0 \cr -1 & 1 \cr} \right ) \ .\qquad
\ea
L'azione sulle strutture di spin \`e data da \cite{igusa}
\be
\left (\matrix{ \tilde \kappa' \cr \tilde \kappa ''\cr}  \right )
=
\left ( \matrix{D & -C \cr -B & A \cr} \right )
\left ( \matrix{ \kappa ' \cr \kappa '' \cr} \right )
+ {1 \over 2} \ {\rm diag} 
\left ( \matrix{CD^T  \cr AB^T \cr} \right )\, ,
\ee
dove diag$(M)$ indica genericamente, per una matrice $M$ $n\times n$, un vettore colonna $1\times n$ i cui valori siano i valori diagonali di $M$. Sulla matrice periodica la trasformazione agisce come
\be
\tilde \Omega = (A\Omega + B ) (C\Omega + D)^{-1} \ ,
\ee
Mentre le funzioni $\vartheta$ trasformano come
\be
\vartheta [\tilde \kappa ] \biggl ( \{(C\Omega +D)^{-1} \}^t v , \tilde
\Omega \biggr ) =
\epsilon (\kappa, M) \det (C\Omega + D) ^{1 \over 2}
e^{ i \pi v ^t (C\Omega +D)^{-1} C v }
\vartheta [ \kappa ] (v, \Omega) \ .
\ee
Il fattore di fase $\epsilon (\kappa, M)$ dipende sia da $\kappa $ che dalla trasformazione modulare $M$, ed \`e tale che $\epsilon (\kappa , M )^8=1$.

Utilizzando le relazioni date, si trovano facilmente le leggi di trasformazione
\ba
d^3 \tilde\Omega _{IJ}
& = & \det (C\Omega+D)^{-3} \ d^3 \Omega_{IJ} \ ,
\nonumber \\
\vartheta [\tilde \delta ] ^4 (0,\tilde \Omega) & = & 
\epsilon ^4 \ \det (C\Omega + D)^2 \ \vartheta [\delta ] ^4 (0,\Omega) \ ,
\nonumber \\
\Xi _6 [\tilde \delta ] (\tilde \Omega) 
& = & 
\epsilon ^4 \ \det (C\Omega + D)^6 \ \Xi _6 [\delta ] (\Omega ) \ ,
\nonumber \\
\Psi _{10} (\tilde \Omega) 
& = & \det (C\Omega + D)^{10} \ \Psi _{10}(\Omega) \ ,
\ea
da cui si vede chiaramente che $\Xi_6[\delta ](\Omega)$ non trasforma come una forma modulare.

La legge di trasformazione modulare per la misura chirale \`e quindi
\be
\label{modmeasure}
d\mu [\tilde \delta] (\tilde \Omega) = 
\det\,(C\Omega+D)^{-5}d\mu [\delta ](\Omega) \ .
\ee
Il peso $-5$ \`e legato alla dimensione critica $D=10$, in maniera analoga a quanto avviene a genere uno. Questo risulta evidente dopo l'integrazione sui momenti interni, che porta alla comparsa di un fattore $\det {\rm Im} \Omega$, la cui legge di trasformazione modulare risulta essere 
\be
\det {\rm Im} \tilde \Omega = |\det (C\Omega +D)|^{-2} \det {\rm Im} \Omega \ .
\ee

Si trova cos\`i che la misura completa per il doppio toro, tenendo conto dei fattori destri e sinistri \`e, come atteso, covariante sotto trasformazioni modulari,
\be
(\det {\rm Im} \tilde \Omega)^{-5} \ d\mu [\tilde \delta] (\tilde \Omega)
\times \overline{d\mu [\tilde {\bar \delta}] (\tilde \Omega)}
=
(\det {\rm Im} \Omega)^{-5} \ d\mu [ \delta] ( \Omega) \times \overline{d\mu [
{\bar \delta}] ( \Omega)} \ .
\ee

\section{Proiezione GSO e funzione di partizione della Tipo II}

Per implementare la proiezione GSO occorre sommare le misure chirali sulle strutture di spin con opportune fasi $\eta_\delta$ come nella (\ref{ampiezzeg2}),
\be
d\mu (\Omega) = \sum_\delta \eta_\delta d\mu [\delta](\Omega) \ ,
\ee
in modo da ottenere una misura completa invariante sotto l'azione di $Sp(4,{\bf Z})$. Considerando le leggi di trasformazione delle misure chirali, di det Im($\Omega$) e della misura $d^3 \Omega _{IJ}$, \`e possibile vedere che per la superstringa di Tipo II tutte le fasi devono essere uguali. 

A questo punto possiamo scrivere il contributo al secondo ordine alla la funzione di partizione della Tipo II e quindi alla costante cosmologica, che risulta essere
\be
\label{ZIIg2}
Z _{{\rm II}}  = 
\int _{{\cal M}_2} {|d^3 \Omega|^2 \over  (\det {\rm Im} \Omega)^5 } \times
 {|\sum _\delta \Xi_6 [\delta ](\Omega ) \vartheta [\delta ]^4 (0,\Omega )|^2  \over 2^8 \pi
^{12} |\Psi _{10} (\Omega)|^2} \ .
\ee

Si pu\`o dimostrare con considerazioni generali sulle propriet\`a delle forme modulari che l'integrando del contributo al secondo ordine alla costante cosmologica si annulla identicamente. 

\section{Degenerazioni delle ampiezze di genere 2}

Nel calcolo delle ampiezze di superstringa un test importante \`e che queste obbediscano alle corrette fattorizzazioni negli stati fisici quando il world-sheet degenera. A genere due esistono due casi inequivalenti, a seconda che la degenerazione separi la superficie in due parti sconnesse o meno. Scegliendo un base di cicli canonici di omologia come in figura \ref{baseomologia}, e usando la seguente parametrizzazione della matrice dei periodi in questa base 
\be
\Omega = \left ( \matrix{\tau_1 & \tau \cr \tau & \tau _2} \right ) \ ,
\ee
si vede facilmente che la \emph{degenerazione separante} corrisponde al limite $\tau \to 0$, per $\tau_1$ e $\tau _2$ fissati, mentre la \emph{degenerazione non separante} avviene nel limite $\tau_2 \to i\infty$, per $\tau _1 $ e $\tau$ fissati (si veda la figura \ref{degenerazioni}).

\begin{figure}
\begin{center}
\includegraphics[width=10cm,height=10cm]{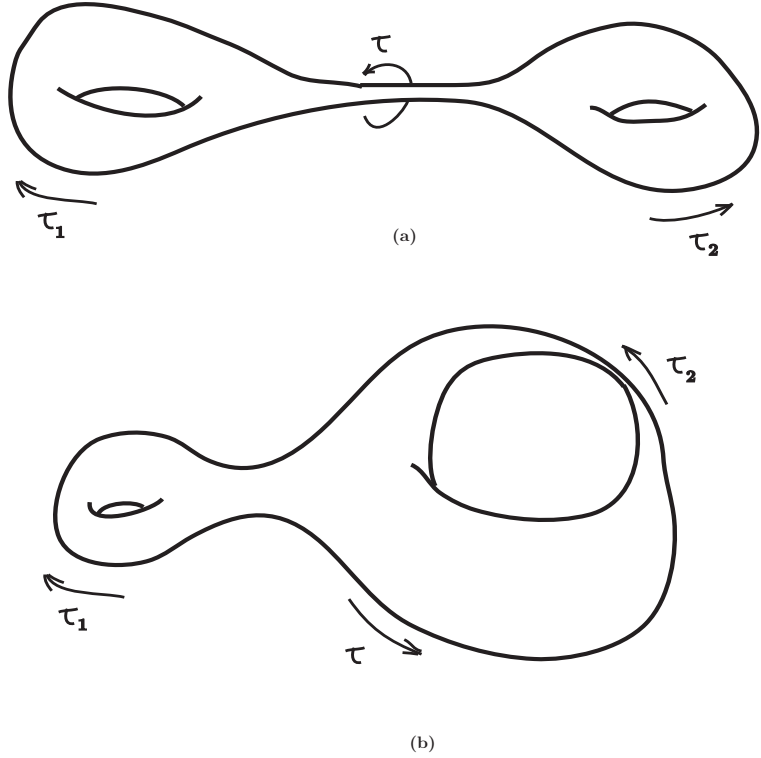}
\end{center}
\caption{Degenerazioni di una superficie di genere due: (a) separante, (b) non separante}
\label{degenerazioni}
\end{figure}

\subsection{Degenerazione separante}

L'andamento asintotico della misura d'integrazione si trova studiando l'effetto delle degenerazioni sulle funzioni $\vartheta$ a genere due. In questo limite \`e utile distinguere due casi. Il primo contiene 9 delle 10 strutture di spin $\delta$ per le quali le strutture di spin $\mu_1$ e $\mu_2$ sulle componeti connesse di genere 1 del doppio toro sono entrambe pari, si tratta del caso NS-NS. Il secondo \`e l'unico caso in cui entrambe le strutture di spin $\nu_0$ sono dispari, si tratta del caso R-R. 

Per studiare la degenerazione si usa l'espansione di Taylor delle funzioni $\vartheta$ di genere 2 intorno a $\tau=0$, data in termini di funzioni $\vartheta$ di genere 1, che indichiamo per semplicit\`a in questa sezione genericamente come $\vartheta_1$. Le espansioni nei due casi risultano essere
\ba
\label{thetaasym}
\vartheta \left [ \matrix{\mu _1 \cr \mu _2 \cr} \right ](0,\Omega )
&=&
\sum _{p=0} ^\infty {(2 \tau) ^{2p} \over (2p)!} 
\partial _{\tau _1}^p \vartheta _1[\mu _1] (0,\tau _1) 
\partial _{\tau _2}^p \vartheta _1[\mu _2] (0,\tau _2)  \ ,
\nonumber \\
\vartheta \left [ \matrix{\nu_0 \cr \nu_0 \cr} \right ](0,\Omega )
&=&
{1 \over 4 \pi i} \sum _{p=0} ^\infty {(2 \tau )^{2p+1} \over (2p+1)!} 
\partial _{\tau _1}^p \vartheta _1 '[\nu_0] (0,\tau _1) 
\partial _{\tau _2}^p \vartheta _1 '[\nu_0] (0,\tau _2) \ .
\ea
I termini dominanti di queste espansioni per le $\vartheta$-costanti sono quindi
\ba
\vartheta \left [ \matrix{\mu _1 \cr \mu _2 \cr} \right ](0,\Omega )
&=& \vartheta _1[\mu _1] (0,\tau _1)\vartheta _1[\mu _2] (0,\tau _2) + {\cal O}(\tau^2) \ ,
\nonumber \\
\vartheta \left [ \matrix{\nu_0 \cr \nu_0 \cr} \right ](0,\Omega )
&=& -2i\pi \tau \eta (\tau_1)^3\eta (\tau_2)^3 + {\cal O}(\tau^3) \ .
\ea
Si trova cos\`i che i limiti di $\Psi_{10}(\Omega)$ e $\Xi_6[\delta](\Omega)$ sono dati da
\ba
\label{xiasym}
\Psi _{10} (\Omega ) & = & 
- (2 \pi \tau )^2 \cdot 2^{12} \cdot \eta (\tau _1)^{24} \eta (\tau
_2)^{24}  +{\cal O}(\tau ^4) \ ,
\nonumber \\
\Xi _6 \left [ \matrix{\mu _1 \cr \mu _2 \cr} \right ] (\Omega)
& = &
- 2^8 \cdot \langle\mu_1 |\nu_0\rangle \langle\mu _2 |\nu_0\rangle \eta (\tau _1)^{12} \eta
(\tau _2)^{12}  +{\cal O}(\tau ^2) \ ,
\nonumber \\
\Xi _6 \left [ \matrix{\nu _0 \cr \nu _0  \cr} \right ] (\Omega)
& = & 
-3 \cdot 2^8 \cdot \eta (\tau _1)^{12} \eta (\tau _2)^{12}  +{\cal O}(\tau ^2) \ .
\ea
Considerando tutti i contributi si trova che il limite della misura \`e
\ba
{\rm NS-NS } \hskip .8in
d\mu \left [ \matrix{  \mu_1 \cr \mu_2  \cr} \right ]
& = &
{ d^3 \tau \over \tau ^2} 
\prod _{i=1,2}   {\langle \mu _i |\nu _0\rangle \vartheta  [\mu_i ]^4 (\tau_i)
\over  32 \pi ^4 \eta (\tau _i)^{12} } + {\cal O}(\tau^0) \ ,
\nonumber \\
{\rm R-R } \hskip 1in
d\mu \left [ \matrix{ \nu_0 \cr \nu_0  \cr} \right ]
& = &
{ 3 \tau ^2 d^3 \tau \over 2^6 \pi ^4} + {\cal O} (\tau ^4) \ .
\ea 
Il caso NS-NS riproduce correttamente i fattori ad un loop, incluse le fasi GSO introdotte dal limite di $\Xi_6[\delta]$. Il prefattore $\tau ^{-2}$ indica la presenza di uno stato intermedio tachionico. Combinando questa misura con la corrispondente misura destra e tenendo conto anche degli effetti dei momenti interni si trova che anche uno stato a massa nulla intermedio \`e presente. Una risommazione GSO parziale in una o nell'altra delle due parti sconnesse ad 1-loop cancella il tachione.

\subsection{Degenerazione non separante}

Il caso non separante si studia in maniera analoga a quanto fatto fin qui. Si trova che 
\ba
d\mu \left [ \matrix{\mu_i \cr 00  \cr} \right ]   
& = & + \
V_i (\tau, \tau_1) {d^3 \tau \over q}  + {\cal O}(q^0) \ ,
\nonumber \\
d\mu \left [ \matrix{\mu_i \cr 0{1\over 2}  \cr} \right ]   
& = & - \
V_i (\tau, \tau_1) {d^3 \tau \over q}  +
{\cal O}(q^0) \ ,
\nonumber \\
d\mu \left [ \matrix{\mu_i \cr {1\over 2} 0  \cr} \right ]  & = & {\cal O}(q^0) \ ,
\nonumber \\
d\mu \left [ \matrix{\nu_0 \cr \nu_0  \cr} \right ] & = & {\cal O}(q^0) \ .
\ea
Nelle prime tre linee, $\mu_i$  rappresenta una delle tre strutture di spin pari di genere 1 e $V_i(\tau, \tau_1)$ \`e la funzione a due punti per il tachione sulla superficie degenere di genere 1. La singolarit\`a $q^{-1}$ corrisponde al tachione che attraverso il ciclo di omologia $A_2$ quando la struttura di spin nel secondo manico \`e o $[00]$ oppure $[0 \ 1/2]$, che corrispondono a condizioni al bordo NS. Una risommazione parziale sulle strutture di spin nel solo settore NS elimina la singolarit\`a tachionica. D'altra parte, come atteso, nessun tachione appare per condizioni al bordo R.

\section{Superstringa di Tipo 0B a genere 2}

\`E possibile generalizzare i risultati ottenuti da D'Hoker e Phong per la superstringa di Tipo II, alla superstringa di Tipo 0 \`e possibile ricordando che $\Xi_6[\delta]$ implementa le corrette fasi GSO per la Tipo II, come si vede chiaramente dalle (\ref{xiasym}). 
Per ottenere la funzione di partizione della Tipo $0B$, che come si ricorder\`a \`e una teoria di superstringa con rottura di supersimmetria esplicita, in analogia con quanto fatto a genere uno si deve prendere il modulo dei diversi oggetti nella sommatoria nella funzione di partizione della Tipo II (\ref{ZIIg2}). La funzione di partizione della Tipo 0B si pu\`o scrivere nella forma
\be
Z _{{\rm 0B}}^{g=2}  = 
\int _{{\cal M}_2} {|d^3 \Omega|^2 \over  (\det {\rm Im} \Omega)^5 } \times
 {\sum _\delta |\Xi_6 [\delta ](\Omega )|^2  |\vartheta [\delta ]^4 (0,\Omega )|^2  \over 2^8 \pi
^{12} |\Psi _{10} (\Omega)|^2} \ .
\ee

Per verificare la correttezza della funzione di partizione proposta si pu\`o studiare la degenerazione separante, nel limite $\tau \to 0$. In questo limite ci si attende di ottenere due funzioni di partizione di genere uno della Tipo 0B, che come si ricorder\`a sono della forma,
\be
{Z}_{0B}^{g=1}\ = \ \int \frac{d^2\tau}{\tau_2^2} \
\frac{|O_8|^2 + |V_8|^2 + |S_8|^2 + |C_8|^2}{\tau_2^4 \ (\eta \ \bar\eta)^8} \ ,
\ee

Scriviamo esplicitamente le strutture di spin, utlizzando una notazione leggermente differente da quella utilizzata da D'Hoker e Phong,
\be
\kappa = \left( \matrix{\alpha_1 & \alpha_2 \cr \beta_1 & \beta_2 \cr} \right )  \ ,
\ee
dove $\alpha_i$ e $\beta_i$ sono riferite alle strutture di spin dell'$i$-esima componente connessa del doppio toro. Le strutture di spin dispari sono,
\ba
\nu_1 &=& \left ( \matrix{0 & 1/2 \cr 0 & 1/2 \cr} \right ) \ ,\qquad
\nu_2 = \left ( \matrix{1/2 & 0 \cr 1/2 & 0 \cr} \right ) \ ,\qquad
\nu_3 = \left ( \matrix{0 & 1/2 \cr 1/2 & 1/2 \cr} \right ) \ ,\qquad \nonumber \\
\nu_4 &=& \left ( \matrix{1/2 & 0 \cr 1/2 & 1/2 \cr} \right ) \ ,\qquad
\nu_5 = \left ( \matrix{1/2 & 1/2 \cr 0 & 1/2 \cr} \right ) \ ,\qquad
\nu_6 = \left ( \matrix{1/2 & 1/2 \cr 1/2 & 0 \cr} \right ) \ ,\qquad 
\ea
mentre quelle dispari sono
\ba
\delta_1 &=& \left ( \matrix{0 & 0 \cr 0 & 0 \cr} \right ) \ ,\qquad
\delta_2 = \left ( \matrix{0 & 0 \cr 0 & 1/2 \cr} \right ) \ ,\qquad
\delta_3 = \left ( \matrix{0 & 0 \cr 1/2 & 0 \cr} \right ) \ ,\qquad \nonumber \\
\delta_4 &=& \left ( \matrix{0 & 0 \cr 1/2 & 1/2 \cr} \right ) \ ,\qquad
\delta_5 = \left ( \matrix{0 & 1/2 \cr 0 & 0 \cr} \right ) \ ,\qquad
\delta_6 = \left ( \matrix{0 & 1/2 \cr 1/2 & 0 \cr} \right ) \ ,\qquad \nonumber \\
\delta_7 &=& \left ( \matrix{1/2 & 0 \cr 0 & 0 \cr} \right ) \ ,\qquad
\delta_8 = \left ( \matrix{1/2 & 0 \cr 0 & 1/2 \cr} \right ) \ ,\qquad
\delta_9 = \left ( \matrix{1/2 & 1/2 \cr 0 & 0 \cr} \right ) \ ,\qquad \nonumber \\
\delta_0 &=& \left ( \matrix{1/2 & 1/2 \cr 1/2 & 1/2 \cr} \right ) \ .\qquad
\ea

Si vede quindi che, nella notazione usuale, la degenerazione delle funzioni $\vartheta$ a genere 2 nel limite $\tau \to 0$ si scrive
\ba
\label{limti1}
\vartheta [\delta_1](0,\Omega ) &=& \vartheta _3(0,\tau _1)\vartheta _3(0,\tau _2) + {\cal O}(\tau^2) \ ,
\nonumber \\
\vartheta [\delta_2](0,\Omega ) &=& \vartheta _3(0,\tau _1)\vartheta _4(0,\tau _2) + {\cal O}(\tau^2) \ ,
\nonumber \\
\vartheta [\delta_3](0,\Omega ) &=& \vartheta _4(0,\tau _1)\vartheta _3(0,\tau _2) + {\cal O}(\tau^2) \ ,
\nonumber \\
\vartheta [\delta_4](0,\Omega ) &=& \vartheta _4(0,\tau _1)\vartheta _4(0,\tau _2) + {\cal O}(\tau^2) \ ,
\nonumber \\
\vartheta [\delta_5](0,\Omega ) &=& \vartheta _3(0,\tau _1)\vartheta _2(0,\tau _2) + {\cal O}(\tau^2) \ ,
\nonumber \\
\vartheta [\delta_6](0,\Omega ) &=& \vartheta _4(0,\tau _1)\vartheta _2(0,\tau _2) + {\cal O}(\tau^2) \ ,
\nonumber \\
\vartheta [\delta_7](0,\Omega ) &=& \vartheta _2(0,\tau _1)\vartheta _3(0,\tau _2) + {\cal O}(\tau^2) \ ,
\nonumber \\
\vartheta [\delta_8](0,\Omega ) &=& \vartheta _2(0,\tau _1)\vartheta _4(0,\tau _2) + {\cal O}(\tau^2) \ ,
\nonumber \\
\vartheta [\delta_9](0,\Omega ) &=& \vartheta _2(0,\tau _1)\vartheta _2(0,\tau _2) + {\cal O}(\tau^2) \ ,
\nonumber \\
\vartheta [\delta_0](0,\Omega ) &=& \vartheta _1(0,\tau _1)\vartheta _1(0,\tau _2) -2i\pi \tau \eta (\tau_1)^3\eta (\tau_2)^3 + {\cal O}(\tau^2) \ ,
\ea
dove  naturalmente nell'ultimo termine $\vartheta _1(0,\tau _1)=0$. Osservando che il modulo di $\Xi _6 [\delta_i] (\Omega)$ cancella le fasi,
\be
|\Xi _6 [\delta_i] (\Omega)|^2 = 
 |2^8 \cdot  \eta (\tau _1)^{12} \eta (\tau _2)^{12}|^2  +{\cal O}(\tau ^4) \ ,
\ee
dove nel caso di $\delta_0$ si \`e operato un riscalamento del fattore 3 di $\Xi _6$, e ricordando che si era trovato
\be
\Psi _{10} (\Omega )  =  
- (2 \pi \tau )^2 \cdot 2^{12} \cdot \eta (\tau _1)^{24} \eta (\tau
_2)^{24}  +{\cal O}(\tau ^4) \ ,
\ee 
si pu\`o calcolare facilmente la degenerazione della funzione di partizione. Nel limite nel limite $\tau \to 0$ si trova in questo modo,  il prodotto di due funzioni di partizione di Tipo 0B a genere 1. Questo dimostra la correttezza della funzione di partizione a genere 2 proposta.

\section{Discendenti Aperti della Tipo 0'B a genere $3/2$}

\subsection{Ampiezze di genere $3/2$}

Come si \`e visto, le funzioni di partizione di stringa aperta ricevono contributi da superfici con bordi e crosscap. Questo tipo di superfici ammettono sempre come ricoprimento doppio superfici chiuse e orientabili. L'immersione nel ricoprimento doppio \`e descritta per mezzo di involuzioni che generalizzano relazioni di riflessione. Queste involuzioni invertono la matrice di intersezione e per questo sono dette anti-conformi, e agiscono per mezzo di matrici che, con un'opportuna scelta di base, possono essere sempre scritte nella forma
\be
\label{periodsempl}
I = \left ( \matrix{I & 0 \cr \Delta & -I \cr} \right ) \ ,
\ee
con $\Delta$ una matrice simmetrica. Il caso di genere 1, discusso in dettaglio, \`e illuminante. L'anello, la bottiglia di Klein e il nastro di M\"obius,  come si \`e visto, ammettono il toro come ricoprimento doppio, via involuzioni anti-conformi con $\Delta=0$ nei primi due casi e $\Delta=1$ nell'ultimo. Questo risultato si estende a tutti i generi e permette un trattamento unificato delle varie superfici di interesse \cite{Bianchi:1989du}.

Partendo da una base canonica di omologia di cicli definita dai cicli $A_I$ e $B_I$, con matrice di intersezione 
\be
J= \left ( \matrix{0 & I \cr -I & 0 \cr} \right ) \ ,
\ee
la matrice di involuzione soddisfer\`a la relazione
\be
\label{cond}
I^TJI \,=\,-J \ ,
\ee
dal momento che inverte l'orientazione della superficie ricoprente. In generale la matrice di involuzione per una superficie di genere $g$ ha forma a blocchi,
\be
I = \left ( \matrix{A & B \cr C & D \cr} \right ) \ ,
\ee
con $A$, $B$, $C$ e $D$ matrici $g\times g$ di interi che soddisfano le condizioni
\be
AB^T=BA^T \ , \qquad CD^T=DC^T \ , \qquad e \qquad AD^T-BC^T=-1 \ , 
\ee
che risultano dalla (\ref{cond}). Non \`e difficile mostrare che una matrice dei periodi $\Omega$ compatibile con l'involuzione I deve soddisfare la condzione
\be
\label{vicoloomeg}
\bar \Omega = I(\Omega)=(C+D\Omega)(A+B\Omega)^{-1} \ .
\ee

Una scelta opportuna della base di omologia semplifica la forma della matrice dei periodi, e si pu\`o dimostrare che \`e sempre possibile porre la matrice di involuzione nella forma (\ref{periodsempl}), \cite{Bianchi:1988fr}. La matrice $\Delta$ \`e molto importante dal momento che determina la forma della parte reale della matrice dei periodi $\Omega$. 

Per ottenere questa forma della matrice di involuzione occorre scegliere dei cicli $a_i$ che siano lasciati invarianti dall'involuzione. A tal fine \`e utile ricordare che l'equivalenza fra tre crosscap e un manico e un crosscap riduce le superfici in tre classi con rispettivamente 0, 1 e 2 crosscap. Nel caso in cui non si abbiano crosscap, l'involuzione lascia fissi un certo numero di bordi. I cicli $a_i$ posssono essere scelti in parte come i cicli lungo questi bordi e in parte come somme di cicli di coppie di manici scambiati dall'involuzione. I cicli $b_i$ sono i cicli lungo i manici che contribuiscono alla $\Delta$, dal momento che trasformano in maniera non banale sotto l'involuzione.  Nel caso di superfici con crosscap, ricordando che un crosscap e un bordo con punti opposti identificati, si capisce che un ciclo che sia identificato con un crosscap induce una rivoluzione intorno a se stesso nella trasformazione del ciclo omologico coniugato. Nel caso della striscia di M\"obius questo argomento porta alla matrice dei periodi nota, non puramente immaginaria.

\`E possibile operare trasformazioni della base in omologia mantenendo la forma triangolare della matrice di involuzione. In generale una trasformazione di questo tipo sar\`a della forma
\be
M = \left ( \matrix{A & 0 \cr C & (A^T)^{-1} \cr} \right ) \ ,
\ee
con $\det A = \pm 1$ e
\be
2C=\Delta A -(A^T)^{-1}\Delta \ ,
\ee
una matrice di interi pari. Trasformazioni di quiesto tipo sono banali a genere 1, ma non a genere  pi\`u alto, dove sostanzialmente scambiano tori vicini.

In questa nuova base di omologia, detta \emph{base identit\`a}, la matrice dei periodi pu\`o essere scritta nella forma particolarmente semplice,
\be
\Omega = \frac{\Delta}{2}+i\Omega_2
\ee
con $\Omega_2$ una matrice a valori reali.

Esistono cinque superfici di genere $3/2$ con ricoprimento doppio di genere due, che possono essere identificate per mezzo di triple di numeri che indicano nell'ordine il numero di manici, di bordi e di crosscap.  Ad esempio, una superficie con tre bordi \`e indicata come $(030)$, mentre un toro con un bordo con (110). Nella base identi\`a le matrici $\Delta$ che definiscono le involuzioni sono
\ba
\Delta_{(110)} &=& \left ( \matrix{0 & 1 \cr 1 & 0 \cr} \right ) \ ,\qquad
\Delta_{(101)} = \left ( \matrix{0 & 1 \cr 1 & 0 \cr} \right ) \ ,\qquad
\Delta_{(030)} = \left ( \matrix{0 & 0 \cr 0 & 0 \cr} \right ) \ ,\qquad \nonumber \\
\Delta_{(012)} &=& \left ( \matrix{1 & 1 \cr 1 & 0 \cr} \right ) \ ,\qquad
\Delta'_{(012)} = \left ( \matrix{1 & 0 \cr 0 & 1 \cr} \right ) \ ,\qquad
\Delta''_{(012)} = \left ( \matrix{0 & 1 \cr 1 & 1 \cr} \right ) \ ,\qquad \nonumber \\
\Delta_{(021)} &=& \left ( \matrix{1 & 0 \cr 0 & 0 \cr} \right ) \ ,\qquad
\Delta'_{(021)} = \left ( \matrix{1 & 1 \cr 1 & 1 \cr} \right ) \ ,\qquad
\Delta''_{(021)} = \left ( \matrix{0 & 0 \cr 0 & 1 \cr} \right ) \ ,
\ea
dove per la bottiglia di Klein con un bordo $(012)$ si distinguono tre casi, collegati da trasfortmazioni modulari in relazione alla posizione del bordo rispetto ai crosscap. Allo stesso nel caso della striscia di M\"obius  con un bordo $(021)$ si distinguono tre casi in relazione alla posizione del crosscap rispetto ai bordi.

\subsection{Contributi alla funzione di partizione a genere $3/2$}

Per studiare la sistematica delle ampiezze di genere $3/2$ in un caso con supersimmetria spazio-temporale rotta, un esempio interessante \`e fornito dal modello non tachionico 0'B, costruito a genere 1 a partire dall'ampiezza di Klein, indicata come ${\cal K}_3$. In particolare si utilizzer\`a questo modello nel caso particolarmente semplice in cui si fissi $n_v=N$, $n_s=\bar N$, $n_o=0$, $n_c=0$, ottenendo nel canale diretto le ampiezze di genere 1,
\ba
{\cal K}_3 &=& \frac{1}{2}(-O_8+V_8+S_8+C_8) \ , \nonumber \\
{\cal A}_3 &=& N\bar N V_8 - \frac{1}{2}(N^2+\bar N^2)C_8 \ , \nonumber \\
{\cal M}_3 &=& -\frac{N+\bar N}{2}\hat C_8 \ ,
\ea
dove come di solito la notazione lascia intesi i gradi di libert\`a bosonici e la misura di integrazione. Nel canale trasverso si ottiene,
\ba
\tilde {\cal K}_3 &=& -\frac{2^6}{2}C_8 \ ,\nonumber \\
\tilde {\cal A}_3 &=& \frac{2^{-6}}{2}[(N^2+ \bar N^2)(V_8-C_8)-(N-\bar N)^2(O_8-S_8)] \ , \nonumber \\
\tilde {\cal M}_3 &=& -(N+ \bar N)\hat C_8 \ .
\ea

Per calcolare le correzioni dovute ai diagrammi di genere 3/2 di questo modello, occorre partire da una generica ampiezza di genere 3/2 che, pu\`o essere ottenuta con semplici considerazioni a partire dall'ampiezza di stringa chiusa di genere 2. In particolare si considerano solo met\`a dei gradi di libert\`a, in analogia con il caso di genere uno, e si introducono dei coefficienti generici $C_\delta^{g=3/2}$ che devono essere opportunamente fissati in relazione alla diverse ampiezze,
\be
\label{32open}
Z^{g=3/2}  = 
\int {d^3 \Omega}\, \frac{1}{\Psi _{10}(\Omega)}\cdot 
 {\sum _\delta C_\delta^{g=3/2}\Xi_6 [\delta ](\Omega )  \vartheta [\delta ]^4 (0,\Omega )} \ .
\ee
Questa \`e un ampiezza di genere 3/2 nel canale di stringhe chiuse (il canale trasverso), dal momento che si \`e scelto di non introdurre potenze della matrice dei periodi al denominatore, ancora in analogia con il canale trasverso di genere 1.

Il caso pi\`u semplice \`e quello dell'ampiezza a tre bordi. Per fissare i coefficienti consideriamo l'ampiezza (\ref{32open}) nel limte $\tau \to i\infty$. In questo limite ci si aspetta che la funzione di partizione fattorizzi in due ampiezze trasverse di anello. Utilizzando il comportamento asintotico per $\tau \to i\infty$ delle funzioni $\vartheta$ di genere 2 (\ref{limti1}), di $\Psi _{10}(\Omega)$ e di $\Xi_6 [\delta ](\Omega )$ si ricava il limite della (ref{32open}) che pu\`o essere confrontato con il contenuto in termini di funzioni $\vartheta$ di genere 1 del generico prodotto $\tilde{\cal A}(\tau_1)\cdot \tilde{\cal A}(\tau_2)$. Si trova cos\`i che
\ba
C_{\delta_1}^{(030)}&=& C_{\delta_5}^{(030)}= C_{\delta_7}^{(030)}=C_{\delta_9}^{(030)}= 4N\bar NM\bar M \ , \nonumber\\
C_{\delta_2}^{(030)}&=& C_{\delta_8}^{(030)}=2N\bar N(M^2+\bar M^2)\ ,\nonumber\\
C_{\delta_3}^{(030)}&=& C_{\delta_6}^{(030)}=2M\bar M(N^2+\bar N^2)\ ,\nonumber\\
C_{\delta_4}^{(030)}&=& C_{\delta_0}^{(030)}= (N^2+\bar N^2)(M^2+\bar M^2) \ ,
\ea

Per verificare la correttezza dell'identificazione si pu\`o studiare il comportamento dell'ampiezza con tre bordi trovata sotto una trasformazione modulare S (\ref{generatori}) e calcolarne il limite di degenerazione separante. Dal momento che questa ampiezza ammette una matrice dei periodi puramente immaginaria, l'azione di $S$ sar\`a
\be
S: \Omega \to -\Omega^{-1}= \frac{-i}{\tau_1\tau_2-\tau^2}\left ( \matrix{\tau_1 & -\tau \cr -\tau & \tau_2 \cr} \right )\ ,
\ee
dove adesso i $\tau_i$ sono reali. Nel limite $\tau_2 \to \infty$ questa trasformazione equivale a due trasformazioni S su ciascuno dei due tori di genere 1,
\be
\Omega \to \left ( \matrix{-\frac{i}{\tau_1} & 0 \cr 0 & -\frac{i}{\tau_2} \cr} \right )\ .
\ee
Ci si aspetta quindi di ottenere la corretta fattorizzazione di due ampiezze di anello nel canale diretto.

Per procedere nel calcolo occorre specializzare le leggi di trasformazione modulari date nei paragrafi precedenti al caso di S. Per le strutture di spin si trova,
\ba
S: \left \{ \matrix{
\tilde \delta_1=\delta_1 & \tilde \delta_2=\delta_5 & \tilde \delta_3=\delta_7 \cr
\tilde \delta_4=\delta_9 & \tilde \delta_5=\delta_2 & \tilde \delta_6=\delta_8 \cr
\tilde \delta_7=\delta_3 & \tilde \delta_8=\delta_6 & \tilde \delta_9=\delta_4 \cr
\tilde \delta_0=\delta_0 & \, & \, \cr}
\right .
\ea
Le trasformazioni per gli altri oggetti che compaiono nell'ampiezza con tre bordi sono
\ba
S: \left \{ \matrix{
d^3 \tilde\Omega _{IJ} & = & \det(-\Omega)^{-3}d^3 \Omega \cr
\vartheta [\tilde \delta ] ^4 (0,\tilde \Omega) & = & \det(-\Omega)^{2}\ \vartheta [\delta ] ^4 (0,\Omega)  \cr
\Xi _6 [\tilde \delta ] (\tilde \Omega) & = & \det (-\Omega)^6 \ \Xi _6 [\delta ] (\Omega ) \cr
\Psi _{10} (\tilde \Omega) & = & \det (-\Omega)^{10} \ \Psi _{10}(\Omega) \cr}
\right. \ .
\ea
Si pu\`o quindi operare una trasformazione S dell'ampiezze e calcolarne successivamente il limite nella degenerazione separante con le regole note,
\ba
Z^{(030)} &\sim & \int \frac{d\tau}{(2\pi)^2\tau^2}\int \frac{d\tau_1d\tau_2}{\tau_1^5\tau_2^5}\,\frac{1}{\eta^8(\tau_1)\eta^8(\tau_2)}\left\{ - N\bar NM\bar M \left[\frac{\vartheta_3^4-\vartheta_4^4}{2\eta^4}\right](\tau_1) \left[\frac{\vartheta_3^4-\vartheta_4^4}{2\eta^4}\right](\tau_2) \right. \nonumber \\
&+& N\bar N\frac{M^2+\bar M^2}{2}\left[\frac{\vartheta_3^4-\vartheta_4^4}{2\eta^4}\right](\tau_1)
\left[\frac{\vartheta_2^4-\vartheta_1^4}{2\eta^4}\right](\tau_2)\nonumber \\
&+& M\bar M \frac{N^2+\bar N^2}{2}\left[\frac{\vartheta_2^4-\vartheta_1^4}{2\eta^4}\right](\tau_1) \left[\frac{\vartheta_3^4-\vartheta_4^4}{2\eta^4}\right](\tau_2) \nonumber \\
&-& \left. \frac{N^2+\bar N^2}{2}\frac{M^2+\bar M^2}{2}\left[\frac{\vartheta_2^4-\vartheta_1^4}{2\eta^4}\right](\tau_1) \left[\frac{\vartheta_2^4-\vartheta_1^4}{2\eta^4}\right](\tau_2) \right \} \ .
\ea
L'espressione trovata corrisponde come atteso al prodotto di due ampiezze di anello nel canale diretto. 

\begin{figure}
\begin{center}
\includegraphics[width=4cm,height=3cm]{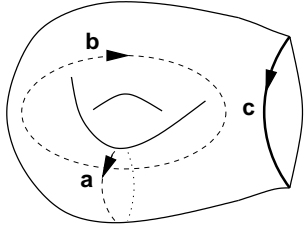}
\end{center}
\caption{Superficie di genere 3/2 con un bordo e un manico.}
\label{doppia1b1h}
\end{figure}

Lo studio delle altre ampiezze risulta pi\`u complesso in ragione della loro matrice dei periodi non puramente immaginaria, ma il metodo per l'identificazione dei coefficienti resta sostanzialmente invariato. 

La superficie con un bordo e un manico (figura \ref{doppia1b1h}), che si ottiene tramite una semplice involuzione di riflessione dal doppio toro (figura \ref{1b1h}) pu\`o essere studiata sulla scorta di quanto gi\`a fatto. La funzione di partizione viene scritta nella forma
\be
\label{(110)}
Z^{(110)}  = 
\int {d^3 \Omega}\, \frac{1}{\Psi _{10}(\Omega)}\cdot 
 {\sum _\delta C_\delta^{(110)}\Xi_6 [\delta ](\Omega )  \vartheta [\delta ]^4 (0,\Omega )} \ .
\ee
In una base canonica di omologia, l'involuzione pu\`o essere scritta nella forma
\be
I= \left ( \matrix{\sigma_1 & 0 \cr 0 & \sigma \cr} \right ) \ ,
\ee
come ci si rende conto facilmente osservando la figura \ref{1b1h}. Il vincolo \ref{vicoloomeg} permette di scrivere la matrice dei periodi nella forma
\be
\Omega = \left ( \matrix{\tau_1 & i\tau \cr i\tau & -\bar \tau_1 \cr} \right ) \ ,
\ee
con $\tau$ \`e reale. La forma della matrice dei periodi \`e tale che, nel limite di degenerazione separante $\tau \to 0$ in cui il digramma si presenta come un toro con un tadpole su un bordo, si ritrovino, come desiderato, i settori di stringa chiusa della funzione di partizione del toro della OB. Per rendersene conto si osservi che in questo limite
\ba
\vartheta\left [ \matrix{\alpha_1 & \alpha_2 \cr \beta_1 & \beta_2 \cr} \right](0, \Omega) &=&
\vartheta \left [ \matrix{\alpha_1 \cr \beta_1 \cr} \right](0, \tau_1) 
\left [ \matrix{\alpha_2 \cr \beta_2 \cr} \right](0,-\bar \tau_1) + {\cal O}(\tau^2)\nonumber \\
&=&  \vartheta \left [ \matrix{\alpha_1 \cr \beta_1 \cr} \right](0, \tau_1)
\vartheta \left [ \matrix{\alpha_2 \cr -\beta_2 \cr} \right](0, \bar \tau_1) + {\cal O}(\tau^2)\nonumber \\
&=& \vartheta \left [ \matrix{\alpha_1 \cr \beta_1 \cr} \right](0, \tau_1)
\overline{\vartheta \left [ \matrix{\alpha_2 \cr \beta_2 \cr} \right](0,\tau_1)} + {\cal O}(\tau^2) \ ,
\ea 
come \`e possibile verificare a partire dalla definizione della funzione $\vartheta$.

\begin{figure}
\begin{center}
\includegraphics[width=7cm,height=6cm]{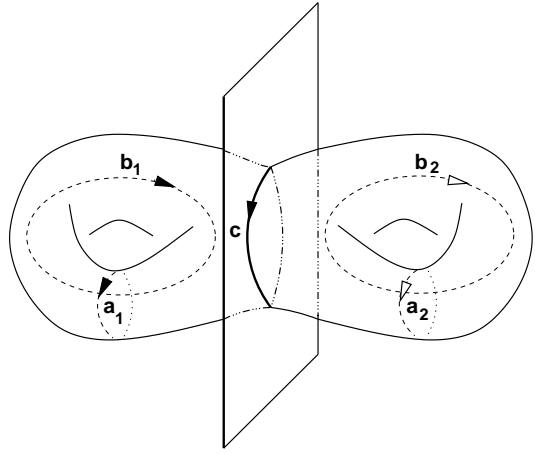}
\end{center}
\caption{Superficie con un bordo e un manico ottenuta da una involuzione di riflessione da una superficie di genere 2.}
\label{1b1h}
\end{figure}

In questo limite ci si aspetta che siano presenti solo i settori NS-NS, dal momento che sul toro per funzioni ad un punto i settori R-R sono zero,
\be
Z \sim (N + \bar N)(V_8^2+O_8^2) \ .
\ee
Si possono fissare in questo modo i coefficienti della ampiezza trovando
\be
C_{\delta_1}^{(110)}=C_{\delta_4}^{(110)}=N+\bar N \ , \qquad {\rm tutti \, gli \, altri}\qquad  C_{\delta_i}^{(110)}=0 \ .
\ee

Con considerazioni analoghe si pu\`o studiare il diagramma con un manico e un crosscap (101), che equivale ad un diagramma con tre crosscap. 

Gli ultimi diagrammi su cui, allo stato attuale della ricerca, sono in grado di fare alcune considerazioni sono la prima e la terza versione del diagramma con due bordi e un crosscap (021) (il caso con un crosscap centrale presenta sottigliezze nella degenerazione separante). Questo tipo di diagramma pu\`o essere studiato in stretta analogia con quanto fatto per il diagramma con tre bordi: se ne fissano i coefficienti studiandone la fattorizzazione nel limite di degenerazione separante e poi se ne verifica la corretta attribuzione passando al canale diretto e studiandone nuovamente la degenerazione separante. 

In questo caso ci si aspetta di trovare il prodotto di una striscia di M\"obius per un anello. Consideriamo il primo caso. Studiando il limite $\tau \to \infty$ nel canale diretto e confrontando con il prodotto $\tilde {\cal M} \times \tilde {\cal A}$, si fissano i coefficienti
\ba
C_{\delta_0}^{(021)}&=& -(N^2+\bar N^2)(M+\bar M) \ ,
C_{\delta_7}^{(021)}= -(N^2+\bar N^2)(M+\bar M) \ , \nonumber \\
C_{\delta_8}^{(021)}&=& (N^2+\bar N^2)(M+\bar M) \ ,
C_{\delta_9}^{(021)}= 2N\bar N(M+ \bar M) \ , \nonumber \\
C_{\delta_1}^{(021)}&=& C_{\delta_2}^{(021)}=C_{\delta_3}^{(021)}=C_{\delta_4}^{(021)}=C_{\delta_5}^{(021)}=0  \ .
\ea
La complicazione di questo diagramma rispetto al caso con tre bordi risiede nella corretta individuzione della matrice che permette il passaggio dal canale trasverso a quello diretto. Ricordando che nel canale trasverso $I(a_i)=a_i $ mentre nel canale diretto $I(a_i)=-a_i$. Si pu\`o individuare la trasformazione che collega le due rappresentazioni che, nella base identit\`a di omologia, risulta essere
\be
M = \left ( \matrix{1 & 0 & 0 & 0 \cr
0 & 0 & 0 & -1 \cr
-2 & 0 & 1 & 0 \cr
0 & 1 & 0 & 0 \cr } \right ) \ .
\ee
Studiando l'azione di questa trasformazione sulle strutture di spin si trova
\ba
M: \left \{ \matrix{
\tilde \delta_1=\delta_1 & \tilde \delta_2=\delta_5 & \tilde \delta_3=\delta_3 \cr
\tilde \delta_4=\delta_6 & \tilde \delta_5=\delta_2 & \tilde \delta_6=\delta_4 \cr
\tilde \delta_7=\delta_7 & \tilde \delta_8=\delta_9 & \tilde \delta_9=\delta_8 \cr
\tilde \delta_0=\delta_0 & \, & \, \cr}
\right .
\ea
Ricordando le leggi di trasformazione di $\Psi _{10}(\Omega)$, di $\Xi_6 [\delta ](\Omega )$ e di $d^3 \Omega$, \`e possibile verificare la correttezza della scelta dei coefficienti.

\section{Un ansatz per i campi magnetici a genere 2}

A titolo di esempio delle possibili applicazioni della tecnica sviluppata consideriamo una proposta di ampiezza con tre bordi con l'introduzione di campi magnetici interni. Questo modello tutt'ora oggetto di studio, presenta sottigliezze non ancora del tutto chiarite. L'ansatz per l'ampiezza, in analogia con il caso noto di genere 1, \`e
\ba
\label{magg2}
Z^{(030)}_{B} &\sim & \sum_{a, b, c} \frac{\pi^2(q_a + q_c)(q_b+q_c)B^2}{\sqrt{1+\pi^2 q_c^2B^2}}\int d^3 \Omega \nonumber \\
&\,& \frac{\partial_{z_1}\vartheta\left [ \matrix{1/2 & 0 \cr 1/2 & 0 \cr} \right](0 | \Omega)
\partial_{z_2}\vartheta\left [ \matrix{0 & 1/2 \cr 0 & 1/2 \cr} \right](0 | \Omega)}
{\vartheta\left [ \matrix{1/2 & 0 \cr 1/2 & 0 \cr} \right](\epsilon_a+\epsilon_c, 0| \Omega)
\vartheta\left [ \matrix{0 & 1/2 \cr 0 & 1/2 \cr} \right](0, \epsilon_b+\epsilon_c | \Omega)} \times \nonumber \\
&\,& \sum_\delta \frac{\Xi_6 [\delta ](\Omega )\vartheta[\delta](\epsilon_a+\epsilon_c, \epsilon_b+\epsilon_c | \Omega)\vartheta^3[\delta](0,0 |\Omega)}{\Psi _{10}(\Omega)} \ .
\ea
Ci si aspetta, nel limite di degenerazione non separante $\tau_2 \to i\infty$, in cui si allontana il bordo $b$ all'infinito, di ottenere un espressione della forma,
\be
\label{magg21}
(q_a+q_c)B\sum_{\alpha, \beta}\frac{\eta_{\alpha \beta} \vartheta \left [ \matrix{\alpha_2 \cr \beta_2 \cr} \right](\epsilon_a+\epsilon_c | \tau_1) \vartheta^3 \left [ \matrix{\alpha_2 \cr \beta_2 \cr} \right](0 | \tau_1)}{\vartheta_1\left [ \matrix{\alpha_2 \cr \beta_2 \cr} \right](0 | \tau_1)\eta^9(\tau_1)}\sqrt{1+\pi^2 q_b^2B^2}
\ee
I calcoli fatti, che risultano molto complicati, e richiedono un utilizzo intenso delle propriet\`a delle funzioni $\vartheta$, mostrano che nel limite $\tau_2 \to i\infty$ si trova una espressione simile alla (\ref{magg21}), insieme per\`o a termini in cui campo magnetico compare in forma esponenziale che risultano piuttosto sottili da interpretare.


\chapter{Conclusioni}

In questo lavoro di Tesi si \`e iniziato il lavoro di generalizzazione dei risultati ottenuti da D'Hoker e Phong per le superstrighe chiuse di Tipo II e Eterotiche agli altri altri modelli di stringhe con supersimmetria rotta (modelli di tipo 0 e modelli con ``brane supersymmetry breaking'' in dieci dimensioni). In particolare si \`e riusciti a studiare la superstringa 0'B individuando tecniche utili alla trattazione degli altri modelli.

Il lavoro svolto, frutto di una collaborazione in atto con il Dr. Carlo Angelantonj dell'Universit\`a di Torino, il Prof. Emilian Dudas dell'Ecole Polytechnique e il mio relatore di Tesi, mostra la consistenza della costruzione di orientifold a genere 3/2 e  rappresenta il punto di partenza per lo studio sistematico delle ridefinizioni di vuoto introdotte a due loops dalla rottura della supersimmetria, in vista anche della costruzione di un'espressione generale per le correzioni di soglia a due loops.


\appendix

\chapter{Funzioni Theta}

\subsubsection{Definizione}
Le funzioni $\vartheta$ di Jacobi \cite{ww} sono definite in termini di una somma infinita
\be
\label{A.1}
\vartheta \left[{\textstyle {\alpha \atop \beta}} \right] (z|\tau) =
\sum_{n\in {\mathbb Z}} \
q^{\frac{1}{2} (n+ \alpha)^2} \ e^{2 \pi i (n + \alpha)(z+\beta)} \ ,
\ee
dove $ \ \alpha, \beta \in {\mathbb R}$ e $q=e^{2\pi i \tau}$, o in maniera equivalente come
\ba
\label{A.2}
\vartheta \left[ {\textstyle {\alpha \atop \beta}} \right] (z|\tau) &=&
e^{2 i \pi \alpha (z+\beta)} \ q^{\alpha^2/2} \prod_{n=1}^\infty \
( 1 - q^n)\prod_{n=1}^\infty \ (1 + q^{n + \alpha - 1/2} e^{2 i \pi (z+\beta)} )\nonumber\\
&& \times \prod_{n=1}^\infty (1 + q^{n - \alpha - 1/2} e^{-2 i \pi (z+\beta)} ) \ ,
\ea
in particolare 
\ba
\label{A.3}
&&\vartheta\left[{\textstyle{0\atop
      0}}\right](z|\tau) \ = \ \vartheta_3(z|\tau) \ = \
\prod_{m=1}^\infty (1-q^m)(1+e^{2\pi iz}\ q^{m-1/2})(1+e^{-2\pi i z} \
q^{m-1/2})
\ ,\nonumber\\
&&\vartheta\left[{\textstyle{0\atop
      1/2}}\right](z|\tau) \ = \ \vartheta_4(z|\tau) \ = \
\prod_{m=1}^\infty (1-q^m)(1-e^{2\pi iz}\ q^{m-1/2})(1-e^{-2\pi i z} \
q^{m-1/2})\ , \nonumber\\
&&\vartheta\left[{\textstyle{1/2\atop
      0}}\right](z|\tau) \ = \ \vartheta_2(z|\tau) \ = \
2q^{1/8}\cos(\pi z)\nonumber\\
&&\qquad\qquad\times\prod_{m=1}^\infty (1-q^m)(1+e^{2\pi iz}\ q^m)(1+e^{-2\pi i z} \
q^m) \ , \nonumber\\
&&\vartheta\left[{\textstyle{1/2\atop
      1/2}}\right](z|\tau) \ = \ -\vartheta_1(z|\tau) \ = \
-2q^{1/8}\sin(\pi z) \nonumber\\
&&\qquad\qquad\times\prod_{m=1}^\infty (1-q^m)(1-e^{2\pi iz}\ q^m)(1-e^{-2\pi i z} \
q^m) \ .
\ea

\subsubsection{Trasformazioni modulari}

\be
\label{A.4}
\vartheta \left[ {\textstyle {\alpha \atop \beta}}
 \right] (z|\tau+1)
= e^{-i \pi \alpha
(\alpha -1)} \vartheta \left[ {\textstyle {\alpha \atop \beta +\alpha - 1/2}}
\right]
(z|\tau) \ ,
\ee

\label{A.5}
\be
\vartheta \left[ {\textstyle {\alpha \atop \beta}} \right]
\left(\frac{z}{\tau}\right|\left.-\frac{1}{\tau}\right) =
(-i \tau)^{1/2} \ e^{2 i \pi \alpha \beta + i \pi z^2/\tau} \ \vartheta
\left[{\textstyle {\beta \atop -\alpha}}\right] (z|\tau) \ .
\ee
in particolare si ha 
\ba
\label{A.6}
&&\vartheta_3(z|\tau+1) \ = \ \vartheta_4(z|\tau+1) \ , \nonumber\\
&&\vartheta_4(z|\tau+1) \ = \ \vartheta_3(z|\tau+1) \ , \nonumber\\
&&\vartheta_2(z|\tau+1) \ = \ q^{1/8} \ \vartheta_2(z|\tau+1) \ , \nonumber\\
&&\vartheta_1(z|\tau+1) \ = \ q^{1/8} \vartheta_1(z|\tau+1) \ ,
\ea
\ba
\label{A.7}
&&\vartheta_3(z/\tau|-1/\tau) \ = \ \sqrt{-i\tau} \ e^{\pi i z^2 /\tau}
\ \vartheta_3(z|\tau) \ , \nonumber\\
&&\vartheta_4(z/\tau|-1/\tau) \ = \ \sqrt{-i\tau} \ e^{\pi i z^2 /\tau}
\ \vartheta_2(z|\tau) \ , \nonumber\\
&&\vartheta_2(z/\tau|-1/\tau) \ = \ \sqrt{-i\tau} \ e^{\pi i z^2 /\tau}
\ \vartheta_4(z|\tau) \ , \nonumber\\
&&\vartheta_1(z/\tau|-1/\tau) \ = \ -i \sqrt{-i\tau} \ e^{\pi i z^2 /\tau}
\ \vartheta_1(z|\tau) \ . \nonumber\\
\ea

\subsubsection{Identit\`a utili}

\textit{Aequatio identica satis abstrusa} di Jacobi
\be
\label{A.8}
\vartheta_3^4-\vartheta_4^4-\vartheta_2^4 \ = \ 0 \ ,
\ee
\be
\label{AB}
\vartheta_3^4(z)+\vartheta_1^4(z)-\vartheta_4^4(z)-\vartheta_2^4(z) \ = \ 0 \ ,
\ee
Altre identit\`a utili sono
\be
\label{A.13}
\vartheta_3(z)\vartheta_3^3-\vartheta_4(z)\vartheta_4^3-\vartheta_2(z)\vartheta_2^3
\ = \ 2\vartheta_1^4(z/2) \ ,
\ee
insieme con le identit\`a per $z=0$ 
\be
\label{A.14}
\vartheta_3^{\prime\prime}\vartheta_3\vartheta_2^2-\vartheta_2^{\prime\prime}\vartheta_2\vartheta_3^2
\ = \ 4\pi^2\eta^6\vartheta_4^2 \ ,
\ee
\be
\label{A.15}
\vartheta_4^{\prime\prime}\vartheta_4\vartheta_3^2-\vartheta_3^{\prime\prime}\vartheta_3\vartheta_4^2
\ = \ 4\pi^2\eta^6\vartheta_2^2 \ .
\ee

\subsubsection{Funzione $\eta$ di Dedekind}

La funzione $\eta$ di  Dedekind \`e definita come
\be
\label{A.11}
\eta(\tau) \ = \ q^{1/24} \ \prod_{n=1}^\infty(1-q^n) \ ,
\ee
e le sue trasformazioni modulari sono
\ba
\label{A.12}
\eta(\tau+1) \ = \ e^{i\pi /12} \ \eta(\tau) \ , \nonumber\\
\eta(-1/\tau) \ = \ \sqrt{-i\tau} \ \eta(\tau) \ .
\ea
Infine
\be
\label{A.9}
\vartheta(0|\tau) \ = \ 0 \ ,
\ee
mentre la derivata prima di $\vartheta_1$ all'origine \`e
\be
\label{A.10}
\vartheta_1^\prime(0|\tau) \ = \ 2\pi \eta^3 \ .
\ee

\chapter[Ampiezze di Superstringa]{Alcuni risultati sulle ampiezze di Superstringa ad albero e ad un loop}

Per generalizzare al caso di Superstringa la teoria perturbativa sviluppata per le stringhe bosoniche \`e utile introdurre il \emph{formalismo dei supercampi}. La supersimmetria locale N=1 sul worldsheet non \`e manifesta nel formalismo in componenti che si \`e utilizzato nel primo capitolo. Al contrario, il formalismo $N=1$ dei supercampi, formulato nel linguaggio $N=1$ della \emph{supergeometria}, pone le trasformazioni di supersimmetria e i diffeomorfismi sullo stesso piano, rendendo entrambe le simmetrie manifeste. La loro azione combinata forma il gruppo dei \emph{super-diffeomorfismi} sulla super-variet\`a del world-sheet. In questo capitolo si introdurranno brevemente i concetti fondamentali della supergeometria, il formalismo dei supercampi e la costruzione delle ampiezze d'interazione di superstringa a partire dall'integrale funzionale (per una rassegna dettagliata sull'argomento si veda \cite{D'Hoker:1988ta}). 

\section{Supergeometria}

Localmente, una supervariet\`a $\Sigma$ ($N=1$) di dimensione $(2|2)$ \`e parametrizzata da coordinate $\xi^M=(\xi^m,\theta^\mu)$, $m=1,2$, $\mu=1,2$ con regole di commutazione
\be
\xi^M\xi^N=(-)^{MN}\xi^N\xi^M
\ee
dove $(-)^{MN}$ \`e $1$ tranne nel caso $M=\mu$, $N=\nu$, in cui vale $-1$. Introduciamo un riferimento (\emph{super-zweibain}) ortonormale locale sulla supervariet\`a, 
\be
E^A\equiv d\xi^M E_M{}^A\qquad\qquad
A=(a,\alpha)\quad a=z,\bar z;\,\,\alpha=+,-\,\,,
\ee
dove le regole di commutazione associate ad indici greci e latini sono le stesse che valgono per le coordinate. Indicando il riferimento inverso come $E_A{}^M$, si ha $E_A{}^M\,E_M{}^B=\delta_A{}^B$ e $E_M{}^A\,E_A{}^N=\delta_M{}^N$.

Si pu\`o introdurre un gruppo di trasformazioni di gauge $U(1)$, sotto il quale i campi del riferimento $E^z$,
$E^{\bar z}$, $E^+$, $E^-$ hanno pesi $-1$, $+1$, $-\frac{1}{2}$, $+\frac{1}{2}$. Un tensore generico $V$ che trasformi come $(E^z)^{\otimes n}$ avr\`a pertanto peso $-n$ nel gruppo $U(1)$. Introducendo un campo di gauge $U(1)$, detto \emp{connessione},
\be
\Omega=d\xi^M\Omega_M
\ee
si pu\`o costruire una super-derivata covariante per il gruppo di simmetria $U(1)$
\be
{\mathcal D}_A^{(n)}V\equiv
E_A{}^M(\partial_M+in\Omega_M)V\,\,.
\ee
I supercampi tensoriali di \emph{torsione} $T_{AB}{}^C$ e di \emph{curvatura} $R_{AB}$ sono definiti dalle relazioni di commutazione (e di anticommutazione per gli indici spinoriali)
\be
[{\mathcal D}_A,{\mathcal D}_B]V=T_{AB}{}^C{\mathcal D}_C V+in\,R_{AB}V\,\,.
\ee
La supergeometria pu\`o essere specificata imponendo alcuni vincoli sui campi di torsione di curvatura,
\be
\label{const1}
T_{\alpha\beta}^\gamma=0 \ ,\qquad
T_{\alpha\beta}^c=2\gamma_{\alpha\beta}^c \ ,\qquad
R_{++}=R_{--}=0 \ ,
\ee 
dove le $\gamma^c$ sono matrici bidimensionali di Dirac. I vincoli (\ref{const1}) possono essere scritti in forma equivalente come
\be
\label{const2}
{\mathcal D}_+^2={\mathcal D}_z,\quad
{\mathcal D}_-^2={\mathcal D}_{\bar z}
\{{\mathcal D}_+,{\mathcal D}_-\}V=in\,R_{+-}V\,\,,
\ee
dove $V$ \`e un generico supercampo tensoriale di peso $n$. Le rimanenti derivate covarianti possono essere trovate utilizzando le identit\`a di Jacobi. L'unica quantit\`a libera nella supergeometria \`e il supercampo di curvatura $R_{+-}=R_{-+}\equiv R$.

La supergeometria \`e invariante sotto trasformazioni che preservino i vincoli sulla torsione (\ref{const1}, \ref{const2}). Le simmetrie della supergeometria sono le trasformazioni locali $sU(1)$, generate da un campo locale $L$,
\ba
E^{\pm}_M &=& e^{\pm(i/2)L}\hat E^{\pm}_M  \ , \qquad {\mathcal D}_+^{(n)}e^{-i(n+1/2)L} {\hat \mathcal D}_+^{(n)} e^{inL} \ , \nonumber \\ 
E^{\phantom{M}z}_M &=& e^{iL}\hat E^{\phantom{M}z}_M  \ , \qquad {\mathcal D}_-^{(n)}e^{-i(n-1/2)L} {\hat \mathcal D}_-^(n) e^{inL} \ , \nonumber \\ 
E^{\phantom{M}\bar z}_M &=& e^{-iL}\hat E^{\phantom{M}\bar z}_M  \ , \qquad \Omega_M=\hat \Omega_M+\partial_M L \ ;
\ea
le super-riparametrizzazioni, che formano il gruppo $sDiff(M)$, la cui versione infinitesima \`e generata da un supercampo vettoriale $\delta V^M$, 
\be
E_A^M\delta E_M^B = {\mathcal D}_A\delta V^B- \delta V^CT_{CA}^{\phantom{CA}B}+\delta V^C\Omega_C E_A^B \ ;
\ee
le super-trasformazioni di Weyl, generate da un supercampo reale scalare,
\be
E_a^M=e^{\Sigma}\hat E_M^a \ , 
E_a^M=e^{\Sigma/2}[\hat E_M^a+\hat E_M^a(\gamma_a)^{\alpha \beta}\hat {\mathcal D}_\beta \Sigma] \ ,
\ee
che per la superconnessione, la supercurvatura e le superderivate risultano essere
\ba
\Omega_M &=& \hat \Omega_M + \hat E_M^a \epsilon_a^{\phantom{a}b}{\hat \mathcal D}_b \Sigma + \hat E_M^\alpha(\gamma_5)_{\alpha}^{\phantom{\alpha}\beta}{\hat \mathcal D}_\beta \Sigma \ , \nonumber \\
R_{+-} &=& e^{-\Sigma}(\hat R_{+-}-2i {\hat \mathcal D}_+ {\hat \mathcal D}_- \Sigma) \ , \nonumber \\
{\mathcal D}_+^{(n)} &=& e^{(n-1/2)\Sigma}{\hat \mathcal D}_+^{(n)}e^{-n\Sigma} \ , \nonumber \\
{\mathcal D}_-^{(n)} &=& e^{-(n+1/2)\Sigma}{\hat \mathcal D}_-^{(n)}e^{+n\Sigma} \ .
\ea
\`E utile combinare le trasformazioni $sWeyl$ e $sU(1)$ in una trasformazione complessa di Weyl, data in termini di un supercampo complesso $\Lambda$, definito come
\be
\Lambda\equiv \Sigma - iL \ , \qquad \Lambda^*\equiv \Sigma + iL \ .
\ee
Le superderivate trasformano sotto $sWeyl\times sU(1)$ come
\be
{\mathcal D}_+^{(n)} =e^{n\Lambda-1/2 \Lambda^*}{\hat \mathcal D}_+^{(n)} e^{-n\Lambda} \ , \qquad
{\mathcal D}_-^{(n)} =e^{-n\Lambda^*-1/2 \Lambda}{\hat \mathcal D}_-^{(n)}e^{n\Lambda^*} \ .
\ee
e la supercurvatura trasforma come
\be
R_{+-}=e^{-\Sigma}\left(\hat R_{+-}-2i{\hat \mathcal D}_+{\hat \mathcal D}_-
\Sigma\right) \ .
\ee

Definiamo le coordinate complesse $(\xi,\bar \xi,\theta,\bar \theta)$ come
\be
\xi=1/\sqrt{2\,\,}\, (\xi^1+i\xi^2) \ , \qquad \theta=1/\sqrt{2\,\,}\,(\theta^1+i\theta^2) \ ,
\ee
con $\bar \xi$ e $\bar \theta$ le rispettive coordinate complesse coniugate. Localmente ogni supergeometria $N=1$, in analogia con quanto succede nel caso bosonico, \`e equivalente a meno di trasformazioni $sWeyl\times sU(1)$ ad una supergeometria piatta (Euclidea) in cui $R_{+-}=0$.  Nel superspazio piatto si ha per il superzweibein
\ba
E_{m}^{a} &=& \delta_{m}^{a} \ , \qquad E_m^{\alpha}=0 \ , \nonumber \\
E_\mu^a &=&(\gamma^a)_\mu^\beta \theta_\beta \ , \qquad E_\mu^\alpha=\delta_\mu^\alpha \ ,
\ea
mentre le superderivate prendono una forma estremamente semplice
\ba
{\mathcal D}_+ &=&{\partial\over\partial\theta} +\theta{\partial\over\partial\xi}\ , \qquad {\mathcal D}_z = {\mathcal D}_+^2 ={\partial\over\partial\xi}\ , \nonumber \\
{\mathcal D}_- &=&{\partial\over\partial \bar \theta}+\bar \theta{\partial\over\partial \bar \xi} \ , \qquad 
{\mathcal D}_{\bar z} ={\mathcal D}_-^2 ={\partial\over\partial \bar \xi} \ .
\ea
La simmetria residua del gruppo $sDiff\times sU(1)\times sWeyl$ che lascia invariante la supergeometria piatta \`e il gruppo delle trasformazioni \emph{superconformi}, che comprende i diffeomorfismi superanalitici
\be
\xi\to\xi'(\xi,\theta) \ ,\qquad
\bar \xi\to\bar \xi'(\bar \xi,\bar \theta) \ ,\qquad
\theta\to\theta'(\xi,\theta) \ ,\qquad
\bar \theta\to \bar \theta'(\bar \xi,\bar \theta) \ ,\qquad
\ee
combinati con trasformazioni di $sWeyl$ e $sU(1)$.

Per ottenere una comprensione pi\`u profonda della supergeometria \`e utile studiare la formulazione \emph{in componenti}, che richiede l'eliminazione di alcune componenti dei supercampi. Per farlo si pu\`o fissare il gauge di Wess-Zumino per il superzweibein, definita dalle condizioni 
\ba
E_\mu^\alpha &\sim & \delta_\mu^\alpha + \theta^\nu  e_{\nu \mu}^{*\alpha} \ , \qquad
E_\mu^a \sim \theta^\nu e_{\nu \mu}^{**a} \ , \nonumber \\
e^{*\alpha}_{\nu \mu} &=& e^{*\alpha}_{\mu \nu} \ , e_{\nu \mu}^{**a} = e_{\mu \nu}^{**a} \ ,
\ea
sull'espansione in potenze di $\theta$, a meno di termini di ordine pi\`u alto. Il superzweibein pu\`o sempre essere ridotto in questa forma con una super-riparametrizzazione e i campi indipendenti rimanenti sono $e_m{}^a$,
$\chi_m{}^\alpha$ e un campo ausiliario $A$. Per le altre componenti si trova
\ba
E_m{}^a &=& e_m{}^a+\theta\gamma^a\chi_m-\frac{1}{2} \theta\bar \theta e_m{}^a A \ ,\nonumber \\
E_m{}^\alpha &=& -\frac{1}{2}\chi_m{}^\alpha- \frac{i}{2} \theta^\beta(\gamma^a)_\beta{}^\alpha e_m{}^a A+\cdots \ ,\nonumber \\
E_\mu{}^a &=& (\gamma^a)_\mu{}^\beta\theta_\beta \ ,\nonumber \\
E_\mu{}^\alpha &=& \delta_\mu{}^\alpha \left(1+\frac{i}{4} \theta \bar \theta A \right) \ .\nonumber \\
\ea

L'integrazione sulla supervariet\`a $\Sigma$ richiede la definzione di una super-misura $d\mu_E$, invariante sotto super-diffeomorfismi, definita come
\be
d\mu_E\equiv s\det\,E_M{}^A d\xi\,d\bar \xi\,d\theta\, d\bar \theta \ ,
\ee
dove il super determinante \`e 
\be
s\det\,E_M{}^A\equiv\det(E_m{}^a-E_M{}^\alpha
(E_\mu{}^\alpha)^{-1}E_\mu{}^a)\cdot
(\det\,E_\mu{}^\alpha)^{-1} \ .
\ee
Il prodotto scalare fra due supercampi $\Phi_1$ e $\Phi_2$ di peso $n$ viene definto come
\be
\left<\Phi_1,\Phi_2\right>=\int\nolimits_{\Sigma}
d\mu_E\,\Phi_1^*\Phi_2\,\,.
\ee

\section{Integrale funzionale di Superstringa}

Per definire l'integrale funzionale di superstringa \`e utile riscrivere l'azione di superstringa e l'azione dei campi di ghost in termini di supercampi. Introduciamo un supercampo di coordinata di stringa, di peso $0$ rispetto all'azione della simmetria $sU(1)$, che definisca l'immersione della supervariet\`a  nello spazio-tempo dieci-dimensionale,
\be
{\mathbf X}: \Sigma \to {\mathbb R}^{10} \ ,
\ee
che in termini di componenti scriviamo come
\be
{\mathbf X}^\mu=X^\mu+\theta^\mu\psi_+{}^\mu+\bar \theta \psi_-^\mu+\theta \bar\theta \,F^\mu\qquad \mu=0,1,\ldots,9\,\,.
\ee
L'azione della superstringa in termini di supercampi si ottiene accoppiando i supercampi di posizione alla supergravit\`a $N=1$ bidimensionale,
\be
S_{\mathbf X}={1\over 4\pi}\int\nolimits_{\Sigma}d\mu_E {\mathcal D}_-{\mathbf X}\cdot{\mathcal D}_+{\mathbf X} \ ,
\ee
che scritta in componenti si riduce all'azione di superstringa nota, a meno di un termine non dinamico $F^2$ che che coinvolge il campo ausiliario. L'azione scritta possiede le simmetrie locali $sU(1)\times sWeyl \times sDiff$, ed \`e evidentemente invariante sotto trasformazioni spazio temporali di Poincar\'e. Si pu\`o dimostrare che l'azione scritta \`e l'unica possibile che possegga tutte le simmetrie volute.

L'azione dei supercampi di ghost $B$ e $C$, il cui peso sotto le trasformazioni di $sU(1)$ pu\`o essere scelto rispettivamente $n$ e $(\frac{1}{2}-n)$, viene scritta come
\be
S_{BC}={1\over 2\pi}\int\nolimits_{\Sigma} d\mu_E(B{\mathcal D}_-^{(n)}C+\bar B{\mathcal D}_+^{(-n)}\bar C)\,\,.
\ee
Per ritrovare l'azione nota del sistema $(b, c)$, scomponiamo i campi $B$, $C$ in componenti
\ba
B &=& \beta+\theta b+\bar \theta b'+\theta\bar \theta \beta' \ , \nonumber \\
C &=& c+\theta\gamma+\bar \theta \gamma'+ \theta\bar \theta c' \ .
\ea
I campi $b'$, $\beta'$, $\gamma'$ e $c'$ sono campi ausiliari, i cui pesi per $sU(1)$ possono essere ricavati ricordando che $\theta$ e $\bar \theta$ hanno pesi $-\frac{1}{2}$ e $\frac{1}{2}$. Si trova che i campi hanno pesi  $\beta$, $3/2$; $b$, $2$; $\gamma$, $-1/2$; $c$, $-1$.

In questo modo si riconosce il sistema $(b, c)$ noto, e un sistema $(\beta, \gamma)$ di campi di ghost commutanti (detti superghost) che ha il giusto peso per essere il patner supersimmetrico dei campi di ghost. I campi $(\beta, \gamma)$ sono spinori sul world-sheet e la loro struttura di spin deve essere la stessa dei campi $\psi_+{}^\mu$ in quanto campi di ghost di una supersimmetria $N=1$. L'azione $BC$ pu\`o essere scritta in componenti nel gauge di Wess-Zumino,
\be
S_{BC}={1\over 2\pi}\int\nolimits_{\Sigma} d^2 z [b\nabla_{(-1)}^zc+\beta\nabla_{(-1/2)}^z\gamma-\chi_{\bar z}{}^+ S_+^{\bar z}+c.c.] \ ,
\ee
e l'azione totale
\be
S = S_{\mathbf X} + S_{BC} \ ,
\ee
\`e invariante superconforme. Procedendo come nel caso bosonico \cite{D'Hoker:1988ta} \`e possibile identificare i supermoduli. Lo spazio delle supergeometrie di genere $g$ che soddisfano i vincoli sulla torsione (\ref{const1}) che siano inequivalenti sotto i gruppi di simmetria $sDiff(M)$, $sWeyl(M)$ e $sU(1)$ definisce lo spazio dei supermoduli
\be
s{\mathcal M}_g\equiv \frac{\{E_M{}^A,\Omega_M\hbox{che soddisfino la (\ref{const1})}\}}{sDiff\times sWeyl\times sU(1)} \ .
\ee
La dimensione dello spazio dei supermoduli si trova essere
\be
dim\,s{\mathcal M}_g=\left\{
\begin{array}{cc}
(0\vert 0) & g=0  \\
(1\vert 0) & g=1  \, \hbox{struttura di spin pari}  \\
(1\vert 1) & g=1  \, \hbox{struttura di spin dispari} \\
(3g-3\vert 2g-2) & g \ge 2 
\end{array}
\right. 
\ee
Lo spazio olomorfo cotangente di $s{\mathcal M}_g$ \`e generato dai differenziali super-olomorfi $3/2$, $\Phi_K$, che soddisfano
\be
{\mathcal D}_-^{(3/2)}\Phi_K=0\qquad\qquad
{K=1,\ldots,\dim\,s{\mathcal M}_g} \ .
\ee
I differenziali di super-Beltrami sono definiti come duali di $\Phi_K$ e possono essere parametrizzati dalla variazione dei supermoduli per $E_M{}^A$
\be
\mu_K\equiv E_-{}^M {\partial E_M{}^z\over
\partial m_K}\qquad K=1,\ldots,\dim\,s{\mathcal M}_g \ ,
\ee
che nel gauge di Wess-Zumino diventano
\be
\mu_K=\bar \theta\left(e_{\bar z}{}^n {\partial e_n{}^z\over\partial m_K}-\theta {\partial\chi_{\bar z}{}^+\over \partial m_K}\right) \ .
\ee
Il primo termine \`e il differenziale di Beltrami consueto, per un modulo pari $m_K$, mentre il secondo termine \`e un differenziale di Beltrami dispari per un modulo dispari $m_K$.

Tralasciando per un momento i problemi di separazione chirale degli spinori, possiamo introdurre la definizione generale delle ampiezze d'interazione. Definendo in analogia con il caso bosonico l'integrale funzionale a partire dall'idea della somma sulle superfici di universo, la forma generale delle ampiezze si scrive come
\be
A_g=\int [dE] [d\Omega] \delta\hbox{(vincoli)}
\int [d{\mathbf X}] V_1\ldots V_N e^{-S_{\mathbf X}} \ ,
\ee
dove i $V_i$ sono operatori di vertice di superstringa da definire. Nella dimensione critica $D=10$ e per operatori di vertice super-conformemente invarianti, si pu\`o dimostrare \cite{D'Hoker:1988ta} che
\ba
A_g=\int\nolimits_{s{\mathcal M}_g}\prod_{K}
dm_K\int [d{\mathbf X}]\int [dBd\bar B]\int [dCd\bar C] V_1\ldots
V_N  e^{-S_{\mathbf X}-S_{BC}} \nonumber \\
 \cdot\prod_{K}\delta(\left<\mu_\kappa,B\right>)
\delta(\left<\bar \mu_\kappa,\bar B\right>) \ .
\ea

\subsection{Quantizzazione BRST}

Come nel caso bosonico, scegliendo un gauge e introducendo i campi di ghost di Faddeev-Popov, la simmetria originaria sopravvive in forma di simmetria BRST. L'azione totale $S = S_{\mathbf X} + S_{BC}$ \`e invariante sotto le trasformazioni dei supercampi
\ba
\delta {\mathbf X}^\mu &=& \lambda C{\mathcal D}_+^2 X^\mu-\frac{1}{2}\,\lambda\,{\mathcal D}_+ C{\mathcal D}_+ {\mathbf X}^\mu+c.c. \ ,\nonumber \\
\delta C &=& \lambda C{\mathcal D}_+^2 C-{1\over 4}\lambda{\mathcal D}_+ C{\mathcal D}_+ C \ , \nonumber \\
\delta B &=& -\lambda T \ ,
\ea
dove $\lambda$ \`e un parametro grassmaniano costante e $T$ \`e il super tensore di energia-impulso. Per il supercampo $\bf X$ si trova 
\ba
T^{({\mathbf X})} &=& -\frac{1}{2}{\mathcal D}_+{\mathbf X}\cdot{\mathcal D}_+^2{\mathbf X}  \nonumber \\ 
&=& -\frac{1}{2}\,\psi_+\cdot\partial_z X+\theta
\left(-\frac{1}{2}\,\partial_z X\cdot\partial_z {\mathbf X}-\frac{1}{2}\,
\partial_z \psi_+\cdot\psi_+\right) \ ,
\ea
mentre per il sistema $(B, C)$ il supertensore $T$ \`e 
\ba
T^{(BC)} &=& -C{\mathcal D}_+^2 B+{1\over2}{\mathcal D}_+ C{\mathcal D}_+B-\frac{3}{2}{\mathcal D}_+^2C B \nonumber \\
&=& -c\partial_z\beta+{1\over2}\gamma b-\frac{3}{2}\partial_z c\beta+\theta(T^{(bc)}+T^{(\beta\gamma)}) \ .
\ea
La corrente BRST associata alla simmetria \`e
\be
J_{BRST}=C(T^{({\mathbf X})}+{\frac{1}{2}}T^{(BC)})-\frac{3}{4}{\mathcal D}_+(C{\mathcal D}_+ CB) \ .
\ee
Come gi\`a visto nel caso bosonico, la carica associata alla corrente $BRST$
\be
Q_{BRST}=\oint dz\,d\theta\,J_{BRST} \ ,
\ee
\`e conservata e nilpotente. 

\section{Teorema di separazione chirale}

In Teoria delle Stringhe esiste un risultato molto potente sulle propriet\`a di olomorfia delle funzioni di correlazione per operatori di vertice fattorizzati di stati fisici (stati \emph{on-shell}), che va sotto il nome di \emph{Teorema di Separazione Chirale}. Un generico operatore di vertice della forma
\be
V_i=\int d^2z_i\,d^2\theta_i\,W_i(z_i,\theta_i,\bar z_i,\bar \theta_i) \ ,
\ee
\`e fattorizzato se $W_i$ \`e prodotto di un fattore olomorfo e di un fattore anti-olomorfo
\be
W_i(z_i,\theta_i,\bar z_i,\bar \theta_i)=W_i(z_i,\theta_i)\tilde W_i
(\bar z_i,\bar \theta_i) 	\ .
\ee

\begin{figure}
\begin{center}
\kern.75true cm\vbox{\epsfxsize=5.00in\epsfbox{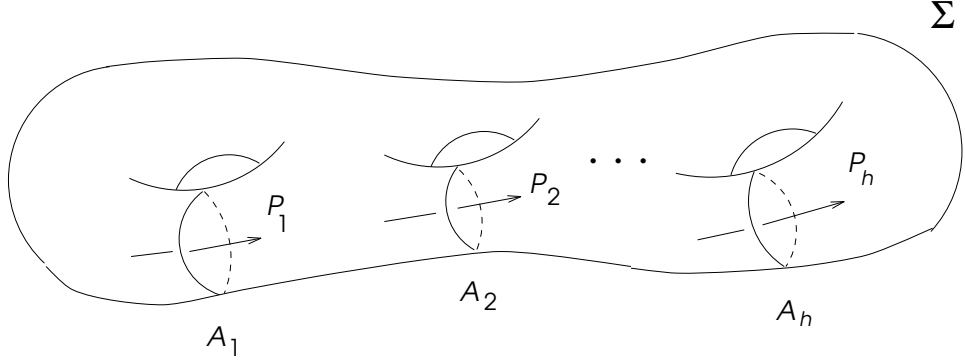}}
\end{center}
\caption{Momenti interi di loop di una Superficie di Riemann di genere $h$ con base canonica in omologia generata dalle curve chiuse $A_I$ e $B_I$}
\label{momentiloop}
\end{figure}

Un generico vertice di interazione pu\`o essere scritto come una combinazione lineare di operatori di vertice fattorizzati. Definendo i $g$ momenti interni $p_I^\mu$, associati agli 1-cicli di una base canonica $A_I$, $B_I$ (\ref{momentiloop}), $I=1, \dots, g$ di una superficie di genere $g$, come
\be
p_I^\mu=\oint_{A_I}dz\,d\theta\,{\mathcal D}_+ {\mathbf X}^\mu \ ,
\ee
le funzioni di correlazione di operatori di vertice non integrati sulla supergeometria, per momenti interni $p_I^\mu$ e supermoduli fissati, sono
\ba
\left<W_1\ldots W_N\right>_E(p_I) &\,&\equiv  \int [dX]\int [dBdC]\,\,W_1\ldots W_N \nonumber\\
&\,&\prod\limits_{I=1}^g\delta\left(p_I^\mu-\int\nolimits_{A_I}
dz\,d\theta\,\,{\mathcal D}_+{\mathbf X}^\mu\right) 
\prod\limits_{K=1}^{\dim\,s{\mathcal M}_g} \vert\left<\mu_K,B\right>\vert^2 e^{-S_{\mathbf X}-S_{BC}} \nonumber\\
\,
\ea

Il Teorema di separazione chirale comprende due risultati. Il primo \`e la fattorizzazione delle ampiezze di superstringa a fissati momenti interni. Le ampiezze risultano scrivibili nella forma
\be
\left<W_1\ldots W_N\right>_E(p_I)=\delta(k)
{\mathcal C}_\nu^F{\bar \mathcal C}_\nu^F \ ,
\ee
con $k$ gli impulsi esterni e dove ${\mathcal C}_\nu{}^F={\mathcal C}_\nu{}^F (z_i,\theta_i;\zeta_i,;m_i; p_I,k_i)$ \`e una funzione analitica complessa dei supermoduli $m_K$, $K=1,\ldots,\dim\,s{\mathcal M}_g$, dei punti di inserzione $(z_i,\theta_i)$, e dei fattori sinistri del tensore di polarizzazione $\zeta_{i\mu}$. Per semplicit\`a si sono considerate strutture di spin uguali per gli spino di chiralit\`a destra e sinistra. ${\bar \mathcal C}_\nu^F$ \`e quindi il complesso coniugato di
${\mathcal C}_\nu{}^F$, con la stessa struttura di spin $\nu$,
\be
{\bar \mathcal C}_\nu^F(\bar z_i,\,\bar \theta_i;\bar \zeta_{i};\,
\bar m_K;\,p_I,\,k_i)=
{\mathcal C}_\nu{}^F(z_i,\,\theta_i;\zeta_i;\,m_K\;\,p_I,\,k_i)^* \ .
\ee 
L'espressione delle funzioni ${\mathcal C}_\nu{}^F$ pu\`o essere trovata in forma esplicita utilizzando le funzioni di Green e il kernel di Szeg\"o.

Il secondo risultato del Teorema di separazione chirale fissa la forma delle ampiezze di superstringa. Si trova per le teorie di tipo II \cite{D'Hoker:1988ta}
\be
A_h=\delta(k)\sum\limits_{\bar \nu}Q_{\nu\bar \nu}
\int\nolimits_{{\mathbb R}^{10g}}d^{10}p_I\int\nolimits_{s{\mathcal M}_h}
dm_K\,d\bar m_K \prod\limits_{i=1}^N
\int\nolimits_{\Sigma}d^2z_i\,
d^2\theta_i\,{\mathcal C}_\nu^F{\bar \mathcal C}_{\bar \nu}^F \ ,
\ee
dove $Q_{\nu\bar \nu}=\pm 1$ sono fattori che dipendono dalla struttura di spin e che realizzano la proiezione GSO.

\section{Operatori di vertice}

Per stringhe chiuse si hanno quattro classi di operatori di vertice: $NS-NS$, $R-NS$, $NS-R$, $R-R$. Iniziamo dagli operatori di vertice $NS-NS$, senza dipendenza dai campi ghost. Le parti destra e sinistra possono essere ricavate utilizzando la separazione chirale, e in analogia con i vertici di stringa bosonica si ha per un vertice generico, in una supergeometria $N=1$ piatta,
\be
V(\epsilon,k)=\int\nolimits_{\Sigma}d\mu_E
P_n(\epsilon,{\mathcal D}_+X,{\mathcal D}_-{\mathbf X},{\mathcal D}_+^2{\mathbf X},\ldots\,) e^{ik\cdot {\mathbf X}} \ ,
\ee
dove $P_n$ \`e una somma di termini con un numero totale di $n$ derivate dei campi ${\mathbf X}$. Nel caso di supergeometrie non piatte, l'operatore di vertice dipender\`a anche dalla supercurvatura $R_{+-}$.  

Per essere invarianti sotto l'azione del gruppo $sU(1)$, i vertici devono avere un ugual numero $n$ di derivate ${\mathcal D}_+$ e ${\mathcal D}_-$. I vertici di questa forma sono invarianti sotto l'azione di $sDiff$ per costruzione, mentre l'invarianza per $sWeyl$ pu\`o essere verificata senza difficolt\`a.

L'operatore di vertice degli stati a massa nulla \`e
\be
\label{masslessNSNS}
V(\epsilon,k)=\epsilon_{\mu\nu}(k)\int\nolimits_{\Sigma} d\mu_E{\mathcal D}_+{\mathbf X}^\mu{\mathcal D}_-{\mathbf X}^\nu e^{ik\cdot {\mathbf X}} \ .
\ee 
Si pu\`o vedere che l'invariaza per $sWeyl$ richiede $\epsilon_{\mu\nu}(k)k^\mu=\epsilon_{\mu\nu}(k)k^\nu=0$, e si ritrovano in questo modo in particolare, i vertici del dilatone, del gravitone e del campo $B_{\mu \nu}$. 

La parte sinistra dell'operatore di vertice (e allo stesso modo la parte destra) pu\`o essere scritta in componenti come
\be
\int d\theta\,{\mathcal D}_+{\mathbf X}_L^\mu e^{ik\cdot {\mathbf X}_L}
=(\partial_z X_L^\mu-i\psi_+^\mu\psi_+^\nu k_\nu)
e^{ik\cdot X_L} \ .
\ee

La definizione dei vertici di Ramond introduce diverse complicazioni \cite{Polchinski, D'Hoker:1988ta}. Non \`e infatti possibile scrivere un operatore di vertice per i fermioni a massa nulla a partire dal solo campo spinoriale, ma si deve opportunamente tener conto del contributo dei campi ghost. Consideriamo la parte sinistra di Ramond di un operatore di vertice, che scriviamo come
\be
W_-(z,u,k)=u^\alpha(k){\mathcal O}(z)S_\alpha(z)e^{ik\cdot X_L(z)} \ ,
\ee
dove $u^\alpha$ \`e uno spinore che costiutisce l'analogo del tensore di polarizzazione $\epsilon$ dei vertici NS. Perch\`e il vertice sia Weyl invariante, l'operatore ${\mathcal O}(z)$, che dipende dai campi ghost $\beta$ e $\gamma$, deve essere un campo primario di peso $3/8$, in modo che $W_-$ abbia peso $1$ per i campi a massa nulla. 

Il sistema di campi di ghost $(\beta, \gamma)$ pu\`o essere rappresentato in termini di due campi bosonici liberi $\phi$ e $\chi$, tali che
\be
\gamma=e^{\phi-\chi} \ , \qquad\qquad \beta=e^{-\phi+\chi}\partial_z \chi \ ,
\ee
con
\ba
\phi(z)\,\phi(w) &\sim & -\ln(z-w) \ , \nonumber \\
\chi(z)\chi(w) &\sim &  \ln(z-w) \ , 
\ea
L'operatore ${\mathcal O}(z)$ pu\`o essre scritto come esponenziale
\be
{\mathcal O}(z)=e^{-\frac{1}{2}\phi(z)} \ .
\ee
L'operatore di vertice destro di Ramond si trova essere
\be
W_+(z,u,k)=e^{\phi/2}u^\alpha(\gamma_\mu)_{\alpha\beta} S'{^\beta}(z)(\partial_z x^\mu-i\psi_+^\mu k\cdot\psi_+)
e^{ik\cdot x}
\ee

\section{Ampiezze ad albero}

Limitandoci alle ampiezze di stringa chiusa, al livello ad albero il world-sheet ha la topologia della sfera $g$, e in questo caso non si hanno supermoduli. Consideriamo lo scattering di stati a massa nulla $NS-NS$. L'operatore di vertice (\ref{masslessNSNS}) pu\`o essere riscritto, fattorizzando $\epsilon_{\mu\nu}(k)=\zeta_\mu(k)\bar \zeta_\nu(k)$ con $\zeta$ e $\bar \zeta$ parametri Grassmaniani, nella forma
\be
V(\zeta,\bar \zeta;k)=\int\nolimits_{\Sigma} d^2 z\,d^2\theta\, e^{ik\cdot {\mathbf X}+\xi\cdot{\mathcal D}_+{\mathbf X}+\bar \xi\cdot{\mathcal D}_- {\mathbf X}} \ ,
\ee
limitandoci a considerare i contributi lineari in $\zeta$ e $\bar \zeta$. Il propagatore di ${\mathbf X}$ si trova essere
\be
\left<X^\mu(z,\theta)X^\nu(z',\theta')\right>=-\ln\,\vert z-z'-\theta\theta'\vert^2\,\eta^{\mu\nu} \ .
\ee
I determinanti associati ai campi $x$, $\psi$, $b$, $c$ e $\beta$, $\gamma$, costituiscono una costante moltiplicativa.

 In questo caso, il teorema di separazione chirale \`e un'ovvia conseguenza della mancanza di momenti interni per la sfera. La funzione di correlazione dei vertici non integrati porta ad un prodotto di una funzione ${\mathcal C}^F$ complessa analitica in $z_i$, $\theta_i$ per la sua complessa coniugata. Consideriamo gli operatori chirali di vertice che sono 
\be
W_L(z_i,\theta_i)=e^{ik_i\cdot X_L(z_i,\theta_i)+ \zeta_i\cdot{\mathcal D}_+ X_L(z_i,\theta_i)} \ ,
\ee
e il corrispondente operatore destro. La contrazione dei campi chirali risulta essere
\be
\left<X_L^\mu(z,\theta)X_L^\nu(z',\theta')\right>=-\ln(z-z'- \theta\theta')\eta^{\mu\nu}\ ,
\ee
da cui si trova per il valore di aspettazione dei vertici chirali
\be
\left<W_L(z_1,\theta_1)\cdots W_L(z_N,\theta_N)\right>= {\mathcal C}^F(z_i,\theta_i) \ ,
\ee
con
\be
\label{contrax}
{\mathcal C}^F=\exp\sum\limits_{i\not=j}^N\left\{ ik_i\cdot\zeta_j\,{\theta_{ij}\over z_{ij}}+\frac{1}{2}
\,\zeta_i\cdot\zeta_j\,{1\over z_{ij}}+\frac{1}{2}\, k_i\cdot k_j\ln\,z_{ij}\right\} \ .
\ee
e l'espressione analoga per i vertici con chiralit\`a opposta. La notazione usata \`e quella standard $\theta_{ij} =\theta_i-\theta_j$, $z_{ij}=z_i-z_j-\theta_i\theta_j$. 

\subsection{Trasformazioni superconformi}

Per la superstringa a livello ad albero si ha invarianza sotto il gruppo di automorfismi superconforme $OSp(1,2)$. La misura di integrazione deve tenere conto di questa simmetria ulteriore. Per definire il gruppo di trasformazioni, consideriamo una terna di coordinate $(v, w, \psi)$ dove le varibili latine (greche) descrivono variabili commutanti (anticommutanti). Su questa tripletta, si ha l'azione naturale di $GL(2 |1)$, $T: W \to TW$,
\be
W=\left(
\begin{array}{c}
v \\
w \\
\psi \\
\end{array}
\right)\ ,
\qquad \qquad
T=\left(
\begin{array}{ccc}
a & b & \alpha \\
c & d & \beta \\
\gamma & \delta & A 
\end{array}
\right) \ ,
\ee
Per stabilire un contatto con il superspazio $N=1$, introduciamo le coordinate proiettive
\be
z=\frac{v}{w} \ , \qquad \qquad \theta=\frac{\psi}{w} \ ,
\ee
sulle quali l'azione di $GL(2 |1)$ \`e
\be
z\to {az+b+\alpha\theta\over cz+d+\beta\theta}\qquad\qquad\qquad
\theta\to {\gamma z+\delta+A\theta\over cz+d+\beta\theta}
\ee
Per ottenere una trasformazione superconforme, l'elemento lineare $dz=dz+\theta\,d\theta$ deve trasformare in se stesso a meno di un riscalamento conforme, in maniera equivalente, la forma quadratica
\be
z_{12}=z_1-z_2-\theta_1\theta_2={v_1w_2-v_2w_1-\psi_1\psi_2
\over w_1w_2}
\ee
deve trasformare in se stessa a meno di un riscalamento conforme. Questo si ottiene imponendo su $T$ il vincolo,
\be
T^T KT=K
\ee
dove K \`e la matrice
\be
\left(
\begin{array}{ccc}
0 & +1 & 0 \\
-1 & 0 & 0 \\
0 & 0 & 1 
\end{array}
\right) \ ,
\ee
La legge di trasformazione della forma quadratica si deriva facilmente, e si trova essere
\be
T: z_{12}\to {z_{12}\over (cz_1+d+\beta\theta_1)(cz_2+d+\beta\theta_2)} \ .
\ee
Allo stesso modo per l'elemento lineare, si trova
\be
dz\to {dz\over (cz+d+\beta\theta)^2} \ ,
\ee
ed infine la legge di trasformazione dell'elemento di volume \`e
\be
dz\wedge d\theta\to{dz\wedge d\theta\over (cz+d+\beta\theta)}\,\,.
\ee
Gli elementi di $OSp(1,2)$ sono in corrispondenza biunivoca con una tripletta di punti del superpiano $(z_1,\theta_1)$,
$(z_2,\theta_2)$, $(z_3,\theta_3)$, vincolata da un equazione a valori grassmaniani. Il conteggio dei parametri funziona perch\`e $OSp(1,2)$ ha 3 parametri commutanti e 2 anticommutanti. Il vincolo \`e una funzione invariante del gruppo $OSp(1,2)$ a valori grassmaniani, che dipende dai 3 punti,
\be
\Delta={z_{12}\theta_3+z_{31}\theta_2+z_{23}\theta_1 +\theta_1\theta_2\theta_3\over(z_{12}z_{23}z_{31})^{1/2}} \ .
\ee
Ponendo $\Delta$ a zero, si fissa uno dei parametri $\theta$. In questo modo si pu\`o vedere che c'\`e una corrsipondenza tra triplette di punti che soddisfino il vincolo e elementi di $OSp(1,2)$. L'elemento di volume invariante indotto su $OSp(1,2)$ si ottiene moltiplicando l'elemento di volume ordinario per una funzione delta che tenga conto del vincolo, 
\be
d\mu={dz_1dz_2dz_3d\theta_1 d\theta_2 d\theta_3\over (z_{12}z_{23}z_{31})^{1/2}}\,\delta(\Delta) \ .
\ee

\subsection{Ampiezze}

Le funzioni a zero-, uno- e due-punti di superstringa sono tutte nulle. Il modo pi\`u rapido per capirlo \`e osservare che il sottogruppo del gruppo superconforme, che lascia 0, 1, o 2 punti fissati ha un volume infinito. 
	
La funzione a tre punti, calcolata a partire dal valore di aspettazione dei vertici (\ref{contrax}), si trova essere
\be
\left<V(\epsilon_1,k_1)V(\epsilon_2k_2)V(\epsilon_3k_3)\right>=
4(2\pi)^{10}\delta(k)
\epsilon_1^{\mu_1\bar \mu_1}\epsilon_2^{\mu_2\bar \mu_2}
\epsilon_3^{\mu_3\bar \mu_3}K_{\mu_1\mu_2\mu_3}
K_{\bar \mu_1\bar \mu_2\bar \mu_3} \ ,
\ee
con
\be
K_{\mu_1\mu_2\mu_3}=
\eta_{\mu_1\mu_2}k_{1\mu_3}+\eta_{\mu_2\mu_3}
k_{2\mu_1}+\eta_{\mu_3\mu_1}k_{3\mu_2} \ .
\ee
Si pu\`o vedere che a causa della trasversalit\`a del tensore di polarizzazione, la funzione a tre-punti per particelle a massa nulla \`e zero.

Per calcolare la funzione a quattro punti, occorre fissare l'invarianza superconforme fissando i verti di interazione. Scegliendo $z=z_1$, $z_2=0$, $z_3=1$, $z_4=\infty$, $\theta_1,\theta_2,\theta_3=\theta_4=0$, si trova dalla (\ref{contrax}),
\ba	
\exp{{\mathcal G}_4^\zeta} &=& \left\{\zeta_1\cdot\zeta_2\zeta_3\cdot
\zeta_4{1\over z_{12}z_{34}}+\zeta_1\cdot\zeta_3\zeta_2
\cdot\zeta_4{1\over
z_{13}z_{24}}+\zeta_1\cdot\zeta_4\zeta_2\cdot\zeta_3
{1\over z_{14}z_{23}}\right\}\nonumber \\
&+& \left\{\zeta_1\cdot\zeta_2\left(k_1\cdot\zeta_3k_2\cdot
\zeta_4{\theta_2\theta_1\over z_{12}z_{13}z_{24}}+k_1
\cdot\zeta_4k_2\cdot\zeta_3{\theta_2\theta_1\over
z_{12}z_{14}z_{23}}\right)+\hbox{perm.}\right\} \ .
\ea
Per trovare l'ampiezza si dovrebbe moltiplicare questa espressione per l'equivalente espressione in $\bar \zeta$, integrare in $z$ e $\theta$ e raggruppare i termini. Il calcolo \`e per\`o enormemente semplificato dalle propriet\`a di fattorizzazione degli integrali di Veneziano. 

Per un integrale ordinario si ha
\be
\int{d^2z\over\pi}z^A\bar z^{\tilde A}(1-z)^B(1-\bar z)^{\tilde B}=
{\Gamma(-1-\tilde A-\tilde B)\over \Gamma(-\tilde A)\Gamma(-\tilde B)}\quad
{\Gamma(1+A)\Gamma(1+B)\over \Gamma(A+B+2)} \ ,
\ee
per  $A-\tilde A$ and $B-\tilde B$ interi. Non \`e difficile verificare che questa formula \`e simmetrica sotto lo scambio $(A,B)\leftrightarrow (\tilde A,\tilde B)$. Una formula analoga pu\`o essere derivata per i super-integrali
\ba
&\,&\int{d^2z_1\over\pi}d^2\theta_2[\theta_1\theta_2]^a
[\bar \theta_1\bar \theta_2]^{\tilde a} z_{12}^A\,\bar z_{12}^{\tilde A}
(1-z_1)^B(1-\bar z_1)^{\tilde B} \nonumber \\
&\,&\qquad \qquad=(-2i)^{1-a}(+2i)^{1-\tilde a}
{\Gamma(-\tilde a-\tilde A-\tilde B)\over
\Gamma(-\tilde A)\Gamma(-\tilde B)}{\Gamma(1+A)\Gamma(1+B)\over
\Gamma(A+B+1+a)}
\ea
con $a$ e $\tilde a$ a valori $0$ o $1$. L'integrale \`e simmetrico per $(aAB)\leftrightarrow (\tilde a\tilde A\tilde B)$. Utilizzando questa propriet\`a, si pu\`o ottenere l'espressione completa delle ampiezze limitandosi a  considerare il contributo dipendende da $z$. 

Per l'ampiezza a quattro punti si trova in questo modo
\ba
\Big< V(\epsilon_1,k_1)
V(\epsilon_2,k_2)V(\epsilon_3,k_3)V(\epsilon_4,k_4)\Big> 
\ea
\ba
&=&(2\pi)^{10}\delta(k)g^4\int d^2z_1d^2\theta_2\vert
z_{12}\vert^{-s}\vert z_1-1\vert^{-u}
e^{{\mathcal G}_4^\zeta+{\mathcal G}_4^{\bar \zeta}} \nonumber \\
&=&\pi(2\pi)^{10}\delta(k)g^4{\Gamma(-s/2)\Gamma(-t/2)
\Gamma(-u/2)\over
\Gamma\left(1+{s\over 2}\right)\Gamma\left(1+{t\over 2}\right)
\Gamma\left(+{u\over 2}\right)}\epsilon^{1\bar 1}\epsilon^{2\bar 2}
\epsilon^{3\bar 3}\epsilon^{4\bar 4}K_{1234}
K_{\bar 1\bar 2\bar 3\bar 4} \ ,
\ea
dove, usando l'abbreviazione $i$ per $\mu_i$ ($K_{\mu_1 \mu_2 \mu_3 \mu_4}=K_{1234}$), si \`e definito il tensore
\ba
K_{1234}
&=&(st\eta_{13}\eta_{24}
-su\eta_{14}\eta_{23}-tu\eta_{12}\eta_{34})\nonumber \\
&-&s(k_1^4k_3^2\eta_{24}+k_2^3k_4^1\eta_{13}-k_1^3k_4^2
  \eta_{23}-k_2^4k_3^1\eta_{14})\nonumber \\
&+&t(k_2^1k_4^3\eta_{13}+k_3^4k_1^2\eta_{24}-
  k_2^4k_1^3\eta_{34}-k_3^1k_4^2\eta_{12})\nonumber \\
&-&u(k_1^2k_4^3\eta_{23}+k_3^4k_2^1\eta_{14}-k_1^4k_2^3\eta_{34}
  -k_3^2k_4^1\eta_{12}) \ .
\ea

\section{Ampiezze ad un loop}

Per calcolare l'ampiezza di superstringa ad un loop, occorre utilizzare il teorema di separazione chirale per determinare le ampiezze chirali ${\mathcal C}_\nu^F$. Cosideriamo per semplicit\`a strutture di spin sinistra e destra uguali. 

Per strutture di spin $\nu$ pari non si hanno moduli e non ci sono zero modi di Dirac. Consideriamo la solita rappresentazione del toro come parallelogrammo di lati $0$, $1$, $\tau$, $1+\tau$, con $\tau$ appartenente al semipiano complesso superiore. La metrica pu\`o quindi essere scelta come $g=2\vert dz\vert^2$. 

L'azione scritta in componenti \`e
\ba
S_X+S_{BC} &=& {1\over 4\pi}\int\nolimits_{\Sigma}
d^2 z \left(\partial_z X\cdot\partial_{\bar z}X-
\psi_+\cdot\partial_{\bar z}\psi_+-\psi_-\cdot
\partial_z\psi_- \right.\nonumber \\
&\,& \left. +2b\partial_{\bar z}c+2\bar b\partial_z\bar c+
2\beta\partial_{\bar z}\gamma+2\bar \beta \partial_z
\bar \gamma\right)\ ,
\ea
e le inserzioni dei campi di ghost e di anti-ghost si riducono a $b\bar b c \bar c$, dal momento che per una struttura di spin pari  non ci sono zero modi. 

Nel settore $NS-NS$ i vertici non contengono campi di ghost. In questo caso quindi l'ampiezza si semplifica e si possono integrare i campi $b$, $c$, $\bar b$, $\bar c$, $\beta$, $\gamma$, $\bar \beta$ e $\bar \gamma$. Tenendo conto anche dei determinanti provenienti dall'integrazione sui campi $X$ e $\psi_{\pm}$ e utilizzando il teorema di separazione chirale si pu\`o scrivere 
\be
\delta(k){\mathcal C}_\nu {\mathcal C}_\nu=M_\nu\bar M_\nu\,{\mathcal F}_\nu {\mathcal F}_\nu \ ,
\ee
dove $M_\nu\,\bar M_\nu$ sono i termini dovuti ai determinanti e ${\mathcal F}_\nu\,{\bar \mathcal F}_\nu$ sono i termini delle funzioni di correlazione degli operatori di vertice. In forma esplicita, si trova,
\be
M_\nu\bar M_\nu =\left({\det'\Delta_0\over [Im(\tau)]^2}\right)^{-5}
\left(\det\,{\notD}_+\right)_\nu^5 \left(\det\,{\notD}_-\right)_\nu^5 
\left({\det'\Delta_{-1}\over [Im(\tau)]^2}\right)^{+1}(\det\,\Delta_{-1/2}^-)_\nu^{-1} \ .
\ee
I determinanti posso essere calcolati esplicitamente, e
\ba
{\det'\Delta_0\over [Im(\tau)]^2} &=& {\det'\Delta_{-1}
\over[Im(\tau)]^2}=\vert\eta(\tau)\vert^4 \ ,\nonumber \\
(\det\,\notD_+\,\notD_-)_\nu &=& (\det\,\Delta_{1/2}^-)_\nu=
\left\vert{\vartheta[\nu](0\vert\tau)\over
\eta(\tau)}\right\vert^2 \ .
\ea
Si ottiene in questo modo
\be
M_\nu={\vartheta[\nu](0\vert\tau)^4\over
\eta(\tau)^{12}} \ ,\qquad\qquad
\bar M_\nu={\overline{\vartheta[\nu](0\vert\tau)^4}\over
\overline{\eta(\tau)}^{12}} \ ,
\ee
a meno di un fattore di fase indipendente da $\tau$, ma che pu\`o dipendere da $\nu$. I contributi dei vertici possono essere similmente calcolati, ottenendo
\ba
{\mathcal F}_\nu(z_i, \theta_i,\zeta,k,p) &=& \exp\{i\pi p^2\tau+2\pi p\cdot \sum\limits_{i}(-\zeta_i\theta_i+ik_iz_i) \nonumber \\
&\,& -\frac{1}{2}\sum\limits_{i\not=j}(k_i\cdot k_jG_\nu(z_i,\theta_i;z_j,\theta_j)+\zeta_i\cdot\zeta_j
  {\mathcal D}_+^i{\mathcal D}_+^jG_\nu \nonumber \\
&\,& +2ik_i\cdot\zeta_j{\mathcal D}_+^jG_\nu)\}\ ,
\ea
dove la funzione di correlazione $G_\nu$ \`e data da
\ba
G_\nu(z_i,\theta_i;z_j,\theta_j)
&=& \left<{\mathbf X}_L(z_i,\theta_i){\mathbf X}_L(z_j,\theta_j)\right> \nonumber \\ 
&=& -\ln\,E(z_i,z_j)+\theta_i\theta_j S_\nu(z_i,z_j) \ .
\ea
$E(z,w)$ \`e la forma prima, e $S_\nu$ \`e il kernel di Szeg\"o. Per il toro queste funzioni sono
\ba
E(z,w) &=& {\vartheta_1(z-w\vert\tau)\over \vartheta'_1(0\vert\tau)}\ , \nonumber \\
S_\nu(z-w) &=& {\vartheta[\nu](z-w\vert\tau) \vartheta'_1(0\vert\tau)\over
\vartheta[\nu](0\vert\tau)\vartheta_1(z-w\vert\tau)}\ .
\ea

Per strutture di spin dispari c'\`e un modulo complesso dispari $\chi$, e $10$ zero modi di Dirac, uno per ogni direzione spazio-temporale. In questo caso, oltre all'azione $S_X+S_{BC}$, c'\`e un'inserzione della supercorrente sul world-sheet, $\chi S_{z+}$,
\be
S_{z+}=\left(\psi_+\cdot \partial_z X-\frac{1}{2}\,b\gamma+{3\over 2}\,\beta\partial_z c+
\partial_z\beta\,c\right)+c.c\ .
\ee
L'integrazione sul modulo $\chi$ porta ad avere un'inserzione di termini $\psi_+\cdot \partial_z x$ e $\psi_-\cdot\partial_{\bar z}x$. Considerando i campi ghost e anti-ghost, l'inserzione totale \`e
\be
b\bar b\,c\bar c\,\delta(\beta)\delta(\bar \beta)\delta(\gamma)
\delta(\bar \gamma)\int S_{z+}\int S_{\bar z-}\ .
\ee

Consideriamo ora le funzioni a uno-, due-, tre-, quattro-punti per stati a massa nulla $NS-NS$. Le strutture di spin dispari contribuiscono nei diagrammi con almeno $5$ operatori di vertice. Quindi per queste ampiezze basta considerare stutture di spin pari, e non c'\`e pertanto differenza fra le stringhe di tipo $IIA$ e $IIB$.

Iniziamo dal determinare i fattori $Q_{\nu\bar \nu}$. Ricordando le propriet\`a delle funzioni $\eta$ e $\theta$, si ha che 
\ba
M_{ab}(\tau+1) &=& -e^{i\pi a}M_{a(b+a+1)}(\tau) \nonumber \\
M_{ab}\left(-{1\over\tau}\right) &=& -{1\over \tau^4} M_{ba}(\tau)\ .
\ea
Le funzioni theta sono definite come
\ba
\vartheta_{00}(z,\tau) &=&\vartheta_3(z,\tau)
  =\vartheta(z,\tau) \ ,\nonumber \\
\vartheta_{01}(z,\tau) &=&\vartheta_4(z,\tau)
  =\vartheta\left(z+\frac{1}{2},\tau\right)\ ,\nonumber \\
\vartheta_{10}(z,\tau) &=&\vartheta_2(z,\tau)
  =e^{i\pi\tau/4+\pi iz}\vartheta\left(z+{\tau\over 2},
  \tau\right)\ ,\nonumber \\
\vartheta_{11}(z,\tau) &=&\vartheta_1(z,\tau)
  =e^{i\pi\tau/4+\pi iz}
  \vartheta\left(z+\frac{1}{2}+{\tau\over 2},\tau\right)\ .
\ea
e la richiesta di invarianza modulare dell'ampiezza, impone
\ba
\sum\limits_{(a,b)}Q_{(ab)\bar \nu}M_{(ab)} (\tau+1) &=&
\sum\limits_{(a,b)} Q_{(a,b)\bar \nu}M_{(ab)}(\tau) \ , \nonumber \\
\sum\limits_{(a,b)}Q_{(ab)\bar \nu}M_{(ab)}(-1/\tau)&=& -
{1\over\tau^4}\sum\limits_{(a,b)}
Q_{(a,b)\bar \nu}M_{ab}(\tau)\ . 
\ea
Si pu\`o in questo modo fissare la fase dell'ampiezza.

Possiamo ora calcolare in maniera diretta la funzione di vuoto. Dal momento che non si hanno vertici l'ampiezza \`e semplicemente
\be
A_1=Vol(M)\int_{{\mathcal M}_1}\, {d^2\tau\over \tau_2^4}\,\sum\limits_{\nu \bar \nu}
Q_{\nu \bar \nu}M_\nu M_{\bar \nu} \ ,
\ee
che risulta annullarsi, per una relazione di identit\`a delle funzioni $\theta$ appartenente alla serie di identit\`a  di Riemann.  In maniera simile, con l'aiuto di queste relazioni, si trova che anche le funzioni a uno-, due e tre-punti si annullano. 

Il significato fisico di questo risultato \`e che lo spazio tempo piatto di Minkowski \`e una soluzione delle equazioni di superstringa al livello ad un loop, e che non ci sono rinormalizzazioni della massa e degli accoppiametni a questo ordine. Gli ultimi due risultati vanno sotto il nome di \emph{teoremi di non rinormalizzazione}.

Rimane da discutere la funzione a quattro-punti per gli stati a massa nulla $NS-NS$. Chiaramente questa ampiezza non pu\`o annullarsi in una teoria interagente. In questo contesto ci limitiamo a fornire il risultato, rimandando alle referenze per i dettagli del calcolo \cite{D'Hoker:1988ta}. Si trova
\be
A_1(k_i,\epsilon_i)=\delta(k)\epsilon^{1\bar 1}\epsilon^{2\bar 2}
\epsilon^{3\bar 3}\epsilon^{4\bar 4}K^{1234}K^{\bar 1 \bar 2\bar 3\bar 4} A_1(s,t) \ ,
\ee
dove
\ba
A_1(s,t)=\frac{1}{2} \int_{{\mathcal M}_1}{d^2\tau\over \tau_2^2}
\int_{\Sigma}{d^2 z_1\over \tau_2}\int_{\Sigma}
{d^2 z_2\over\tau_2}\int_{\Sigma} {d^2 z_3\over\tau_2}
&\,& e^{_{s\over 2}(G_{12}+G_{34}-G_{13}-G_{24})} \nonumber \\
&\,& e^{+{t\over 2} (G_{23}+G_{14}-G_{13}-G_{24})} \ .
\ea
I propagatori dei campi $X$ ad un loop sono
\be
G_{ij}=G(z_i-z_j)\vert\tau)
\equiv -\ln\,\left\vert{\vartheta_1(z_i-z_j\vert\tau)\over
\theta'_1(0\vert\tau)}\right\vert^2+{2\pi
\over \tau_2}\,Im(z_i-z_j)^2 \ .
\ee


\end{document}